\documentclass[
11pt, 
english, 
singlespacing, 
headsepline, 
]{MastersDoctoralThesis} 

\usepackage[utf8]{inputenc} 
\DeclareUnicodeCharacter{2212}{-}
\DeclareUnicodeCharacter{03B3}{$\gamma$}
\newcommand{\arcsec}{\ensuremath{^{\prime\prime}}}

\usepackage[T1]{fontenc} 
\usepackage{graphics}
\usepackage{mathpazo} 

\usepackage{amsmath}	

\usepackage{amssymb}	

\usepackage{algorithm}
\usepackage{algpseudocode}

\usepackage[backend=bibtex,style=authoryear,natbib=true]{biblatex} 

\usepackage[autostyle=true]{csquotes} 

\usepackage{acronym} 

\usepackage{units} 

\usepackage{longtable}
\usepackage{booktabs}
\usepackage{changepage}
\usepackage{multirow}

\usepackage{lscape}
\usepackage{pdflscape}

\usepackage{amsbsy}

\usepackage{rotating}

\usepackage{tablefootnote}
\usepackage{threeparttable}

\usepackage[svgnames,table]{xcolor}

\thesistitle{Title} 
\supervisor{Supervisor 1 \\ Supervisor 2} 
\examiner{} 
\degree{Doctor of Philosophy in Physics} 
\author{Author} 
\addresses{} 

\subject{Physics and astrophysics} 
\university{UNIVERSITAT AUTÒNOMA DE BARCELONA \\ DEPARTAMENT DE FÍSICA} 
\department{\href{}{Institute of Space Sciences (ICE-CSIC/IEEC)}} 

\AtBeginDocument{
\hypersetup{pdftitle=\ttitle} 
\hypersetup{pdfauthor=\authorname} 
\hypersetup{pdfkeywords=\keywordnames} 
}

\begin{document}

\frontmatter 

\pagestyle{plain} 


\begin{titlepage}
\begin{center}

\vspace*{.06\textheight}
{\scshape\large \univname\par}\vspace{1cm} 

\includegraphics[scale=0.16]{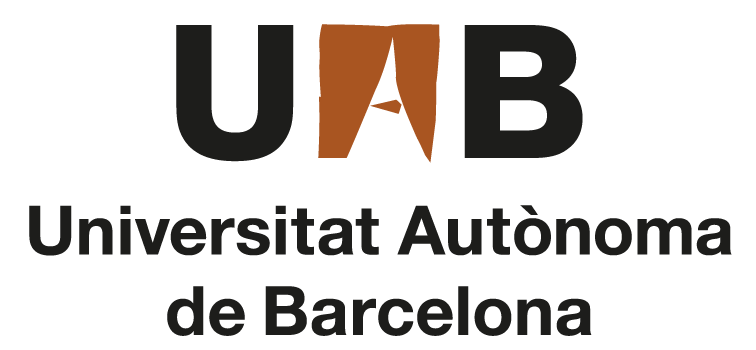}\vspace{0.5cm} 

\textsc{\Large Doctoral Thesis}\\[0.5cm] 

\HRule \\[0.4cm] 
{\huge \bfseries Application and development of advanced mathematical tools for population and time series analysis
in pulsar astrophysics\par}\vspace{0.4cm} 
\HRule \\[1.5cm] 
 
\begin{minipage}[t]{0.4\textwidth}
\begin{flushleft} \large
\emph{Author:}\\
\href{}{Carlos R. García} 
\end{flushleft}
\end{minipage}
\begin{minipage}[t]{0.4\textwidth}
\begin{flushright} \large
\emph{Advisors:} \\
\href{}{Prof. Diego F. Torres}\\ 
\href{}{Dr. Alessandro Patruno}
\end{flushright}
\end{minipage}\\[1cm]
\begin{minipage}[t]{0.4\textwidth}
\begin{flushleft} \large
\end{flushleft}
\end{minipage}
\begin{minipage}[t]{0.4\textwidth}
\begin{flushright} \large
\emph{Tutor:} \\
\href{}{Prof. Lluís Font Guiteras} 
\end{flushright}
\end{minipage}\\[3cm]
 
\vfill

\large A thesis submitted for the degree of \\ \textit{\degreename}\\[0.3cm] 

\deptname\\[2cm] 
 
\vfill


\vfill
\end{center}
\end{titlepage}

\vspace*{0.2\textheight}

\noindent\enquote{\itshape Para mi mamá y mi abuela.}\bigbreak



\begin{abstract}
\addchaptertocentry{\abstractname} 

In this thesis, we introduce novel methods for analyzing pulsar populations using a variety of mathematical techniques.  

These tools—particularly graph theory—have been thoroughly validated in advanced mathematics, enabling us to overcome some of the constraints (even dimensional) inherent in conventional visualization approaches.  

This exploration benefits from dimensionality reduction techniques, which not only lessen computational demands but also highlight potential for describing physical characteristics.  
The resulting structures encode information about pulsar similarities that extend beyond standard spin parameters, revealing relationships that are not readily apparent in traditional diagrams.
With a physically motivated topological perspective, we leverage the strengths of these methods and present results that span from prospective source classification and the emergence of new classes to catalog comparison, among other applications.  

This new approach enables fresh interpretations of longstanding problems, laying a new foundation for visualizing the pulsar population and categorizing sources.  

Building on this, we identify several sources as likely members of specific binary subclasses and investigate the potential transitional nature of others. We also introduce an algorithm capable of quickly determining the position of newly discovered pulsars within the resulting structure, which we apply to known transitional candidates and accreting systems. By evaluating their positions under reasonable assumptions about currently unknown parameters, we demonstrate that this method can effectively narrow parameter spaces, guiding future targeted observations and classification efforts.

Furthermore, we extend the use of graph theory to the boundary of machine learning, demonstrating its capability for binary separation in an unsupervised context.  

We demonstrate how these topological structures offer transparent and efficient representations in high-dimensional spaces, thereby avoiding opaque, black-box models that render interpretations difficult or questionable.  

Finally, we introduce and apply an innovative, flexible time-series alignment technique—adapted from applications in economics, medicine, and speech recognition—to the field of gamma-ray astrophysics.
We delve into he mathematical framework behind dynamic time warping and apply it to the study of gamma-ray pulsar emission processes from the perspective of their light curve morphological similarity.
The method identifies notable similarities among light curves, unaffected by their flux levels, and frequently irrespective of their timing, physical, or spectral properties. Very different pulsars can produce extremely similar light curves. 
The results presented here are promising, offering a refreshing direction for the field and new pathways for rigorous mathematical analysis, ultimately providing meaningful alternatives to traditional approaches in high-energy astrophysics.

\newpage
\textbf{This thesis is based on the following publications}:

\begin{itemize}

\item \citet*{MST-1}\\
{\it Visualizing the pulsar population using graph theory}\\
{Monthly Notices Of The Royal Astronomical Society (MNRAS), 515, 3, 3883 - 3897}

\item \citet*{MST-2}\\
{\it Quantitative determination of minimum spanning tree structures: using the pulsar tree for analyzing the appearance of new classes of pulsars}\\
{Monthly Notices Of The Royal Astronomical Society (MNRAS), 520, 1, 599 - 610}

\item \citet*{MST-MSPs}\\
{\it Millisecond pulsars phenomenology under the light of graph theory}\\
{Astronomy and Astrophysics (A\&A), 692, A187}

\item \citet*{MST-FRBs}\\
{\it Separating repeating fast radio bursts using the minimum spanning tree as an unsupervised methodology}\\
Astrophysical Journal (ApJ), 977, 273

\item \citet*{DTW_I}\\
{\it Quantitative exploration of the similarity of gamma-ray pulsar light curves}\\
Astrophysical Journal Letters (ApJL),  982, L51\\

and briefly commented here, also: \\

\item \citet*{MST-PWNs} \\
{\it Analysis of the possible detection of the pulsar wind nebulae of PSR J1208-6238, J1341-6220, J1838-0537, and J1844-0346} \\
{Astronomy and Astrophysics (A\&A), 691, A332.}

\end{itemize}

\end{abstract}


\begin{acknowledgements}
\addchaptertocentry{\acknowledgementname} 

I want to thank Diego, who has not only been my thesis director but also someone I turn to for discussion, debate, and, above all, learning. Nothing in this thesis would have been possible without him. I also thank Alessandro for the opportunity he gave me to embark on this path. I appreciate Dani's efforts in making this journey more comfortable.
Thanks to the entire MAP group; it has been truly inspiring to work so closely with such highly skilled people. 
Thanks to everyone at ICE-CSIC for creating a motivating, healthy, and supportive environment for personal growth. 
Thanks to my friends for being my anchor and helping me stay awake. 
Thank you, mami, for always being there for me. I would like to thank my family for being my home from afar, for understanding each of my absences, and for accompanying me throughout this journey. 
I would like to thank my sister, Isa, for insisting on staying in my life and appearing in this thesis. 
Thank you, Coriss, for being my strength and withstanding the test of time.
Thank you, t, for walking with me through the darkness, never letting go of my hand.
"I've lost my mind on this game like Vincent van Gogh dedicated his life to his art, and lost his mind in the process, that's happened to me, but foock it..then it will pay, then I'll be happy to lose my mind.”- C.M.
\end{acknowledgements}


\tableofcontents 

\listoffigures 

\listoftables 


\newcommand\aap{A\&A}                
\let\astap=\aap                          
\newcommand\aapr{A\&ARv}             
\newcommand\aaps{A\&AS}              
\newcommand\actaa{Acta Astron.}      
\newcommand\afz{Afz}                 
\newcommand\aj{AJ}                   
\newcommand\ao{Appl. Opt.}           
\let\applopt=\ao                         
\newcommand\aplett{Astrophys.~Lett.} 
\newcommand\apj{ApJ}                 
\newcommand\apjl{ApJ}                
\let\apjlett=\apjl                       
\newcommand\apjs{ApJS}               
\let\apjsupp=\apjs                       
\newcommand\apss{Ap\&SS}             
\newcommand\araa{ARA\&A}             
\newcommand\arep{Astron. Rep.}       
\newcommand\aspc{ASP Conf. Ser.}     
\newcommand\azh{Azh}                 
\newcommand\baas{BAAS}               
\newcommand\bac{Bull. Astron. Inst. Czechoslovakia} 
\newcommand\bain{Bull. Astron. Inst. Netherlands} 
\newcommand\caa{Chinese Astron. Astrophys.} 
\newcommand\cjaa{Chinese J.~Astron. Astrophys.} 
\newcommand\fcp{Fundamentals Cosmic Phys.}  
\newcommand\gca{Geochimica Cosmochimica Acta}   
\newcommand\grl{Geophys. Res. Lett.} 
\newcommand\iaucirc{IAU~Circ.}       
\newcommand\icarus{Icarus}           
\newcommand\japa{J.~Astrophys. Astron.} 
\newcommand\jcap{J.~Cosmology Astropart. Phys.} 
\newcommand\jcp{J.~Chem.~Phys.}      
\newcommand\jgr{J.~Geophys.~Res.}    
\newcommand\jqsrt{J.~Quant. Spectrosc. Radiative Transfer} 
\newcommand\jrasc{J.~R.~Astron. Soc. Canada} 
\newcommand\memras{Mem.~RAS}         
\newcommand\memsai{Mem. Soc. Astron. Italiana} 
\newcommand\mnassa{MNASSA}           
\newcommand\mnras{MNRAS}             
\newcommand\na{New~Astron.}          
\newcommand\nar{New~Astron.~Rev.}    
\newcommand\nat{Nature}              
\newcommand\nphysa{Nuclear Phys.~A}  
\newcommand\pra{Phys. Rev.~A}        
\newcommand\prb{Phys. Rev.~B}        
\newcommand\prc{Phys. Rev.~C}        
\newcommand\prd{Phys. Rev.~D}        
\newcommand\pre{Phys. Rev.~E}        
\newcommand\prl{Phys. Rev.~Lett.}    
\newcommand\pasa{Publ. Astron. Soc. Australia}  
\newcommand\pasp{PASP}               
\newcommand\pasj{PASJ}               
\newcommand\physrep{Phys.~Rep.}      
\newcommand\physscr{Phys.~Scr.}      
\newcommand\planss{Planet. Space~Sci.} 
\newcommand\procspie{Proc.~SPIE}     
\newcommand\rmxaa{Rev. Mex. Astron. Astrofis.} 
\newcommand\qjras{QJRAS}             
\newcommand\sci{Science}             
\newcommand\skytel{Sky \& Telesc.}   
\newcommand\solphys{Sol.~Phys.}      
\newcommand\sovast{Soviet~Ast.}      
\newcommand\ssr{Space Sci. Rev.}     
\newcommand\zap{Z.~Astrophys.}       

\def\xmm {\emph{XMM--Newton}}
\def\cxo {\emph{Chandra}}
\def\nustar {\emph{NuSTAR}}
\def\rst {\emph{ROSAT}}
\def\swift {\emph{Swift}}
\def\nicer {\emph{NICER}}
\def\hxmt {\emph{Insight}-HXMT}
\def\pks {Parkes}

\def\gleamfirst {\mbox{GLEAM-X\,J162759.5-523504.3}}
\def\gleam {\mbox{GLEAM-X\,J1627}}
\def\mtp{\mbox{PSR J0901-4046}}
\def\gpm {\mbox{GPM J1839–10}}
\def\rcw{\mbox{1E\,161348-5055}}
\def\lowbsgr{\mbox{SGR\,0418+5729}}
\def\sgrfrb{SGR\,1935+2154}
\def\xte{XTE\,J1810$-$197}

\def\msun{M$_{\odot}$\,}

\acrodef{ATNF}[ATNF]{Australia Telescope National Facility}
\acrodef{PHL}[PHL]{Parkes High-Latitude}
\acrodef{PMB}[PMB]{Parkes Multi-Beam}
\acrodef{SMB}[SMB]{Swinburne Intermediate-latitude}
\acrodef{MLP}[MLP]{multi-layer perceptron}
\acrodef{CNN}[CNN]{convolutional neural network}
\acrodef{MDN}[MDN]{mixture density network}
\acrodef{MAF}[MAF]{masked autoregressive flow}
\acrodef{NPE}[NPE]{neural posterior estimation}
\acrodef{SBC}[SBC]{simulation-based calibration}
\acrodef{DM}[$DM$]{dispersion measure}
\acrodef{SNR}[SNR]{Supernova Remnant}
\acrodef{GRBs}[GRBs]{Gamma-Ray Bursts}
\acrodef{FRBs}[FRBs]{Fast Radio Bursts}
\acrodef{ANNs}[ANNs]{artificial neural networks}
\acrodef{Adam}[Adam]{Adaptive Moment Estimation}
\acrodef{RMSE}[RMSE]{root-mean-square error}
\acrodef{MRE}[MRE]{mean relative error}
\acrodef{SKA}[SKA]{Square Kilometer Array}
\acrodef{ML}[ML]{machine learning}
\acrodef{RPPs}[RPPs]{rotation-powered pulsars}
\acrodef{SGRs}[SGRs]{soft gamma-ray repeaters}
\acrodef{AXPs}[AXPs]{anomalous X-ray pulsars}
\acrodef{RRATs}[RRATs]{rotating radio transients}
\acrodef{CCOs}[CCOs]{central compact objects}
\acrodef{XDINSs}[XDINSs]{X-ray dim isolated neutron stars}
\acrodef{CCSN}[CCSN]{core-collapse supernova}
\acrodef{ReLU}[ReLU]{rectified linear unit}
\acrodef{MWA}[MWA]{Murchison Widefield Array}
\acrodef{ATCA}[ATCA]{Australia Telescope Compact Array}
\acrodef{ASKAP}[ASKAP]{Australian Square Kilometre Array Pathfinder}
\acrodef{VLA}[VLA]{Very Large Array}
\acrodef{VLITE}[VLITE]{Low Band Ionospheric and Transient Experiment} \acrodef{GMRT}[GMRT]{Giant Metrewave Radio Telescope}

\newcommand{\dft}[1]{{\color{cyan}  dft: #1}} 
\newcommand{\ch}[1]{{\color{green} ch: #1}} 
\def\red#1{\textcolor{red}{ #1}}

\mainmatter 

\pagestyle{thesis} 


\chapter{Pulsars}
\label{chapter1}

\section{Introduction}
\label{chapter1: introduction}
The idea that neutron stars might exist preceded their observational confirmation by several decades. Baade and Zwicky made the first proposal for the existence of neutron stars in 1934 \citep{Baade1934}, an idea already suggested by \citet{Landau1938}: the existence of stars composed mainly of neutrons\footnote{In \citet{Shapiro1983, Landau_discussion1932, Haensel2007}, based mainly on to recollections of Rosenfeld (1974, Proc. 16th Solvay Conference on Physics, p. 174), note that Lev Landau in (Phys. Z. Sowjetunion, 1, 285, 1932) already suggests the possibility of stable structures within stars composed of neutrons (see also \cite{Landau1932, Brecher1999}), as credited in \citet{Cameron1970}, after the discovery of the neutron by \citet{Chadwick_1932}.}. 
The former work suggested that a neutron star may be the final stage of a normal star's collapse following a supernova explosion.
After that, \citet{Tolman1939, Oppenheimer1939} independently performed the first calculations of neutron star models and analyzed the structure based on a star composed of degenerate neutron gas. This approach showed that the degeneration was such that temperature was irrelevant and that everything depended on the relationship between density and pressure. Therefore, the focus of attention from that point onward was on determining this relationship, known as the Equation of State (EoS) \citep{Lyne2012}. 

In the following years, Harrisson, Wakano, and Wheeler 1958\footnote{Wheeler, J. A. 1958, paper read at the Solvay Conference, Brussels. Published with B. K. Harrison and
M. Wakano, in La Structure et la Révolution de l'Univers (Stoops, Brussels, 3100 pp., 1959).},
\citet{Cameron1959, Ambartsumyan1960, Hamada1961} conducted research on EoS and neutron star models (see \cite{HTWW_book} for extensive discussions of these models).
However, interest in these objects remained low, primarily among astronomers, as work on neutron stars had focused mainly on understanding how the cores of massive stars, composed solely of neutrons, could be a source of stellar energy.
In addition, observing neutron stars with the optical telescopes of the time was complicated due to their small size and residual thermal energy, which meant that even less attention was paid to them \citep{Shapiro1983}.

This changed in 1967 when the PhD student Jocelyn Bell Burnell (see \cite{Burnell1977} for her role in the discovery), who was part of the group of astronomers led by Antony Hewish\footnote{Antony Hewish received the Nobel Prize in 1974 for his role in the pulsars' discovery.}, detected the first pulsar. 
The find originated from data collected at the Mullard Radio Astronomy Observatory, where an unusual, stable radio signal in the form of pulses was observed.
In an initial explanation, they proposed white dwarfs or neutron stars as the sources of these unusual stable oscillations\footnote{Recorded as CP1919, the signals observed were so surprising that they persuaded astronomers to consider that an extraterrestrial civilization was responsible for them (initially called \textit{Little Green Men} or "LGM"), see \citet{Burnell1977, Penny_SETI}.
} \citep{Hewish_Bell1968}.

\begin{figure}
  \includegraphics[width=1\textwidth]{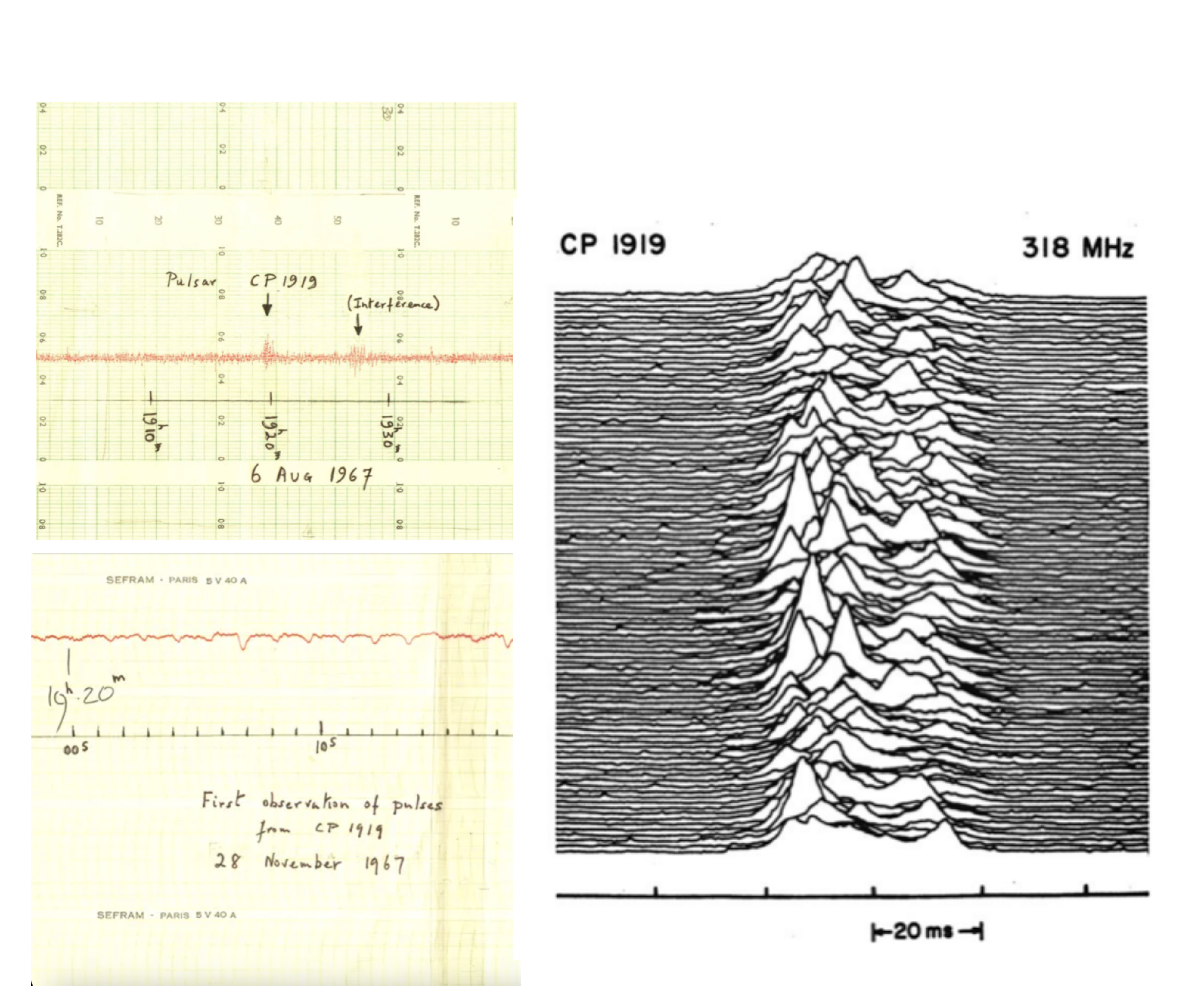}
  \centering
  \caption[First pulsar detection: PSR B1919+21.]{First column: shows the actual first detection charts taken from Jocelyn Bell (figure adopted from \url{https://www.cam.ac.uk/stories/journeysofdiscovery-pulsars}). The first panel displays the initial record of CP1919, dated August 6, 1967, in which the anomalous regular pulse was observed. The second panel presents the record dated November 28, 1967, which consolidated the previous observations and, as a result, confirmed the detection of the first pulsar: PSR B1919+21.
  Second column: Stacked plot on CP1919, also known as a plot illustrating successive mean pulse spectra, as referenced in Harold Dumont Craft Jr, who titled it as "Many consecutive pulses from CP1919" seen in his PhD dissertation "Radio observations of the pulse profiles and dispersion measures of twelve pulsars", Cornell University, 1970, p. 214 (figure adopted from \cite{Turchetti_2023}, see this publication for more information).
  }
  \label{chapter_1_fig: CP1919}
\end{figure}

\begin{figure}
  \includegraphics[width=1\textwidth]{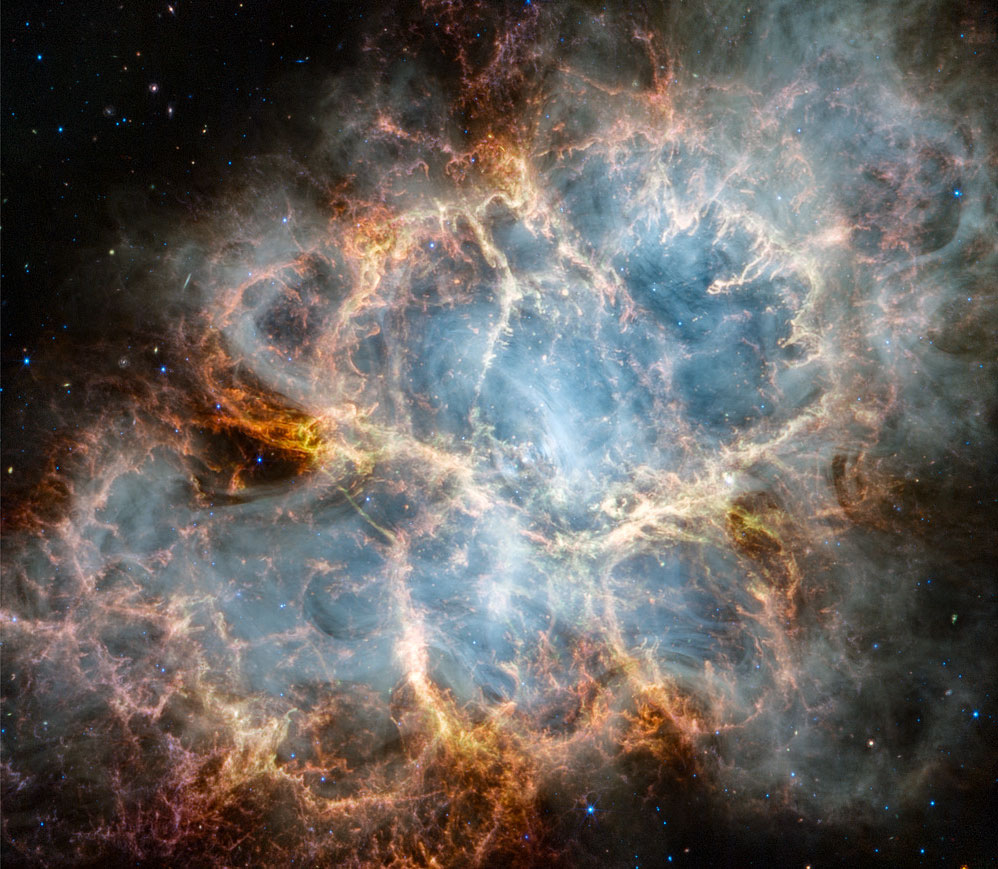}
  \centering
  \caption[The Crab Nebula as seen by the James Webb Space Telescope.]{The Crab Nebula as seen by the James Webb Space Telescope in infrared light. At the center of this ring-shaped structure is a bright white spot: the Crab Pulsar (Credits: NASA, ESA, CSA, STScI, Jeff Hester (ASU), Allison Loll (ASU), Tea Temim (Princeton University)).
  }
  \label{chapter_1_fig: Crab_Webb_998}
\end{figure}

This discovery contributed to research proposing the existence of stable stars denser than white dwarfs, which would be rapidly rotating and have strong magnetic fields \citep{Hoyle1964, Woltjer1964, TsurutaCameron1966}.
It had even been conjectured that the energy source seen in the Crab nebula\footnote{The Crab Nebula is widely accepted as the remnant of a supernova explosion observed by Chinese astronomers in 1054 AD, who referred to this event as a "guest star". This identification was first proposed by \citet{Duyvendak1942, Mayall1942, Baade1942, Minkowski1942}.} came from this type of object \citep{Wheeler_1966}, see Figure~\ref{chapter_1_fig: Crab_Webb_998}.

In the following years, both the Cambridge team and the National Radio Astronomy Observatory groups at Green Bank (West Virginia) and Arecibo (Puerto Rico), two of its facilities, detected more pulsars, which helped determine that these stable oscillations with regular patterns were due to rapidly rotating neutron stars.
\citet{GOLD1968} then explained that the pulses observed were rapidly rotating neutron stars with a strong magnetic field, which would generate relativistic velocities in the plasma surrounding their magnetosphere. 
\citet{Pacini1967, PACINI_1968}, in turn, explained how a simple dipolar magnetic model could be capable of carrying out this process, connecting it with the activity seen in the Crab Nebula. 
This leads to the simple model that a rapidly rotating neutron star would behave like a rotating beacon, or a lighthouse, generating pulsed signals with a stable pattern, which is why the name "pulsar" was coined.
Subsequent observations and studies of the energy source of the Crab Nebula \citep{Staelin1968, COMELLA1969, Cocke1969} as well as the slowdown of what would be the Crab pulsar, showed that the energy lost by the latter would be the energy used to activate the Crab Nebula \citep{Richards1969}. 
Similarly, the observed Vela supernova remnant \citep{LARGE1968}, together with the detection of the Vela pulsar at its center, offered one of the earliest observational links between neutron stars and supernova explosions. 
This interpretation was supported by the works of \citet{Radhakrishnan_Cooke_1969, Radhakrishnan_etAl1969, Radhakrishnan_Manchester_1969}, who collectively showed that the Vela pulsar was a rotating neutron star with a well-defined magnetic geometry, whose alignment with the remnant and energy loss properties were consistent with theoretical expectations.
These findings were observational support for the hypothesis initially proposed by Baade and Zwicky, suggesting that neutron stars are the remnants left behind by supernova explosions (see also \cite{Ronchi_thesis} for a comprehensive review).

Over the decades, advances in observational technology and instrumentation have enabled the discovery of $\sim$3800 pulsars \citep{ATNF-Catalog} detected through their radio emission by The Australian Telescope National Facility (ATNF\footnote{\url{https://www.atnf.csiro.au/research/pulsar/psrcat/}}). 
Additionally, it was found that pulsars are also bright in X-rays, optical, and gamma rays. In particular, the Large Area Telescope (LAT) on board the \textit{Fermi} Gamma-ray Space Telescope\footnote{\url{https://fermi.gsfc.nasa.gov/}} has detected $\sim$300 pulsars released in the {\it Fermi}-LAT Third Pulsar Catalog (3PC, \cite{Fermi3PC}).

Most known pulsars reside within our Galaxy and are concentrated along the Galactic plane. In addition, about $\sim$345 pulsars have been discovered in globular clusters\footnote{A comprehensive and regularly updated list of pulsars in globular clusters: \url{https://www3.mpifr-bonn.mpg.de/staff/pfreire/GCpsr.html}} (see e.g., \cite{Hessels2015}), where stellar densities could favor binary evolution.

In this regard, it should be noted that while most known pulsars are isolated neutron stars, a significant fraction reside in binary systems (see \cite{Postnov2014} for an extensive review). 
The first radio pulsar in a binary system\footnote{PSR B1913+16 was detected in July 1974 and confirmed in August as a result of observing a significant change in its spin period \citep{Haensel2007}.} was followed by \citet{Hulse_Taylor_BinaryPulsar1975}\footnote{Russell A. Hulse and Joseph H. Taylor were awarded with the Nobel prize in 1993 for this discovery.} using the Arecibo radio telescope. 
Before 1974, neutron stars in X-ray binaries (such as Sco X-1 \cite{Giacconi1962}\footnote{This detection was key to the model proposed by \citet{Shklovsky1967}, where the X-ray emission detected originated from a close binary system due to an accreting process.}, Her X-1 \cite{Tananbaum1972}, and Cen X-3 \cite{Schreier1972}) had already been discovered. 
These systems supported the idea that a neutron star could accrete matter from a companion, leading to high-energy emissions across the X-ray and gamma-ray regimes.
These discoveries provided crucial observational support for binary evolution models of neutron stars, and established the foundation for the development of the recycling scenario, in which neutron stars in close binaries are spun up through accretion of material from the companion to form millisecond pulsars (MSPs) \citep{Alpar1982, Bhattacharya1991}.
The MSPs constitute a distinct subpopulation characterized by rapid rotation and low magnetic fields.
MSPs offer unique insights into stellar evolution, the interaction between magnetic fields and plasma transferred by the donor star, and particle acceleration from compact objects, particularly in binary systems (see e.g., \citealt{Manchester2017, Papitto2022} for reviews).

Throughout this chapter, we will show the formation and phenomenological processes through observed and derived properties that define the population of pulsars as a whole, and MSPs in particular. 
We will show the evolution of binary systems and the relevance of their classification based on their behavior. 
Additionally, we will briefly introduce Fast Radio Bursts, which are events of uncertain origin but are increasingly linked to compact objects.

\subsection{Compact object formation}
\label{chapter1: compactObject_formation}

To describe what a pulsar is and its characteristics, we must revisit its origin, i.e., its progenitor. 
When a normal star consumes most of its nuclear fuel, it is considered to have reached the end of its stellar evolution and therefore dies as we know it.
At that moment, what is known as a compact object is born: a white dwarf, neutron star, or black hole. 
Two fundamental characteristics differentiate compact objects from normal stars \citep{Shapiro1983}:
\begin{itemize}
    \item Compact objects no longer sustain thermonuclear reactions in their cores, so the thermal pressure that previously counterbalanced gravity disappears. However, this collapse can be prevented by other processes, such as electron degeneracy pressure in the case of white dwarfs, or primarily neutron degeneracy pressure and nuclear interactions in the case of neutron stars. 
    This does not happen in the case of black holes, which are the outcome of complete gravitational collapse, where no known force stops the contraction once a critical mass threshold is exceeded. 
    \item Compact objects are significantly smaller in size compared to normal stars of comparable mass, leading to extreme compactness and resulting in much stronger surface gravitational fields.
    
\end{itemize}

The mass of a normal star is a significant factor in determining whether it will ultimately evolve into a white dwarf, a neutron star, or a black hole in its final stage. 
White dwarfs\footnote{White dwarfs, which emit primarily their residual thermal energy, are characterized by having higher effective temperatures than normal stars, despite having lower luminosity. This is mainly due to their small radius, since according to the blackbody relation, for an object with temperature $T$ and radius $R$, its radiative flux varies with $T^{4}$, and its luminosity as $L\sim R^{2}T^{4}$. White dwarfs can be observed directly through optical telescopes during their cooling period, which spans billions of years.} are believed to originate from stars with masses $10\pm 2M_{\odot}$ \citep{Althaus2010}. 
These progenitors of white dwarfs are expected to form a planetary nebula through the controlled ejection of matter in their final stage. 
According to the Chandrasekhar limiting mass, once converted into a white dwarf, they can only reach a maximum of $1.4M_{\odot}$ \citep{Chandrasekhar1931}.
Note that massive progenitors $>20M_{\odot}$ may lead to black holes\footnote{Black holes do not produce any direct emissions, as they do not permit light or matter to escape their gravitational pull. Their observation is primarily indirect, based on the influence they exert on their surroundings, accretion disk emission detectable in X-ray and optical emission, and gravitational wave signals from mergers (see, e.g., LIGO/Virgo, which detected the first gravitational wave from the merger of stellar-mass binary black holes \cite{FirstGWsdetected}). The Event Horizon Telescope (\url{https://eventhorizontelescope.org/}) produced the first image of a black hole shadow in the galaxy M87$^{\star}$ \citep{1stBHimage}.} \citep{Fryer_1999, Heger_2003}. 

However, neutron stars are stellar remnants produced in Type II, Ib, or Ic supernova explosions, caused by the gravitational collapse of ordinary stars with initial masses between $8M_{\odot}$ and $25M_{\odot}$.
A supernova explosion occurs when the progenitor star can no longer produce energy from nuclear fusion. As a result, the pressure gradient from radiation is not sufficient to balance gravitational attraction, causing the star to collapse.
This event in the inner layers of the star releases gravitational energy, which is transferred to the outer layers, causing them to be expelled.

Neutron stars derive their name from the predominance of neutrons in their interior, whose degeneracy pressure prevents the neutron star from collapsing entirely.
Neutron stars can have masses of 1-2$M_{\odot}$ and a radius of 10-12 km \citep{Burgio2020}.

This occurs due to the process of mutual elimination through $\beta$-decay between protons and electrons that they undergo. 
This process determines the equilibrium of the neutron fluid that comprises the central part of the star, which is $\sim$5\% protons and electrons.
As a result, their density reaches nuclear values since they essentially function as a giant nucleus held together by self-gravity \citep{Shapiro1983}.

Assuming the predicted onion structure (see Figure 1 and further details in \cite{Burgio2020}) for the interior of a neutron star, as well as the parameters\footnote{See \citet{Ozel2016} for a comprehensive summary of proposed EoSs, as well as observational constraints on neutron star masses and radii.} typically accepted for describing it, with a radius of 10 km and a mass of 1.4$M_{\odot}$, we observe how specific values are comparable to those of nuclear matter. 
For example, neutron stars reach, in their core, a density of 6.7$\times 10^{14}$g cm$^{-3}$, close to 2.7 $\times 10^{14}$g cm$^{-3}$ calculated for nuclear matter. 
Even under normal conditions, the mean number density of neutrons of a neutron star\footnote{1 femtometre (fm) = 1$0^{-15}$ m.}, 0.4 fm$^{-3}$, is close to the number density of neutrons and protons seen for nuclear matter of 0.16 fm$^{-3}$.
The crust of neutron stars also appears to be remarkably formed.
So much so that their outer crust, $\sim$1 km thick and composed of iron near its surface, has a composition of heavy nuclei that become increasingly rich in neutrons as one descends toward its core. 
Assuming the number of neutrons $N$ and the atomic number $Z$ of normal matter are similar, and $N\sim2Z$ for heavy nuclei, the neutron fluid has $N\gg Z$ \citep{Lyne2012}.

Neutron stars are multi-wavelength emitters that can be observed across the entire electromagnetic spectrum, from radio to gamma bands. 
More details on radio and gamma detectability on neutron stars, which behave as pulsars, are provided in later sections. 
Additionally, neutrino\footnote{For instance: The Sudbury Neutrino Observatory (SNO): \url{https://sno.phy.queensu.ca/}, and the Ice Cube Neutrino Observatory
(IceCube): \url{https://icecube.wisc.edu/}} and gravitational wave\footnote{For instance: Laser Interferometer Gravitational-Wave
Observatory (LIGO): \url{https://www.ligo.caltech.edu/}, European Gravitational Observatory (VIRGO detector): \url{https://www.virgo-gw.eu/},
and the Large Interferometer Space Antenna (LISA): \url{https://lisa.nasa.gov/}.} observatories can also detect these objects \citep{Haensel2007}.

These fundamental principles governing the formation of compact objects, ranging from white dwarfs to neutron stars and black holes, offer a natural starting point for understanding how pulsars emerge from a specific progenitor and how they differ from other possible endpoints of stellar evolution.

\section{Pulsar phenomenology}
\label{chapter1: phenomenology}

\begin{figure}
  \includegraphics[width=0.5\textwidth]{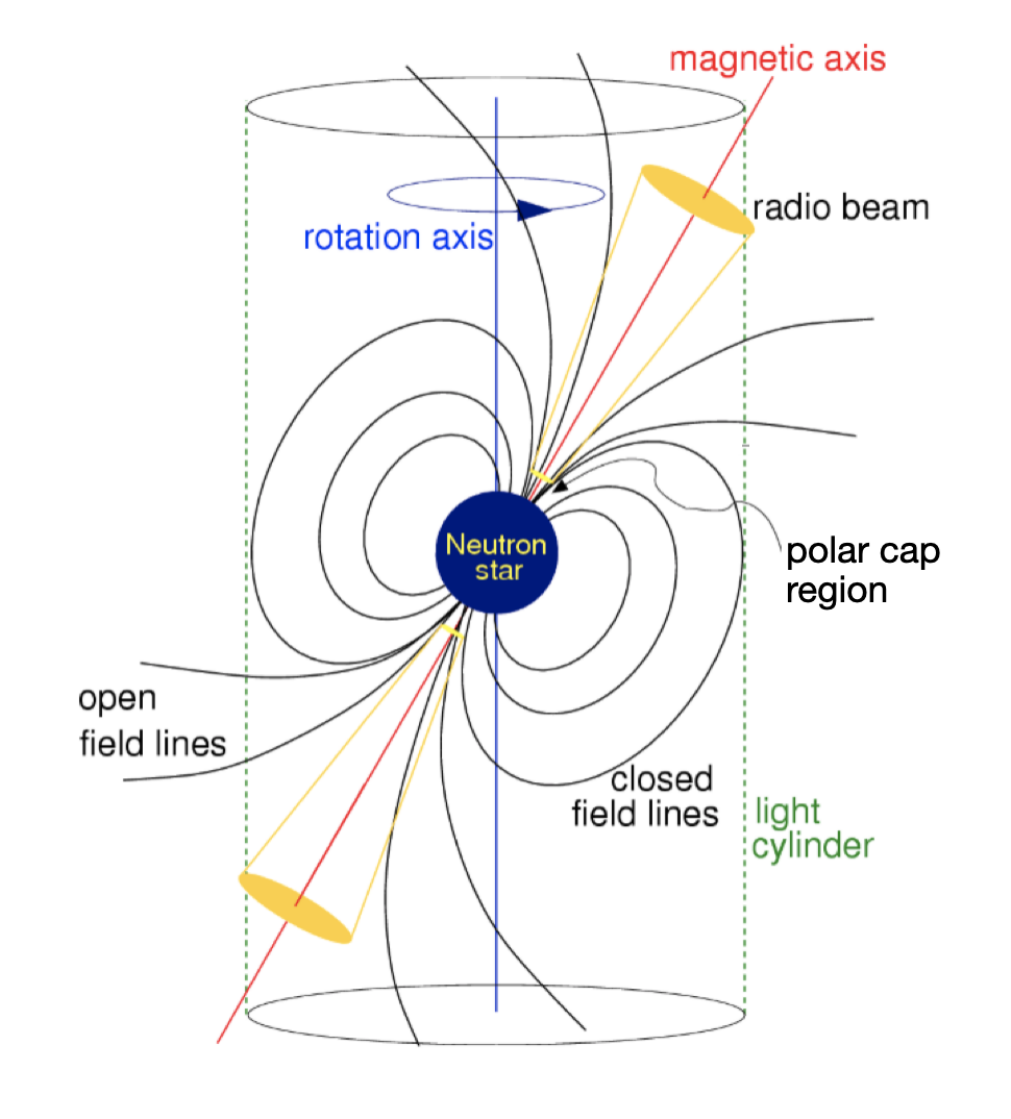}
  \centering
  \caption[Scheme of the pulsar lighthouse model.]{Scheme of the pulsar "lighthouse" model showing its emission geometry (not drawn to scale). The rotation axis (blue) is inclined concerning the magnetic axis (red), producing a beam (yellow) of radiation that sweeps across the observer's line of sight.  
  The description of the rest of the scheme can be found in Section~\ref{chapter1: magnetosphere} (figure adopted from \cite{Lorimer2012}).
  }
  \label{chapter_1_fig: NS_emission_geometry}
\end{figure}

As described in Section~\ref{chapter1: introduction}, following the discovery of the first pulsar \citep{Hewish_Bell1968}, a large number of observations were conducted, establishing key facts about them. Primarily, and even considering open questions based on reasonable doubts, a basic model \citep{GOLD1968} has long been established, stating that pulsars are highly magnetized neutron stars that rotate rapidly and are formed during the supernova explosions of massive stars.
The "lighthouse" model, also known as the dipole model \citep{Pacini1967, PACINI_1968, Ostriker1969}, describes a rotating neutron star with a misaligned magnetic axis relative to its rotation axis as shown in Figure~\ref{chapter_1_fig: NS_emission_geometry}.

The neutron star rotates, accelerating charged particles that are distributed along field lines to form a beam (radio beam in Figure~\ref{chapter_1_fig: NS_emission_geometry}). These accelerated particles emit electromagnetic radiation, which is typically detected in radio frequency through the observed pulses. Each pulse is produced when the radio beam, aligned with the magnetic axis, sweeps across the observer's line of sight during each rotation. 
The pattern seen in the pulses, defined by their repetition period, corresponds to the rotation period of the neutron star \citep{Lorimer2008}.

\subsection{Spin Periods and Slowdown Rates}
\label{chapter1: Spin_periods}

\begin{figure}
  \includegraphics[width=0.5\textwidth]{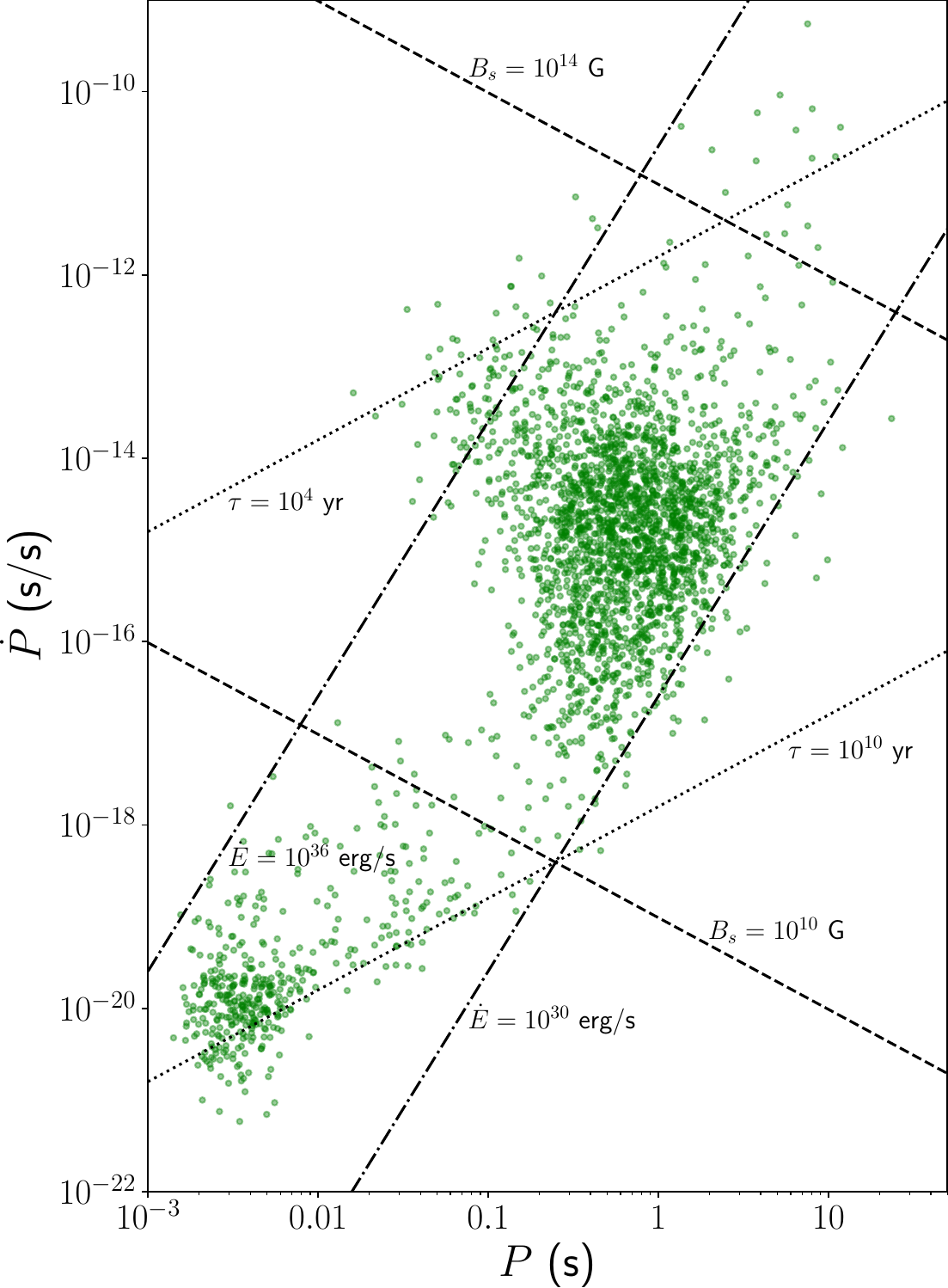}
  \centering
  \caption[$P\dot{P}$ diagram.]{The $P\dot P$ diagram based on the the ATNF v2.6.1 restricted to $\dot P>0$ resulting in 2752 pulsars.
  Lines of constant magnetic field at the surface ($B_s$) in dashed, characteristic age ($\tau_c$) in dotted, and spin-down energy loss rate ($\dot{E}_{sd}$) in dash-dotted are also shown.
  }
  \label{chapter_1_fig: PPdot_diagram_constant_lines}
\end{figure}

We will focus on the radio emission of pulsars and the properties of these that can be described through this process, adopting the formalism seen in \citet{Lorimer2012} and the references cited.
Timing analyses in radio, through the periodicity of the arrival time of the pulses, allow us to determine the spin period ($P$) of pulsars and their corresponding rate of spin-down, also known as the spin period derivative ($\dot{P}$).
These measurements place the population of radio-emitting pulsars on a plane called the $P\dot P$ diagram, which enables us to explore various characteristics that define them. 
Figure~\ref{chapter_1_fig: PPdot_diagram_constant_lines} shows the population of pulsars, as seen in the ATNF v2.6.1, that contains 2752 pulsars.

A remarkable differentiation comes from $P$, where the so-called "normal pulsars" are located from $P\sim 0.5$ s which increases at rates between $\dot{P}\sim10^{-15}$ ss$^{-1}$ and $\dot{P}\sim10^{-10}$ s, while the MSPs with $P<10$ ms and $\dot{P}\sim10^{-20}$ ss$^{-1}$.
Assuming a pulsar behaves as a rotating magnetic dipole, its observed $P$ and $\dot{P}$ provide insight into key physical properties. 
Variations in these quantities imply differences in derived properties considering the spin evolution, such as the magnetic field at the surface, spin-down energy loss, characteristic age, and other quantities related to the pulsar's magnetosphere, including the magnetic field at the light cylinder, the Goldreich–Julian charge number density, and the surface electric voltage (see Section~\ref{chapter1: magnetosphere}).
Note that in Figure~\ref{chapter_1_fig: PPdot_diagram_constant_lines}, we restrict the sample to pulsars with $\dot{P} > 0$.
This selection is essential to ensure the physical validity of the derived properties, which are based on mathematical expressions involving positive values of $P$ and $\dot{P}$.
For instance, observed negative values of $\dot{P}$ are often associated with pulsars in globular clusters, due to the acceleration in the cluster's gravity well (see Section~\ref{chapter6: significance_branches}). 
These values do not reliably reflect the intrinsic spin evolution of the pulsar and are therefore excluded from the analyses presented in this work.
In the following sections, we examine the theoretical relationships that link these observables to the derived properties of the pulsars and explore how they influence the pulsar population, as depicted in the $P\dot{P}$ diagram.
We will assume a dipole model, with $P(\text{s})$ and $\dot{P}(\text{s/s})$, where a neutron star behaves like a sphere of uniform density rotating in a vacuum, following \citet{Lorimer2012} and references therein.
In this way, its moment of inertia $I=kMR^{2}$ will be calculated with the values $M=1.4M_{\odot}$\footnote{1$M_{\odot}$=1.989$\times10^{33}$ g.}, $R=10^{6} \text{ cm}$, and $k=0.4$, resulting in the canonical value of $I=10^{45}\text{ g cm}^{2}$.

\subsection{Spin evolution}
\label{chapter1: Spin_evolution}

The evolution of the pulsar's spin can be observed through the pulse period, which increases over time.
Let $\dot{P} = \mathrm{d}P/\mathrm{d}t$ denote the increase in $P$ and $E_{rot}=I\Omega^{2}/2$ the rotational kinetic energy, with $\Omega=2\pi/P$ the angular frequency,
we can relate the change in $P$ to the loss of $E_{rot}$ as:

\begin{equation}
\dot{E}_{sd}\equiv -\frac{\mathrm{d}E_{rot}}{\mathrm{d}t}=-\frac{\mathrm{d}(I\Omega^{2}/2)}{\mathrm{d}t}=-I\Omega\dot\Omega=4\pi^{2}I\dot{P}P^{-3}~.
\label{chapter_1_eq: Edot_relation}
\end{equation}

The quantity $\dot{E}_{sd}$ is known as the spin-down energy loss rate (also referred to as the spin-down luminosity or spin-down power) and quantifies the rate at which the neutron star loses rotational energy.
By substituting the canonical value of $I=10^{45} \text{ g cm}^{2}$ (see Section~\ref{chapter1: Spin_periods}) into Equation~(\ref{chapter_1_eq: Edot_relation}), we obtain:

\begin{equation}
\dot{E}_{sd}\simeq 3.95\times10^{46} \dot{P}P^{-3} \text{ erg s}^{-1}~.
\label{chapter_1_eq: Edot_model}
\end{equation}

From Equation~(\ref{chapter_1_eq: Edot_model}) comes the value of $\dot{E}_{sd}$ seen in the population of pulsars observed at the ATNF. If we look at Figure~\ref{chapter_1_fig: PPdot_diagram_constant_lines}, the most energetic pulsars are young pulsars (see the discussion below on age), unlike older ones, which are usually less active.

Furthermore, not all of $\dot{E}_{sd}$ is converted into radio emission; in fact, only a small part of it is obtained in that frequency band. The majority of the energy loss is emitted in the form of high-energy radiation, including X-rays and gamma rays, as well as through pulsar winds and, undetectable yet, gravitational waves \citep{Lasky_2015, Glampedakis2018}.

To model how a pulsar loses its rotational energy over time, the braking index $n$ is introduced. 
To do this, we assume that a rotating magnetic dipole with a particular moment $|\vec{m}|$, with a separation $\alpha$ between its magnetic axis and its axis of rotation, emits electromagnetic radiation at a given $\Omega$: 

\begin{equation}
\dot{E}_{\text{dipole}}=\frac{2}{3c^{3}}|\vec{m}|^{2}\Omega^{4}\sin^{2}\alpha~.
\label{chapter_1_eq: Edot_dipole}
\end{equation}

This energy loss, seen in Equation~(\ref{chapter_1_eq: Edot_dipole}), is assumed to be equal to the lost rotational energy described in Equation~(\ref{chapter_1_eq: Edot_relation}). In this way, we can evaluate the evolution of its rotational frequency over time, described by:

\begin{equation}
\dot{\Omega}=-\bigg (\frac{2|\vec{m}|^{2}\sin ^{2}\alpha}{3Ic^{3}}\bigg )\Omega^{3}~.
\label{chapter_1_eq: Omega_dot}
\end{equation}

When considering the rotational frequency $\nu = 1/P$, Equation~(\ref{chapter_1_eq: Omega_dot}) can be generalized as a power-law in the following form:

\begin{equation}
\dot{\nu} = - K\nu^{n}~,
\label{chapter_1_eq: braking_index}
\end{equation}
 
where $\dot{\nu}$ is its rotational frequency over the time, and $K$ is a constant. 
The $n$ reflects the dominant mechanism by which the pulsar spins down. In other words, measuring $n$ allows us to investigate the processes that influence the long-term spin evolution of pulsars and explore potential evolutionary connections among the different pulsar populations. 

This is because $n$ can also be measured whenever $\dot \nu$ can be observed, so that by differentiating Equation~(\ref{chapter_1_eq: braking_index}) and not considering $K$ for simplifying, we have:

\begin{equation}
n = \nu \ddot{\nu}/\dot{\nu}^{2}~.
\label{chapter_1_eq: braking_index_measuring}
\end{equation}

The measurement of $n$ can be affected by timing noise and glitches in some instances. 

Glitches are sudden increases in the spin frequency of pulsars and are thought to offer insights into the internal structure of neutron stars. 

Timing noise, on the other hand, refers to irregularities in rotational behavior that cause deviations in pulse arrival times from a steady state of spin-down.
While measuring $\nu$ and $\dot{\nu}$ using standard timing models is relatively straightforward, determining $\ddot{\nu}$ is significantly more challenging due to its minimal magnitude. 

This follows from Equation~(\ref{chapter_1_eq: braking_index_measuring}), which shows that accurately measuring $n$ requires a reliable estimate of $\ddot{\nu}$. 
Since younger pulsars tend to spin down more rapidly, their $\ddot{\nu}$ are more pronounced, making them more favorable targets for such studies (\cite{Parthasarathy_2019_I, Parthasarathy_2020_II} and references therein).
\citet{Archibald_2016_brakingIndex} showed that pulsars with $n > 3$ exist, as exemplified by PSR J1640-4631, which was the first observed case where other physical mechanisms, such as magnetic quadrupoles, are essential for explaining pulsar spin-down.
Also, it can be observed that there is compelling evidence for a common trend among young glitching pulsars, with several exhibiting values of $n \leq 2$ \citep{Espinoza2017}.
Furthermore, as shown in \citet{Parthasarathy_2019_I} as well as in \citet{Parthasarathy_2020_II}, $n$ can vary significantly across pulsars. 

Estimations of $n=5$ suggest that the pulsar behaves as a pure emitter of gravitational waves, characterized by a quadrupole moment, to explain specific observations made as the case of the high $n$ seen in J1640-4631 \citep{Araujo2016}.

The value of $n = 3$ corresponds to the case of pure magnetic dipole radiation in vacuum, the canonical model for pulsar spin-down proposed initially in early theoretical work \citep{PACINI_1968, Ostriker1969}, and adopted in this study. 
This choice is motivated by its ability to link observable quantities to physical properties of the neutron star, such as age or magnetic field strength, through simplified and interpretable expressions.

Thus, we can estimate the pulsar's age from this starting point. 
This can be obtain through Equation~(\ref{chapter_1_eq: braking_index}), taking $\nu=1/P$, can be expressed as $\dot{P} = K P^{2-n}$, and integrating the latter while assuming $K$ constant and $n\neq 1$, we have:

\begin{equation}
T=\frac{P}{(n-1)\dot{P}}\Bigg[ 1- \bigg( \frac{P_0}{P} \bigg )^{n-1}  \Bigg]~,
\label{chapter_1_eq: age_pulsar}
\end{equation}

where $P_{0}$ denotes the spin period at birth. 
Under the model's considerations, $n=3$, so the dipolar magnetic radiation produces the spin-down, 
and $P_{0}\ll P$, Equation~(\ref{chapter_1_eq: age_pulsar}) can be seen as: 

\begin{equation}
\tau_{c}\equiv\frac{P}{2\dot{P}}\simeq 1.58\times 10^{-8} \frac{P}{\dot{P}} \text{ yr}~.
\label{chapter_1_eq: characteristic_age_pulsar}
\end{equation}

The quantity in Equation~(\ref{chapter_1_eq: characteristic_age_pulsar}) is denoted as the characteristic age, and it is the value expressed in Figure~\ref{chapter_1_fig: PPdot_diagram_constant_lines}. 
This figure shows the normal pulsars, the bulk of the population, with $\tau_c \sim 10^7$ yr, while young pulsars are those with $\tau_c \leq 10^{5}$ yr, and MSPs present a $\tau_{c}\sim 10^{10}$ yr.

Even so, it must be considered that, due to the conditions imposed on $\tau_c$, it only serves as an estimate, especially when true ages are not accessible, for locating pulsars within a population (see Section~\ref{chapter6: variables_PCA} for cases where it is discarded). 
Thus, it can be verified that $\tau_c$ does not provide the true age of the pulsar, for example, using the Crab pulsar (see Section~\ref{chapter1: introduction}).
According to Equation~(\ref{chapter_1_eq: characteristic_age_pulsar}), and taking $P=0.033$ s and $\dot P=4.2\times10^{-13} \text{ ss}^{-1}$  from the Crab pulsar, we see a $\tau_c = 1260 $ yr. 
However, the supernova explosion of the Crab Nebula is dated to 1054 AD, so the true age of the Crab pulsar is estimated to be less than 1000 years.

Among the physical properties associated with the immediate surface of the neutron star, the magnetic field at the surface ($B_s$) plays a central role. Assuming a purely dipolar behavior, the magnetic field strength is related to the magnetic moment through:

\begin{equation}
B = \frac{|\vec{m}|}{r^{3}}~.
\label{chapter_1_eq: magnetic_field_moment}
\end{equation}

Substituting $\Omega = 2\pi /P$ and introducing Equation~(\ref{chapter_1_eq: magnetic_field_moment}) into Equation~(\ref{chapter_1_eq: Omega_dot}), we obtain:

\begin{equation}
B_{s}\equiv B(r=R) =\sqrt{\frac{3c^{2}IP\dot{P}}{8\pi^{2}R^{6}\sin^{2}\alpha}}.
\label{chapter_1_eq: Bs_relation}
\end{equation}

Taking the canonical values of $I=10^{45}\text{ g cm}^{2}$, $R=10^{6} \text{ cm}$, and assuming\footnote{With $\alpha = 90^{\circ}$ the magnetic axis is perpendicular to the rotation axis, which equates to the maximum dipole torque.} $\alpha = 90^{\circ}$, by Equation~(\ref{chapter_1_eq: Bs_relation}) we have:

\begin{equation}
B_{s}\simeq3.2\times10^{19}\sqrt{P\dot{P}} \text{ G}.
\label{chapter_1_eq: Bs_quantity}
\end{equation}

Equation~(\ref{chapter_1_eq: Bs_quantity}) provides a widely used estimate of the magnetic field strength at the surface of the neutron star, and its behavior is shown in Figure~\ref{chapter_1_fig: PPdot_diagram_constant_lines}.
We can distinguish pulsars with high $B_s \sim 10^{14}G$, usually referred to as magnetars, from normal pulsars that are around $B_s \sim 10^{12}G$, and, for instance, MSPs with $B_s \sim 10^{8}G$.
However, it is essential to note that this value is derived from model assumptions. 
Thus, the true magnetic field strength may vary from this estimation, in particular due to deviations from ideal dipole geometry, contributions from multipole fields, or the presence of plasma in the magnetosphere \citep{Spitkovsky_2006, Petri_2015, Gralla2017}. 

In some cases, the magnetic field strength can be measured directly by detecting cyclotron line features in X-ray spectra \citep{Truemper1978, Wheaton1979, Coburn2002, Tiengo2013}, or even in isolated pulsars \citep{Bignami2003, Suleimanov2010}.
The observed values are generally consistent in order of magnitude with $B_{s}$.
Furthermore, as seen in \citet{Shapiro1983}, Equation~(\ref{chapter_1_eq: Bs_quantity}) can be interpreted as the magnetic field strength at the magnetic equator, but the field at the magnetic poles is approximately twice as strong: $B_{\text{pole}} = 2 B_{s}$.

This distinction becomes important when comparing theoretical models with observational diagnostics, particularly in studies of X-ray emission \citep{Zavlin_2000}.

\subsection{Magnetosphere}
\label{chapter1: magnetosphere}

The realistic evolution of the model assumed for the rotating neutron star in a vacuum is based on the idea that it behaves like a highly magnetized, rotating conductive sphere. This is because the Lorentz forces acting on the charges inside the neutron star are stronger than the gravitational forces. 
\citet{Deutsch1955} was the first to study the structure of the electromagnetic field around an isolated, rotating, magnetized neutron star in vacuum that has a magnetic axis misaligned with the rotation axis.
Then, \citet{Goldreich1969} shows that a neutron star with aligned magnetic axes and rotation generates an external electric field that extracts plasma from its surface. 
Although this model is an approximation with some deviation from a realistic case, it contributes to our understanding of the pulsar's behavior and facilitates its observation.
Thus, the pulsar magnetosphere is the region surrounding the neutron star, filled with magnetized plasma and shaped by interactions with the intense magnetic field.
The magnetosphere spans from the stellar surface to distances of several thousand kilometers and may be shaped by interactions with the interstellar medium, surrounding accretion material, or binary companions.
The complexity of such an environment has led to the assumption that analytical solutions are only applicable in simple cases, providing a convincing answer.
In Section~\ref{chapter1: emission_models}, we present studies conducted on the description of the magnetosphere through models that will aid in understanding gamma emission.

To describe the physical properties of magnetospheric proxies relevant to our work, we follow the treatment presented in \citet{Lorimer2012}, as also outlined in \citet{Goldreich1969}.
We start from the premise that the neutron star behaves like a perfectly conductive, rotating, and magnetized sphere. 
Therefore, we expect an induced electric field to be produced by the magnetic field of this rotating magnetized sphere, balanced by the electric field generated by the distribution of charges, since it is a conductive sphere.
If the space outside the star were a true vacuum, these induced surface charges would generate an external electric field, which is quadrupolar and associated with a net charge of zero. 
This field can be described by the electric potential:
\begin{equation}
\Phi(r, \theta)= \frac{R^{5}\Omega B_s}{6cr^{3}}(3\cos^{2}\theta-1)~, 
\label{chapter_1_eq: quadropolar_field}
\end{equation}
expressed in polar coordinates ($r, \theta$) centered on the star.

Charged particles are greatly influenced by the electric field on the surface and are expelled, making it impossible to maintain a vacuum around the neutron star. 
This flow of charged particles, combined with the neutron star's high magnetic field, forms the dense plasma that surrounds it.
This creates a charge distribution, visible at the magnetic poles and the equator, where charges of opposite signs will be positioned respectively. 
To measure the maximum value of the charge separation at the magnetic pole, we use the Goldreich-Julian charge number density ($\eta_{GJ}$):

\begin{equation}
\eta_{GJ}= \frac{\Omega B_s}{2\pi c e}\simeq \frac{B_s}{c e P}=2.2\times10^{18}\sqrt{\frac{\dot{P}}{P}} \text{ cm}^{-3}~.
\label{chapter_1_eq: GJ_density}
\end{equation}

The quantity referred to in Equation~(\ref{chapter_1_eq: GJ_density}) sets a baseline for magnetospheric plasma densities and serves as a reference for comparing different emission regimes.

The plasma contained in the envelope surrounding the neutron star is in a state of rigid co-rotation with it, influenced by an electromagnetic field that affects all charged particles uniformly. 
The plasma's dynamic behavior can only be maintained up to a certain distance, beyond which such co-rotation would require particles to move faster than the speed of light. 
This maximum distance can be calculated if we consider the co-rotation velocity, $v =\Omega R$, which defines the radius of an imaginary surface around the neutron star. 
This distance, where $v=c$ is known as the radius of the light cylinder, taking again $\Omega=2\pi/P$, can be described as: 

\begin{equation}
R_{lc}= \frac{c}{\Omega}=\frac{cP}{2\pi}=4.77\times10^{4}P \text{ km}~.
\label{chapter_1_eq: R_lc_P}
\end{equation}

Now, assuming a dipolar behavior of the magnetic field at a distance $r$ as we saw in Equation~(\ref{chapter_1_eq: magnetic_field_moment}), where $B(r) \propto 1/r^{3}$, this can be defined using $B_s$, also seen as the magnetic field at the equator, and $R$: 

\begin{equation}
B(r)= B_s\bigg(\frac{R}{r}\bigg)^{3}~.
\label{chapter_1_eq: B_r_Bs}
\end{equation}

Then, we can define the magnetic field at the light cylinder ($B_{lc}$) by introducing Equation~(\ref{chapter_1_eq: R_lc_P}) in Equation~(\ref{chapter_1_eq: B_r_Bs}):

\begin{equation}
B_{lc}= B_s\bigg(\frac{R}{R_{lc}}\bigg)^{3}=B_s\bigg(\frac{2\pi R}{cP}\bigg)^{3}~.
\label{chapter_1_eq: B_l_relation}
\end{equation}

Assuming the canonical values seen in Equation~(\ref{chapter_1_eq: Bs_quantity}) for Equation~(\ref{chapter_1_eq: B_l_relation}):

\begin{equation}
B_{lc}\simeq 3 \times 10^{8} \sqrt{\frac{\dot{P}}{P^{5}}} \text{ G}~.
\label{chapter_1_eq: B_l_quantity}
\end{equation}

The quantity of $B_{lc}$ seen in Equation~(\ref{chapter_1_eq: B_l_quantity}) gives the value taken from the ATNF to define the population of pulsars used in this work, and captures magnetospheric activity (see Chapter~\ref{chapter4} for more information on its impact on population separation).

To understand the role of the light cylinder, we refer to Figure~\ref{chapter_1_fig: NS_emission_geometry}. 
The light cylinder is defined by its $R_{lc}$ and has a significant influence on the structure of the magnetic field lines. 
Inside the light cylinder lie the so-called closed field lines, which loop back to the surface of the neutron star. 
In contrast, open field lines extend beyond the light cylinder and do not return; instead, they carry relativistic plasma into the pulsar wind. 
These open field lines originate from a small region around the magnetic poles known as the polar cap (see Figure~\ref{chapter_1_fig: NS_emission_geometry}). 
In other words, the "last open field line", which touches the light cylinder tangentially, defines the edge of the polar cap over the neutron star surface.
This region plays a crucial role in shaping both the radio emission beam and the structure of the particle outflow.

From that region, we aim to estimate the potential drop between the magnetic pole and the edge of the polar cap, referred to as surface electric voltage ($\Delta{V}$). 

First, we need to define the polar cap region, which can be described by ($R, \theta_p$), where $\theta_p$ is the angle between the last open field line and the magnetic axis. More specifically, we have to compute the polar cap radius ($R_p$).

Considering these dipolar magnetic field lines as constant, under the boundary conditions $(R,\theta_p)$ and $(R_{lc}, \theta = \pi/2)$ due to aligned axes, we can see:

\begin{equation}
\frac{\sin^{2}\theta}{r} =  \frac{\sin^{2}\theta_p}{R}= \frac{1}{R_{lc}}~.
\label{chapter_1_eq: magnetic_line}
\end{equation}

As explained above, the region of the polar cap is a circle on the neutron star surface delimited by the last open field lines.
Therefore, $R_p$ can be seen as the projected distance from the magnetic axis to the "last open field line" under $\theta_p$.

Thus, using Equation~(\ref{chapter_1_eq: magnetic_line}) and Equation~(\ref{chapter_1_eq: R_lc_P}) to describe $\theta_p$ , we can compute $R_p$ as:

\begin{equation}
R_p = R\sin \theta_p = \sqrt{\frac{2\pi R^{3}}{cP}}\simeq 144 \times \sqrt{P} \text{ m} ~.
\label{chapter_1_eq: Rp_relation_quantity}
\end{equation}

where $\Omega=2\pi /P$, and $R=10^{6} \text{ cm}$.

Now, we can obtain $\Delta V$ by applying Equation~(\ref{chapter_1_eq: quadropolar_field}) at the magnetic pole $(R_{lc}, \pi/2)$ and polar cap $(R_p, \theta_p)$:

\begin{equation}
\Delta V = \frac{B_s \Omega^{2}R^{3}}{2c^{2}} \simeq 2.1\times 10^{20}\sqrt{\frac{\dot{P}}{P^{3}}} \text{ V}~.
\label{chapter_1_eq: SEV_relation_quantity}
\end{equation}

For the calculation of Equation~(\ref{chapter_1_eq: SEV_relation_quantity}), in addition to the canonical values set in Section~\ref{chapter1: Spin_periods}, we recall that the CGS system\footnote{$c=3\times10^{10}\text{ cm/s}$; $e=4.8032\times 10^{-10} \text{ cm}^{3/2} \text{ g}^{1/2} \text{ s}^{-1}$} is assumed throughout this chapter except when the magnitude is indicated. 
Thus, the result in the first part of the above expression is given in Statvolts, whose equivalence to Volts is seen as $1 \text{ Statvolt} \simeq 300 \text{ V}$, giving the final result. 
The $\Delta{V}$ quantity provides an estimate of the pulsar's ability to accelerate particles, which may be relevant for emission modeling.

\section{Alternative emission models}
\label{chapter1: emission_models}

In this section, we revisit the main theoretical models developed to describe pulsar magnetospheres, with a particular focus on their ability to reproduce the gamma-ray light curves observed by instruments such as \textit{Fermi}-LAT \citep{Fermi3PC}. 
Although pulsars emit across a broad range of wavelengths, only a limited subset of $\sim$300 pulsars have been detected in gamma-rays (see Section~\ref{chapter1: MSPs} for more details about the population). 
This emission is highly anisotropic and originates from particles accelerated along open magnetic field lines by strong electric fields induced by the rotation of the magnetic field. 
As such, gamma-ray light curves offer constraints on the geometry and physical conditions within the magnetosphere.
This emission is highly anisotropic and originates from particles accelerated along magnetic field lines opened by intense electric fields generated by the rotation of its magnetic field. The pulsar's radiation is explained based on assumptions regarding the location and geometry of the emission zones, as described by \citet{Brambilla2015}. 
See also \citet{Philippovreview} for a comprehensive review. 

Thus, models such as the Polar Cap (PC; see, e.g., \cite{Daugherty1996}), Outer Gap (OG; see, e.g., \cite{Cheng1986, Romani1995}), and Slot Gap (SG; see, e.g., \cite{Arons1983, Muslimov2004}) were built upon the magnetic dipole geometry assumed in the vacuum model discussed in Section~\ref{chapter1: Spin_periods}. 
However, the PC emission model, which assumes that particle acceleration occurs near the neutron star due to strong parallel electric fields, was ruled out due to the sensitivity achieved by \textit{Fermi}-LAT, for example, in measuring the Vela spectrum in \citet{Fermi2PC}. 
However, none of these models can fully explain the complete phenomenology of gamma-ray light curves (see, e.g., \cite{RomaniWatters2010, Pierbattista2012}), nor are they consistent with the global properties of the magnetosphere, such as current closure. 
In particular, see \citet{Vigano2015a, Vigano2015b} for a study of the approximations and assumptions behind OG models.

A more representative view of the geometry of the external field than that assumed by the magnetic dipole model in a vacuum has advanced the study of pulsar magnetospheres. 

This progress is facilitated by the development of, for instance, Force-Free Electrodynamics (FFE; see, e.g., \cite{Spitkovsky_2006, Kalapotharakos2009, Petri2012a, Harding2017, Petri2021}),
and Particle-In-Cell models (PIC; see, e.g., \cite{Chen2014, Philippov_Spitkovsky_2014, Cerutti2015, Cerutti2025}).
The former assumes a perfectly conducting plasma and therefore cannot account for particle acceleration due to the absence of parallel electric fields ($E_\parallel$). 
The latter incorporates kinetic microphysics and self-consistent particle acceleration.
Each of these, and others such as Magnetohydrodynamics models (MHD; see, e.g., \cite{Komissarov2006, Tchekhovskoy2013}) or 
resistive or dissipative magnetospheres models (see, e.g., \cite{Kalapotharakos2012, Kalapotharakos2014, Brambilla2015, Cao2024}) serve their aims and have their limitations.
Also mention the Effective Synchro-curvature model, which considers the motion of charged particles along curved magnetic field lines and calculates their radiation through synchro-curvature processes \citep{Vigano2015c}, and tries to 
reproduce, using only three effective parameters that act as order parameters, both the gamma-ray spectra across the entire population of known gamma-ray pulsars \citep{Vigano2015c, Vigano2015d} and the subset with observed non-thermal X-ray pulsations \citep{Torres2018, Torres2019}. 
Recent advancements have further showcased its versatility, with successful applications in fitting gamma light curves of \textit{Fermi}-LAT pulsars (see \cite{Iniguez2024, Iñiguez2025} and Section~\ref{chapter9: intro} for more details).

\section{Millisecond pulsars}
\label{chapter1: MSPs}

As mentioned at the beginning of this chapter, neutron stars can also be members of binary systems, which make up a significant fraction of the population shown in Figure~\ref{chapter_1_fig: PPdot_diagram_constant_lines}. 
These systems offer a distinct perspective on the formation and evolution of neutron stars. 

Observations allow us to establish constraints that provide information about the properties of their companions. 
Most binary systems include white dwarfs, main-sequence stars, or other neutron stars as companions, although a few systems exhibit unusual companions. 
For instance, PSR B1257+12 is accompanied by up to three terrestrial-mass bodies (see Table~\ref{chapter_6_tab: BW_RW_tMSP} for more details). 
At the same time, PSR B1620−26 is part of a triple system that includes a young white dwarf and a Jupiter-mass third body, located in the globular cluster M4 \citep{Thorset1993, Backer1993, Thorsett1999, Sigurdsson2003}.
Noteworthy is J1719-1438, whose system produced interest due to an uncertain companion that poses planetary mass but a remarkably high density \citep{Bailes2011, van_Haaften2012, Horvath_2012}.

Most of the binary systems observed are composed of MSPs with $P<10 \text{ ms}$, and are considered older neutron stars. 
The first MSP discovered was PSR B1937+21, as reported by \citet{Alpar1982}, who proposed that these were merely the result of accretion.
As we will see below, the MSPs are formed due to the spin-up that the pulsar experiences when it accretes mass from its companion, known as the recycling scenario \citep{Bhattacharya1991, Tauris_2003, Tauris_2012}.

However, a subset of MSPs is found to be isolated, which likely formed in binary systems but later lost their companion through mechanisms such as ablation or disruption. 
Ablation, in particular, occurs when the pulsar's energetic wind evaporates the companion over time, a scenario supported by observations of so-called "black widow" and "redback" systems, also denominated spider pulsars, transitioning into isolated MSPs (see, eg., \cite{Fruchter1988, Roberts2013, Chen2013}, and a further explanation about these systems in Section~\ref{chapter1: SpiderPulsars}).
Disruption, in contrast, refers to the dynamical breakup of a binary system, especially in dense stellar environments, resulting from gravitational interactions with passing stars. 
This is especially relevant in globular clusters, where the high number of isolated pulsars is a consequence of the increased frequency of such encounters and disruption events.
It is worth noting that several alternative formation scenarios have been proposed to explain this observation, including accretion following mergers of neutron stars and main-sequence stars, as well as mergers of massive white dwarf binaries and tidal disruption events of main-sequence stars. 
In this regard, it is essential to note that the overall proportion of MSPs in globular clusters is significantly higher than that observed in the galactic plane (see, e.g., \cite{Yin_FAST_2024} and references therein).

Another approach to characterizing the MSP population is through their gamma-ray emission (see, e.g., the Third \textit{Fermi}-LAT Pulsar Catalog (3PC) or dedicated studies such as \cite{Torres2017}). 
In the 3PC, approximately 143 gamma-ray millisecond pulsars (gMSPs) have been identified.

\subsection{Formation and evolution of binary systems to millisecond pulsars}
\label{chapter1: MSPs_Form_Evo}

To explain a plausible formation pathway for MSPs, we focus on the evolution of binary systems, following the general framework described in \citet{Lorimer2008}. 
The process begins with the formation of a neutron star in a core-collapse supernova within a binary system. 
The binary can remain bound only if the mass ejected during the explosion does not exceed a critical fraction of the system's pre-supernova mass, typically, less than half. 
Otherwise, the sudden loss of gravitational binding leads to system disruption, leaving behind an isolated, rapidly rotating neutron star (see, e.g., \citealt{Blaauw1961, Hills1983}).
If the binary survives the explosion, the system may acquire a high orbital eccentricity due to the kick imparted to the neutron star at birth \citep{Tauris2017}.
Over time, if the companion star is sufficiently massive, it will evolve off the main sequence and expand. 
Once it fills its Roche lobe, mass transfer begins, marking the initial point of the recycling scenario \citep{Bhattacharya1991, Tauris_2003, Tauris_2012, Patruno_Watts}.
In this process, the neutron star accretes matter and angular momentum from its companion, gradually spinning up to millisecond periods \citep{Alpar1982, Radhakrishnan1982}.
During this accretion phase, the system becomes a low-mass X-ray binary (LMXB), characterized by strong X-ray emission produced as the infalling material heats the surface of the neutron star. Once accretion ceases, the pulsar reactivates in the radio band, now observable as a recycled MSP.
Statistical analyses of pulsar spin distributions provide empirical evidence in support of this evolutionary scenario. 
\citet{PapittoTorres2014} showed that different classes of accreting and rotation-powered MSPs exhibit distinct spin periods.  
They note that accreting MSPs and eclipsing rotation-powered MSPs seem to occupy an intermediate stage, reinforcing the notion of spin-down during the late stages of mass accretion and reflecting the diversity of evolutionary channels that can produce MSPs.  
The last two types of MSPs, as we will see in Section~\ref{chapter1: SpiderPulsars}, can sometimes be adopted by the same transitional system as different states, referred to as transitional millisecond pulsars.

\subsection{Characterization of binary systems}
\label{chapter1: BinaryParameters}

In addition to the properties described in Section~\ref{chapter1: phenomenology}, which will continue to be of interest for characterizing the binary population, except $\tau_{c}$ (see discussion in Section~\ref{chapter6: variables_PCA}), we will use the binary period ($P_B$), the projected semi-major axis ($A_1$), and the minimum mass of the companion ($M_C$). 

These last parameters enable us to describe the binary nature of the MSP in question, providing a new and distinct perspective when classifying the population. 
These quantities will be obtained from the ATNF and analyzed in depth in Chapter~\ref{chapter6} on their application. 

It should be noted that $P_B$ represents the orbital period observed in the system and will be expressed in days. It is a quantity that can be directly measured from the periodic variations in pulse arrival times caused by the pulsar's motion around the center of mass of the binary system.

On the other hand, $A_1$ represents the projected semi-major axis that describes the elliptical orbit of the system as it moves. This quantity is derived directly from timing measurements and is called projected because we only see the component along our line of sight:

\begin{equation}
A_1 = \frac{a_1 \sin i}{c} \text{ lt-s}~.
\label{chapter_1_eq: projected_semi_major_Axis}
\end{equation}

In Equation~(\ref{chapter_1_eq: projected_semi_major_Axis}), $a_1$ denotes the semi-major axis of the pulsar's orbit and $i$ represents the orbital inclination angle. An inclination of $i = 90^\circ$ corresponds to an edge-on view, while $i = 0^\circ$ describes a face-on orientation.

Assuming the third law of Kepler and the relationship of center of mass \citep{FundamentalAstronomy}, we can explore the allowed range of $M_C$ from the binary mass function, which relates the latter to the observables $P_B$, $a_1$, and $i$, alongside $M$ and the gravitational constant $G$ \citep{Tauris2006, Bailes2011, Podsiadlowski2012}:

\begin{equation}
f(M_C) = \frac{4\pi^{2}}{G}\frac{(a_1\sin i)^{3}}{P_{B}^{2}}=\frac{(M_{C}\sin i)^{3}}{(M_{C}+M)^{2}}\text{ }M_{\odot}~.
\label{chapter_1_eq: mass_function}
\end{equation}

For its calculation, assumptions such as $M\simeq1.4 M_{\odot}$ and $i=90^{\circ}$  will give the minimum value of the companion's mass, $M_C$. 

The steps carried out for Equation~(\ref{chapter_1_eq: mass_function}) are also applicable in cases where some observables are measured over a range of values or must be estimated through other parameters (see Section~\ref{chapter_6_figures: MST_3FGL1544} for a specific case).

\subsection{Spider pulsars}
\label{chapter1: SpiderPulsars}

Spider pulsars (see, e.g., \citealt{Eichler1988, Roberts2013, Roberts20172018, DiSalvo2023}), including black widows and redbacks, as well as a special state of the latter called transitional millisecond pulsars, form a special class of MSPs that reside in binary systems characterized by tight orbits and low-mass companions that are actively ablated by the pulsar wind, as we introduced in Section~\ref{chapter1: MSPs}.
They show irregular radio eclipses due to the presence of intrabinary material. For a comprehensive review of these objects and their phenomenology, see \citet{Papitto_tMSPSectionBook}.
The first black widow, B1957+20, discovered by \citet{Fruchter1988}, is an eclipsing pulsar with a $P=$1.6~ms and a binary period of 9.1~hr, orbiting a low-mass companion of 0.02~$M_{\odot}$.
This discovery supported the idea that the companions of isolated MSPs were ablated by energetic particles and/or $\gamma$-rays produced by the pulsar wind.
Although the recycling scenario (see Section~\ref{chapter1: MSPs_Form_Evo}) was increasingly validated by observations of fast millisecond pulsations in X-rays in transient LMXBs and X-ray bursts, it remained to be explained how the change from accretion to quiescence occurred to produce radio emission. 
This problem was described as the "missing link" between MSPs and LMXBs, and introduced the idea of a transitional millisecond pulsar. 
This is a class of redback systems in which a neutron star can alternate between accretion-powered and rotation-powered states.
During the accretion stage, the system is observed to emit bright X-rays. When the mass transfer rate decreases in the final evolutionary stages, these binaries come to contain an MSP whose radio emission is enhanced by the rotating magnetic field of the neutron star.
This scenario was favoured by the detection of X-ray millisecond pulsations from accreting neutrons stars, as SAX J1808.4−3658 \citep{Chakrabarty1998, Wijnands1998}, and evidence of the existence of an accretion disk in the past of an MSP with the discovery of PSR J1023+0038, classified initially as a MSP and later found to exhibit characteristics of an accreting system \citep{Archibald2009, Archibald2010, Patruno2014, Stappers2014}. 
Although the latter was an early step in confirming this scenario, the first confirmed observation of a pulsar transitioning in real time between these two states was reported by \citet{Papitto2013}, who detected X-ray pulsations from IGR J18245–2452 during an accretion episode followed by radio pulses a few days later, thus establishing it as the first confirmed transitional millisecond pulsar.
PSR J1023+0038, now classified as a transitional millisecond pulsar, remains the best-studied system of its kind. It was the first to exhibit simultaneous optical and X-ray pulsations during the high-intensity X-ray mode, which disappear during the low-intensity mode, suggesting a shared underlying physical mechanism \citep{Papitto_2019ApJ}.
State transitions from XSS J12270–4859 were observed, resulting in the third confirmed transitional millisecond pulsar to date. The unusual properties seen in XSS J12270–4859 were determined to be characteristic of the sub-luminous state of tMSPs. The disappearance of optical emission lines and the significant dimming across multiple wavelengths indicated that XSS J12270–4859 underwent a transition from an accretion disc state to a radio pulsar state during a short time interval in late 2012 \citep{Saitou2009, deMartino2010, Hill2011, deMartino_2013A&A, Bassa2014, deMartino2014}.

As a summary, we must note that redbacks are eclipsing radio pulsars in tight binary systems, either with a non-degenerate main-sequence companion with a mass in the range $\sim$0.1–0.8 M$_\odot$. 
In contrast, black widows have a $\lesssim0.06$ M$_\odot$ semi-degenerate companion.
While a transitional millisecond pulsar exhibits dramatic state changes: transitioning from rotation-powered when they behave as redbacks, to accretion-powered and vice versa, on timescales as short as a few weeks.
It is worth noting that all three confirmed transitional millisecond pulsars exhibit a peculiar state of accretion disk, characterized by an X-ray luminosity that is lower than that of outbursting AMXPs but brighter than that of both rotation-powered MSPs and quiescent AMXPs. A defining feature of this sub-luminous disk state is the presence of distinct X-ray intensity modes, typically classified as "high" (or active) and "low" (or passive), along with sporadic flaring episodes \citep{Papitto_tMSPSectionBook}.
One of the most efficient ways to identify candidate tMSPs has turned out to be searching for this enigmatic variability in the X-ray emission between the two intensity levels, high and low modes, in the sub-luminous disc state \citep[see, e.g.,][]{deMartino_2013A&A, Papitto2013, Patruno2014, Archibald_2015ApJ, Linares_2014MNRAS, Bogdanov_2015ApJ, Papitto_2019ApJ, Baglio_2023A&A}. 
This method has proven successful in identifying a few candidates \citep[see, e.g.,][]{Bogdanov_2015, CotiZelati_2019}.
In Section~\ref{chapter6: MST}, we will show a more individual analysis of these classes of pulsars.

\section{Fast radio bursts}
\label{chapter1: FRBs}

Fast Radio Bursts (FRBs) were first detected in 2001 by the Parkes Telescope and reported in 2007 by \citet{Lorimer2007}, denoted as FRB~20010724 (also called FRB~010724 or "Lorimer Burst", see \citet{Lorimer2024FRBs} for a review of the discovery and highlights of the FRBs topic). 
They are transient radio pulses, typically lasting on millisecond timescales (see \cite{Zhang2023} for a comprehensive review).
Although there is consensus on its astronomical origin based on all observations carried out and all events detected, this was not the case from the outset. 
An artificial origin occurring on Earth had to be ruled out ("perytons"; see, e.g., \cite{Burke-Spolaor2011, Petroff2015}). 
Additionally, there were speculations that it could be the pulse of other classes of isolated neutron stars observed at X-ray and gamma-ray wavelengths \citep{McLaughlin2006}. 

However, the detection reported by \citet{Thornton2013} not only confirmed an extragalactic origin but also pinpointed its origin as an astronomical event.

The detectability of events originating from what is believed to be the same source establishes a distinction between two categories, at least from an observational point of view: repeaters and non-repeaters, albeit it is unclear whether all non-repeaters will eventually repeat.
This is because, until \citet{Spitler2016} detected several bursts from the same source, named FRB~20121102A (also called FRB~121102, "R1", or "Spitler burst"), which demonstrated that the source survives the process that produces the bursts, all detected FRBs had been attributed to different sources. The repetition established that the FRBs do not originate in cataclysmic astrophysical events, contrary to previous hypotheses.

After that, the number of detected repeating sources increased due to observations made by the Canadian Hydrogen Intensity Mapping Experiment (CHIME, \cite{CHIME2019a, CHIME2019b}), the Australian Square Kilometre Array Pathfinder (ASKAP) \citep{Kumar2019}, and the Five-hundred-meter Aperture Spherical radio Telescope (FAST) in China (\cite{LuoFASTRepeater, NiuFASTRepeater}). Furthermore, most of these sources have been detected at cosmological distances, leading to the expectation that they generate extreme, coherent radio emissions.

The repetitive behavior of FRB~20121102A prompted specific observation campaigns using the Karl G. Jansky Very Large Array (VLA) in conjunction with the Arecibo telescope, focusing on detecting possible additional bursts and estimating the source's location using the interferometric technique \citep{Chatterjee2017}.
Noteworthy are the observations via interferometry carried out by several groups, such as the ASKAP collaboration and Deep Synoptic Array (DSA) collaboration, which led to the location of both repeating and non-repeating sources in different types of host galaxies \citep{Bannister2019, Prochaska2019, Ravi2019, Macquart_et_al2020, Marcote2020,  Bhandari2022, Xu2022}.

Regarding its origin, several theoretical models were proposed (see, e.g., \cite{Platts2019}, for a summary) before the first discovery. 
It has been suggested that their emission could be powered by the dissipation of magnetic fields of a magnetar \citep{Popov2013, Lyubarsky2014, Katz2016, Murase2016, Kumar2017, Nicholl2017, Margalit2018, Lu2020, Yang2021}. 
This has been substantiated after CHIME \citep{CHIME} and STARE-2 radio array \citep{STARE2} discovered FRB~200428 in association with a hard X-ray burst from the Galactic magnetar, SGR 1935+2154 \citep{HXMT-FRB, Mereghetti2020, Zhong2020}. 
In the magnetar interpretation, the leading source model for FRBs, some differences may arise from age, with young magnetars producing a higher repetition rate than older ones \citep{Beloborodov2017, Metzger2017, Margalit2020, Levin2020, Feng2022}.
However, this may not represent the entire FRB population; see, e.g., the analysis observed in FRB~20200120E in \citet{Pearlman2025}. 
They argue against typical ultraluminous X-ray bursts, magnetar-like giant flares, or a SGR 1935+2154-like intermediate flare as associations for this repeater. Additionally, they suggested that it is unlikely to be an ultraluminous X-ray source or a young extragalactic pulsar embedded in a Crab-like nebula acting as the engine of FRB~20200120E. Other scenarios proposed include a giant radio pulse-emitting pulsar or a magnetar formed through a delayed channel.
\citet{Zhang2023} also offers an extensive review of FRB source models.
Although initially unexpected in the context of pulsars, the behavior of FRBs immediately suggested a compact object origin. Among the most compelling models are those involving neutron stars, particularly magnetars, as progenitors. This connection, along with the discovery of repeating FRBs, has prompted the inclusion of these events in this work.

\section{Conclusions}
\label{chapter1: Conclusions}

This section outlines the fundamental observables of pulsars, $P$ and $\dot{P}$, and how they provide a first approach to the pulsar population. 
From these quantities, several derived quantities such as the spin-down energy loss, characteristic age, and surface magnetic field can be inferred, as discussed in Section~\ref{chapter1: Spin_evolution}. 
These parameters describe (as a proxy) the rotational energy loss and provide a first estimate of the pulsar's evolutionary stage and magnetic properties. 
However, they are derived under assumptions that the pulsar behaves as an isolated, rotating magnetic dipole in a vacuum. 
As such, they do not account for the presence of a magnetosphere and do not describe the detailed mechanisms by which radiation is produced. 
Some physical parameters discussed in Section~\ref{chapter1: magnetosphere} relate directly to the environment where pulsar emission originates. 
These include the Goldreich–Julian charge number density, magnetic field at the light cylinder, and the surface electric voltage.
They are linked to the conditions under which particle acceleration and radiation can occur. 
In particular, the structure of the magnetosphere and the associated electric and magnetic fields determine whether a pulsar can accelerate particles to the energies required for producing coherent radio waves. 
It can be observed that the presence of open field lines, a well-defined polar cap, and a sufficient potential drop are relevant for sustaining radio emission.

While all these derived properties are model-dependent, they serve as a proxy within the complete set of physical parameters, providing a framework for understanding how pulsars evolve, emit, and interact with one another.
The $P\dot{P}$ diagram, as a visual tool, has played a central role in the pulsar population. However, it has a limitation when it comes to displaying all the information in the population's feature space. This can be seen, for example, in Figure~\ref{chapter_1_fig: PPdot_diagram_constant_lines}, where the addition of constant lines from the physical property relationships would render the diagram useless.
For this reason, which we shall further comment on later, relying solely on the $P$–$\dot{P}$ diagram may be misleading, especially when trying to separate populations or infer connections between classes of pulsars, as it requires all available information.
We will present approaches that enable the feature space to be expanded by leveraging similarity relationships, such as graphs —specifically, the Minimum Spanning Tree (see Chapter~\ref{chapter2}) —in conjunction with dimension reduction and feature space analysis techniques that rely on the Principal Component Analysis (see Chapter~\ref{chapter3}).

Particular attention is paid to MSPs, especially those in binary systems. 
These objects represent a key phase in the evolutionary process of neutron stars, offering insight into recycling processes, mass transfer, and spin-up mechanisms. 
Binary MSPs also introduce additional observables into the analysis, such as orbital period, the projected semi-major axis, and companion mass estimates, thereby enriching the feature space and enabling a more detailed classification.

In Chapters~\ref{chapter4}, \ref{chapter5}, and \ref{chapter6}, we will see specific applications of the described pulsar populations.

Furthermore, in Chapter~\ref{chapter7}, we will discuss FRBs from these approaches to offer a new form of separation based on their repetition rate.
\chapter{Graph theory}
\label{chapter2}

\section{Introduction}
\label{chapter2: introduction}

Graph theory has become central to advanced mathematics and data science, offering a flexible framework for modeling relationships between objects. Originating from Euler's solution to the Königsberg bridge problem in the 18th century (see, e.g., \cite{Euler1736, Shields2012}), it has since evolved into a foundational tool across various fields, including computer science, engineering, biology, linguistics, and economics.

\begin{figure}
  \includegraphics[width=0.5\textwidth]{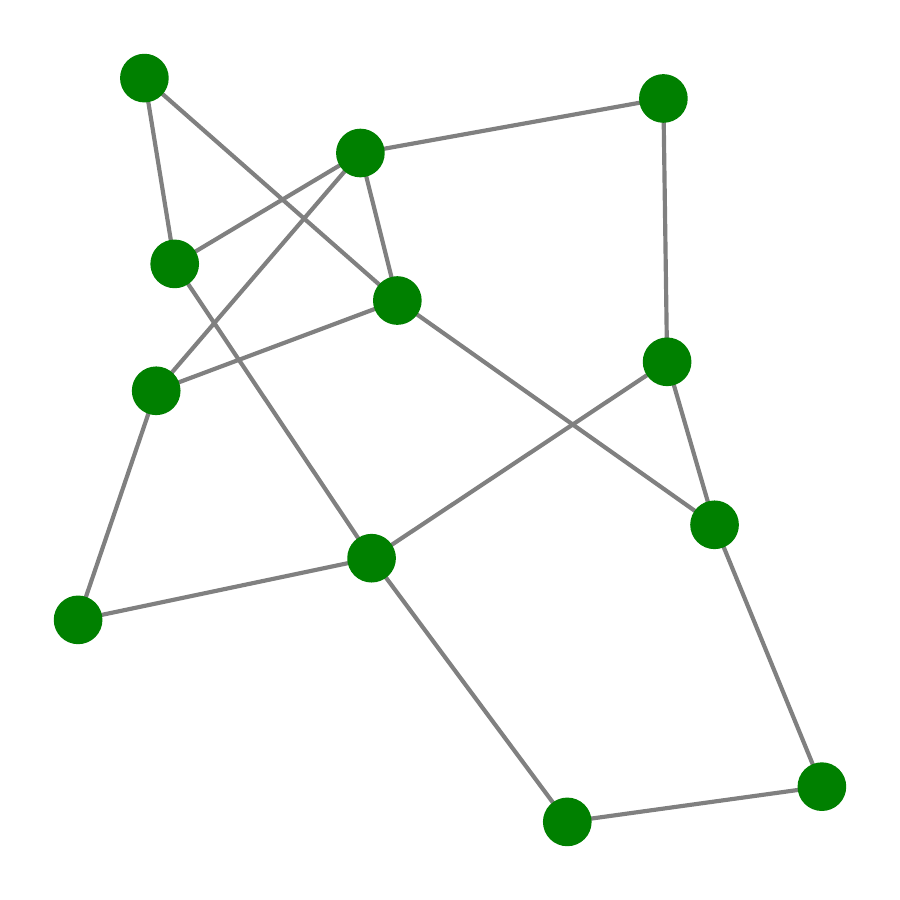}
  \centering
  \caption[Example of a graph.]{
Representation of a graph composed of nodes (green) and edges (black). An edge indicates a direct relationship between two nodes. This relationship can be quantified, for example, by assigning a weight to the corresponding edge, which represents a distance or similarity. 
  }
  \label{chapter_2_fig: graph_theory_example}
\end{figure}

Its applications are wide-ranging: in social networks, it is used to identify influential nodes \citep{Granovetter1973, Rashidi2024}; in biology, to map protein-protein interactions \citep{Ghosh2015}; in finance, to study risk propagation through interconnected asset networks \citep{Smerlak2015}; in transportation, to optimize routes and infrastructure \citep{Henzinger1997, Roughgarden2002}; in neuroscience, it supports the analysis of brain connectivity and cognitive processes \citep{Bullmore2009}; in genomics, it plays a central role in genome sequencing and the reconstruction of DNA sequences \citep{Chikomana2023}. The list is not exhaustive. 
Graph theory is particularly well-suited for analyzing systems of discrete elements with pairwise interactions, as these are often represented as nodes connected by edges, as seen in Figure~\ref{chapter_2_fig: graph_theory_example}.

In particular, the Minimum Spanning Tree (MST) is a useful structure for uncovering relationships in multidimensional data. 

The MST is a graph that connects points in a multidimensional space. Each point (or node) is linked to at least one other by an edge whose length is associated with a given distance, aiming to minimize the total edge weight while forming a connected, acyclic graph.

Unlike methods that rely on Euclidean space, the MST operates without imposing assumptions about the space's global geometry. 
This flexibility allows it to capture complex patterns in the data that might be missed by conventional techniques. 
Moreover, the MST is firmly rooted in graph theory, which provides a mathematically rigorous and conceptually rich framework. 

In addition, graph theory demonstrates that the MST is unique as long as the distances are distinct \citep{Kleinberg2005, Cormen2009, Goodrich_2014}, and its definition intuitively suggests that it is an optimization technique. 

The MST was widely used in engineering problems, starting from its original application developed by Borůvka in 1926, for electricity distribution in Moravia, as seen in \citet{Nesetril2021}. Currently, the MST is employed for analyzing cognitive impairment \citep{Simon2021} to risk in financial markets \citep{Pozzi2013}.
Early applications in scientific problems include describing the interrelationship of species or genetics (see the work of Florik in the 1950s and Edwards in the 1960s, as commented in \cite{Hartigan1981} and \cite{Winther2018}, respectively), disciplines in which this technique is widely exploited. 
This non-exhaustive reference list demonstrates growing interest in utilizing the MST across various fields.

In astronomy, the MST has been used for finding high-energy sources for photon-starved imaging techniques \citep{Campana2013}, establishing differences between cluster and field stars \citep{Sanchez2018}, detecting filaments \citep{Pereyra2020}, galaxy clustering \citep{Barrow1985}, and cosmology \citep{Bonnaire2020, Naidoo2020, Naidoo2022, Naidoo2024}.
This approach has also been implemented to analyze event samples in particle colliders \citep{Rainbolt2017} or cosmic rays \citep{Harari2006}. Despite this interest, as noted, the MST has barely been used for pulsars. To our knowledge, the first related publication is \citep{Maritz2016}, which uses 11 handpicked objects. The goal was to demonstrate that an MST could distinguish between binaries and isolated pulsars using the dispersion measure as a proxy for distance. In a more recent work, the MST has been utilized for classifying pulsars based on their profiles \citep{Vohl2024}.

Graph theory, when applied through the MST, also offers various centrality estimators, such as betweenness and closeness, that serve as tools for structurally analyzing populations of objects.
For example, in \citet{Baron}, the authors present an algorithm that identifies the main trend in a dataset by constructing graphs and exploiting the characteristics of their structures to reveal global sequences, with applications in astronomy, geology, and natural image data.
This chapter introduces the fundamental concepts necessary to build and work with weighted graphs, with a focus on the MST. We aim to establish it as a robust, practical framework for extracting meaningful insights from complex astronomical data. 
In addition, we demonstrate the utility of betweenness and closeness centrality as structural metrics for a deeper understanding of the relative importance of individual nodes within the graph.

\section{Basic definitions for building a weighted graph}
\label{chapter2: basics_graph}

Graph theory seeks to establish relationships among objects based on their connections, ultimately representing these relationships in a graph. 
Let $G(V, E)$ denote a graph of a set of nodes $V$ and a set of edges $E$, where each edge $e = \{v_n, v_m\}$ connects a pair of nodes. To represent the relationship between the nodes, a weight $w(e)$ (e.g., a distance value), is assigned to each edge, making $G$ a weighted graph \citep[see, e.g.,][]{Wilson2010}.

When the edges have no orientation imposed, $e=\{v_n,v_m\}=\{v_m,v_n\}$, then $G$ can be considered an undirected graph.

A possibility is to assume the $w(e)$ as the Euclidean distance calculated as the straight-line distance between $(v_n,v_m)$ in an $N$-dimensional space. This can be calculated as follows:

\begin{eqnarray}
d_{nm}=\sqrt{ \sum_{j=1}^{N} (v_{jn}-v_{jm})^{2} }
~.\label{chapter_2_eq: d_eucl}
\end{eqnarray}

In Equation~(\ref{chapter_2_eq: d_eucl}), $d_{nm}$ represents the $w(e)$, where $v_{jn}$ and $v_{jm}$ denote the $j$-th variable associated with nodes $v_n$ and $v_m$, respectively. A normalization process is typically required when variables have different units or scales.
The total weight of $G$ will be obtained from the sum of each specific $w(e)$ as follows:

\begin{eqnarray}
w(G)=\sum_{e\in E(G)} w(e)
~.\label{chapter_2_eq: weights}
\end{eqnarray}

The resulting value in Equation~(\ref{chapter_2_eq: weights}) is particularly useful in optimization problems where minimizing $w(G)$ is of central interest.

\subsection{General properties of graphs}
\label{chapter2: graph_props}

The following properties are general and applicable to any graph, as seen in \citet{Wilson2010}. The notation introduced in Section~\ref{chapter2: basics_graph} for the specific case of $G$ is retained here.
A \textit{path} in a graph is a sequence of nodes in which no node is repeated. If the path starts and ends at the same node, it is called a \textit{cycle}.
A graph is said to be \textit{connected} if a path exists between every pair of nodes.
The \textit{degree} of a node $v$, denoted $g(v)$, is the number of edges incident to it, that is, the number of adjacent nodes directly connected to $v$.
The overall structure of a graph can also be encoded in its \textit{adjacency matrix} $A$, which has dimensions $|V| \times |V|$\footnote{The notation $|\cdot|$ represents the cardinality (or size) of the specified set, indicating the number of elements within it.}. The element $A_{nm}$ indicates whether nodes $v_n$ and $v_m$ are adjacent. For unweighted graphs, $A_{nm} = 1$ if the edge $e=\{v_n, v_m\}$ exists; for weighted graphs, $A_{nm} = w(e)$. The main diagonal of $A$ is zero when the graph contains no loops (i.e., no node is connected to itself).
Finally, we introduce the notion of a \textit{cut}, particularly relevant in the algorithms discussed in Section~\ref{chapter2: MST_graph}. A cut is a partition of $V$ into two disjoint subsets, $S$ and $V-S$. The associated \textit{cut set} consists of all edges $e = \{v_n, v_m\}$ such that $v_n \in S$ and $v_m \in V-S$.

\section{The minimum spanning tree}
\label{chapter2: MST_graph}

Since $G$ has an edge between every pair of nodes, it is a complete, undirected, and weighted graph. 
Let $T(V', E')$ be a subgraph of $G(V, E)$, where $V'\subseteq V$ and $E'\subseteq E$. We define $T$ as a spanning tree of $G$ if it connects all nodes of $G$, which implies $V'=V$, and contains no cycles. 
The following statements regarding $T$ are equivalent (see e.g., \cite{Tarjan1983})

 \begin{itemize}
     \item $T$ is a spanning tree of $G$.
     \item $T$ is connected and contains no cycles.
     \item $T$ is connected and $E'$ contains $|V|-1$ edges.
     \item $T$ contains no cycles and $E'$ contains $|V|-1$ edges.
     \item $T$ is minimally connected: removing an edge would disconnect it.
     \item $T$ is maximally acyclic: adding an edge would form a cycle.
 \end{itemize}

An MST is a spanning tree $T$ of a graph $G$ for which the sum of all edge weights is the smallest possible, according to Equation~(\ref{chapter_2_eq: weights}).
Thus, the MST connects all nodes in $V$ using locally minimal distances, subject to the global constraint of minimizing the total distance, resulting in a connected and acyclic structure.
The MST is obtained via greedy algorithms, which select the best possible local choice in each iteration to pursue a globally optimal solution (see, e.g., \cite{Roughgarden2019} for more details). 

\subsection{The Kruskal's algorithm}
\label{chapter2: Kruskal}

We employ a specific greedy algorithm, known as Kruskal's algorithm \citep{Kruskal1956}, to compute the MST of $G(V,E,w)$. 
The algorithm constructs an MST by iteratively adding the lowest-weight edge that does not form a cycle. 
Its basic steps are:

\begin{itemize}
  \item Sort the edges in $E$ by increasing $w(e)$.
  \item Initialize an empty set of edges $E'$ and let $T = (V, E')$.
  \item Choose the edge $e$ with the smallest $w(e)$ and add it to $E'$.
  \item Make sure that the chosen $e$ does not produce any cycle in the structure of $T$.
  \item Repeat until $T$ spans all nodes in $V$, such that $|E'| = |V| - 1$ and the total weight $w(T)$ is minimized (see Equation~\ref{chapter_2_eq: weights}).
\end{itemize}

The pseudocode for Kruskal's algorithm, as presented by \citet{Erickson2019}, is shown below.

\begin{algorithm}
\renewcommand{\thealgorithm}{}
\caption{Kruskal($G(V,E,w)$)}\label{pseudo_code_kruskal}
\begin{algorithmic}[1]
\Require{sort $E$ in increasing order according to $w(e)$}
\State $T \gets (V, \varnothing)$
\For{each $v \in V$} 
    \State \textit{MakeSet}($v$)
\EndFor

\For{each $e = \{v_n, v_m\} \in E$ (in order)}
    \If{$\textit{Find}(v_{n}) \neq \textit{Find}(v_{m})$}
        \State \textit{Union}($v_{n},v_{m}$)
        \State $T \gets T \cup e$
    \EndIf
\EndFor
\State \Return $T$
\end{algorithmic}
\end{algorithm}

Furthermore, the implementation of Kruskal's algorithm relies on a disjoint-set data structure (see, e.g., \cite{Cormen2009}), which operates on a collection of disjoint subsets of $V$.
This structure supports three key operations: 

\begin{itemize}
    \item \textit{MakeSet}($v$): Generates a new subset containing only $v$.
    \item \textit{Find}($v$): Operation returns an identifier for the subset containing $v$.
    \item  \textit{Union}($v_n,v_m$): Merges the subsets containing $v_{n}$ and $v_{m}$.
\end{itemize}

\begin{figure}
  \includegraphics[width=1\textwidth]{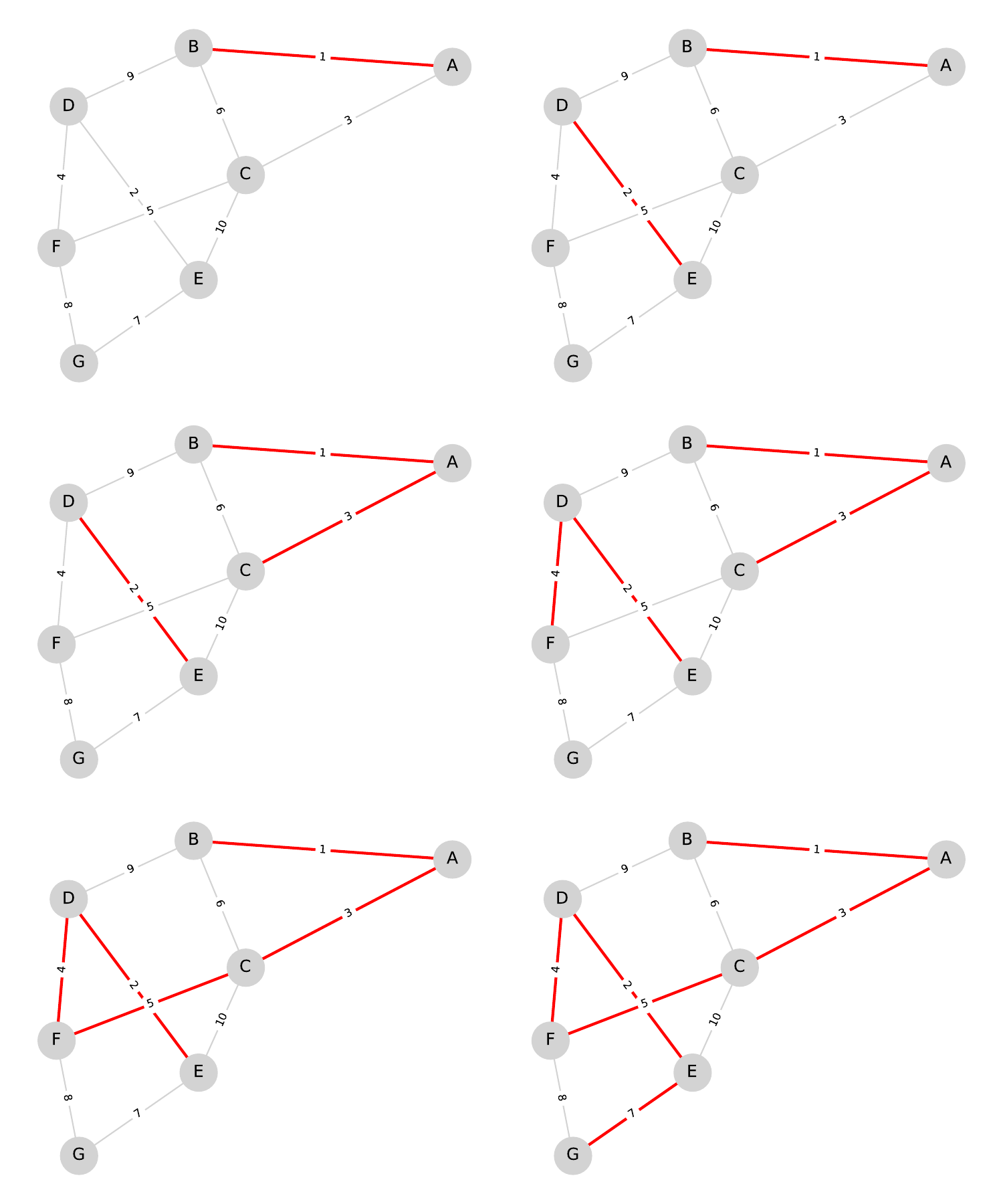}
  \centering
  \caption[Kruskal's algorithm application.]{
From left to right and top to bottom, we illustrate the application of Kruskal's algorithm. Each panel represents an iteration of the algorithm, adding an edge (marked in red) while avoiding the creation of cycles. 
Let $G(7,10)$ be an undirected and weighted graph, whose edges are colored grey, except in the first panel where the $e=\{A, B\}=1$ edge starts in red. 
The last panel displays the resulting MST, $T(7,6)$, with all edges in red. 
  }
  \label{chapter_2_fig: Kruskal_application}
\end{figure}

Figure~\ref{chapter_2_fig: Kruskal_application} illustrates the application of Kruskal's algorithm to a weighted and undirected graph, yielding an MST.
Let $G(7,10)$, creating a $T(7,E')$ with $E'=0$ the edges are included into the $E'$ in increasing order starting from $e=\{A,B\}=1$.
Then, the edges are added to $E'$ to connect all the nodes, avoiding cycles, forming an MST as $T(7,6)$.
Note in the last panel that $e=\{B, C\}=6$ would create a cycle, so it is discarded and the $e=\{E, G\}=7$ is added. 

\subsection{The Prim's algorithm}
\label{chapter2: Prim}

Prim's algorithm \citep{Prim1957} is an alternative method for computing the MST of $G(V, E,w)$ and is equally valid. 
This algorithm grows the tree from an initial node by successively attaching the nearest node not yet in the tree, always selecting the lightest available edge that connects to the existing structure without cycles.
Its basic steps are:
\begin{itemize}
\item Initialize the tree $T$ with an arbitrary starting node $v_{0} \in V$ and an empty set of edges $E'$.
\item Create a priority queue storing all edges of $E$ connecting $T$ to nodes not yet in $T$.
\item Choose the edge $e$ with the smallest $w(e)$ that connects a visited node to an unvisited one and add it to $E'$.
\item Repeat until $T$ spans all nodes in $V$, such that $|E'| = |V| - 1$ and the total weight $w(T)$ is minimized (see Equation~\ref{chapter_2_eq: weights}).
\end{itemize}

The pseudocode for Prim's algorithm presented below is a hybrid adaptation of the versions seen in \citet{Erickson2019, Cormen2009}.

\begin{algorithm}
\renewcommand{\thealgorithm}{}
\caption{Prim($G = (V, E, w), v_0$)}\label{pseudo_code_prim_clrs}
\begin{algorithmic}[1]
\For{each $v \in V$}
    \State $priority(v) \gets \infty$
    \State $parent(v) \gets \text{None}$
\EndFor
\State $priority(v_0) \gets 0$
\State $Q \gets$ PriorityQueue($V$, keyed by $priority$)
\While{$Q \neq \emptyset$}
    \State $v_n \gets \textit{ExtractMin}(Q)$
    \For{each neighbor $v_m$ of $v_n$}
        \State $e \gets \{v_n, v_m\}$
        \If{$v_m \in Q$ and $w(e) < priority(v_m)$}
            \State $parent(v_m) \gets v_n$ 
            \State $priority(v_m) \gets w(e)$ 
            \State $\textit{DecreaseKey}(v_m, w(e))$ 
        \EndIf
    \EndFor
\EndWhile
\State $E' \gets \{\{v, parent(v)\} \mid v \in V \setminus \{v_0\}\}$ 
\State \Return $T = (V, E')$ 
\end{algorithmic}
\end{algorithm}

This implementation of Prim's algorithm is based on the following standard definitions and operations as described in \citet{Cormen2009}:

\begin{itemize}

\item \textit{priority}($v$): Represents the minimum weight of any edge connecting $v$ to the growing MST.
\item \textit{parent}($v$): Stores the node through which $v$ will be attached to the final MST.
\item \textit{PriorityQueue}: A data structure that maintains the set of candidate nodes ordered by their priority values.
\item \textit{ExtractMin}: Selects and removes the node with the lowest priority from the queue, corresponding to the next node added to the MST.
\item \textit{DecreaseKey}: Updates the priority of a node in the queue when a lighter connection is found, preserving the correct ordering.

\end{itemize}

\begin{figure}
  \includegraphics[width=1\textwidth]{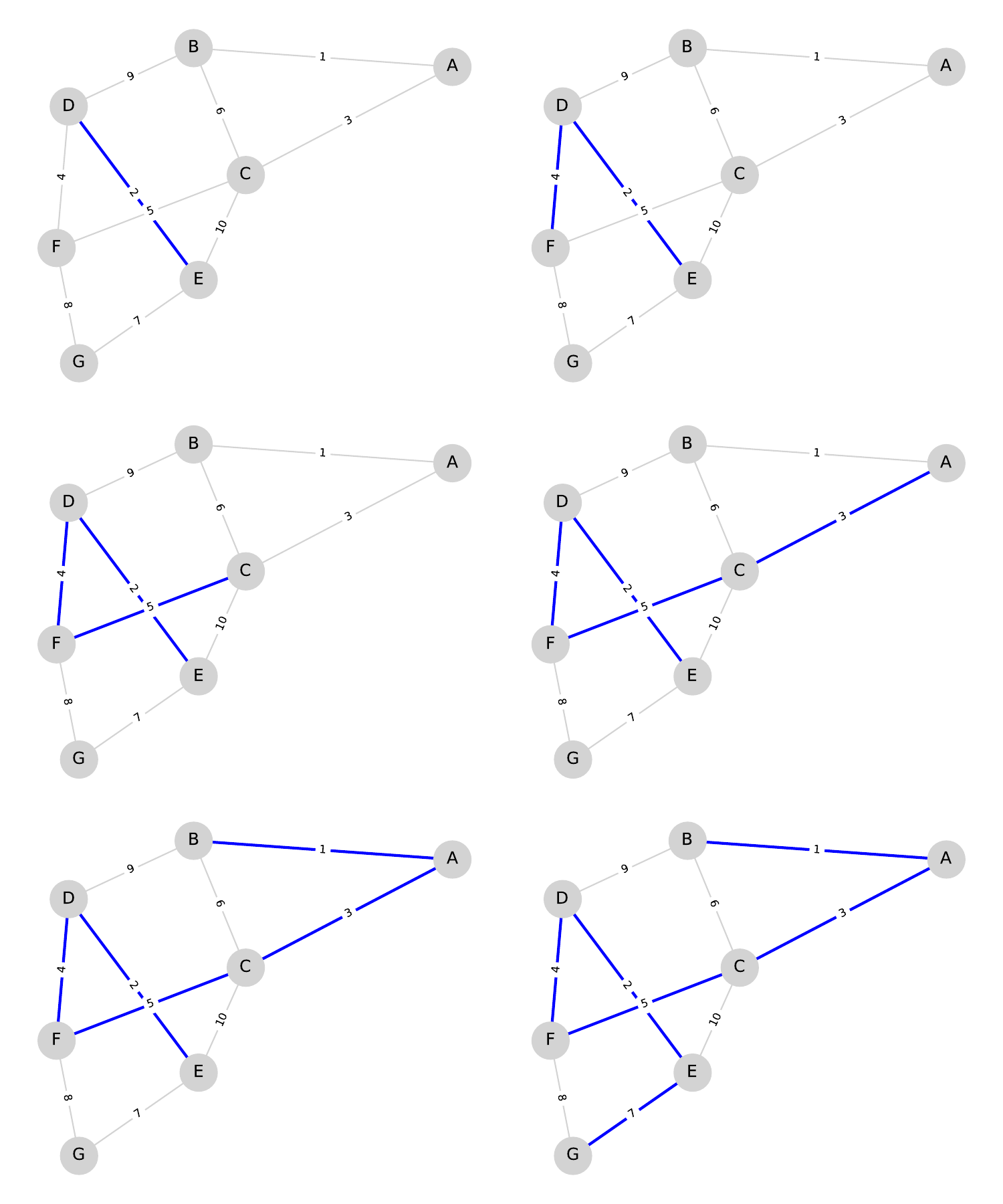}
  \centering
  \caption[Prim's algorithm application.]{
From left to right and top to bottom, we illustrate the application of Prim's algorithm. Each panel represents an iteration of the algorithm, adding an edge (marked in blue) while avoiding the creation of cycles. 
Let $G(7,10)$ be an undirected and weighted graph, whose edges are colored grey, except in the first panel where the $e=\{E, D\}=2$ edge starts in blue. 
The last panel displays the resulting MST, $T(7,6)$, with all edges in blue. 
  }
  \label{chapter_2_fig: Prim_application}
\end{figure}

Figure~\ref{chapter_2_fig: Prim_application} shows the application of Prim's algorithm on a weighted and undirected graph, resulting in an MST.
Let $G(7,10)$, and $v_0=E$ (this selection is completely random), create a $T(v_0,E')$ with $E'=0$. We create a priority queue of the edges connected to $E$, 
$e=\{E,C\}=10$,
$e=\{E,D\}=2$, and 
$e=\{E,G\}=7$, from where the smallest one is selected, $e=\{E,D\}=2$, and the node $D$ is added to $T$. This step is repeated until all the nodes are included in $V$, without creating any cycle, forming an MST as $T(7,6)$.
Note in the last panel that $e=\{B, C\}=6$ is on the priority queue, so $B$ and $C$ are already part of $T$, but would create a cycle, so it is discarded and the $e=\{E, G\}=7$ is added.

\subsection{MST properties}
\label{chapter2: MST_properties}

\begin{figure}
  \includegraphics[width=1\textwidth]{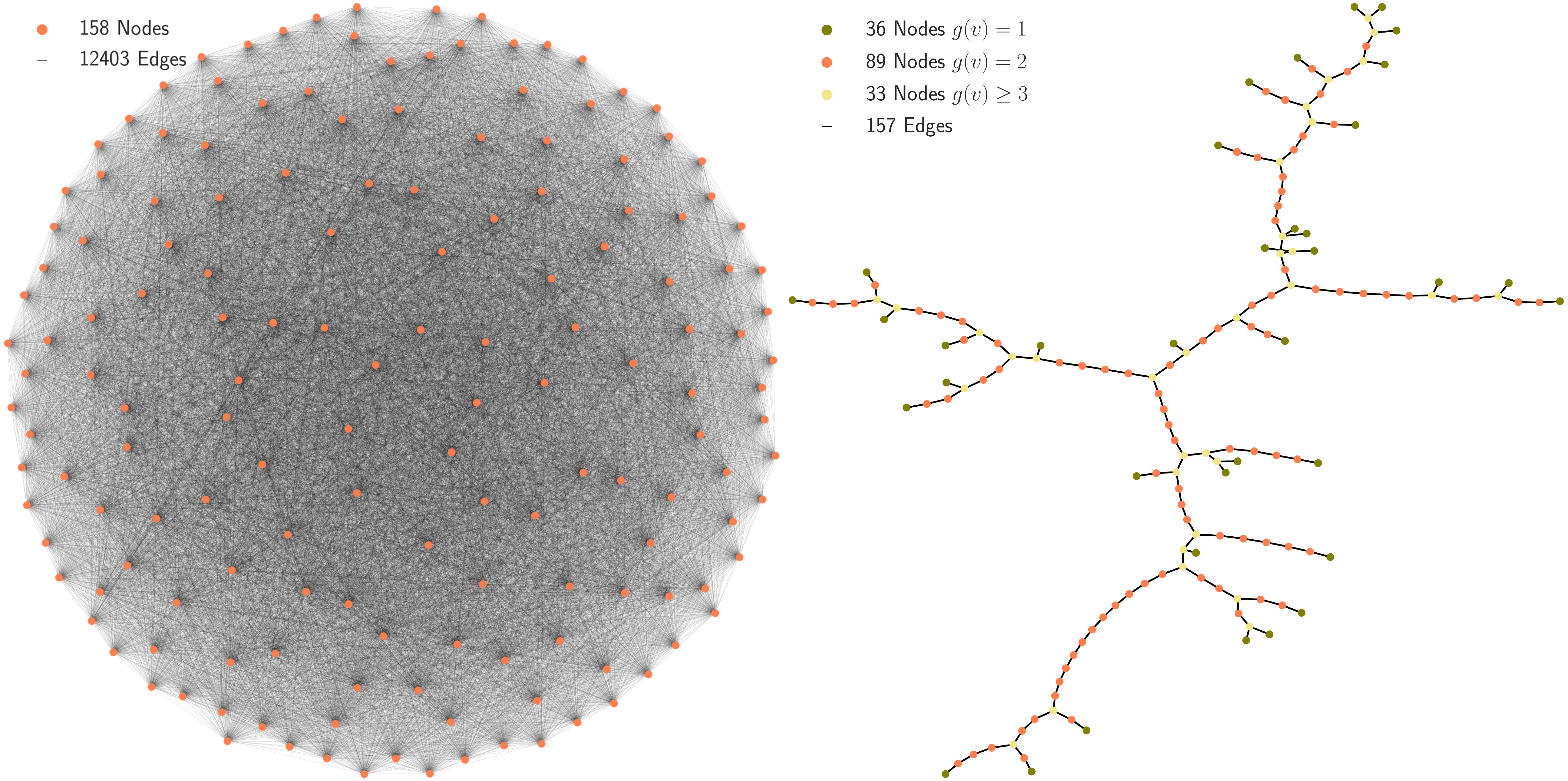}
  \centering
  \caption[Complete, undirected and weighted graph, and its MST.]{Left: Complete, undirected and weighted graph $G(158,12403)$. 
  Right: Corresponding MST, $T(158, 157)$, from the defined $G$.
  The nodes are marked according to the degree: $g(v)=1$ --olive, $g(v)=2$ --coral, and $g(v)\geq 3$ --yellow.}
  \label{chapter_2_fig: figure11}
\end{figure}

Below, we list key properties associated with the MST; see, for example, \citet{Kleinberg2005, Goodrich_2014}, where proofs can also be found.
Considering that $T(V,E')$ is the MST of $G(V,E)$, the principal properties are:

\begin{itemize}

\item Uniqueness:
The resulting MST will be unique if all $w(e)$ values in $E(G)$ are distinct. This implies that choosing any other greedy algorithm instead of Kruskal's as a solution to search for the MST (e.g., Prim's algorithm) would give the same results. This is the case, for instance, seen in Figure~\ref{chapter_2_fig: figure11}.

\item Cycle property:
From the construction of $T$ itself, any edge $e\in E$ such that $E' \cup \{e\}$ creates a cycle $C$ in $T$. On the other hand, if for some $e_{c}\in C$ it is determined that T($V, E'\cup\{e\}-\{e_{c}\}$) is a spanning tree. Thus, if $T$ is an MST, then $w(e_{c})>w(e)$. Otherwise, the edge with the largest weight in $C$ will not be contained in $E'$ for $T$ to remain an MST. Therefore, every node $v\in T$ has among its incident edges the $e$ with the smallest value $w(e)$. 

\item Cut property:
From the construction of $T$ itself, any edge $e\in E'$ such that $T(V, E'-\{e\})$ will cause a cut in $T$ separating it into two connected components. For the resulting cut set, denoted as $D$, it is observed that for any $e_{D}\in D$ such that $T(V, E'-\{e\}\cup\{e_{D}\})$ is a spanning tree. Thus, if $T$ is an MST, then $w(e_{D})>w(e)$; otherwise, the lightest edge contained in $D$ must be included in $E'$ for $T$ to remain an MST.

\end{itemize}

Figure~\ref{chapter_2_fig: figure11} shows an example to exemplify the structures of $G$ and its derived MST $T$, together with the complexity of working with a high number of nodes $|V|$.
Due to the uniqueness property described above, $T$ is a unique MST because all weights of $G$ are distinct. 
Therefore, this result is achieved by applying both the Kruskal and Prim's algorithms.
A similar result is seen for the examples in Figures \ref{chapter_2_fig: Kruskal_application} and \ref{chapter_2_fig: Prim_application}.

\section{Centrality estimators in graphs}

In graph theory, centrality measures quantify the importance of individual nodes within a graph's structure. 
These estimators capture how well-positioned a node is to control flow, connect different regions, or act as a structural bridge. 
This section focuses on two centrality measures relevant to the MST: betweenness centrality and closeness centrality. 
Although betweenness centrality was ultimately chosen for our analysis, closeness centrality is presented as a complementary reference to illustrate the methodological contrast and clarify the reasoning behind our selection; this discussion is included in Section~\ref{chapter2: appl_astro}. We provide a step-by-step example of their application in Appendix~\ref{Appendix_2} to facilitate understanding.

\subsection{Betweenness centrality estimator}
\label{chapter2: bet_cen}

Betweenness centrality is an estimator that defines how central a node is, or, put otherwise, how many times a given node in a graph is situated between any two others \citep{original, Moxley1974}.

Let $G(V, E, w)$, with $v_s \in V$, $v_t \in V$ and $v_j \in V$, this estimator measures centrality from the ratio between the number of times a $v_j$ appears on the shortest path between any two other nodes $(v_s, v_t)$, and the number of possible shortest paths that could occur between these.
This can be seen via (see, e.g., \cite{Brandes}):

\begin{eqnarray}
C_{B}(v_j) = \frac{2}{(N-1)(N-2)}\sum_{s\neq j\neq t \in V} \frac{\sigma_{s,t}(v_j)}{\sigma_{s,t}}
~.\label{chapter_2_eq: betweenness_centrality}
\end{eqnarray}

Here, $\sigma_{s,t}(v_j)$ is assigned to one when any shortest path between the nodes $(v_s,v_t)$ passes through the node $v_j$, or zero if it does not.  
Note that in unweighted graphs, shortest paths are defined by the minimal number of edges; in weighted graphs, they are determined by the paths with minimal total edge weight.
However, as explained in Section~\ref{chapter2: MST_graph}, the MST is acyclic, so only one path is possible between any two nodes when we work on $T$.
This means that for all $s\neq t$, $\sigma_{s,t} =1$; as such, the calculation of $C_B(v_j)$ in an MST becomes independent of $w(e)$ and reflects only the topological position of the node within the tree.
Let $G$ (or $T$ in case of the MSTs) is undirected, $\sigma_{s,t} = \sigma_{t,s}$ will be taken into account only once in Equation~(\ref{chapter_2_eq: betweenness_centrality}), which will be normalized by multiplying it by the factor $2/(N-1)(N-2)$ where $N=|V|$. 
This allows us to compare the results obtained between graphs of different $|V|$. 
Betweenness centrality helps formalize an intuitive mental concept of how central a node is in a given graph, allowing us to establish mathematical definitions based on its distribution.
An explicit example in Appendix \ref{Appendix_2: practical_example} with an MST and the graph from which we obtain that MST of a few nodes will help clarify the computation.

\subsection{Closeness centrality estimator}
\label{chapter2: clo_cen}

Closeness centrality measures the proximity of a node to all other nodes in a graph, based on the average length of the shortest paths connecting it to the rest \citep{Sabidussi_1966, original}. 
Using the same notation and assumptions introduced in Section~\ref{chapter2: bet_cen}, this estimator quantifies the proximity of $v_j$ to every other $v_t$ by computing the inverse of the shortest path distance $d_G(v_j,v_t)$. This is defined as the sum $w(e)$ along the path from $v_j$ to $v_t$ (with $d_G(v_j,v_t) = d_G(v_t,v_j)$ by definition). 
The closeness centrality of $v_j$ is given by (see, e.g., \cite{Brandes}):

\begin{eqnarray}
C_{C}(v_j) = (N-1)\sum_{j\neq t \in V} \frac{1}{d_{G}(v_j,v_t)}
~.\label{chapter_2_eq: closeness_centrality}
\end{eqnarray}

The factor $(N-1)$ in Equation~(\ref{chapter_2_eq: closeness_centrality}) ensures that the measure is normalized and comparable across graphs of different $|V|$. 

As discussed below, this estimator relies on spatial distance in the MST approach, since this graph is weighted. 
This dependence makes it unappealing when searching for a quantifier to highlight the different structures observed in an MST from a topological perspective, such as identifying the tree's main trunk.

An example of the calculation of this estimator can be found in Appendix \ref{Appendix_2: practical_example2}.

\section{Conclusions}
\label{chapter2: appl_astro}

This chapter establishes the theoretical foundations and practical utility of graph theory, focusing on the MST to represent multidimensional data structures. The MST offers a novel visualization of complex relationships between objects in a straightforward and intuitive manner. Subject to the framework of graph theory, the MST allows us to explore different populations by leveraging robust mathematical tools. 

Its simplicity and robustness allow both interpretability and flexibility for various scientific applications.
In the broader context of this thesis, the MST framework plays a central role in the exploration and interpretation of pulsar populations and fast radio bursts. 

In this work, we selected Kruskal's algorithm for computing the MSTs primarily over Prim's algorithm because of its ease of implementation and conceptual alignment with the problem setup. 
Since the complete set of pairwise distances between nodes was available in advance, Kruskal's edge-sorting strategy offered a direct and efficient means to construct the MST, without requiring dynamic key updates characteristic of Prim's approach.
Although the latter can scale well in sparse and dense settings, depending on the data structures used, Kruskal's clarity and edge-centric design made it better suited to our study's conditions and assumptions.

Regarding the applications, as shown in Chapter~\ref{chapter4}, the MST can be used as a descriptive and alerting tool to visualize the pulsar population and their relationships in high-dimensional space, revealing patterns that are not easily captured by traditional representations, such as the $P\dot{P}$ diagram. 

One can also utilize the MST to define possible subpopulations of pulsars that are distinguishable from each other by considering their physical properties through purely mathematical tools, such as centrality estimators. See cases of this approach in Chapter~\ref{chapter5} and Chapter~\ref{chapter7}.

Specifically, we adopt betweenness centrality to quantify the relative structural importance of nodes within the MST. 
Rather than using closeness centrality, this choice is motivated by the desire for a measure less directly affected by the geometry of the feature space from which the edge weights are derived.
Closeness centrality, computed over these weights as we see in Equation~(\ref{chapter_2_eq: closeness_centrality}), is inherently sensitive to the distribution of distances in that space. 
In contrast, betweenness centrality, under the acyclic attribute of the MST, reflects how often a node serves as a bridge along shortest paths between other nodes, independent of the problem geometry, providing a more topologically robust and intuitive representation of what visually appears to be the tree's main trunk.

On the other hand, it is worth noting the MST's predictive and inferential potential described in Chapter~\ref{chapter6}. 
It helps us characterize millisecond pulsars and promotes possible candidates for spider pulsars based on the structural position. 
Similarly, it enables us to suggest plausible ranges for undetermined parameters in some sources, considering their proximity in the MST to better-known neighbors. 
Finally, Chapter~\ref{chapter7} presents MST as an unsupervised methodology that separates repeaters from non-repeaters of fast radio bursts, helping to identify repeater candidates and the most relevant variables for their separation. 
These applications underscore the value of MST as a visualization technique and framework for exploratory, descriptive, prescriptive, and even predictive analysis. 
Its integration into astrophysical problems provides a fresh, robust, and interpretable alternative to usual classification and clustering approaches.

\chapter{Principal Component Analysis}
\label{chapter3}

\section{Introduction}
\label{chapter3: introduction}

Principal Component Analysis (PCA) is a widely used technique for reducing the dimensionality of complex datasets while preserving as much information as possible, while also retaining the importance of each original variable. Comprehensive discussions can be found in \citet{Mardia1979, Jolliffe2002, Izenman2008, Rencher2012}; see also the reviews by \citet{Shlens2014, Jolliffe2016}.

Initially introduced by \citet{Pearson1901} in the context of the principal axes of ellipsoids in geometry, the method was independently developed by \citet{Hotelling1933}, known as the "Hotelling transform," for statistical analysis. It is closely related to the Kosambi-Karhunen-Loève transform \citep{Kosambi1943, Karhunen1947, Loeve1948}, and belongs to a family of classical dimensionality reduction techniques, alongside Independent Component Analysis \citep{Herault1984, AnsHerault1985, Herault1985, Comon1994}, and Non-negative Matrix Factorization \citep{Lee1999, Lee2000}.

In astronomy, PCA has proven especially valuable for extracting the dominant sources of variation in high-dimensional datasets \citep{WallJenkins2003, Ivezic2014}. 
Early applications in astronomy (see, e.g., \cite{Francis1999}) include spectral analyses of stars \citet{Deeming1964, Whitney1983}, galaxies \citep{Faber1973, Bujarrabal1981, Efstathiou1984}, and quasars \citep{Mittaz1990, Francis1992, Boroson1992}. 
It has also been applied to imaging studies of the interstellar medium \citep{Heyer_1997, Brunt2009} and to the analysis of X-ray binaries (see, e.g., \cite{Malzac2006, Koljonen2013, Koljonen2015}), blazars \citep{Gallant2018}, and symbiotic stars \citep{Danehkar2014a}. 
A notable area of use is the study of X-ray variability in Active Galactic Nuclei (AGN; see, e.g., \cite{Vaughan2004, Miller2008, Parker2014a, Parker2014b, Gallo2015}).
A broader overview of PCA applications, including those in AGN studies, is given by \citet{Danehkar_review}.

Beyond astrophysical research, PCA is a fundamental tool in machine learning for data preprocessing, visualization, or unsupervised classification (e.g., \cite{Bishop2007, Raschka2015, Muller2016, Witten2017, Geron2019}).

This Chapter introduces PCA as a complementary technique to the framework discussed in Chapter~\ref{chapter2}. 
While MST captures the topological relationships within the data, PCA helps identify the dominant physical variables, enabling a more comprehensive interpretation of the problem.

\section{Understanding PCA} 

The core objective of PCA is to reduce the number of variables in a dataset while preserving as much of its variance as possible. 
Variance quantifies the extent to which data spreads around its mean; it's calculated as the average of the squared deviations from the mean and acts as a proxy for the information content in the dataset.
Keep in mind that variance is not uniformly distributed across variables; some directions in the feature space exhibit more variability than others.
PCA also aims to clarify the contribution of each original variable to the variance. To understand how PCA accomplishes this, it is helpful to approach it from two complementary perspectives:

\begin{itemize}
\item A geometric perspective explains PCA as a projection and rotation in a new coordinate system.
\item A mathematical perspective shows how this projection is derived from an optimization problem.
\end{itemize}

\subsection{Geometric Interpretation of PCA}
\label{chapter3: geometric_interpretation}

\begin{figure}
    \centering
    \includegraphics[width=0.9\linewidth]{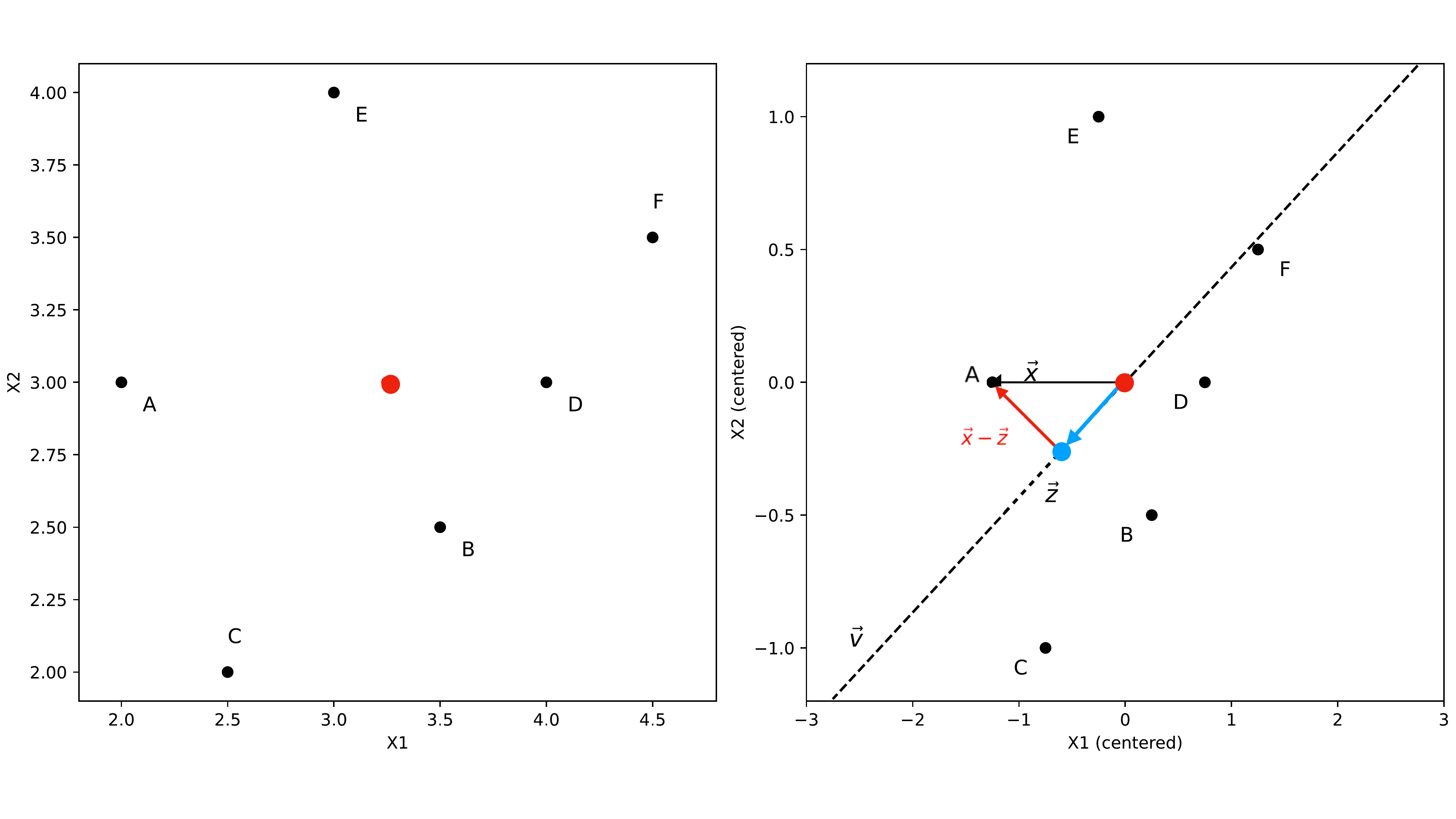}
    \caption[Geometry behind PCA.]{Left panel. Data points in a feature space of two variables, x1 and x2. The red dot represents the mean of the data. 
    Right panel. Data points centered (mean subtracted) on the red dot (origin), illustrating the PCA geometry considering Equation~(\ref{chapter_3_PCA: Pythagorean_th}). The data point A represented by $\vec{x}$ is projected onto the direction of maximum variance $\vec{v}$, resulting in the projection $\vec{z}$. The squared distance from $\vec{z}$ to the origin is the variance contribution of that point in $\vec{v}$; the distance to the axis is the error $\vec{x}-\vec{z}$.}
    \label{chapter_3_PCA_fig: pca_projection}
\end{figure}

PCA finds new axes (principal components, PCs) for the data space such that the first axis ($PC_1$) points in the direction of maximum variance. Each subsequent axis ($PC_2$, $PC_3$, \dots) is orthogonal to the previous ones and captures the following highest remaining variance.
The maximum number of PCs that can be obtained is limited by the number of variables. 
Figure \ref{chapter_3_PCA_fig: pca_projection} helps to explain why PCA seeks to maximize variance: it ensures the projected data best preserves the structure of the original dataset.
The process consists of projecting the data points onto these new axes. 
To preserve the structure of the original data set as much as possible, the distance from the projected point to the original point should be minimized.
This quantity is referred to as the reconstruction error (hereafter error). 
Therefore, this new axis, denoted by the unit vector $\vec{v}$, will minimize the sum of all squared errors across all data points. 
On the other hand, the variance along $\vec{v}$ is equal to the average distance squared of the projection of the points to the center of the new axis. 
This can be interpreted using the Pythagorean theorem:

\begin{equation}
\|\vec{x}\|^2 = \|\vec{z}\|^2 + \|\vec{x} - \vec{z}\|^2,
\label{chapter_3_PCA: Pythagorean_th}
\end{equation}

where $\vec{x}$ is an original data point, $\vec{z}= (\vec{v}^T \vec{x}) \vec{v}$ is its orthogonal projection onto $\vec{v}$, and $\vec{x}-\vec{z}$ is the error. 
Since $\vec{x}$ is constant, minimizing $\|\vec{x}-\vec{z}\|^{2}$ is equivalent to maximizing $\|\vec{z}\|^{2}$, which corresponds to maximizing the variance along the direction along $\vec{v}$.
In other words, the new axis, which captures the maximum variance in the data, is analogous to the PC; it is the axis that least deforms the original data.

\subsection{Mathematical Interpretation of PCA}
\label{chapter3: mathematical_interpretation}

The variance of a projection in direction $\vec{v}$, which is a unit vector, is given by:

\begin{equation}
\text{Var}(\vec{v}) 
= \frac{1}{n} \sum_{i=1}^{n} \|\vec{z}_{i} \|^{2} 
= \frac{1}{n} \sum_{i=1}^{n} (\vec{v}^{T}\vec{x}_{i})^{2} 
= \frac{1}{n} \| \tilde{X} \vec{v} \|^{2} 
= \frac{1}{n} \vec{v}^{T} \tilde{X}^{T}\tilde{X} \vec{v} 
= \vec{v}^{T} C \vec{v}~.
\label{chapter_3_eq: var}
\end{equation}

Formally, $\tilde{X}\in \mathbb{R}^{n\times N}$ is the mean-centered data matrix, where $n$ represents the number of data points, rows in $\tilde{X}$, and $N$ represents the variables of the data, columns in $\tilde{X}$. Note $ \| \vec{z}_i \|^2 = (\vec{v}^T \vec{x}_i)^2 $ represents the variance contribution of data point $\vec{x}_i$ in $\vec{v}$.
In addition, the covariance matrix ($C$) of the dataset is defined as:

\begin{equation}
C = \frac{1}{n-1} \tilde{X}^T \tilde{X}~.
\label{chapter_3_eq: cov_matrix_EVD}
\end{equation}

As a note, while PCA is traditionally applied to mean-centered data, it can also be performed using median-centered variables. 
The chosen type of normalization will manage this situation; for instance, data standardization involves subtracting the mean and then dividing by the standard deviation of the data, whereas scalarization consists of subtracting the median and dividing by the range between the first and third quartiles.
This approach is often more robust when the data distribution is non-Gaussian or has skewed tails, as the presence of outliers can affect the mean as a proper measure of central tendency (see, e.g., \cite{RobustScaler} or the following Chapters \ref{chapter4}, \ref{chapter6}, and \ref{chapter7}, for an idea about this discussion in practical cases).
Regardless of the above, the directions of the PCs are found by solving the following constrained quadratic optimization problem, considering the maximization of Equation~(\ref{chapter_3_eq: var}):

\begin{equation}
\max_{\vec{v}} \quad \vec{v}^T C \vec{v} \quad \text{subject to} \quad ||\vec{v}|| = 1~.
\label{chapter_3_eq: var_max}
\end{equation}

Introducing a Lagrange multiplier $\lambda$ in Equation~(\ref{chapter_3_eq: var_max}), we define:

\begin{equation}
\mathcal{L}(\vec{v}, \lambda) = \vec{v}^T C \vec{v} - \lambda (\vec{v}^T \vec{v} - 1)~.
\label{chapter_3_eq: var_max_lagrange}
\end{equation}

Taking the derivative concerning Equation~(\ref{chapter_3_eq: var_max_lagrange}) and setting it to zero leads to the eigenvalue decomposition (EVD, see Section \ref{chapter3: methods}):

\begin{equation}
\nabla_{\vec{v}} \mathcal{L} = 2C\vec{v}-2\lambda=0 \implies C\vec{v} = \lambda \vec{v}~.
\label{chapter_3_eq: EVD_derived}
\end{equation}

Equation~(\ref{chapter_3_eq: EVD_derived}) shows that the PCs are the eigenvectors $\vec{v}$ of $C$ defined by Equation~(\ref{chapter_3_eq: cov_matrix_EVD}). At the same time, the amount of variance captured by each $\vec{v}$ is given by the corresponding eigenvalue $\lambda$.

The transformation of the dataset seen through $\tilde{X}$ into the PC plane is:

\begin{equation}
PC_m = \vec{v}_m \tilde{X}^T~,
\label{chapter_3_eq: PC_space}
\end{equation}

where each resulting PC is a linear combination of the original variables. 
The result of this projection, the PC scores, represents the coordinates of the original data points in the new basis defined by the PCs.
Moreover, as shown in Equation~(\ref{chapter_3_eq: PC_space}), the PCs are uncorrelated because they are obtained by projecting the dataset onto the $\vec{v}_{\mathrm{s}}$ of $C$, which are orthogonal by construction.
In addition to measuring the variance that each PC retains, we define the explained variance:

\begin{equation}
\text{Explained Variance}\text{ }(PC_m) = \frac{\lambda_m}{\sum_{j=1}^{N} \lambda_j}~.
\label{chapter_3_eq: explained_variance}
\end{equation}

Through the Equation~(\ref{chapter_3_eq: explained_variance}), we can establish an order of importance of the PCs by quantifying the dimensionality reduction.
\\
\\

It is essential to note that PCA fundamentally assumes linearity in the data, as outlined in Equation~(\ref{chapter_3_eq: PC_space}). Specifically, it identifies directions that are linear combinations of the original variables and maximizes variance along those directions. This assumption works well when the variables are linearly correlated, but it may be limiting when relationships are nonlinear. 
For instance, in the case of pulsars (see Chapter~\ref{chapter4}), variables derived from the dipole model, such as the characteristic age, magnetic field at the surface, and spin-down luminosity, are nonlinear functions of the spin period and its spin period derivative. 
However, logarithmic transformations make these relationships linear.
On the other hand, PCA may not fully capture the data's variance when non-linearity is present, even if statistical dependencies exist. 
Nevertheless, at that point, PCA can be employed primarily as an exploratory method to uncover patterns, reduce dimensionality, and assess the relative contributions of variables (see, e.g., Chapters \ref{chapter6} and \ref{chapter7}).

\subsection{Visualizing Principal Components}
\label{chapter3: visual_tools}

Once the PCs have been computed, several standard plots are used to visualize and interpret their usefulness and relevance. 
These visualizations are crucial for understanding how many PCs to retain and how the original variables contribute to each PC. The most common visual tools include:

\paragraph{Explained Variance Plot (Scree Plot).} This plot displays the variance each PC explains according to Equation~(\ref{chapter_3_eq: explained_variance}). 
The $\lambda_{\mathrm{s}}$ (or the squared singular values $\sigma^{2}_{\mathrm{s}}$, see the following section for further details) are plotted in descending order. 
It helps identify how many PCs account for most of the variance. 

\paragraph{Cumulative Explained Variance.} The cumulative sum of explained variance seen in Equation~(\ref{chapter_3_eq: explained_variance}) is plotted as a function of the number of PCs. 
This plot illustrates the number of PCs required to retain a specified percentage of the total variance in the data.

\paragraph{Variable Contribution Heatmap.} A heatmap of the PCA application corresponds to the coefficients (or "weights") of the original variables, also known as loadings, which indicate the extent to which each original variable contributes to each PC. Each row represents a variable, and each column represents a PC. This visualization highlights the variables that contribute most to each PC. 
\paragraph{Plane of Principal Components (PC plane).} A PC plane is a scatter plot that displays the PC scores in the space defined by the selected PCs. 
\\
\\
As examples of these visual tools, see Figures \ref{chapter_4_figure: figure2}, \ref{chapter_4_figure: figure3}, \ref{chapter_6_figures: PCA_10v}, and \ref{chapter_7_figures: PCA}, generated on real data of pulsars, millisecond pulsars, and fast radio bursts, respectively.

\section{Methods for Principal Components Computation}
\label{chapter3: methods}

The optimization approach seen in Equation~(\ref{chapter_3_eq: EVD_derived}) leads to an eigenvalue problem whose solution yields the PCs and their associated variances, denoted by $\vec{v_{\mathrm{s}}}$ and $\lambda_{\mathrm{s}}$, respectively.
However, different mathematical strategies can be used to apply the PCA. The most commonly used methods are the EVD and the Singular Value Decomposition (SVD), which are mathematically equivalent in their outcomes but differ in their computational approaches.
In Appendix~\ref{Appendix_3}, we show the application of both methods for the same example.

\subsection{Eigenvalue Decomposition}
\label{chapter3: EVD}

This method seen in Equation~(\ref{chapter_3_eq: EVD_derived}) is applied directly to $C$ considering Equation~(\ref{chapter_3_eq: cov_matrix_EVD}), giving rise to Equation~(\ref{chapter_3_eq: PC_space}).
This approach is conceptually straightforward and well-suited for datasets with a relatively small $N$. However, for high-dimensional data where $N$ is large or comparable to $n$, working directly with $C$ can make this method computationally intensive.
Alternatives, such as SVD, which do not strictly deal with $C$ as shown below, are preferred for computational efficiency, numerical precision, and stability.

\subsection{Singular Value Decomposition}
\label{chapter3: SVD}

The SVD method computes the PCs directly from $\tilde{X} \in \mathbb{R}^{n \times N}$, factorizing it as:

\begin{equation}
\tilde{X} = U \Sigma V^T~.
\label{chapter_3_eq: SVD_methodology}
\end{equation}

Equation~(\ref{chapter_3_eq: SVD_methodology}) shows $U \in \mathbb{R}^{n \times n}$ that contains the left singular vectors, $\Sigma \in \mathbb{R}^{n \times N}$ which is a diagonal matrix with non-negative singular values $\sigma_1 \ge \sigma_2 \ge \dots \ge \sigma_N \ge 0$, and $V \in \mathbb{R}^{N \times N}$ that contains the right singular vectors. 
The columns of $V$ correspond to the $\vec{v}_{\mathbf{s}}$ of $C$. 
To see how this connects with the PCA, consider Equation~(\ref{chapter_3_eq: cov_matrix_EVD}), where introducing Equation~(\ref{chapter_3_eq: SVD_methodology}), and since $U^{T}U= I$ due to $U$ is orthogonal, we get:

\begin{equation}
\tilde{X}^T \tilde{X} = V \Sigma^T \Sigma V^T \quad \Rightarrow \quad C = \frac{1}{n-1} \tilde{X}^T \tilde{X} = V \left( \frac{\Sigma^T \Sigma}{n-1} \right) V^T~,
\label{chapter_3_eq: cov_SVD_methodology}
\end{equation}

which is the eigendecomposition of $C$. 
Thus, the $\lambda_{\mathrm{s}}$ of $C$, considering $\left( \frac{\Sigma^T \Sigma}{n-1} \right)$ in Equation~(\ref{chapter_3_eq: cov_SVD_methodology}), can be obtained as:

\begin{equation}
\lambda_m = \frac{\sigma_m^2}{n-1}~.
\label{chapter_3_eq: lambdas_SVD_methodology}
\end{equation}

Finally, the left singular vectors $U$ can be computed multiplying both sides of Equation~(\ref{chapter_3_eq: SVD_methodology}) by $V$, taking $V^{T}V=I$ by orthogonality, obtaining:

\begin{equation}
\tilde{X}V = U \Sigma 
~.
\label{chapter_3_eq: PCs_SVD_methodology}
\end{equation}

Equation~(\ref{chapter_3_eq: PCs_SVD_methodology}) corresponds to projecting the original data onto the $\vec{v}_{\mathrm{s}}$, where each row of $U \Sigma$ corresponds to the coordinates of a data point in the PC plane.

\section{Conclusions}

PCA offers a rigorous and intuitive framework for reducing dimensionality and analyzing the internal structure of multivariate data.
The following chapters apply PCA to datasets that include physical and observational parameters from various astrophysical sources, such as pulsars, millisecond pulsars, and fast radio bursts.
In Chapter \ref{chapter4}, we apply PCA to the pulsar population to unravel whether the intrinsic properties, spin period, and spin period derivative, are sufficient to determine how they relate. Additionally, we observe a significant reduction in the number of necessary variables (now, just two PCs) to collect the whole variance without losing information, even when the population is defined by up to eight variables, enabling manageable visualization.
This feature space is extended to define the population of millisecond pulsars in Chapter \ref{chapter6}. We can observe the contributions of binary parameters versus intrinsic properties through PCA, which helps in analyzing clustering methods.
In Chapter \ref{chapter7}, we utilize the PCs derived from the PCA application over the population of fast radio bursts, described by both observed and derived properties, as an alternative to these. Using population separation methods, we investigate whether PCs can reveal hidden patterns that distinguish classes of fast radio bursts.

\chapter{Visualizing the pulsar population using graph theory}
\label{chapter4}

\section{Introduction}
\label{chapter4: intro}

Ever since the discovery of the first pulsar \citep{CP1919}, the $P\dot{P}$ diagram has been used to summarize our knowledge and guide our research on the pulsar population. 
Classes of pulsars and possible links among them are referred to in this diagram, as it is what we know about their possible evolutionary tracks along the pulsar's lifetime (see, e.g., \cite{Enoto2019} for a review). 
However, we have looked at the same diagram for over five decades, so a fresh appraisal may be healthy. 
Is the $P\dot{P}$ diagram the most practical or complete way to visualize the pulsars we know? 
Does it introduce any unwarranted bias in what we consider to be similar pulsars? 

In this chapter, building on the work presented in \citet{MST-1}, we apply PCA (see Chapter~\ref{chapter3}) to show that even when all the variables of the rotating dipole model used to describe a pulsar depend on $P$ and $\dot{P}$, the population's variance is not fully captured by these two physical properties alone. 
Therefore, any classification or visualization of similarity based solely on $P$ and $\dot{P}$ is potentially misleading. To address this multidimensional problem, which arises from the need to utilize additional properties to complete our vision, we introduce the MST (see Chapter~\ref{chapter4}), also referred to as the Pulsar Tree, and discuss its potential applications. It is computed based on a properly normalized distance that measures the closeness between pulsars.

The Pulsar Tree hosts information about pulsar similarities that go beyond $P$ and $\dot{P}$ and are thus naturally challenging to read from the $P\dot{P}$ diagram. 
Additionally, we will introduce an online tool that encompasses all our results, enabling users to focus on user-defined problems.

\section{The pulsar variance}
\label{chapter4: variance}

\subsection{Variables definition}
\label{chapter4: variables}

We consider pulsars listed in the Australia Telescope National Facility catalog, version 1.67 \citep{ATNF-Catalog} (ATNF v1.67, as of March 2022), which includes radio pulsars, X-ray and/or gamma-ray pulsars, and magnetars for which coherent pulsations have been detected. 
Accretion-powered pulsars, such as e.g., SAX J1808.4-3658, are not considered. 
The number of pulsars listed in ATNF v1.67 is 3282, of which 2509 have a known spin period and spin period derivative (larger than 0). 
From the latter, 2242 are isolated pulsars, and 267 are pulsars residing in binary systems. 
All the methods considered in this chapter will be applied to this set without distinction. 
For characterizing the pulsar population and ultimately defining a distance from one pulsar to another to establish a relationship, we consider the following physical set of pulsar variables (see Section~\ref{chapter1: Spin_evolution})\footnote{In \citet{MST-1}, the equations for the $\tau_{c}$, $\Delta \Phi$, and $\eta_{GJ}$ contain incomplete constant factors. However, their dependencies on $P$ and $\dot{P}$ were correctly stated, and all numerical computations used the proper expressions.}:

\begin{itemize}
  \item Spin period:\\ $P$ [s],
  \item Spin period derivative:\\ $\dot{P}$ [s s$^{-1}$],
  \item Magnetic field at the surface: \\ $B_{s} = (3c^{3}I)^{1/2}/(8\pi^{2} R^{6} \sin^{2}\alpha)^{1/2}
  \sqrt{P\dot{P}}  \simeq 3.2 \times 10^{19}  P^{1/2}\dot{P}^{1/2} {\rm G}$,
  \item Magnetic field at the light cylinder:\\ $B_{lc}=B_{s}(\Omega R)^{3}/c^{3} \simeq 3\times 10^8 P^{-5/2} \dot{P}^{1/2} {\rm G}$,
  \item Spin-down energy loss rate:\\ $\dot{E}_{sd} = 4\pi ^{2}I\dot{P} P^{-3}\simeq  3.95 \times 10^{46} P^{-3}  \dot{P} \;\;\; {\rm erg} \; {\rm s^{-1}}$,
 \item Characteristic age: \\ $\tau_c = {{P}}/{2\dot{P}} \simeq 15.8 \times 10^{-9} P \dot{P}^{-1} {\rm yr}$,
  \item Surface electric voltage: \\ $\Delta \Phi = ({B_s 4\pi^2 R^3})/({2 c P^2}) \simeq
  2.1 \times 10^{20} P^{-3/2} \dot{P}^{1/2} {\rm V}$,
  \item Goldreich-Julian charge number density: \\ $\eta_{GJ} = (\Omega B_{s})/(2\pi ce)\simeq  
  2.21 \times 10^{18}P^{-1/2}  \dot{P}^{1/2} {\rm cm^{-3}}$.
\end{itemize}

As a summary of the above, the moment of inertia $I$ was assumed as $10^{45}$ g cm$^{2}$, the radius of the star $R$ was assumed as $10$ km, and the inclination $\alpha$ between the magnetic and rotation axes as $90^o$.
The remaining constants ($c$ and $e$) have their usual meanings.
The measurable quantities $P$ and $\dot{P}$ are the leading magnitudes in this set of variables, from which all others are calculated using the rotating dipole model, as is usual for pulsar estimations. 
The surface magnetic field and spin-down power are fundamental magnitudes that characterize the energetics and magnetospheres of pulsars. 
In our set, others suggested that we incorporate the idea that dissimilar pulsars (e.g., those with millisecond and regular periods) can have similar magnetospheres. 
The magnetic field at the light cylinder partly describes this, which may be similar for both.
The surface electric voltage gives the potential drop between the magnetic pole and the edge of the polar cap \citep{Lorimer2012}. 
It is thought to represent the variety introduced by the electromagnetic configuration, for which another parameter of interest is the Goldreich-Julian charge number density, $\propto B_s / P$. 
Note that these magnitudes, being all functions of $P$ and $\dot{P}$, can have a relationship between themselves, as just noted.
Ideally, the mass and radius of the neutron stars would also be considered variables of interest in our study. However, this information is only available for a tiny percentage of the sample.

Other variables of interest are those related to the birth properties of pulsars, such as the initial spin-down power, the initial magnetic field, or the spin-down timescale. However, these are not known for most pulsars in our sample. 
They all depend on the unknown (except for a few) pulsars' actual age (for which the characteristic age $\tau_c$ is only a proxy). Similarly, the braking index is measured for only a handful of pulsars (and it is known that it may vary significantly). 
Estimates from it, using $\ddot{P}$, would dramatically reduce the sample size. Finally, other measurable quantities exist unrelated to intrinsic properties (transverse velocities, DM, distances, $\ldots$) and/or are known for a limited number of objects. 
Using luminosities and other properties at different frequencies (e.g., fluxes, pulse shapes, peak separation, $\ldots$) would also significantly reduce the sample, as it would be affected by extrinsic conditions (absorption, distances, $\ldots$), and/or would incorporate parameters that are difficult to compare for the population as a whole.

\subsection{Treating variables}
\label{chapter4: treating_variables}

\begin{figure}
  \includegraphics[width=1\textwidth]{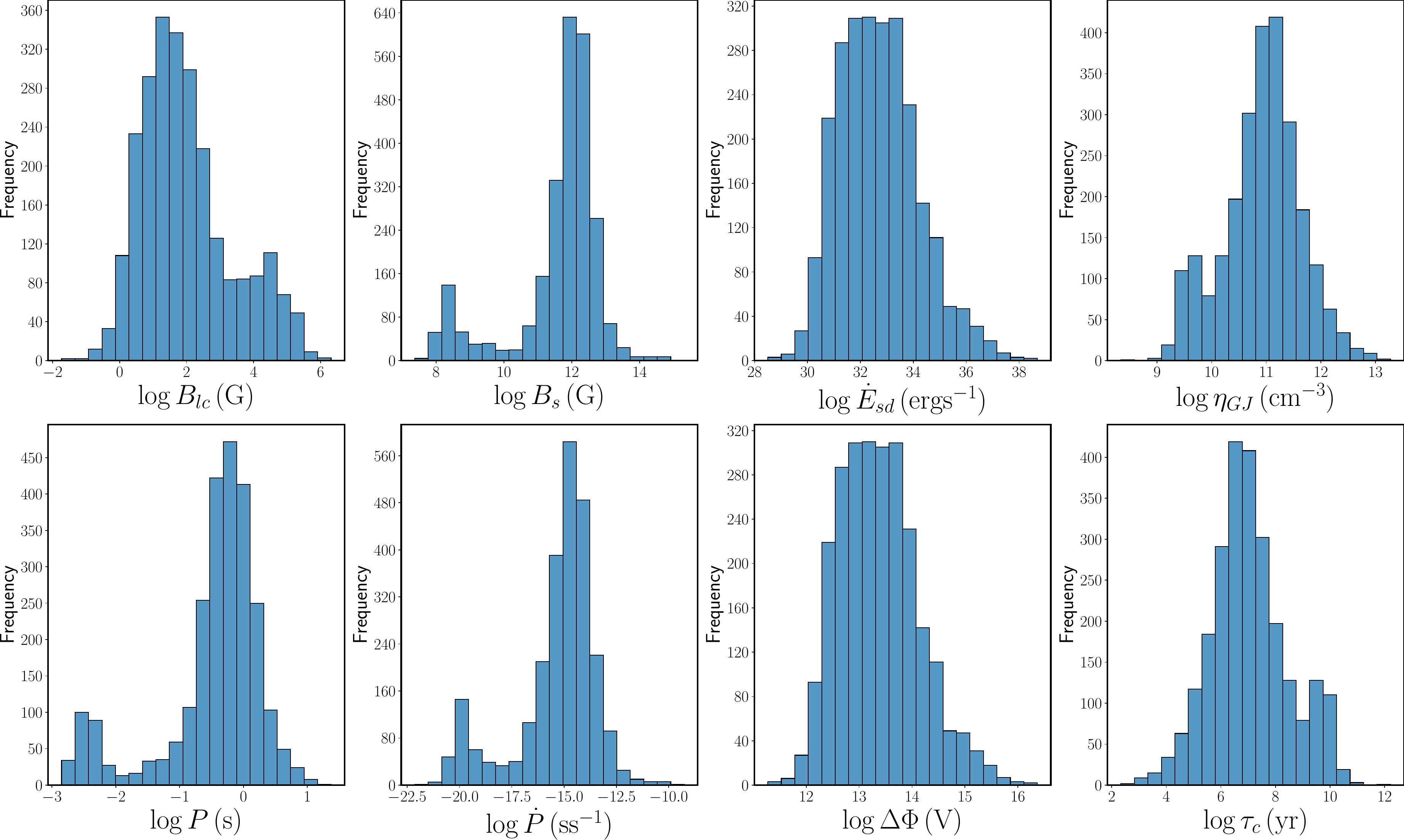}
  \centering
  \caption[Distribution of the properties considered.]{Distribution of the logarithm of the eight variables used for the complete set of 2509 pulsars. }
  \label{chapter_4_figure: figure1}
\end{figure}

The magnitudes' values may differ by several orders of magnitude for different pulsars, so we consider the logarithm of them to mitigate this as much as possible. 
The distributions of the logarithm of these variables are shown in Figure~\ref{chapter_4_figure: figure1}. 

These distributions are not normal (as is the case for the original set of variables without logarithms). 
Two clear populations of millisecond and normal pulsars appear separately in all variables except those related to the $\dot{E}$ and $\Delta \Phi$. 
As is evident from their centralization (mean and median) and dispersion (standard deviation and interquartile range (IQR)), the log variables are orders of magnitude closer together. 
Given that the distributions are not normal, we use the {\it robust scaler} (see, e.g., \cite{RobustScaler}) for scaling the log variables:

\begin{eqnarray}
x_{i}^{\dag} = \frac{x_{i}-Q_{2}}{IQR}
~.\label{chapter_4_eq: robust_scaler}
\end{eqnarray}

The $\dag$-symbol represents that the quantity $x_{i}$ has been scaled, $Q_{1}$, $Q_{2}$ and $Q_{3}$ represent the 1st quartile, median, and 3rd quartile of the distribution, respectively, with the IQR as $(Q_{3}-Q_{1})$. 
After being scaled, the variables' distributions have a median equal to zero and an IQR equal to one.

Note that if we take the logarithm of the variables and then apply Equation~(\ref{chapter_4_eq: robust_scaler}) to scale them, the relations between variables are more clearly revealed. 
For instance, $\dot{E}_{sd}$ and $\Delta \Phi$, lead to $\log \dot{E}_{sd}^\dag = \log \Delta \Phi^\dag$, that is also visible in the corresponding distributions of Figure~\ref{chapter_4_figure: figure1}. 
Considering both at once in defining the nearness of two given pulsars is adopted to represent that the physical meaning of the two original magnitudes is different.

\section{Principal components analysis}
\label{chapter4: pca-pulsars}

PCA is especially suitable for identifying the main factors that introduce variance in a population, particularly when the variables involved are linear (see Section~\ref{chapter3: mathematical_interpretation} for further explanations). 

In this case, since six of the variables that describe the intrinsic properties of the pulsar population (see Section~\ref{chapter4: variables}) are derived from $P$ and $\dot{P}$, one might intuitively conclude that two PCs would suffice to describe the population's variance. 

\begin{figure}
  \centering
  \includegraphics[width=1\textwidth]{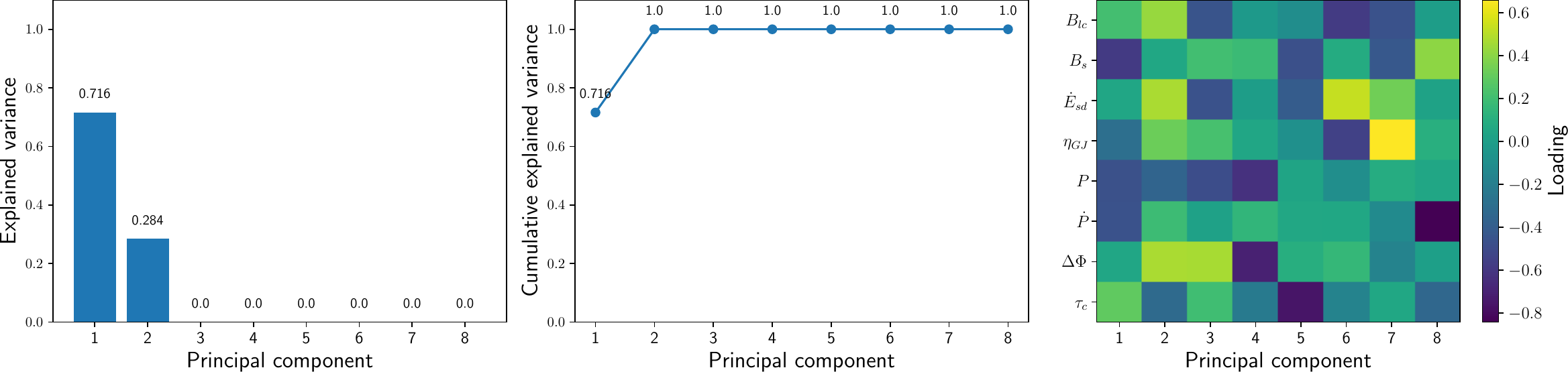}
  \caption[PCA results of the properties considered.]{
  PCA results for the logarithm of the set of variables for the population of 2509 pulsars, see Section~\ref{chapter3: visual_tools} for further explanations about PCA definitions.
  The left panel, referred to as the scree plot, displays each PC's explained variance based on the eigenvalues of the covariance matrix.
  Note that the covariance matrix calculates the relationships between pairs of variables, showing how changes in one variable are associated with changes in another. 
  It represents the amount of information contained in each PC.
  The central panel shows the cumulative explained variance by the new variables defined through the PCA analysis. 
  The right panel shows the "weight", also called loading, each variable has concerning each PC, indicating its contribution to the variance captured by that PC. 
  This value is the coefficient held in each eigenvector of the covariance matrix.
  Negative values imply that the variable and the PC are negatively correlated. 
  Conversely, a positive value shows a positive correlation between the PC and the variable.
  }
  \label{chapter_4_figure: figure2}
\end{figure}

However, Figure~\ref{chapter_4_figure: figure2} illustrates that the PCs identified to reproduce the whole variance, $PC_1$ and $PC_2$, are not simply $P$ and $\dot{P}$ themselves, but rather linear combinations of all the variables.

In other words, the variance of the population is not entirely contained within the individual variances of $P$ and $\dot{P}$ alone.

Thinking only in terms of $P$ and $\dot{P}$ to compare pulsars may thus be misleading, except in the extreme case where these values are the same. 

The first two PCs, containing 71.6\% and 28.4\% of explained variance (see left panel of Figure~\ref{chapter_4_figure: figure2}), respectively, can be seen as (see Equation~(\ref{chapter_3_eq: PC_space})):

\begin{align}
PC_1 &=\; 0.21 B_{{lc}_l}^\dag - 0.29 \eta_{{GJ}_l}^\dag + 0.05 \Delta \Phi_l^\dag + 0.05 \dot{E}_{{sd}_l}^\dag - 0.46 \dot{P}_l^\dag \nonumber \\
     &\quad - 0.59 B_{s_l}^\dag - 0.47 P_l^\dag + 0.29 \tau_{c_l}^\dag 
     \hfill \nonumber \\
PC_2 &=\; 0.43 B_{{lc}_l}^\dag + 0.32 \eta_{{GJ}_l}^\dag + 0.47 \Delta \Phi_l^\dag + 0.47 \dot{E}_{{sd}_l}^\dag + 0.19 \dot{P}_l^\dag \nonumber \\
     &\quad + 0.05 B_{s_l}^\dag - 0.36 P_l^\dag - 0.32 \tau_{c_l}^\dag~.
     \hfill 
\label{chapter_4_eq: PC1_PC2_8V}
\end{align}

In Equation~(\ref{chapter_4_eq: PC1_PC2_8V}) the variables are scaled following Equation~(\ref{chapter_4_eq: robust_scaler}), the sub-index $l$ shows the logarithm of them. 
We note that all physical properties have a non-zero loading associated with them, as shown in the right panel of Figure~\ref{chapter_4_figure: figure2}.

After some computations (see Appendix \ref{Appendix_4: algebra_PC1_PC2} for more details), Equation~(\ref{chapter_4_eq: PC1_PC2_8V}) can be reformulated as:

\begin{align}
PC_1 &= -8.471 - 1.178 \log P - 0.832 \log \dot{P} \nonumber \\
PC_2 &= 14.182 - 2.931 \log P + 1.105 \log \dot{P}~.
\label{chapter_4_eq: PCA_2V}
\end{align}

Equation~(\ref{chapter_4_eq: PCA_2V})\footnote{Note that no dag marking is needed, as the corresponding IQR and median of each variable is absorbed into the coefficients, and that the units of $P$ and $\dot P$ are as in Equation~(\ref{chapter_4_eq: PC1_PC2_8V}).} corroborates the direct non-equivalence between $P$ and $\dot P$, and $PC_1$ and $PC_2$, as discussed above. 
These equations also indicate that several variables are equally important (with similar loadings) in defining the PCs.

\begin{figure}
  \includegraphics[width=1\textwidth]{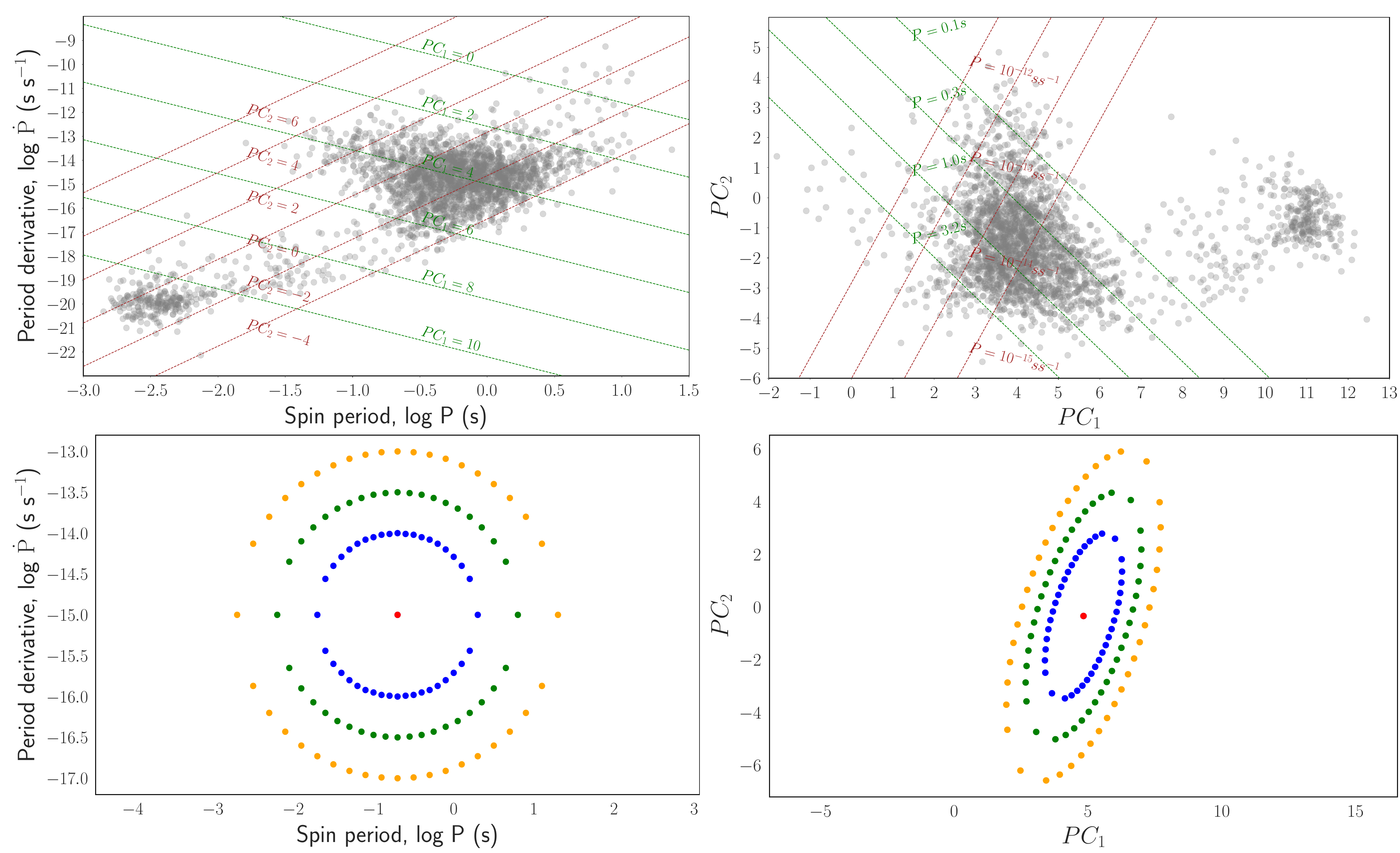}
  \centering
\caption[The pulsar population seen through $P$ and $\dot{P}$, and $PC_{1}$ and $PC_{2}$.]{
Top-Left panel: $P\dot{P}$ diagram of the pulsar population. Constant values of $PC_{1}$ and $PC_{2}$ are shown in green and brown lines, respectively. 
Top-right panel: $PC_1PC_2$ plane of the pulsar population containing the PC scores. Constant values of $P$ and $\dot{P}$ in green and brown lines, respectively. 
Bottom-left panel: Synthetic pulsars positioned circularly at different radii from a given center in the $P\dot{P}$ diagram.
Bottom-right panel: Transformation of the circle into the $PC_1PC_2$ plane.
In all panels, the values are obtained from Equation~(\ref{chapter_4_eq: PCA_2V}).
}
       \label{chapter_4_figure: figure3}
\end{figure}

The left panel of Figure~\ref{chapter_4_figure: figure3} shows the pulsar population in the $P$$\dot P$ diagram together with lines representing equal values of the PCs, i.e., constant PC scores.
The right panel of Figure~\ref{chapter_4_figure: figure3} shows the same pulsars but directly in the $PC_1PC_2$ plane with the PC scores derived from Equation~(\ref{chapter_4_eq: PC1_PC2_8V}).
Note that nearness in one plane does not have the same meaning as in another. 
To exemplify this, we plot a circle in the $P\dot{P}$ diagram and see how the circle transforms to the $PC_1PC_2$ plane via Equation~(\ref{chapter_4_eq: PCA_2V}).
This is illustrated in the second row of Figure~\ref{chapter_4_figure: figure3}, where the difference in relative distances can be as much as a factor of 3 or more. This change of shape advances the idea that any nearness ranking will be affected if considering the PCs instead of $P$ and $\dot{P}$ (further comments about this can be found in Appendix \ref{Appendix_4: nearness-comment}). 

\section{Minimum Spanning Tree of the pulsar population}
\label{chapter4: mst-pulsars}

Chapter~\ref{chapter2} presents all the concepts needed to understand and compute an MST, and we take this for granted in what follows. We define a Euclidean distance (see Equation~(\ref{chapter_2_eq: d_eucl})) using the eight-scaled (via Equation~(\ref{chapter_4_eq: robust_scaler})) logarithm of the variables introduced in Section~\ref{chapter4: variables}. 
Equivalently, after analyzing the data in Section~\ref{chapter4: pca-pulsars}, we can use just $PC_1$ and $PC_2$, which contain the population variance in the ATNF v1.67. Both choices produce the same MST (and thus, the results from the study that follows are the same), but the latter is less demanding due to the reduced dimensionality of the problem. 

\begin{figure}
\centering
\includegraphics[width=\textwidth]{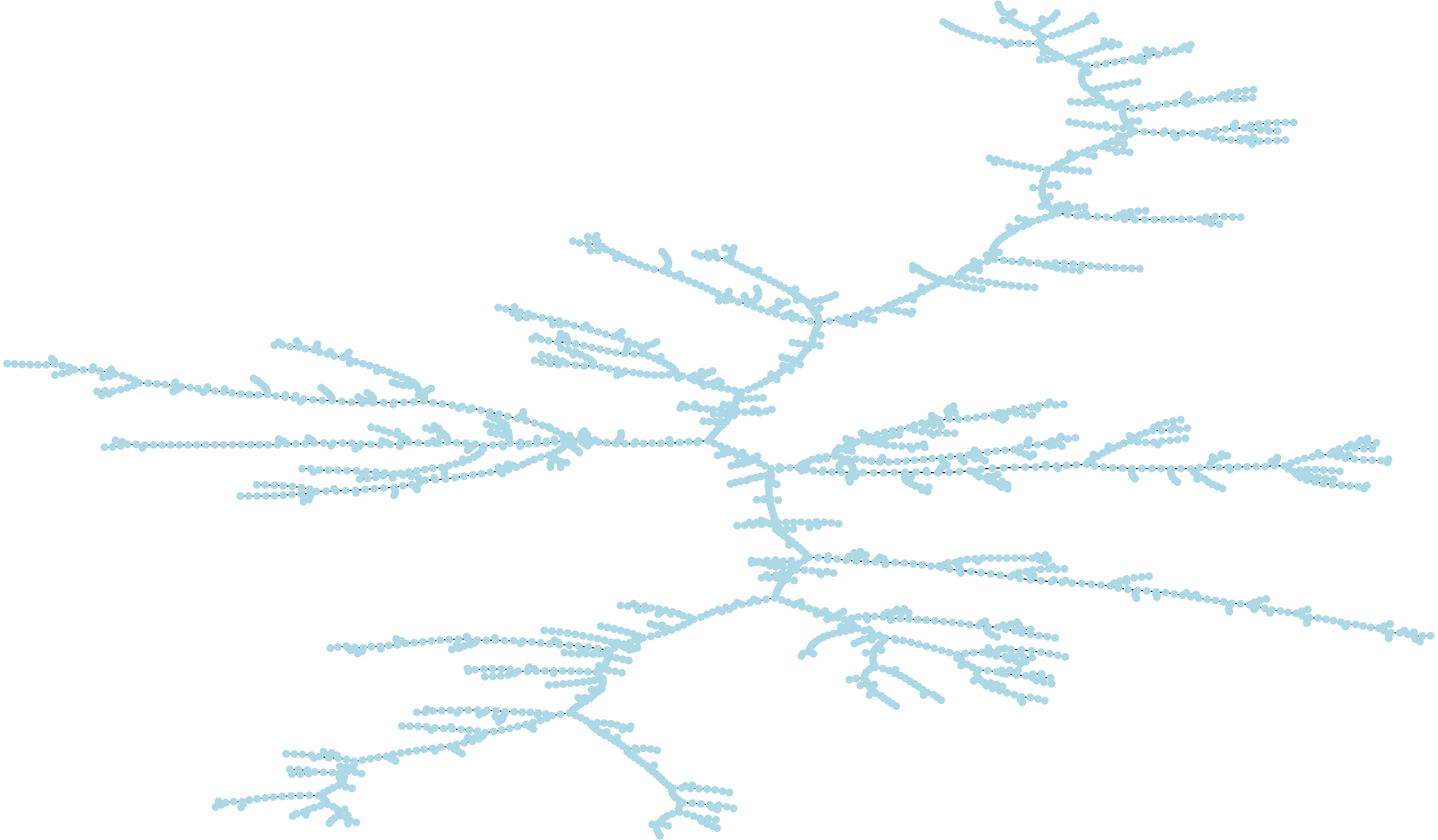}
\caption[The Pulsar Tree of the pulsar population.]{A different look at the population. MST, $T(2509, 2508)$, based on the complete, undirected, and weighted graph $G(2509,3146286)$ for the 2509 pulsars (each node represents a pulsar) from the ATNF v1.67 and their full combination of weights computed from their Euclidean distance among eight scaled variables (or the equivalent 2 PCs). Each node in the MST represents a pulsar. 
}
       \label{chapter_4_figure: figure4}
\end{figure}

With the Euclidean distance defined over the population, we first obtain a complete, undirected, and weighted graph $G(V, E)=G(2509,3146286)$.
From that, we calculate the MST, $T(2509, 2508)$, which we call the Pulsar Tree, which is shown in Figure~\ref{chapter_4_figure: figure4}. 

\subsection{Branch analysis and pulsar classification in the MST}
\label{chapter4: branching}

\begin{figure}
  \includegraphics[width=1\textwidth]{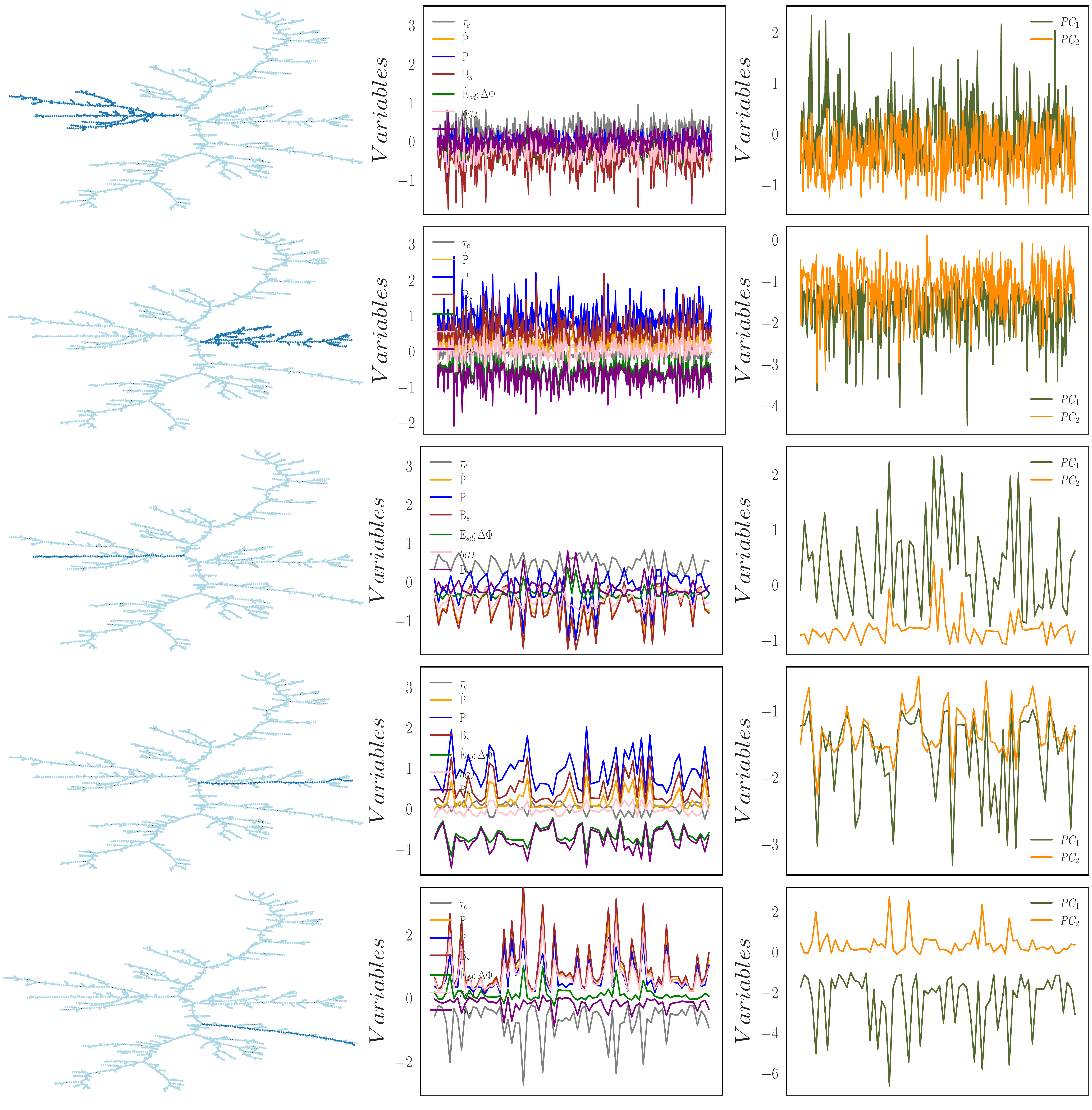}
  \centering
\caption[Track of the properties following an arbitrary mixing of the nodes.]{
Top two rows: On the left, nodes from different branches are highlighted in dark blue in the MST. The middle and right panels display the behavior of the eight scaled variables and the two PCs when following an arbitrary mixing of the nodes in the highlighted part of the MST, as shown in the corresponding panel in the left column. 
The third, fourth, and fifth rows are taken from a unique branch, respectively; however, the nodes of the selected path are now arbitrarily mixed. 
For visualization purposes, we subtracted the corresponding mean for each PC in the right column for $PC_1$ and $PC_2$.
}
       \label{chapter_4_figure: figure5}
\end{figure}

\begin{figure}
  \includegraphics[width=1\textwidth]{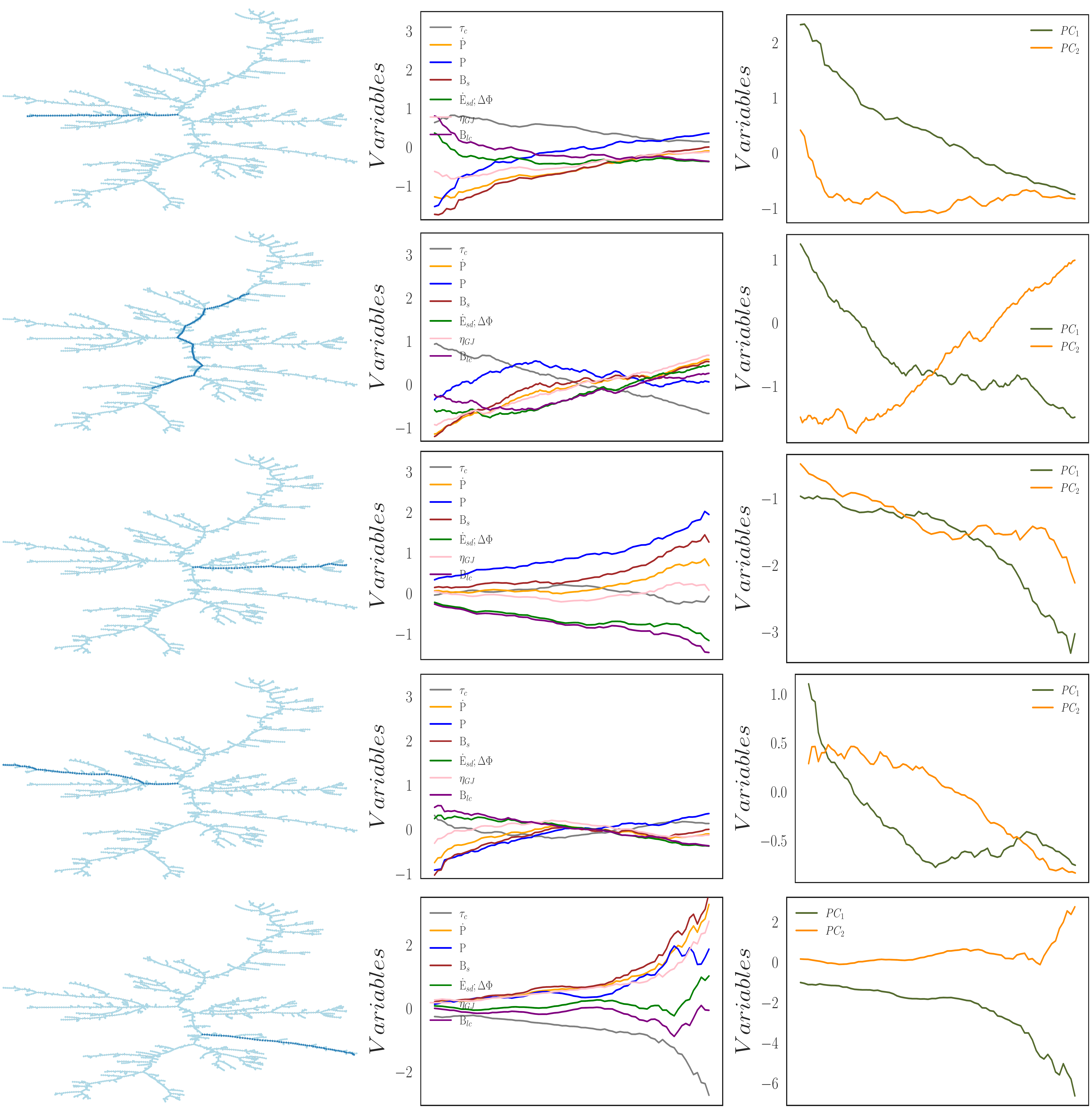}
  \centering
\caption[Track of the properties following an arbitrary path.]{Similar to the panels in Figure~\ref{chapter_4_figure: figure5}, but following the path nodes as appearing along the marked branch of the MST in its corresponding panel in the left column. The path is read from left to right (except in those branches that appear mostly vertical, where they are read from top to bottom).
The first, third, and fifth panels in the left column mark the same nodes as the third, fourth, and fifth panels, respectively, in the left column in Figure~\ref{chapter_4_figure: figure5}.
}
       \label{chapter_4_figure: figure6}
\end{figure}

\begin{figure}
  \includegraphics[width=1\textwidth]{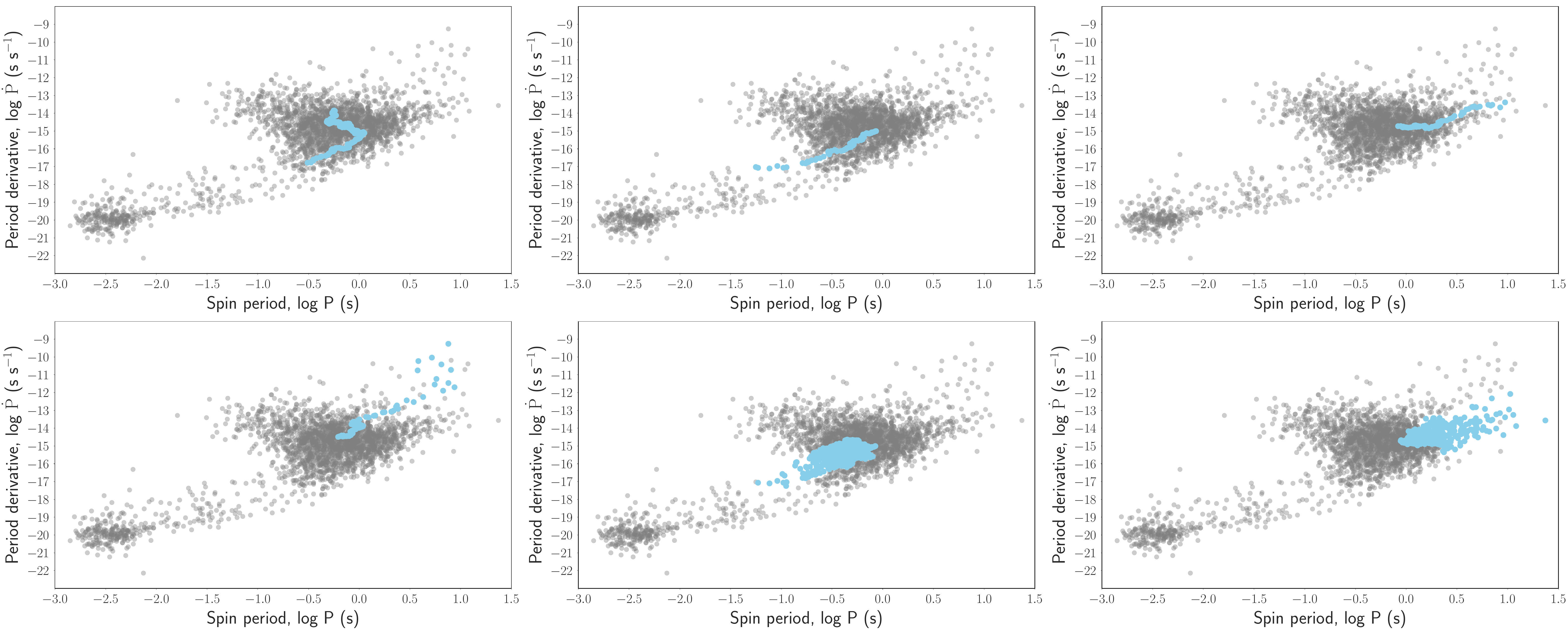}
  \centering
\caption[Positions of the branches in the $P\dot{P}$-diagram.]{Positions of the branches and paths analyzed in Figures \ref{chapter_4_figure: figure5} and \ref{chapter_4_figure: figure6} according to the $P\dot{P}$ diagram. From left to right, and top to bottom: second row of Figure~\ref{chapter_4_figure: figure5}; and, third, fourth, fifth, first, and second rows, respectively, following Figure~\ref{chapter_4_figure: figure6}.
}
       \label{chapter_4_figure: figure7}
\end{figure}

As shown in Figure~\ref{chapter_4_figure: figure5}, mixing nodes from different branches of the MST produces a scattered distribution of variables.  
This generic behavior occurs when mixing branches in the MST and any blending of nodes, even within a path (see the panels in the last three rows of Figure~\ref{chapter_4_figure: figure5}). If we read the MST in a disordered manner, nothing is learned from it. Instead, Figure~\ref{chapter_4_figure: figure6} shows that if we choose one of the paths at a time and run along with the nodes in it in an orderly manner, a smooth behavior of the variables naturally appears. Mixing branches or paths of the Pulsar Tree is equivalent to grouping pulsars by their nearness in the $P\dot{P}$ diagram, as shown in Figure~\ref{chapter_4_figure: figure7}. The Pulsar Tree hosts information that is challenging to read from the $P\dot{P}$ diagram.

\subsection{The MST as a descriptive tool}
\label{chapter4: MST_descriptive_tool}

\begin{figure}
  \includegraphics[width=1\textwidth]{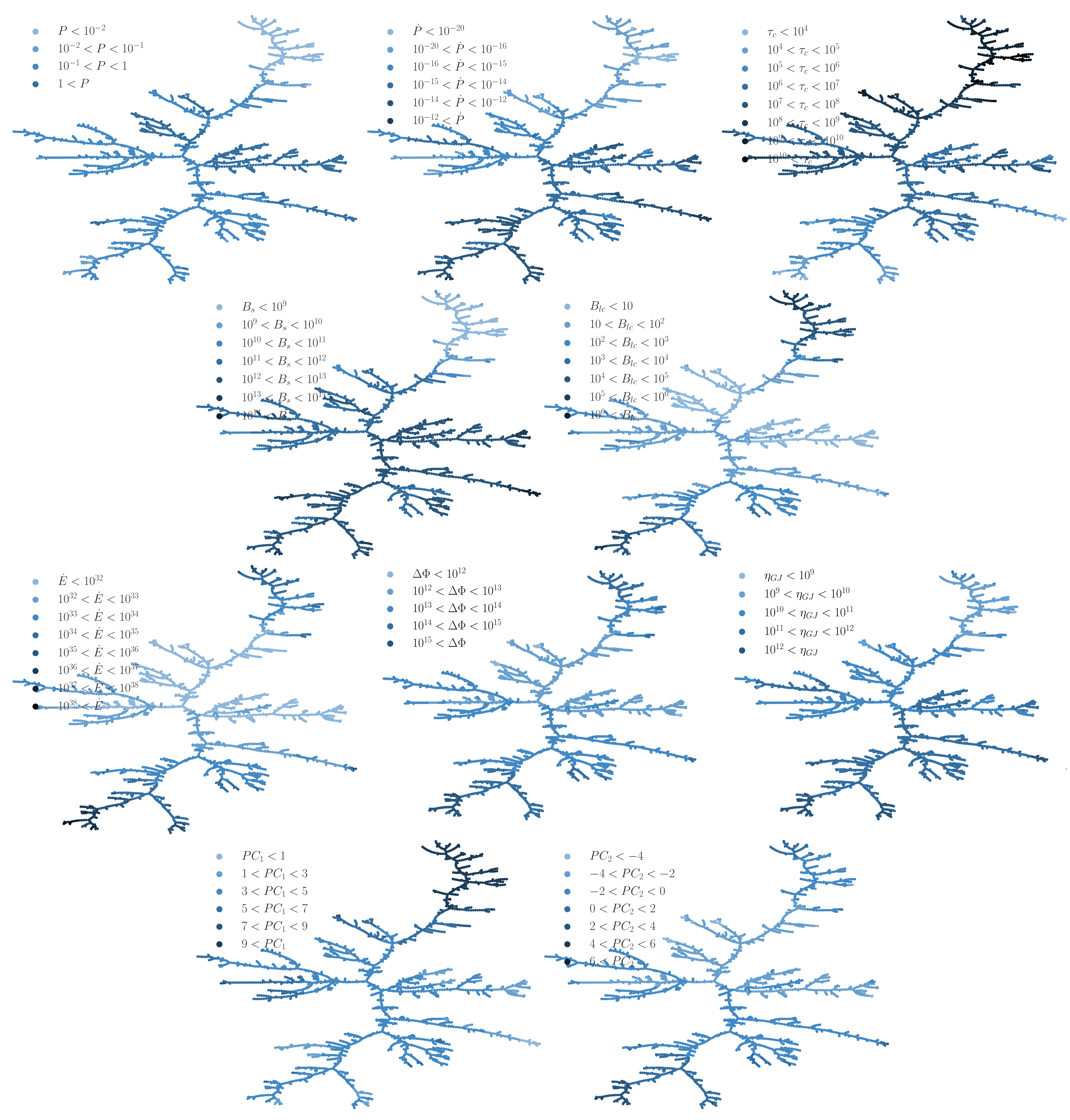}
  \centering
\caption[Representation of the properties considered onto the Pulsar Tree.]{
Representation of the variables considered directly onto the MST. The first row shows $P$, $\dot{P}$, and $\tau$; the second row shows the magnetic field at the surface and at the light cylinder; the third row shows the spin-down power, the surface electric voltage, and the Goldreich-Julian current, and finally, the last row shows directly the $PC_{1}$ and $PC_{2}$.}
       \label{chapter_4_figure: figure8}
\end{figure}

The ordering introduced by each branch and path indicates an internal physical grouping in the MST, as shown in the variations of the variables seen in Figure~\ref{chapter_4_figure: figure8}.  
These variations illustrate the physical properties of different pulsar classes, and understanding them may reveal physical connections among pulsars or insights into their evolution. To emphasize this, we shall observe how some known groups of pulsars are located in the MST.

\begin{figure}
  \includegraphics[width=1\textwidth]{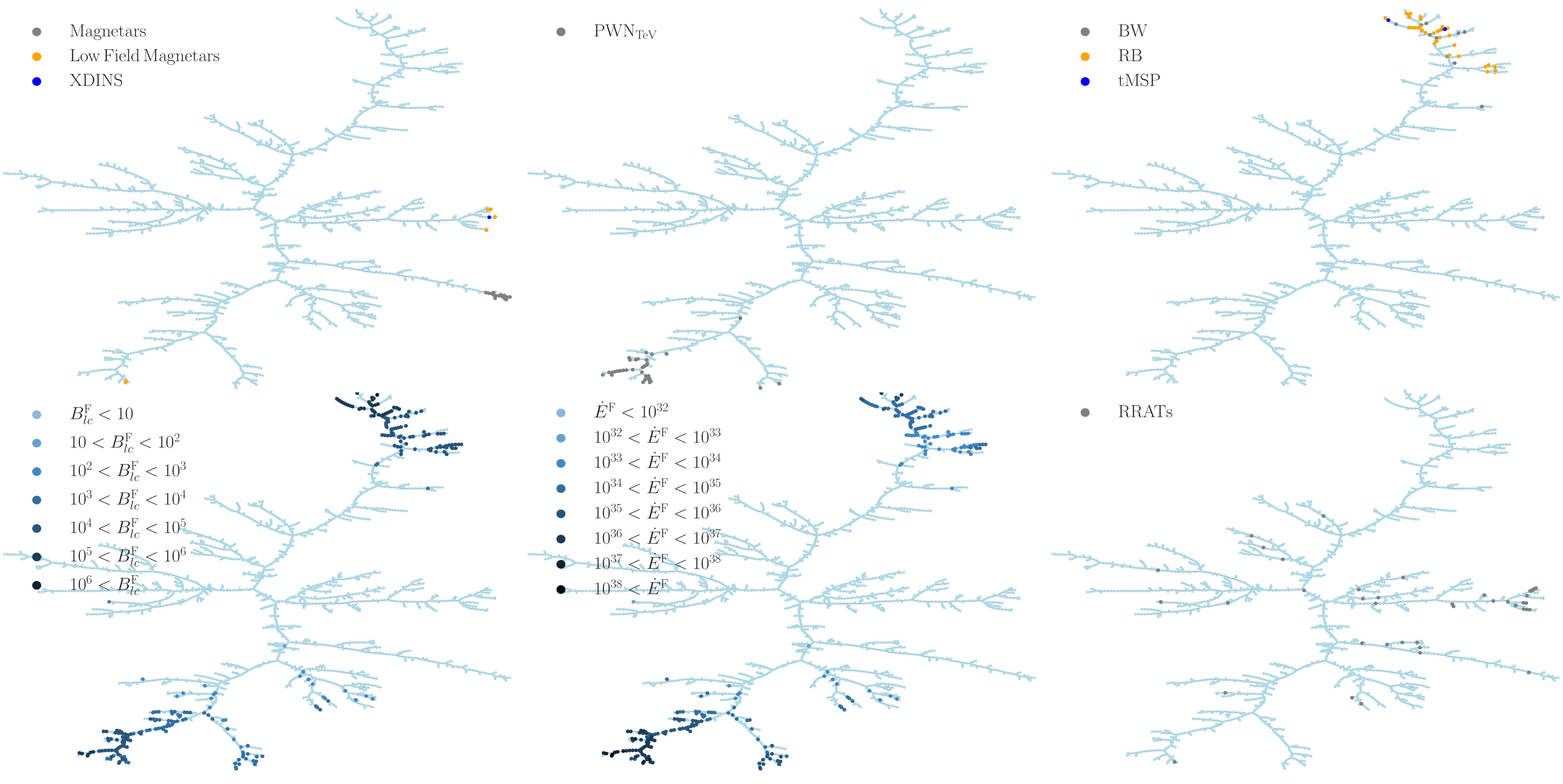}
  \centering
\caption[Representation of different pulsar classes onto the Pulsar Tree.]{
Representation of different pulsar classes on the MST. The first row shows magnetars, low field magnetars, X-ray Dim Isolated Neutron Stars (XDINS), pulsars associated with TeV pulsar wind nebulae (PWN in TeV), redbacks (RB), black widows (BW), and transitional millisecond pulsars (tMSPs). The first two panels of the second row show the {\it Fermi}-LAT pulsars (thus the superscript F), further characterized by their magnetic field at the light cylinder and spin-down power in radio, as depicted in Figure~\ref{chapter_4_figure: figure8}. The third panel in the second row shows the Rotation Radio Transients (RRATs). 
}
       \label{chapter_4_figure: figure9}
\end{figure}

The main tree trunk travels from young and energetic pulsars at the bottom\footnote{
In what follows, we use "bottom", "top", "right", or "left" to refer to a particular position in the MST shown. However, we emphasize that (in the absence of axes) what matters is not the representation but the edges among nodes, see Appendix \ref{Appendix_4: representing_MST} for a further explanation.} 
to millisecond pulsars (MSPs) in the top (see panels -referred from left to right and top to bottom- first, second, and fifth of Figure~\ref{chapter_4_figure: figure8}). However, other than representing the overall separation of the binary pulsars from the rest of the sample, the variables used in this MST do not allow us to delve deeper into the sub-population of binaries, as they are heavily influenced by the evolutionary processes that occur for MSPs during the accreting (recycling) phase. 
As an example, consider BWs, RBs, and tMSPs, see \citet{Papitto2022} for reviews, and also \citet{swihart_blacwidows, Linares2023, Freire2017, Lynch_2012, Douglas2022, Strader_2019} for identifying the sources.

As their physical properties are similar, no clear grouping appears for these binary sub-samples (see Figure~\ref{chapter_4_figure: figure9}). In Chapter~\ref{chapter6}, we explore supplementing intrinsic variables with others that represent the companion, orbital parameters, and the environment of binary pulsars to address these issues. 

The branches departing from the main trunk cover deviations are better represented by the variability in one or a few of the magnitudes considered. Thus, the extremes of each branch are extreme pulsars of the population in a particular way. 

For instance, the most extended rightward branch (moving along it towards the rightmost node) groups pulsars with increasing $\tau_{c}$, $P$, and $\dot{P}$. Pulsars in this branch are not particularly energetic, nor have significant $B_{lc}$; instead, they have an increased $B_s$, reaching extreme values. The second-longest rightward branch has similar behavior but is formed by less magnetized objects at their $B_s$ and is also less energetic, slower, and older. Not surprisingly, magnetars and the only XDIN, J1856-3754, quoted in the ATNF v1.67, are located in both branches (see \cite{Olausen_2014, Zelati_2018}). Figure~\ref{chapter_4_figure: figure9} shows this in more detail. Interestingly, the XDIN and the magnetars do not share the same branch. 

Some of the low-field magnetars, since they are more energetic, less magnetized objects that have nevertheless shown magnetar-flaring behavior, appear pretty separate from the rest. The low-field magnetars depicted in the MST having $P< 1$ s are J1846-0258 \citep{Gavriil2008} and J1119-6127 \citep{Archibald2016}, and they appear in Figure~\ref{chapter_4_figure: figure9} in the bottom leftmost branch.
This location differs from that of J2301+5852, J1647-4552, J1822-1604, and J0418+5732, the other low-field magnetars, which are located on the second-longest rightward branch above the magnetars.

The different branches at the bottom of the MST contain all energetic pulsars. Despite being very different in almost every aspect, they share the same $B_{lc}$ as MSPs at the top of the MST. The small branches in which the energetic pulsars divide at the bottom of the plot also separate them into those having a significantly higher value of different variables, like $\dot E$, $B_{lc}$, $B_s$. 
Figure~\ref{chapter_4_figure: figure9} consistently shows how the pulsars associated with TeV confirmed or candidate pulsar wind nebulae (PWNe, \cite{HESSPWNe}) are essentially all located in these branches. Only three TeV PWNe, B1742-30(1)/J1745-3040, J1858+020/J1857+0143 and CTA 1/J0007+7303 are somewhat outliers. The former two are TeV PWN candidates, the oldest and less energetic PWNe in the population (see Table 4 of \cite{HESSPWNe}). The MST alone cannot judge the reality of the proposed association, but it emphasizes how distinct these two are from the rest (see Section~\ref{chapter4: MST_PWNe} for more details about the MST in this context).
The case of CTA 1 has been studied in detail as a possible PWN in the reverberation phase and/or due to having a higher magnetization \citet{Martin2016}; see also \citet{Torres2014}. 
Its peculiarity has already been noted from a physical standpoint. However, it is less of an outlier than the rest of the PWNe population (both in the MST and in comparing PWNe models). 

Another interesting example of the MST view on pulsars is to note where the {\it Fermi}-LAT detected gamma-ray pulsars that are part of the ATNF v1.67 fall on it (see \cite{2fpc, GLAMCOG}). Figure~\ref{chapter_4_figure: figure9} shows two panels to this effect, where $B_{lc}$ and $\dot{E}$ are noted. The detected gamma-ray pulsars cluster at specific locations, with the most empty of gamma-ray emission, corresponds to the fact that gamma-ray pulsars have $B_{lc}>100$ G, with most having $B_{lc}>10^3$ G, and relatively high $\dot{E}$. Some other magnitudes are less decisive; for example, there are gamma-ray pulsars across the full range of $\eta_{GJ}$ values. Along the branches where the {\it Fermi}-LAT pulsars lie, there might be a close sequence of detected and non-detected pulsars despite the MST clearly showing the similarity of their intrinsic properties. This suggests that extrinsic features, such as distance, environment, or geometry, play a role in {\it Fermi}-LAT detectability at an individual level.

The {\it Fermi}-LAT pulsar isolated in the central part of the MST is J2208+4056, the only one depicted with $B_{lc}<100$ G. This pulsar has been noted by \citet{Smith2019} as having a $\dot{E}$ ($\sim 8 \times 10^{32}$ erg s$^{-1}$) about three times lower than the previously observed gamma-ray emission death-line. The outlier {\it Fermi}-LAT pulsar in the left part of the MST is J1231-5113 and has an even lower $\dot{E}$ ($\sim 5 \times 10^{32}$ erg s$^{-1}$), while its other magnitudes $B_{lc}, \tau, \eta_{GJ}$ are similar to the rest of the gamma-ray population. 

We also investigate the position in the MST of the 40 RRATs (taken from \cite{Abhishek2022, RRATalog}) with known $P$ and $\dot{P}$ appearing in the ATNF v1.67. RRATS are pulsars showing extreme radio variability, as most are discovered through their single, isolated pulses, and only one of the RRATs considered is located in the main trunk. 
This pulsar is located near the degree 3 node J1828-1336, which separates the main trunk into the two central branches. 

The MST can also be used to analyze any other pulsar population, for instance, those with known glitches or the intermittent and nulling pulsar group. 

The online tool, the Pulsar Tree web (see Section~\ref{chapter4: pulsartreeweb}), provided with this work, promotes this kind of analysis.

\subsection{The MST view on PWNe}
\label{chapter4: MST_PWNe}
A dedicated application of the MST methodology to the study of PWNs is presented in the work by \citet{MST-PWNs}.
This analysis, focused on the potential detection of PWNs around PSR J1208−6238, J1341−6220, J1838−0537, and J1844−0346, provides a thorough implementation of the method, along with insightful results that validate its significance beyond population studies. 
Considering the Pulsar Tree, this study identifies potential TeV PWNe based on their locations, utilizing a one-zone PWN leptonic model to predict their possible energy spectrum characteristics.  
As illustrated in Figure~1 in \citet{MST-PWNs}, most pulsars with detected TeV PWNe \citep{HESSPWNe} are positioned in the lower-left branch of the Pulsar Tree.  
The noteworthy quantity of TeV-detected PWNe in a specific part of the Pulsar Tree is not a result of random selection, as obtaining a non-trivial number of a particular type of pulsar is not straightforward.  
The MST relates equally energetic and relatively young sources that are likely to exhibit TeV PWN behavior. 
However, there is no correlation with the environment, or the progenitor is included in the distances on which the Pulsar Tree is computed.

\subsection{The MST view of evolutionary tracks}
\label{chapter4: evolutionary_tracks}

\begin{figure}
    \centering
    \includegraphics[width=\textwidth]{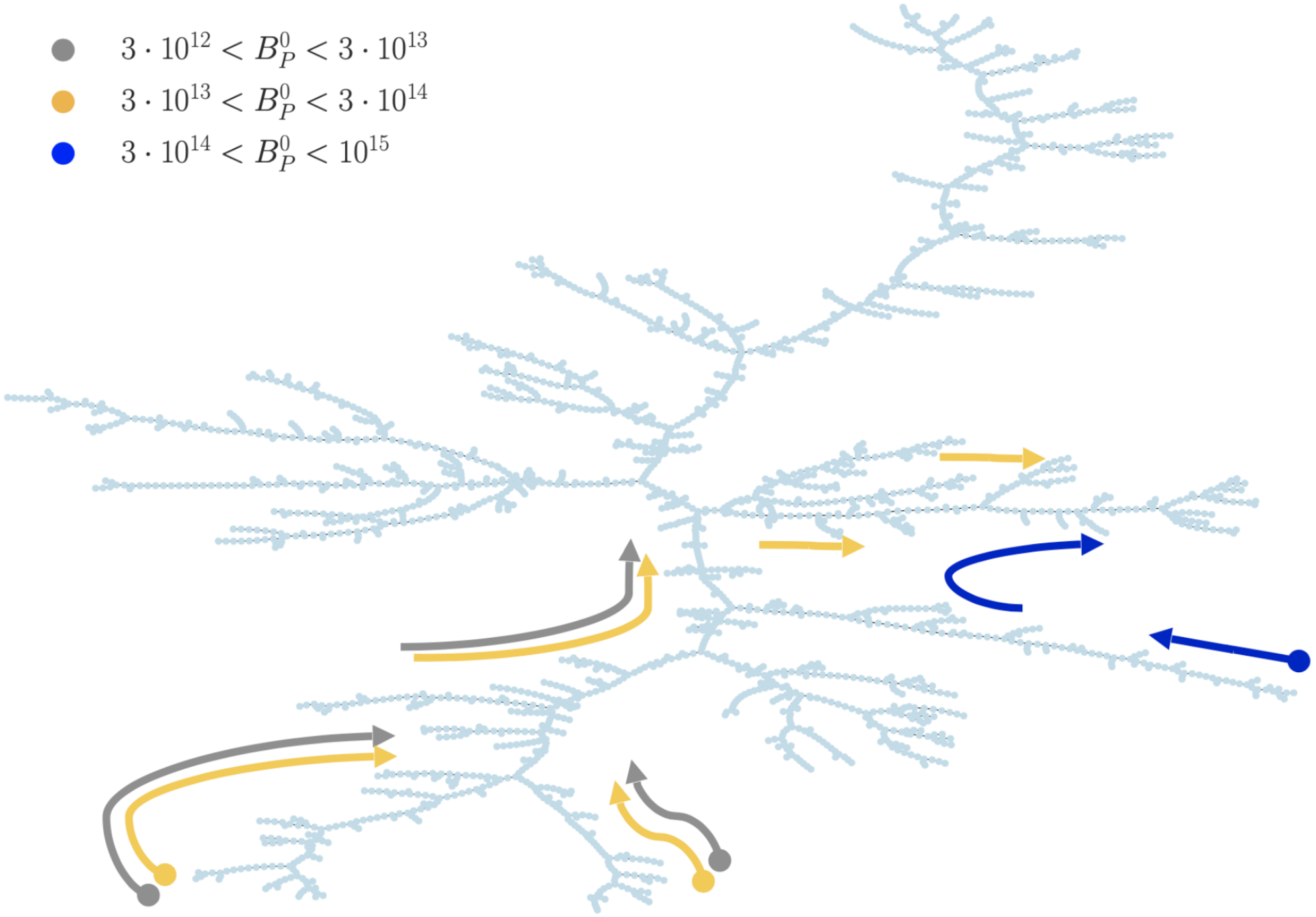}
\caption[Representation of evolutionary tracks into the Pulsar Tree.]{
Representation of evolutionary tracks described in \citet{Vigano2013} into the MST, according to their initial magnetic field ($B_{P}^{0}$).
}
       \label{chapter_4_figure: figure10}
\end{figure}

While pulsars evolve, they change their timing parameters and move across the $P \dot{P}$ diagram. Evolutionary models are then constructed following the fully coupled evolution of temperature and magnetic field in neutron stars (e.g., see \cite{Vigano2013}). To simulate evolutionary tracks, we have created synthetic pulsars based on the theoretical tracks presented in Figure~10 of \citet{Vigano2013} and individually studied where they would fall in the MST should they be part of our sample. This is shown in Figure~\ref{chapter_4_figure: figure10}, where the arrows show increasing age at a fixed initial magnetic field ($B_{P}^{0}$), and the rounded cap point in the origin of the arrows shows possible birthplaces. In agreement with what was already discussed in Figure~\ref{chapter_4_figure: figure6}, we find that tracks are not randomly found in the MST.
There are two birth zones: at the bottom part, where we find all energetic pulsars, and at the rightmost branch, where we see the magnetars. Then, for low $B_{P}^{0}$, pulsars die on the main trunk of the MST. The pulsars populate the middle branches when the $B_{P}^{0}$ increases. At the same time, this is high enough to shift the birthplace from the energetic pulsar zone to the classical magnetar range; the evolution is mainly confined to the two rightmost branches in Figure~\ref{chapter_4_figure: figure10}. In these branches, we find pulsars evolving in two directions (from the main trunk to the extreme and vice versa) depending on their origin. 

\subsection{The MST as an alerting tool}
\label{chapter4: MST_alerting_tool}

In addition to providing a descriptive perspective, the MST may serve as an alerting tool for pulsars of interest (see Chapter~\ref{chapter6}). These examples consider the MST location and the ranking of distances. Implied connections are not always obvious using $(P, \dot{P})$ only (see the discussion on the PCA and the distance ranking above and in the Appendix \ref{Appendix_4: nearness-comment}). We use The Pulsar Tree web (see Section~\ref{chapter4: pulsartreeweb}) to note the following:

\begin{itemize}

\item Based on the location of the energetic low-field magnetars J1846-0258, J1119-6127 at the bottom part of the MST, and due to its nearness ranking, other pulsars with essentially the same characteristics are noted, particularly J1208-6238. It has been suggested as a possible low-field magnetar in the literature (\cite{Clark2016}) and is second (first) in the distance ranking of J1846-0258 (J1119-6127) after J1119-6127 (followed by J1846-0258). PSR J1513-5908 (in the composite SNR MSH 15-52), J1640-4631, and J1930+1852 follow in the ranking of J1846-0258, and J1640+4631, J1614-5048, and J1513-5908 do so in the ranking of J1119-6127.

\item Panel 5 of Figure~\ref{chapter_4_figure: figure8} shows that few locations of the MST showing $B_{lc} \sim 100$ G or beyond and no detected gamma-ray pulsars yet. These regions become of special interest for future searches. In particular, those near J1231-5113 (already detected) in the MST appear as promising potential targets.
A few neighboring pulsars to the outlier J1231-5113, at the end of this branch, also show a relatively large $B_{lc}$ with a similar range of spin-down power and other variables, in comparison to {\it Fermi}-LAT pulsars. Likewise, PSR J1915+1616 and J2129+1210B, located at the end of the nearby branch, are of interest. Again, note that both have $B_{lc}>10^3$ G, and $\dot E>10^{33}$ erg s$^{-1}$. 

\item The detection of the radio pulsar PSR J2208+4056 by the {\it Fermi}-LAT despite its low $\dot{E}$ has been ascribed by \citet{Smith2019} to a possible case of favorable geometry. If this is the case, it may remain isolated in the MST, which appears close to the main trunk. Its closest neighbors (J0532-6639, J0502+4654, and J1848-0123) call for attention to test this.

\item Others pulsars-of-interest regarding their possible detection in gamma-rays maybe J1818-1607 and J1550-5418. These lie in the magnetar branch, where no other detected {\it Fermi}-LAT pulsar is located \citep{Li2017}. The latter authors established a {\it Fermi}-LAT integrated upper limit for J1550-5418 and attempted folding, finding no signal. However, these pulsars have similar properties to other pulsars already detected by {\it Fermi}-LAT (ordered by distance, J1208-6238 occupies the third position, followed by J1119-6217, both of which were detected by {\it Fermi}-LAT).
In the case of the pulsar J1550-5418, among its ten closest pulsars according to the calculation of the Euclidean distance, the pulsars J1734-3333, J1746-2850, and J1726-3530 are in the sixth, eighth, and ninth positions, respectively. They are located in the lower part of the MST, with a high density of {\it Fermi}-LAT pulsars, although they have not yet been detected in gamma rays. The distance ranking of these two magnetars is uncommon compared to others: other magnetars in the branch have neither a {\it Fermi}-LAT pulsar nor other pulsars located in high-density areas of {\it Fermi}-LAT detections in the first positions of their distance ranking.
\item The pulsars J1842-0905 and J1457-5902, and J1413-6141 and J1907+0631 are the closest neighbours  to the pulsar wind nebulae J1745-3040/PWN B1742-30(1) and J0007+7303/PWN CTA 1, respectively. The latter is an outlier of the PWN population (see above). Thus, their neighbors are interested in testing whether this pulsar parameter space region is prone to producing observable nebulae.

\item Finally, nodes with higher degrees, particularly those connected to the main trunk or located at the ends of significant branches, are suggested for individual examination. While their positions in the MST are partly determined by the spanning nature of the tree, some of these nodes (as identified in Chapter~\ref{chapter5}) also serve as key structural points, potentially revealing subclass distinctions.

\end{itemize}

We remind that the MST is constructed solely from intrinsic pulsar properties and does not incorporate spatial position or distance. 
As a result, it may cluster together sources with similar emission characteristics regardless of their actual location. 
For instance, one of the pulsars highlighted (J0532–6639) is located in the Large Magellanic Cloud. This suggests that, despite its extragalactic origin, its emission properties could be intrinsically similar to those of some Galactic pulsars included in the analysis.

\section{The Pulsar Tree web}
\label{chapter4: pulsartreeweb}

The Pulsar Tree web\footnote{\url{http://www.pulsartree.ice.csic.es}} accompanying this paper contains visualization tools and data to produce all plots and go beyond what has been presented in this paper. It allows the readers to gather information regarding MST localization, $P\dot{P}$ comparison, and distance ranking. Among the functionalities included already, it can locate a given pulsar of the sample in the MST, $P \dot{P}$, $PC_1 PC_2$-diagrams and on any other diagram using the variables adopted in this paper; identify all properties of the given pulsar and all neighboring nodes both in the MST; zoom around a given portion of the MST (or on any of the other diagrams); obtain tables of the properties of the nodes in the region of interest; obtain tables of the distance ranking for any pulsar, and more. 
The continuation of the research led to the development of new functionalities for the Pulsar Tree web, specifically related to the binary pulsar population (see Chapter~\ref{chapter6} for details).

\section{Discussion}
\label{chapter4: discussion}

We have examined the pulsar population in a manner different from the usual $P\dot{P}$ diagram. Instead of considering just $P$ and $\dot{P}$, we used a set of 8 variables as proxies for the intrinsic physical properties of all pulsars. While these variables all depend on $P$ and $\dot{P}$, the variance of the population is not fully contained by the variance of the latter quantities. $P$ and $\dot{P}$ are not equivalent to $PC_1$ and $PC_2$. Distance ranking (or visualizations) based only on $P$ and $\dot{P}$ may hide interesting connections and mislead our intuition. We subsequently computed the Pulsar Tree using an adequately scaled Euclidean distance, discussing its properties and how the different classes of pulsars find their ordered place within it.

Based on a more comprehensive set of variables, this more accurate source definition yields a multidimensional problem that is challenging to solve using traditional techniques. We have seen how objects such as the MST can be robust solutions within the framework of graph theory. 
Additionally, we have demonstrated that, under suitable conditions, the MST provides a meaningful visualization of the pulsar population, reflecting the underlying physical relationships.

The MST approach offers applications beyond what we have described above. For instance, advanced analysis of the MST, including clustering, centralization, betweenness, and closeness, can illuminate physical connections and link pulsars to one another. 
In Chapter~\ref{chapter5}, we show a methodology for separating the MST using the betweenness centrality estimator (discussed in Section~\ref{chapter2: bet_cen}). The goal is to cluster the pulsar population from a graph theory context and provide a technique to predict the emergence of new classes of pulsars.
The method used here can also be generalized to consider other variables in the distance definition and the PCA, and then in the MST done with them, allowing different problems to be treated. For instance, where the focus is on binary pulsars, see Chapter~\ref{chapter6}. Orbital parameters (orbital period, the projected semi-major axis, mass of the companion, etc.) are considered part of the distance definition. 
Chapter~\ref{chapter7} demonstrates the scalability of the MST as a separation technique and its predictability when dealing with various events, such as fast radio bursts.
Although the MST technique is not as widely used in astrophysics as in other fields, it can offer new perspectives in classification, clustering, source identification, and cross-correlation of source properties, including those of pulsars.

\chapter{Quantitative determination of minimum spanning tree structures: Using the Pulsar Tree for analyzing the appearance of new classes of pulsars}
\label{chapter5}

\section{Introduction}
\label{chapter5: intro}

In this chapter, based on the work \citet{MST-2}, we introduce a quantitative methodology to define relevant parts of the Pulsar Tree, primarily the trunk and what we will refer to as significant branches.
This method of clustering is generalizable to any spanning tree (and, of course, to any MST), as we will see in Chapter~\ref{chapter6}, where its application is demonstrated on an MST based on the MSPs within binary systems.
In particular, we aim to introduce a qualifier to identify which parts of the MST exhibit statistically significant differences, allowing us to observe them as independent clusters. This enables the analysis of these clusters based on their underlying physical properties, which are used to calculate the MST, and helps to identify potential subpopulations.
Furthermore, once we establish the trunk and significant branches, we can determine whether the resulting groups have evolved in previous iterations of the pulsar population.
To do this, we start with the ATNF v1.67 pulsar population and the physical properties we used in Chapter~\ref{chapter4} for its characterization (see Section~\ref{chapter4: variables} for a more detailed explanation of the variables used to calculate the Pulsar Tree and the catalog content). Note that this version contains 2509 pulsars with $\dot{P}>0$, 2242 of which are isolated, and 267 belong to binary systems.
To delimit these incarnations, we use the date label from the ATNF catalog as a filter, which refers to the date of pulsar discovery, and consider the historical evolution of the pulsar population. As a derivative of the latter study, we analyze how to determine, through defined evolutionary metrics, whether a new class of pulsars emerges in new data, future surveys, or new incarnations of pulsar catalogs.

\section{Clustering algorithm: Significant Branches}
\label{chapter5: method}

We introduce a clustering algorithm based on significant branches using the Pulsar Tree, the MST of the pulsar population of the ATNF v1.67, as seen in Figure~\ref{chapter_4_figure: figure4}. 
Its purpose is to separate the population on which the MST is built into distinct groups or clusters by dividing the graph into separate structures.

This approach includes a topological stage, where the most central part of the MST structure, denoted as the trunk due to its resemblance to a real tree, is identified using betweenness centrality (see Section~\ref{chapter2: bet_cen}) as an indicator of node centrality under certain topological conditions. Once the trunk is established, we can locate the branches as the components that extend from it. 

We then incorporate a statistical stage that considers the physical meaning underlying these branches. This approach enables us to select the branches that demonstrate a statistically significant difference among them, thereby positioning the clusters as potential distinct classes of pulsars. 

\subsection{Definition of the main trunk and branches}
\label{chapter5: conditions}

\begin{figure}
\centering
\includegraphics[width=\textwidth]{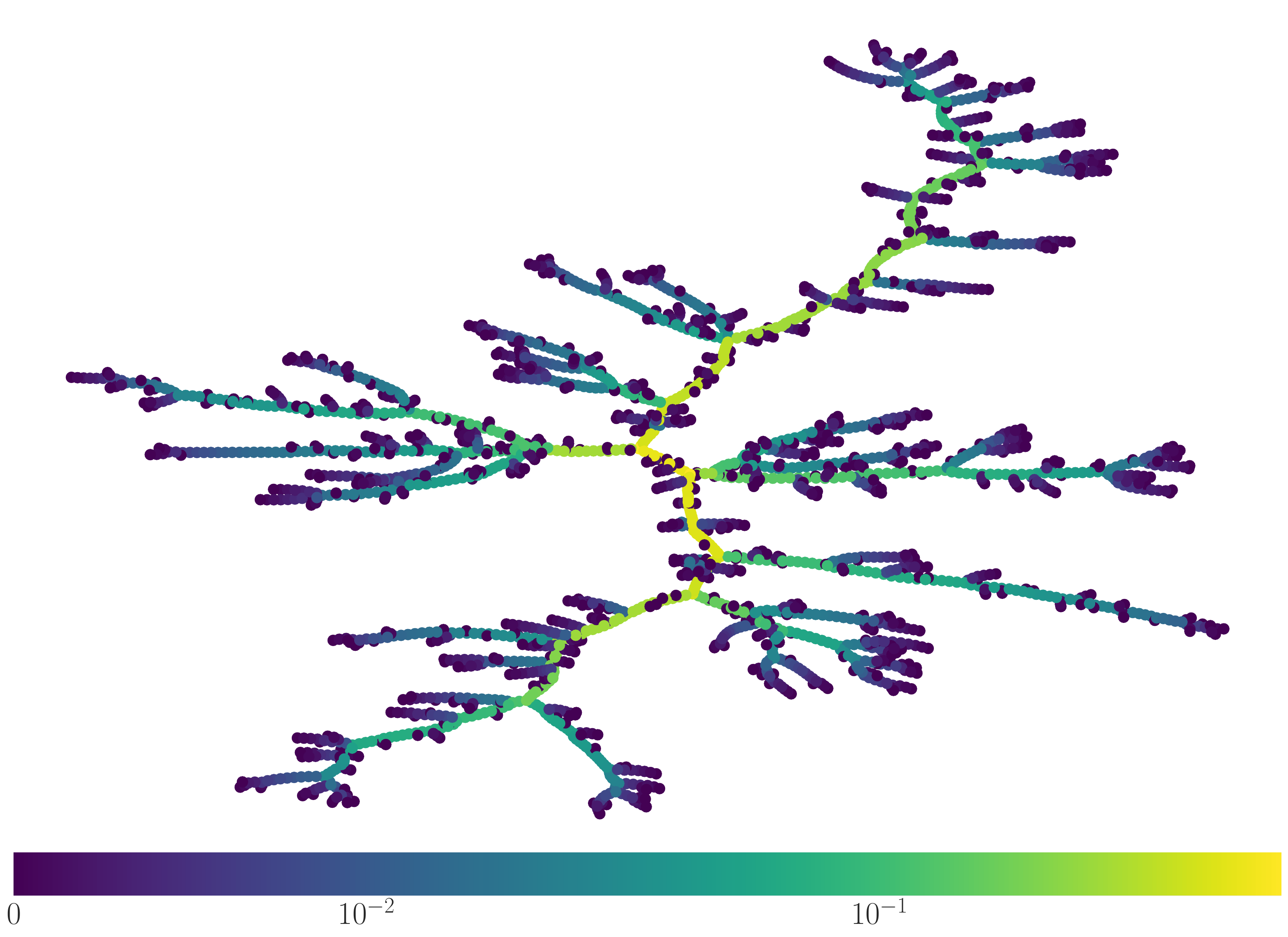}
\caption[A color-coded representation of MST according to the values of the betweenness centrality coefficient.]{A color-coded representation of MST(2509, 2508) according to Equation~(\ref{chapter_2_eq: betweenness_centrality}). The color intensity varies from zero (with the strongest color) for nodes located at the termination of the graph, which are nodes of degree 1, to nodes in the central part of the MST, where $C_B$ reaches its highest values, around 0.6. 
}
       \label{chapter5_fig: BetCen_ModelVersion}
\end{figure}

Figure~\ref{chapter5_fig: BetCen_ModelVersion} shows the Pulsar Tree after the application of Equation~(\ref{chapter_2_eq: betweenness_centrality}).
As explained in Section~\ref{chapter2: bet_cen}, the value of the betweenness centrality estimator ($C_B$) enables us to establish a hierarchical structure in the MST by identifying nodes of higher centrality.
In other words, the higher the $C_B$, the greater the centrality in the MST, and therefore, making the node eventually part of the trunk.

\begin{figure}
\centering
  \includegraphics[width=0.35\textwidth]{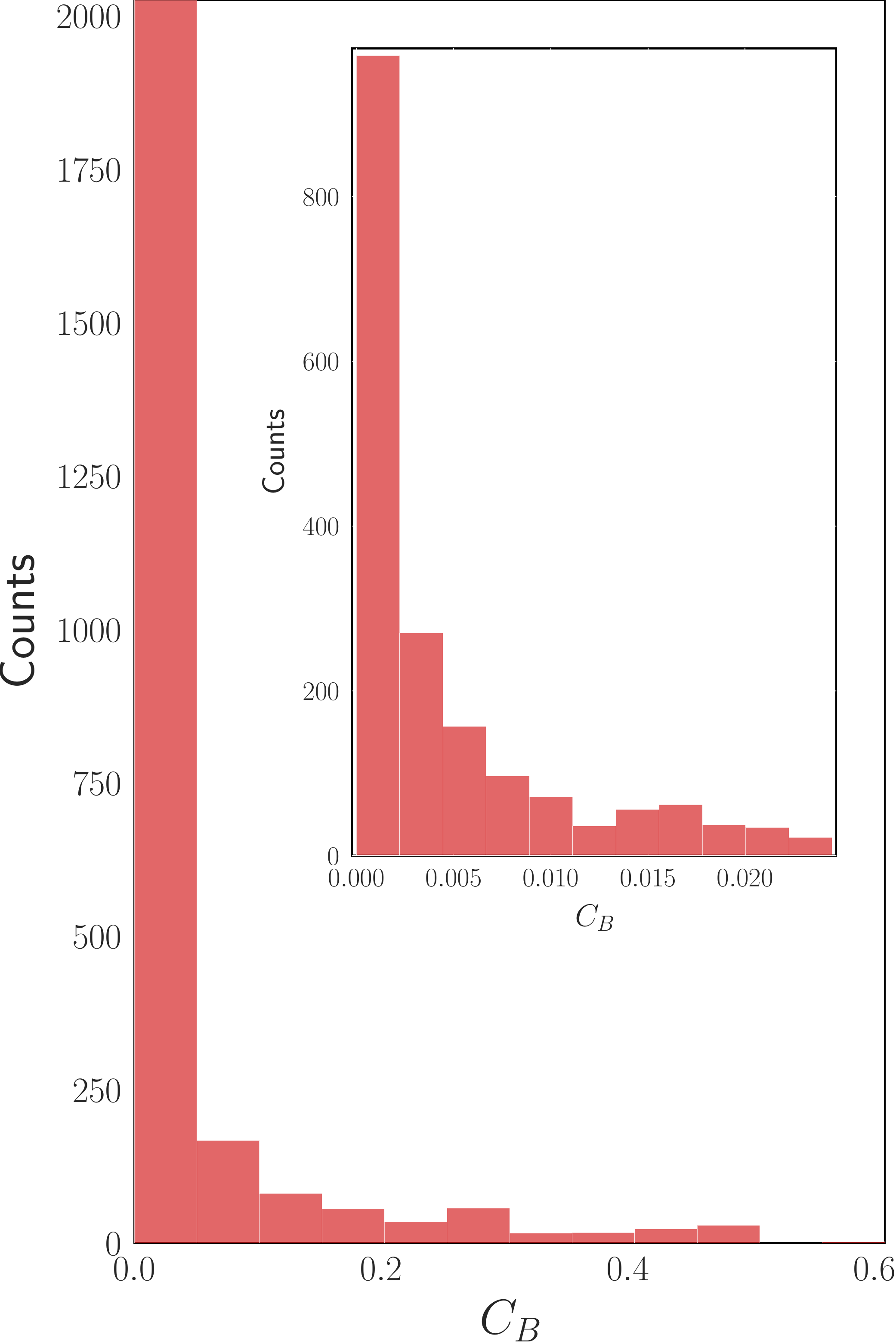}
  \caption[Distribution of the betweenness centrality values.]{Distribution of the $C_B$ shown in Figure~\ref{chapter5_fig: BetCen_ModelVersion}. 
  A zoom-in of the lower 70\% of the data is displayed in the top right.
}
  \label{chapter5_fig: Distributions_betweenness}
\end{figure}

Accordingly, Figure~\ref{chapter5_fig: Distributions_betweenness} shows the distribution of the $C_B$ computed for all nodes in the Pulsar Tree.
The distribution is strongly right-skewed, with a few outliers exhibiting extreme values. As shown in Figure~\ref{chapter5_fig: BetCen_ModelVersion}, and consistent with the definition of betweenness centrality, these outliers correspond to the most central nodes of the graph. 
We interpret that from these will come the trunk of the Pulsar Tree.

We shall use an appropriate technique for asymmetric distributions to delimit the nodes that will form the trunk (see, e.g., \cite{Tukey}):

\begin{eqnarray}
[Q_{1}-k\times IQR];[Q_{3}+k\times IQR]
~.\label{chapter_5_eq: outliers}
\end{eqnarray}

The use of quartiles (${Q_{1}}$, ${Q_{3}}$) to identify central values and outliers is validated for both symmetric and asymmetric distributions: They do not assume anything about the mean or the standard deviation, and their use is compatible with distributions of positive rightward skew, as ours. Furthermore, any asymmetric distribution is more robustly defined by the median as a measure of central tendency and the interquartile range (IQR= $Q_{3}-Q_{1}$) as a measure of its dispersion, as both are less sensitive to extreme values. 

For our case, the right-hand side in Equation~(\ref{chapter_5_eq: outliers}), together with the usual value of 3 for $k$, sets the condition for a node to be considered belonging to the trunk (i.e., an outlier of the distribution of $C_B$). We designated each of these as a potential trunk node (pTN).
This condition is necessary, but not sufficient, for a node to be considered part of the trunk. In addition to being formed by pTNs (something that relates to a bare-eye identification as the trunk in Figure~\ref{chapter5_fig: BetCen_ModelVersion}), we need to provide a criterion to establish where the trunk starts and ends.
To that aim, we shall request a topological condition: the trunk must be a path, i.e., a sequence of consecutive nodes containing no duplicates (see Section~\ref{chapter2: graph_props}), and we require that it starts and ends at nodes, called contour nodes (CNs), whose degrees are greater than 2. 
In this way, we ensure that nodes give rise to substructures known as contour branches (CBs) at the terminations of the trunk. These can be related to the mental image of branches opening from the trunk in a real tree.
To prevent CBs formed just by a few nodes (noise), we define a significant threshold ($\alpha_s$) so that the number of nodes in these branches must exceed this lower limit. 
Under the validation of a given $\alpha_s$ for the consideration of CBs, all the possible CNs are identified, and the potential trunk\footnote{The number of potential trunks: $\# \mathrm{pTs} = \mathrm{CNs} \times (\mathrm{CNs}-1)/2$} (pTs), as the path that pair these CNs, can be computed.
Concluding, for each pT, we have a set of branches starting from every node of degree larger than 2 (including the CBs) and having $\alpha_{s}$ as the minimum number of nodes. 

\subsection{Significant branches}
\label{chapter5: conditions-branch}

Once the branches are obtained, we adopt a general conceptual definition: A significant branch is a group of nodes departing from the main trunk containing at least a $\alpha_{s}$ of the graph and can be statistically distinguished from other branches. 
To differentiate the branches, i.e., to select the significant branches, we will use the Kolmogorov-Smirnov (KS) statistics to define the significance threshold and immediately measure whether one branch distinguishes itself statistically from the others.
The KS test compares two distributions under study by measuring the maximum distance between their empirical cumulative distribution functions, as described in \citet{KS-test3, KS-test4, KS-test5}.
The KS test does not assume any form of distribution beforehand, making it a non-parametric test that can be used for any distribution.
The aim here will be to determine if we can reject the null hypothesis ($H_0$): the distribution of the properties of the nodes of two given branches is consistent with them being drawn from the same parent distribution.
We will seek to reject this $H_0$ at a 95\% confidence level (CL) or better.
When this happens, we shall establish that whatever distance is used to compute the weights between nodes and form the MST separates branches whose nodes are drawn from statistically distinct parent populations.

Therefore, each set obtained will consist of a trunk and its significant branches. 
Depending on the conditions, the significant attributes of the branches can be adaptable to each case study. 
This will allow us to focus our work on a given clustering and discard those that do not accord with the proposed objectives (see, e.g., Section~\ref{chapter6: significance_branches}).

\subsection{Significant Branches in the Pulsar Tree}
\label{chapter5: branches_pulsarTree}

To compute the significant branches derived after applying the clustering algorithm to the Pulsar Tree, we consider the distribution of $C_B$ shown in Figure~\ref{chapter5_fig: BetCen_ModelVersion} and the outlier condition defined in Equation~(\ref{chapter_5_eq: outliers}). 
From this, we identify that approximately 10\% of the nodes in the Pulsar Tree qualify as pTNs.

\begin{figure}
\centering
\includegraphics[width=\textwidth]{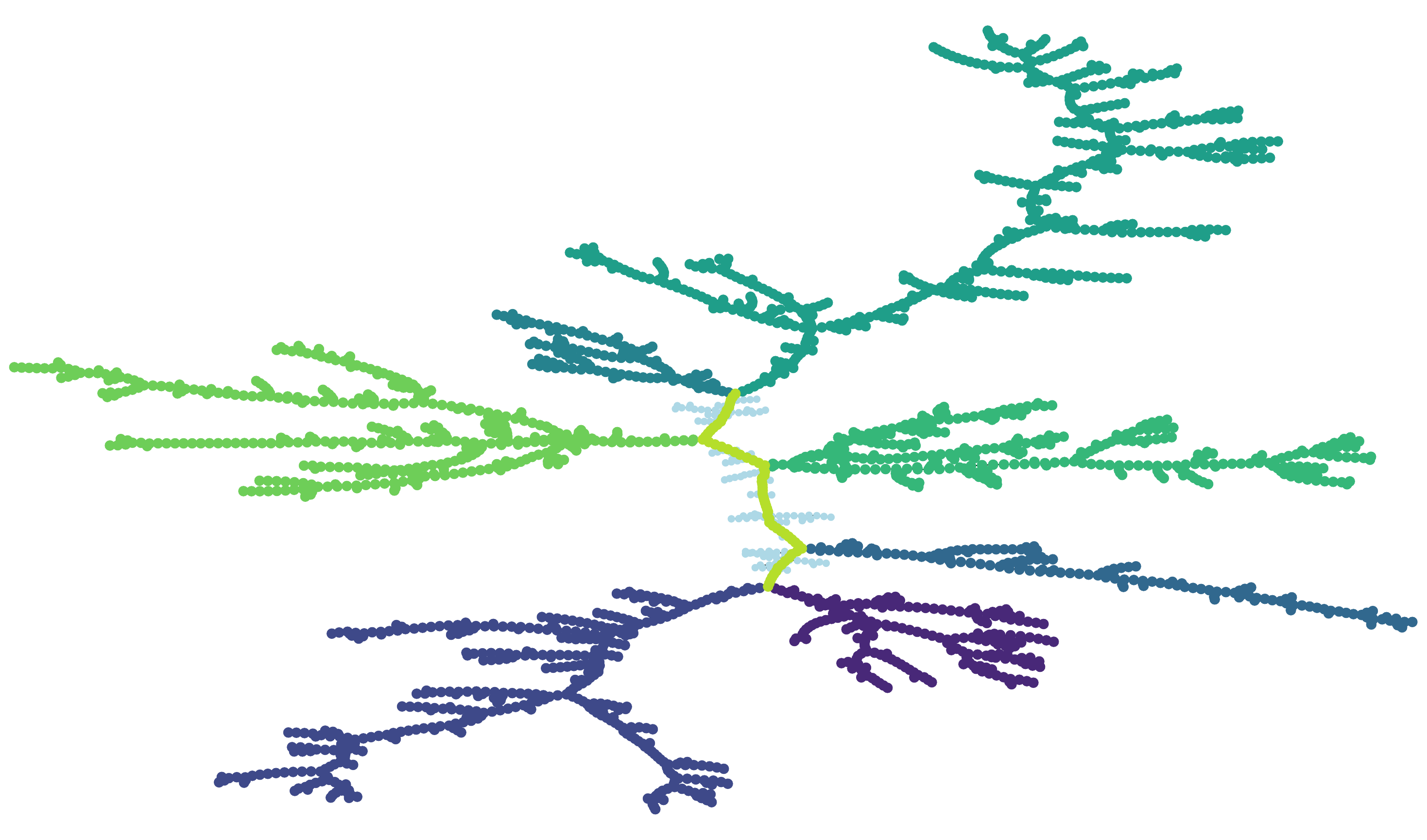}
\caption[MST according to the significant branches.]{MST according to the significant branches of the pulsar population of the ATNF v1.67 after applying the clustering algorithm described in Section~\ref{chapter5: conditions}. 
These have at least 5\% of the total population size; following the branches clockwise (from 12:00), they have the following pulsars: 585, 364, 157, 224, 444, 427, and 119, respectively. 
The MST nodes depicted in the main part are 64 and represent the trunk. Note that the remaining light blue (125) nodes belong to non-significant branches and are considered the noise of the trunk. 
}
       \label{chapter5_fig: MST_Branches_V167}
\end{figure}

To ensure statistical robustness in comparing the branches obtained, we impose an $\alpha_s\sim5\%$ of the population. This value ensures that each branch contains enough values for the KS test to operate reliably.
Additionally, we prioritize a set of significant branches with comparable sizes to improve the interpretability of the KS test results and minimize the impact of sample size heterogeneity. 
The seven significant branches of the Pulsar Tree are shown in Figure~\ref{chapter5_fig: MST_Branches_V167}.

\begin{figure}
  \centering
  \includegraphics[width=1\textwidth]{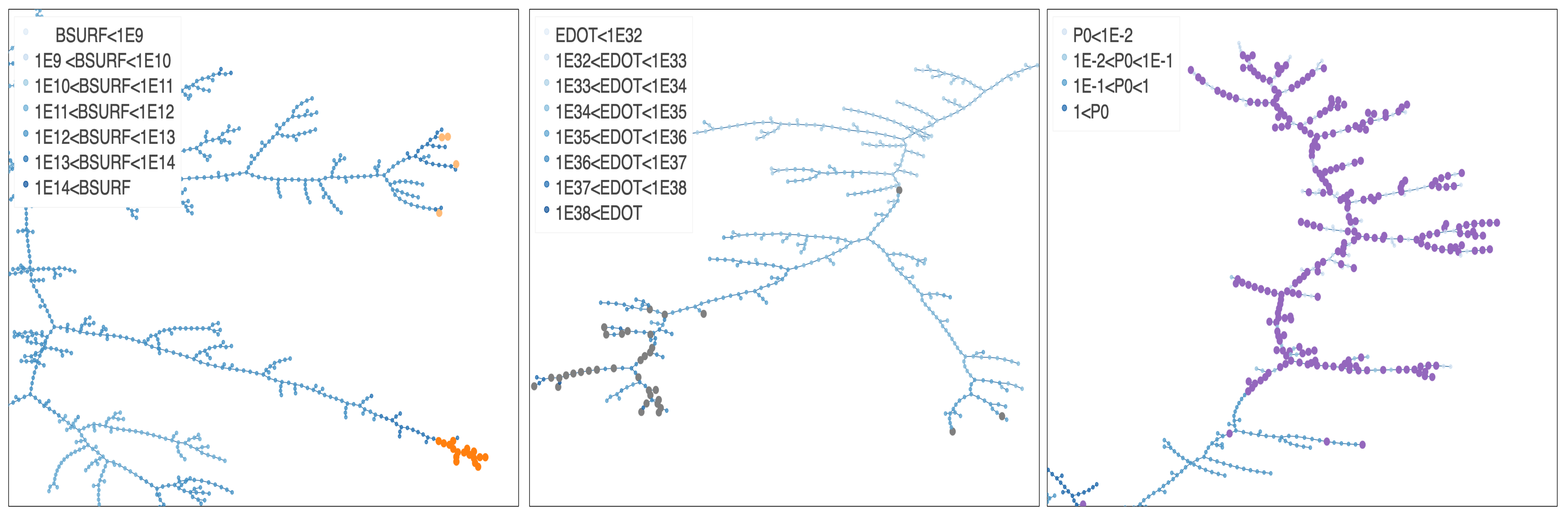}
  \caption[Three examples of the ranges of variables for some of the significant branches of the Pulsar Tree]{
Three examples of the distribution of variables for some of the significant branches of the Pulsar Tree can be extracted from the online tool at {\url{http://www.pulsartree.ice.csic.es}}. 
We suggest examining these examples directly to access additional functionalities, zoom in, and view individual values for each pulsar. 
The left panel shows the distribution of the surface magnetic field in the middle branches. It increases towards the outskirts of the MST and ends in low-field magnetars (light orange) and classical magnetars (orange). 
The central panel shows the spin-down power distribution of the bottom branch, with those having a pulsar wind nebula detected in TeV \citep{HESSPWNe} (grey) noted. 
The right panel shows the spin period distribution in the upper branches of the MST, where binaries (purple) and long-period pulsars are located. 
  }
  \label{chapter5_fig: pt-examples}
\end{figure}

The significant branches identified resemble those one would point by hand if looking at the Pulsar Tree.
When commenting on the MST as a descriptive tool, some branches were already used as examples in Section~\ref{chapter4: MST_alerting_tool}.
The top and bottom branches of Figure~\ref{chapter5_fig: MST_Branches_V167} roughly correspond to binary pulsars and to the more energetic isolated pulsars of the sample (the youngest and those with the highest magnetic field at the light cylinder pulsars are located towards the end of this branch), respectively. 
Rightwards departing branches are characterized by increasing values of the surface magnetic field, ending with magnetars at their extremes.
The outgoing leftward branch contains the oldest isolated pulsars.
Figure~\ref{chapter5_fig: pt-examples} shows examples of the distribution of variables for some of the significant branches of the Pulsar Tree, as can be extracted from the Pulsar Tree web\footnote{\url{http://www.pulsartree.ice.csic.es}} introduced in Section~\ref{chapter4: pulsartreeweb}.

\begin{figure}
  \includegraphics[width=1\textwidth]{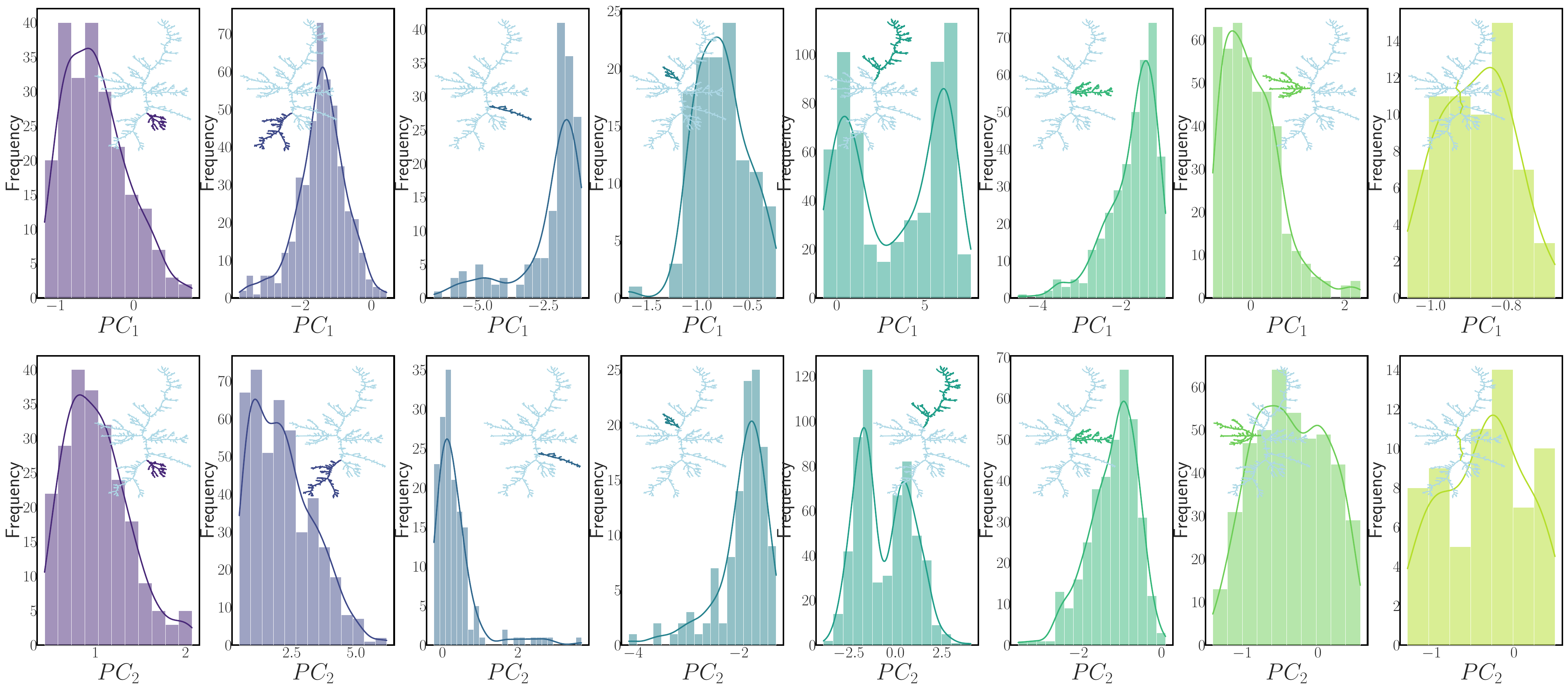}
  \centering
\caption[The distributions of the $PC_1$ and $PC_2$ according to the significant branches.]{The distributions of the $PC_1$ and $PC_2$ according to the significant branches seen in Figure~\ref{chapter5_fig: MST_Branches_V167}, as noted in the inset. 
All panels are plotted in the corresponding scale for $PC_1$ and $PC_2$. 
The last panel in each row shows the distributions for the trunk.}
       \label{chapter5_fig: dist-var1}
\end{figure}

The significant branches, by definition, separate different physical properties.
To emphasize this point, Figure~\ref{chapter5_fig: dist-var1} compares the $PC_1$ and $PC_2$ (see Section~\ref{chapter4: pca-pulsars}) corresponding to each significant branch, showing differences in their distribution. 

We also note the case of the fifth panel from the left in Figure~\ref{chapter5_fig: dist-var1}, showing the distribution of the significant branch containing the binary pulsars. 
This significant branch is large and encompasses both pulsars in binaries and isolated ones. As such, it exhibits a double-peak structure in the distribution, which is not present in any other significant branch of the MST.
Looking at the latter, the long-period, isolated pulsars are located at the bottom part of the significant branch.
This significant branch can also be separately examined by focusing on the two smaller branches (73 and 45 pulsars), which are notably located in its lower part.
None of these branches exceeds $\alpha_s$ and, according to the definition, cannot yet be classified as individually significant branches.
However, analyzing them through the KS test, they reject $H_0$ concerning the rest of the significant branches as well as each other, over the distributions of all the variables used in this work.  
We expect that in further studies, a higher number of pulsars will likely join these branches into one, generating an additional significant branch, or that the number of nodes on each branch will increase to make them individually significant.

\begin{figure}
  \includegraphics[width=\textwidth]{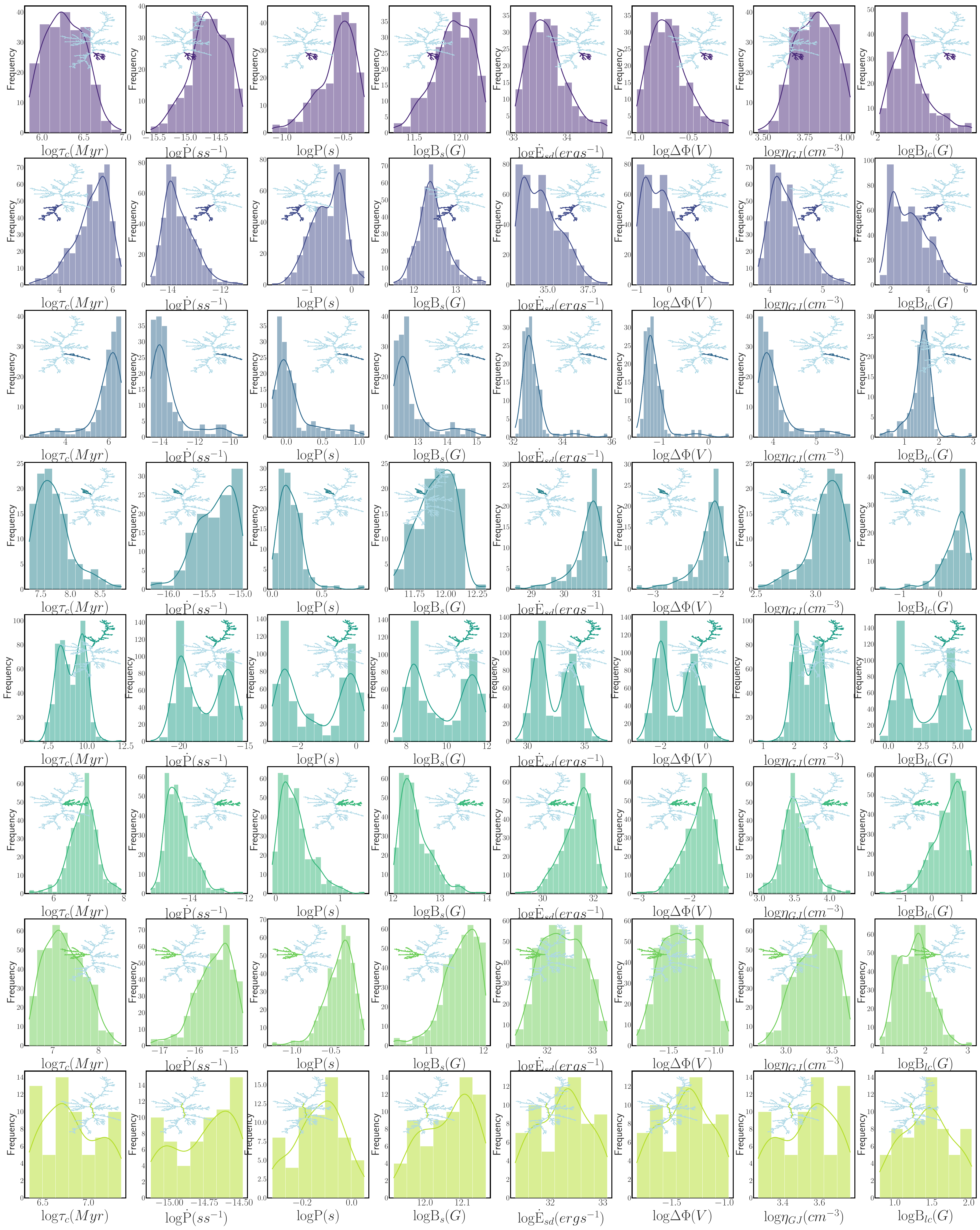}
  \centering
\caption[The distributions of the variables considered according to the significant branches.]{The distributions of the logarithm of the eight intrinsic variables considered to compute the significant branches (1 for each row) seen in Figure~\ref{chapter5_fig: MST_Branches_V167}, as noted in the inset. 
The variables shown are the spin period ($P$) and spin period derivative ($\dot P$), the characteristic age ($\tau_c$), magnetic field at the surface ($B_{s}$), the magnetic field at the light cylinder ($B_{lc}$), spin-down energy loss rate ($\dot{E}_{sd}$), surface electric voltage ($\Delta \Phi$), and Goldreich-Julian charge number density ($\eta_{GJ}$).
All panels are plotted on the corresponding scale for each magnitude. The last row shows the distributions for the trunk.
}
       \label{chapter5_fig: dist-var2}
\end{figure}

Figure~\ref{chapter5_fig: dist-var2} shows the distributions of each underlying physical magnitude considered according to the significant branches of the Pulsar Tree seen in Figure~\ref{chapter5_fig: Branches_trunk_v167}.

\subsection{Monte Carlo validation of significant branches}
\label{chapter5: simulations}

To evaluate the statistical robustness of the observed branches, we conducted a series of Monte Carlo simulations to test whether a similar level of separation could arise purely by chance. Specifically, we generated synthetic groupings by randomly partitioning the pulsar population into seven groups, ensuring that their sizes match those of the original significant branches. These random groupings are entirely independent of the MST structure and serve as a null model for comparison. 
We repeated this process $10^{5}$ times, and for each simulation, we applied the KS test to assess the separability of the random "branches" based on the physical properties under consideration. 
The results show that achieving comparable separation by chance is highly unlikely. 
In none of the $10^{5}$ simulations were all seven random branches simultaneously distinguishable using the KS test. 
In fact, in 55\% of the simulations, even a separation between just two of the seven random groups could not be statistically supported; the $H_0$ could not be rejected in those cases.

\section{Evolution of the MST with the pulsar population }
\label{chapter5: epochs}

\begin{figure}
  \includegraphics[width=1\textwidth]{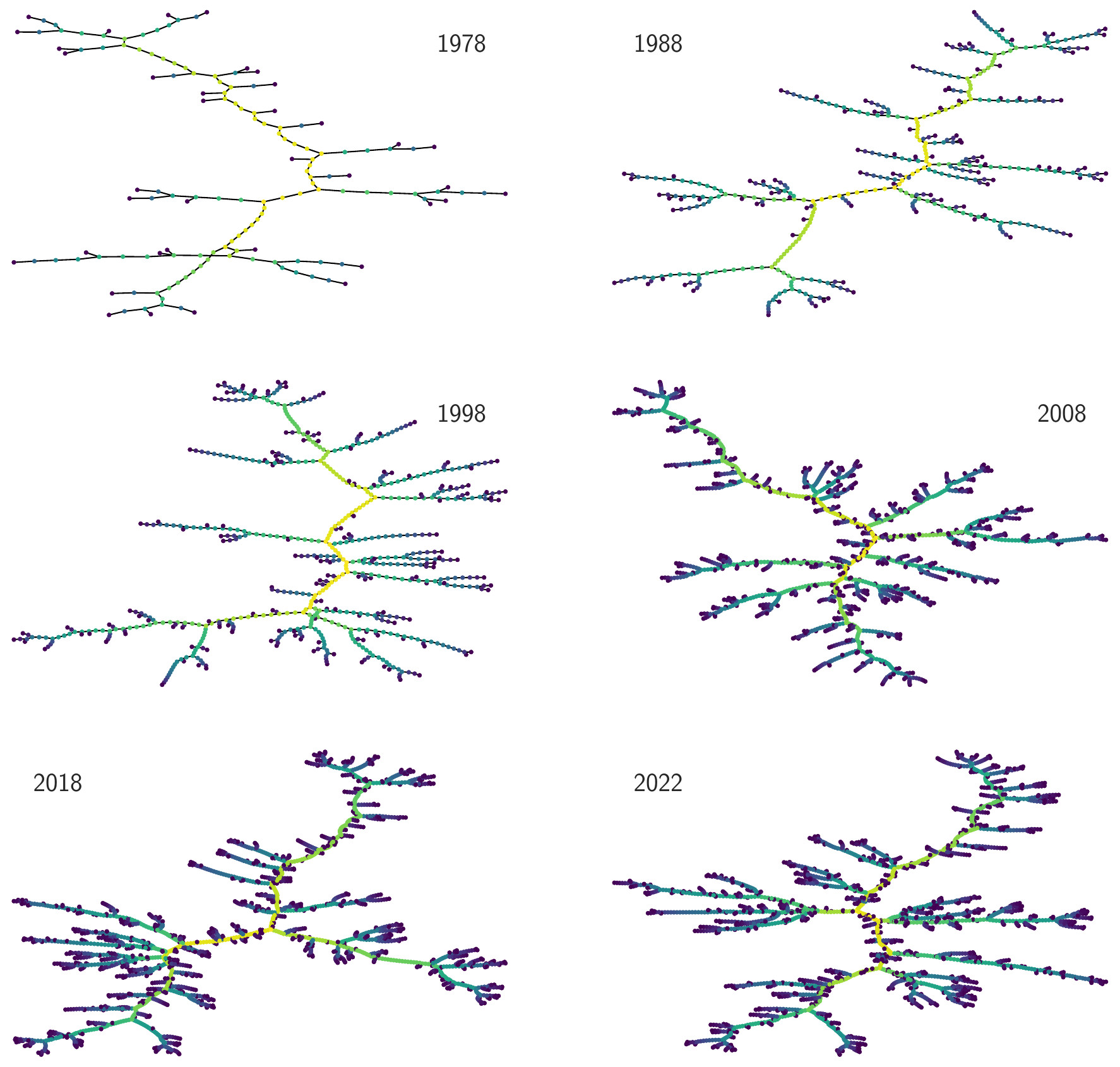}
  \centering
\caption[Representation of the $C_B$ values in the MST]{Representation of the $C_B$ values found after the application of Equation~(\ref{chapter_2_eq: betweenness_centrality}) in the MSTs of the pulsar population throughout history, with nodes representing all pulsars known up to the different years according to the legend (up to December 31st of the prior year, and up to the date of version 1.67 of the ATNF catalog in March 2022). 
Similar to Figure~\ref{chapter5_fig: BetCen_ModelVersion}, and on the same scale, the color map represents the highest $C_B$ values in yellow colors and increases the darkness of the paint to dark purple for those values that reach zero.
}
       \label{chapter5_fig: Betweenness_MST_versions}
\end{figure}

We now consider how the clustering algorithm applies to the evolution of the pulsar population throughout history.
According to ATNF v1.67, we shall consider the pulsars known up to 1978, 1988, 1998, 2008, 2018, and 2022, resulting in sets containing 147, 439, 662, 1660, 2267, and 2509 pulsars.
The clustering algorithm is applied individually to these subpopulations.
Figure~\ref{chapter5_fig: Betweenness_MST_versions} shows the MST computed based on each subpopulation, considering the physical variables described for this work, with the $C_B$ values noted as in Figure~\ref{chapter5_fig: BetCen_ModelVersion}.

\begin{figure}
  \includegraphics[width=1\textwidth]{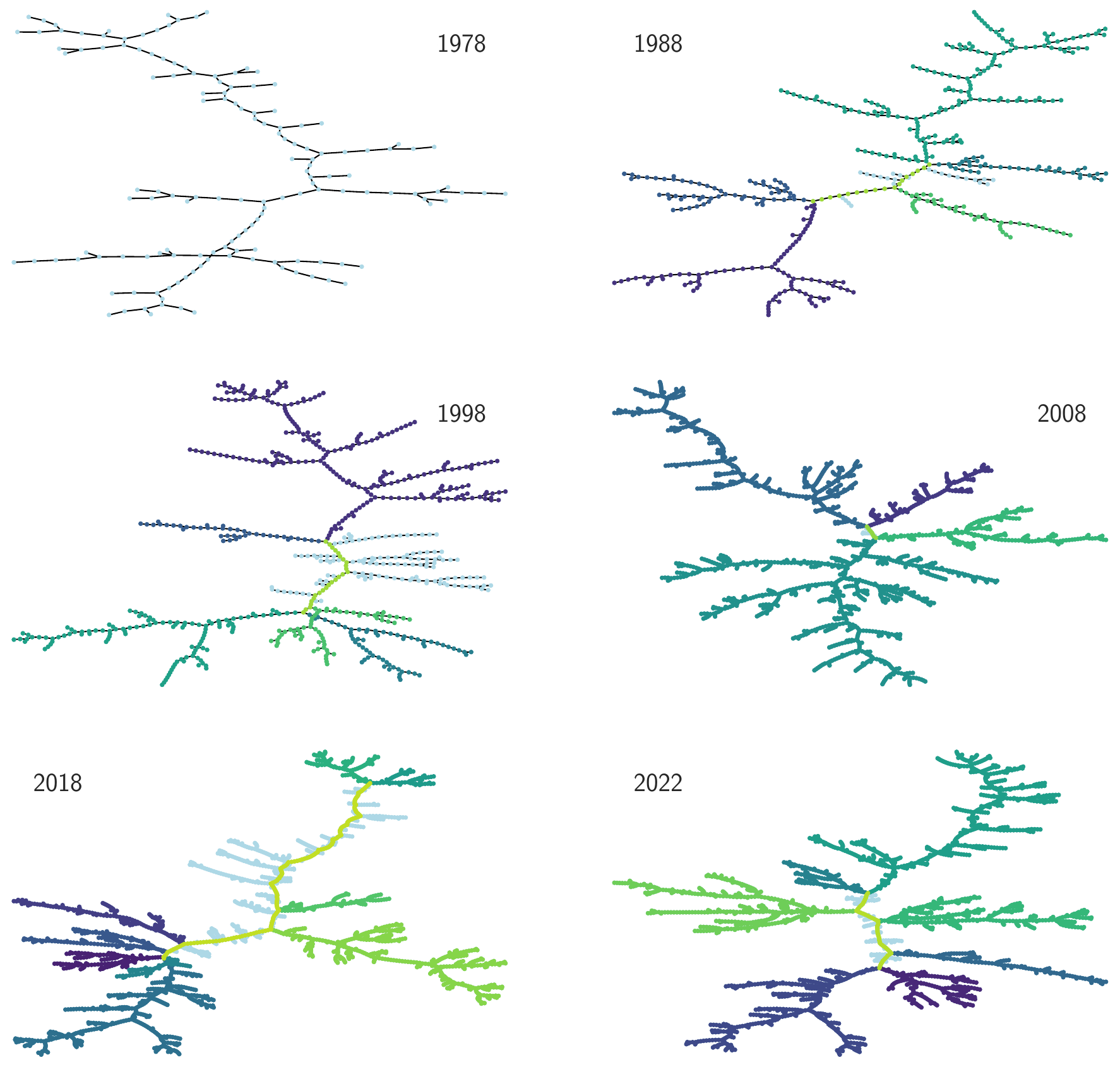}
  \centering
\caption[Significant branches of the pulsar population throughout history.]{Significant branches and trunks for each MST corresponding to each subpopulation filtered by the time indicated in the legend. 
No part of the MST in 1978 is indicated because the application of Equation~(\ref{chapter_5_eq: outliers}) does not find nodes that can represent the trunk (there are no outliers in the distribution of $C_B$). 
The different branches for each epoch are assigned a random color (MSTs are not color-coded for intercomparison). 
The trunk always shows the most intense green color. From 1988 to 2022, the number of significant branches was 5, 5, 4, 9, and 7, respectively.
}
       \label{chapter5_fig: Branches_trunk_versions}
\end{figure}

Equation~(\ref{chapter_5_eq: outliers}) establishes the following pTNs for each subpopulation: 0\%, 7\%, 9\%, 9\%, 13\%, and 10\%.
Due to the size of the subpopulations, we reduce the minimum branch size to consider it significant. 
That is, we set $\alpha_s\sim3\%$ to obtain significant branches in all subpopulations. 
Thus, the number of significant branches extracted from the population of pulsars throughout history, that is, for each subpopulation, is: 0, 5, 5, 4, 9, and 7. 
It should be noted that the number of significant branches observed in each subpopulation is comparable despite introducing up to 20 times more pulsars than those known until 1978. 
The significant branches for each subpopulation are shown in Figure~\ref{chapter5_fig: Branches_trunk_versions}.
Interestingly, due to the small sample size, the MST of the population of pulsars known up to 1978 does not reveal any outliers in its distribution of $C_B$. Correspondingly, we observe that the MST for this subpopulation has fewer substructures than those with more pulsars.

As stated before, these MSTs are still frames of the knowledge gathered from the pulsar population known up to each of the years quoted, where even whole classes of pulsars were undiscovered.

\section{Pulsar classes appearance: \textit{togetherness} and \textit{growth} of the Pulsar Tree's significant branches}
\label{chapter5: history}

\begin{figure}
  \includegraphics[width=1\textwidth]{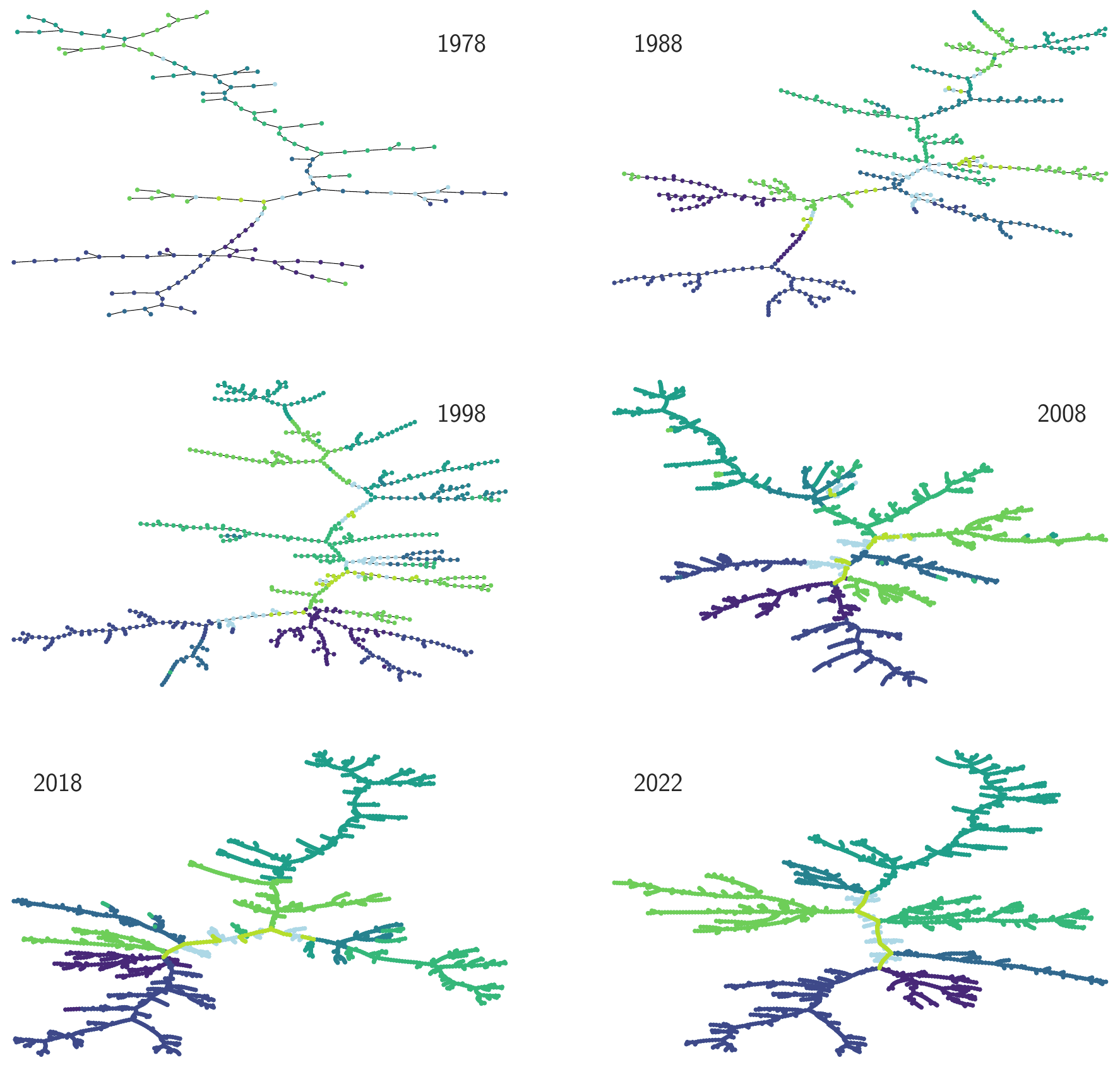}
  \centering
\caption[Projection of the branches into the MSTs.]{Projection of the significant branches seen in the Pulsar Tree (the last panel corresponds to that shown in Figure~\ref{chapter5_fig: MST_Branches_V167}) into the MSTs corresponding to each epoch, following a chronological order as indicated within each panel. 
}
       \label{chapter5_fig: Branches_trunk_v167}
\end{figure}

Figure~\ref{chapter5_fig: Branches_trunk_v167} illustrates how the nodes of each significant branch observed in the Pulsar Tree were distributed in previous versions of the MSTs according to the pulsars known at the time. 
Over time, all the significant branches of the Pulsar Tree converge, as most of the nodes in these branches were also previously connected. 
To quantify this evolutionary process, we define a measure called {\it togetherness}, which shows the percentage of nodes belonging to a given branch (or the trunk) that were together at earlier times, considering the significant branches produced in the MSTs of the other subpopulations.
In cases where the given branch cuts into smaller pieces (e.g., its nodes populate two significant branches of the previous incarnation of the catalog), we shall also consider the group containing the most substantial number of nodes, as this is necessarily more representative of the final significant branch. 
This involves starting from a significant branch in the pulsar population of the ATNF v1.67, where we shall count the number of nodes that existed and were located together in a branch of the previous version (2018). 
Starting from the latter set, we checked how many existed and were already together in the catalog before it (1998), and so on.

\begin{figure}
  \includegraphics[width=0.8\textwidth]{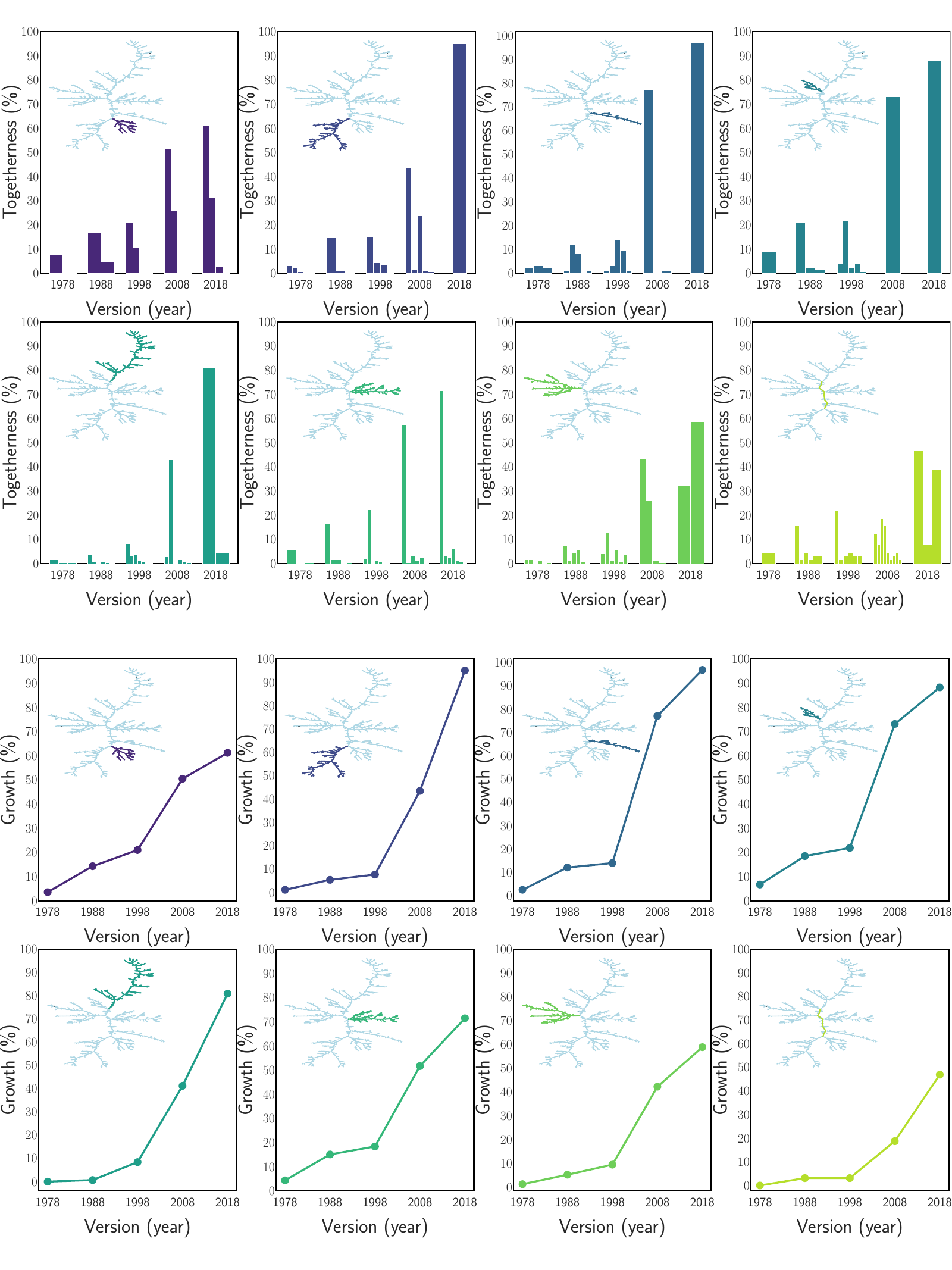}
  \centering
  \caption[Percentage of nodes concerning the number of nodes of each significant branch: \textit{togetherness} and \textit{growth}.]{
  First/Second row: The \textit{togetherness} is shown, representing the percentage of nodes related to each significant branch of the Pulsar Tree, according to the marking in the inner MST depicted in each panel, which was already together in an MST at earlier times in 2022. When the given significant branch of the Pulsar Tree appeared in different branches in the earlier versions of the Pulsar Tree (see Figure~\ref{chapter5_fig: Branches_trunk_v167}), different bars are shown; whereas, when the nodes in a significant branch were located altogether in a single group, only a bar appears.
  Third/Fourth row: The \textit{growth} is shown, representing the percentage of nodes related to each significant branch of the Pulsar Tree, according to the marking in the inner MST depicted in each panel, that remain together throughout history after appearing. 
  To obtain these percentages, we have taken into account the best representation (the largest bar for each branch in 2018, as seen in each panel of the first row) and analyzed how this group has grown from the previous time, repeating the process until 1978.
}
  \label{chapter5_fig: togetherness_growth}
\end{figure}

We also introduce a measure called \textit{growth}, which, unlike \textit{togetherness}, focuses solely on the pulsars within the significant branch under analysis. 
Specifically, \textit{growth} tracks whether these pulsars remain grouped over time across all subsequent MSTs, as additional pulsars are incrementally added until the branch reaches its final size.
Figure~\ref{chapter5_fig: togetherness_growth} shows that all branches in the Pulsar Tree not only join groups of pulsars that were earlier separated (\textit{togetherness} in the first row) but also have a core group of pulsars that stay together along all subpopulations (\textit{growth} in the second row).

\section{Conclusions}
\label{chapter5: conclu}

A new class of pulsars is conceptually a new set with distinct properties. 
In this chapter, we introduce a methodology based on the graph theory-motivated pulsar tree and the underlying physical properties of the pulsar population, which enables the study of this phenomenon from a clustering perspective. 
In general terms (since it may find applications for other problems beyond pulsars), we have introduced a methodology to utilize the MST to qualify the nodes that form it. 
The logic is as follows:

\begin{enumerate}
\item Produce the MST based on the Euclidean distance of the variables containing the full variance of the population as determined from PCA.
\item Compute betweenness centrality for all nodes, produce its distribution, and determine which nodes are outliers. 
\item Set the significance threshold, defined as the minimum percentage of the total population of pulsars for which the branches are statistically distinct (measured by a KS test).
\item If having former incarnations of the sample, with fewer nodes (in our cases, those provided by the historical discovery of pulsars), compute \textit{togetherness} and \textit{growth} to see convergence stability and establishment of significant branches.
\end{enumerate}

The former methodology enables us to quantitatively define the trunk and significant branches of any tree, for which the nodes have statistically different distributions of the selected variables. 
In the Pulsar Tree, using the intrinsic variables, the significant branches can be directly associated with "classes" or, at the very least, with connections among the respective nodes, which quantitatively distinguish them from others in different locations within the tree.
Studying population evolution throughout history allows us to see how our nodes, which were initially non-connected, become connected, while others remain linked all the time. 
In the future, we may see new branches emerge, with little to no projection onto the ATNF v1.67 pulsar catalog, indicating a new class.
We may also see, as described, non-significant sub-branches develop into full-fledged significant ones.
The MST will provide a quantitative perspective on whether we see something "new" or just "more of the same" when new pulsars are discovered and whether currently unknown pulsars connect dislocated groups. 

This methodology has been applied to the population of MSPs in binary systems (see Chapter~\ref{chapter6}) to identify clusters that may suggest the existence of new pulsar classes or provide further insight into the behavior of known systems, such as spider pulsars. 
This is characterized not only by physical properties but also by binary parameters.

\chapter{Millisecond pulsars phenomenology under the light of graph theory}
\label{chapter6}

\section{Introduction}
\label{chapter6: intro}

This chapter is based on \citet{MST-MSPs} and continues the study that integrates graph theory (see Chapter~\ref{chapter2}) and the application of PCA (see Chapter~\ref{chapter3}) to the field of pulsar astrophysics.

Here, we specifically consider the millisecond pulsars (MSPs) population, i.e., weakly magnetized and rotating neutron stars with spinning periods shorter than 10 ms. They are usually hosted in tight binary systems in a low mass ($< 1 \, \mathrm{M_\odot}$) companion star. MSPs offer unique insights into stellar evolution, the interaction between magnetic fields and plasma transferred by the donor star, and particle acceleration from compact objects, particularly in binary systems (see Section~\ref{chapter1: MSPs} for more details and \citealt{Manchester2017, Papitto2022} for comprehensive reviews). By treating each MSP as a node, we shall compute the MST (see Section~\ref{chapter2: MST_graph}) of the MSP population and use it to describe the population as a whole and to identify individual pulsars that may warrant further investigation due to their unique attributes or positions within the graph. We shall specifically examine locations in the MST in which spider pulsars reside (e.g., \citealt{Eichler1988, Roberts2013, Roberts20172018, DiSalvo2023}), i.e., eclipsing radio pulsars in tight binary systems either with a non-degenerate main-sequence companion with a mass in the range $\sim$0.1–0.8 M$_\odot$ (redbacks, RBs) or with a $\lesssim0.06$ M$_\odot$ semi-degenerate companion (black widows, BWs).  We also consider transitional millisecond pulsars (tMSPs, \citealt{Papitto_tMSPSectionBook}) that exhibit dramatic state changes, i.e., go from rotation-powered to accretion-powered and vice versa, on timescales as short as a few weeks (see Section~\ref{chapter1: SpiderPulsars} for more details). Driven by the method introduced in Chapter~\ref{chapter5}, which allows for dividing an MST into distinct parts based on specific variables, we explore how to classify MSPs into different groups. Finally, we shall implement an algorithm to search for the position of pulsars within the graph, even when some properties have not yet been measured. This will allow us to predict ranges for specific variables and their characteristic phenomenology.

\section{The sample variance and the MST}
\label{chapter6: sample_variables}

\subsection{Sample selection}
\label{chapter6: sample}

The sample used is taken from the Australia Telescope National Facility catalog, v2.1.1 (ATNF v2.1.1, see \cite{ATNF-Catalog}) imposing that pulsars have a spin period in the millisecond range ($P<10^{-2}$ s) and a positive spin period derivative ($\dot P>0 \ \, \mathrm{s \, s^{-1}}$). The latter allows us to calculate the intrinsic variables derived from $P$ and $\dot P$ as described in Section~\ref{chapter6: variables_PCA}. A total of 218 pulsars result from these cuts. Of these, 43 are confirmed as BWs, 9 of them in globular clusters (see, e.g., \citealt{swihart_blacwidows, Linares2023, Freire2017, Lynch_2012, Douglas2022}), 16 as RBs, 2 of them in globular clusters (see, e.g., \citealt{strader_redbacks, Linares2023}), and two as tMSPs, J1023+0038 \citep{Archibald2009} and J1227-4853 \citep{Bassa2014}. Note that tMSP J1824-2452I \citep{IGRJ1824_INTEGRAL} is excluded from the sample as it does not have a $\dot P$ measurement, although it will be analyzed in Section~\ref{chapter6: IGRJ1824}. These pulsars are listed in Table~\ref{chapter_6_tab: BW_RW_tMSP}. According to the third pulsar catalog by the Fermi Large Area Telescope (\textit{Fermi}-LAT, see \citealt{Fermi3PC}), 103 pulsars out of the 218 are found to be gamma-ray emitters.

\subsection{Variables and PCA}
\label{chapter6: variables_PCA}

\begin{figure}
  \includegraphics[width=\textwidth]{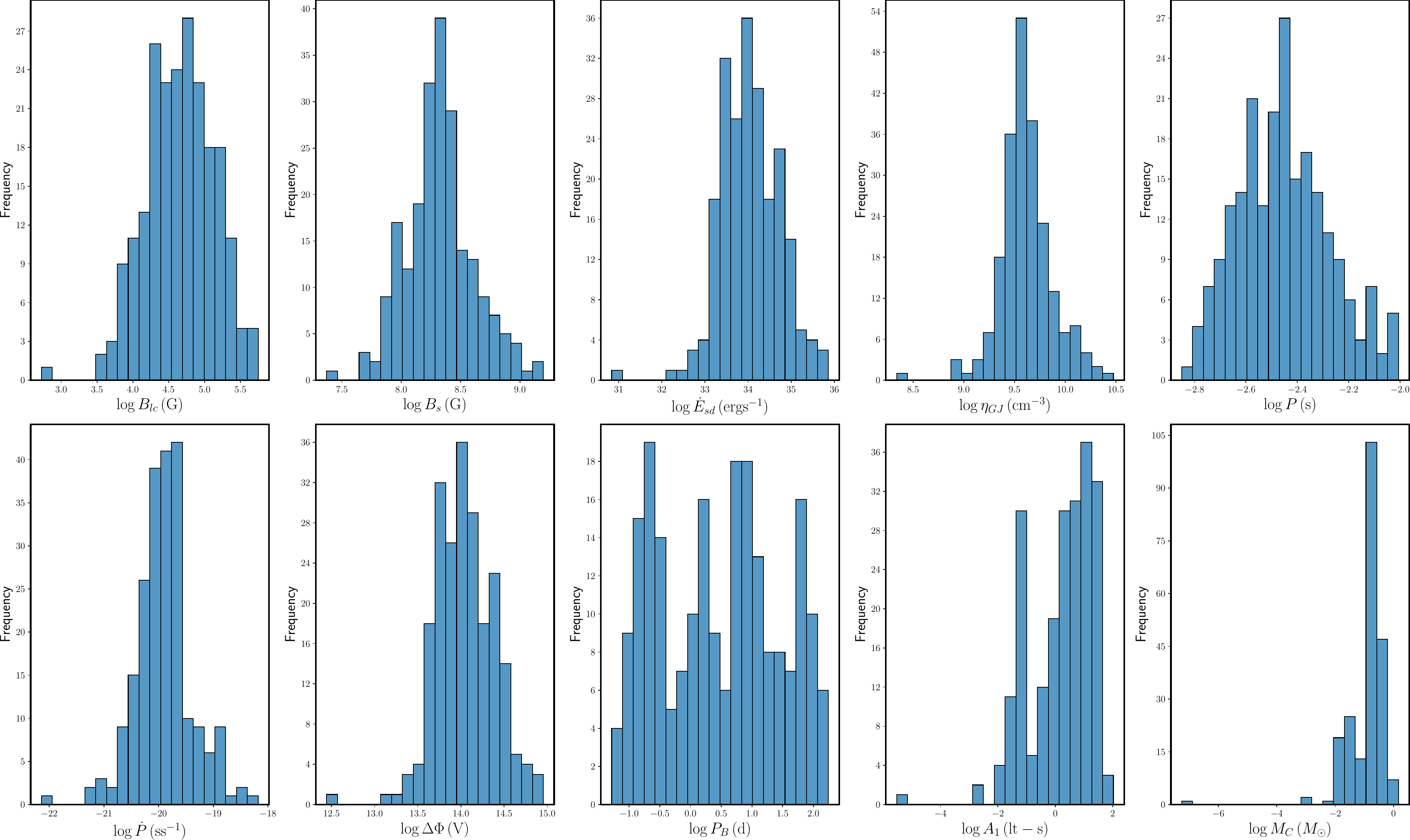}
  \centering
  \caption[Distribution of the properties considered for the MSPs population.]{Distribution of the logarithm of the 10 variables considered for the sample of 218 pulsars.}
  \label{distr_var}
\end{figure}

\begin{figure}
  \includegraphics[width=1\textwidth]{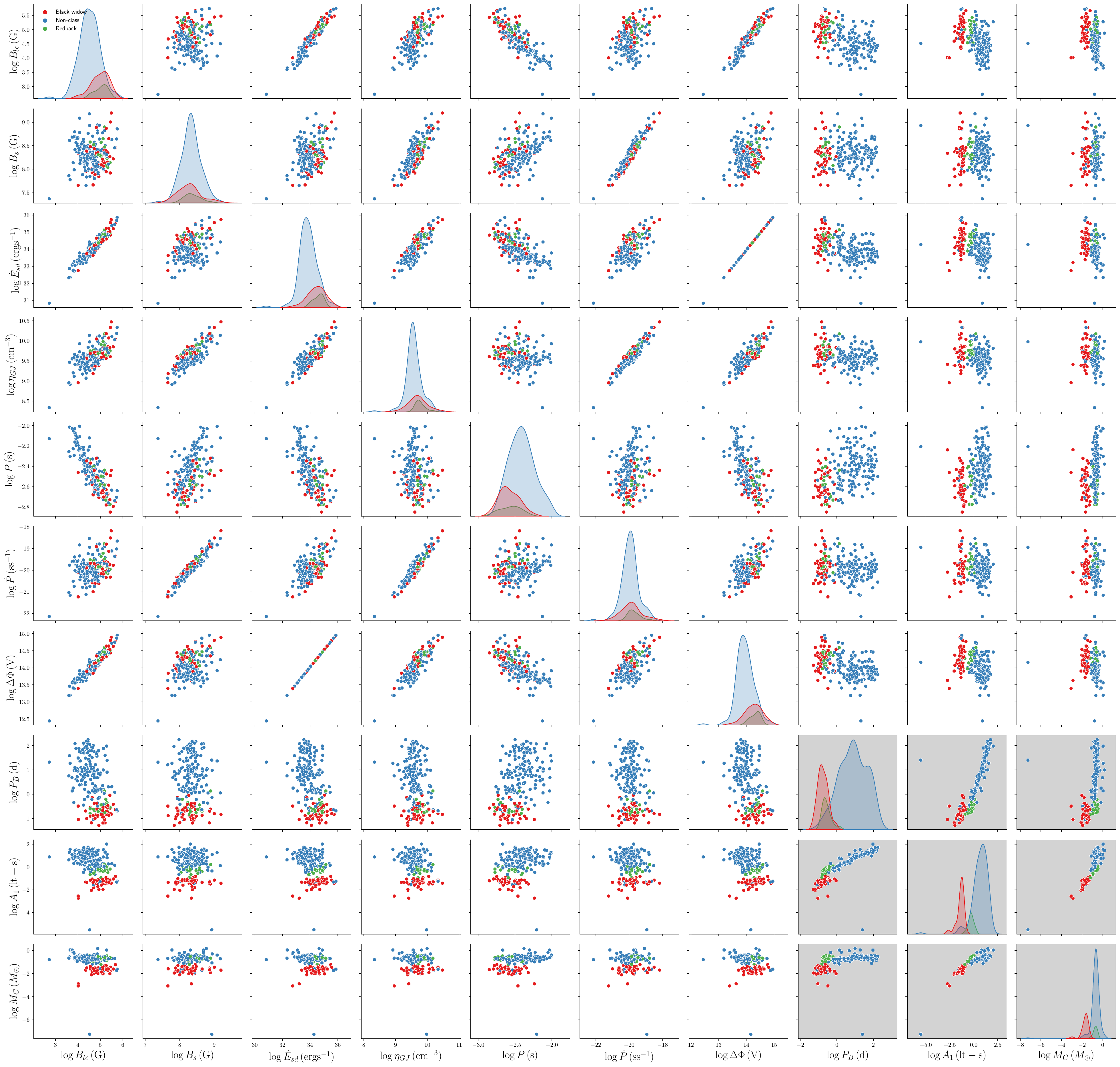}
  \centering
\caption[Cross-correlation of the properties considered for the MSPs population.]{
Cross-dependence of the 10 variables considered. BWs (in red) and RBs (in green) confirmed are separately noted. For visualization purposes, no distinction is made between MSPs residing in globular clusters, and the tMSPs are marked in green due to their behavior as RB. The pulsars not yet assigned to any of these classes are shown in blue. The main diagonal shows the distribution for each variable. The panels with the shaded background show the pairs formed with $P_{B}$, $M_C$, and $A_{1}$, which best differentiate the BWs from the rest.
}
  \label{chapter_6_figures: pair_plotclasses11v}
\end{figure}

Figure~\ref{distr_var} shows the logarithmic distribution of the spin period ($P$) and spin period derivative ($\dot P$), magnetic field at the surface ($B_{s}$), the magnetic field at the light cylinder ($B_{lc}$), spin-down energy loss rate ($\dot{E}_{sd}$), surface electric voltage ($\Delta \Phi$), Goldreich-Julian charge number density ($\eta_{GJ}$), binary period ($P_B$), projected semi-major axis of the orbit ($A_1$) and the median mass of the companion star for each system ($M_C$). We do not consider here the characteristic age ($\tau_c = P/ 2 \dot P$) because, in binary systems, additional torques imparted on the pulsar during accretion phases can render this estimate unreliable (see, e.g., \citealt{Kiziltan_2010, Tauris_2012, Jian2013}). The distributions are not normal (which is also the case for the set of distributions of the original variables, without logarithm), and due to this fact, we use the robust scaler to scale them (see Equation~(\ref{chapter_4_eq: robust_scaler})). Figure~\ref{chapter_6_figures: pair_plotclasses11v} shows how the $P_{B}$, $A_{1}$, and $M_C$ distribution distinguishes BWs from other classes. However, when none of these variables are used in the analysis, the distinction is no longer evident, and spider pulsars are seen as a more homogeneous group.

\begin{figure}
  \centering
  \includegraphics[width=\textwidth]{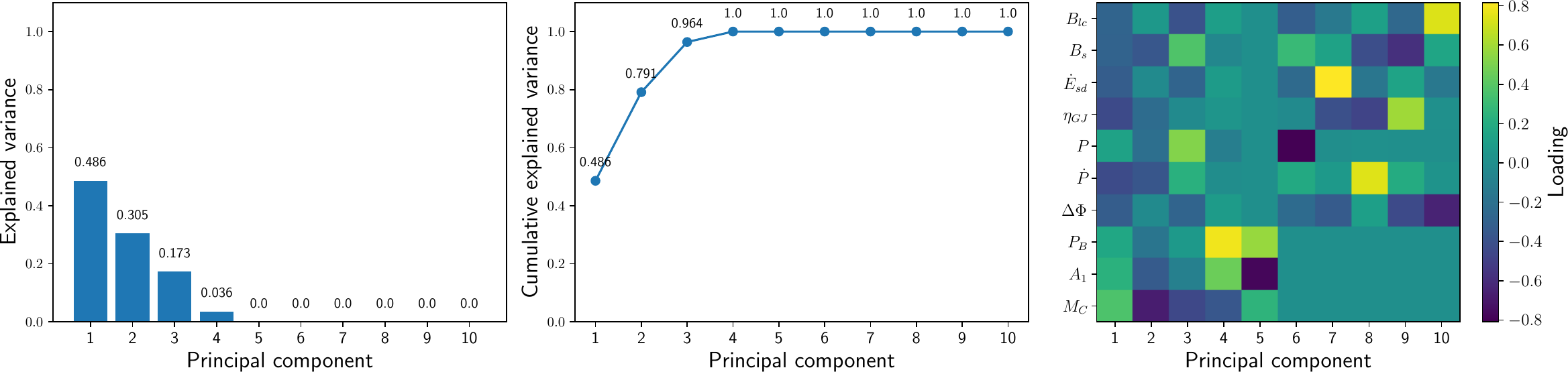}
  \caption[PCA results of the properties considered for the MSPs population.]{
  PCA results for the logarithm of the set of variables for the population of 218 MSPs, see Section~\ref{chapter3: visual_tools} for further explanations about PCA definitions.
  The left panel, called the scree plot, shows each PC's explained variance according to the covariance matrix's eigenvalues.
  Note that the covariance matrix calculates the relationships between pairs of variables, showing how changes in one variable are associated with changes in another. 
  It represents the amount of information contained in each PC.
  The central panel shows the cumulative explained variance by the new variables defined through the PCA analysis. 
  The right panel shows the "weight", also called loading, each variable has concerning each PC, indicating its contribution to the variance captured by that PC. 
  This value is the coefficient held in each eigenvector of the covariance matrix.
  Negative values imply that the variable and the PC are negatively correlated. 
  Conversely, a positive value shows a positive correlation between the PC and the variable.
  }
  \label{chapter_6_figures: PCA_10v}
\end{figure}

Figure~\ref{chapter_6_figures: PCA_10v} shows the results of the PCA analysis applied to these variables (see Chapter~\ref{chapter3} for more details).
In the left panel of Figure~\ref{chapter_6_figures: PCA_10v}, we observe how most of the variance is distributed in the first two PCs.
Likewise, the central panel shows that the whole variance is represented in a 4-dimensional space, i.e., we need the first 4 PCs to cover 100\% of the variance of the sample.
This relatively low number of PCs results from the fact that within the dipolar model used as a proxy (see Section~\ref{chapter4: variables}), all physical variables depend on $P$ and $\dot P$, and also that $A_{1}$ of a binary system is related to the $P_{B}$ and the $M_{C}$ (in this case through the total mass of the system, i.e., the sum of the masses of the pulsar and its companion) via Kepler's third law.
The PCA, as shown in the right panel of Figure~\ref{chapter_6_figures: PCA_10v}, does not highlight the dominant influence of key variables for the separation between pulsar spiders, such as the $M_C$, but provides a more complex classification where its dominance is balanced with other variables.
Technically, the four PCs needed to describe all the variance have similarly significant loadings in several variables (see the right panel of Figure~\ref{chapter_6_figures: PCA_10v}).

\subsection{MST}
\label{chapter6: MST}

\begin{figure}
\centering
\includegraphics[width=\textwidth]{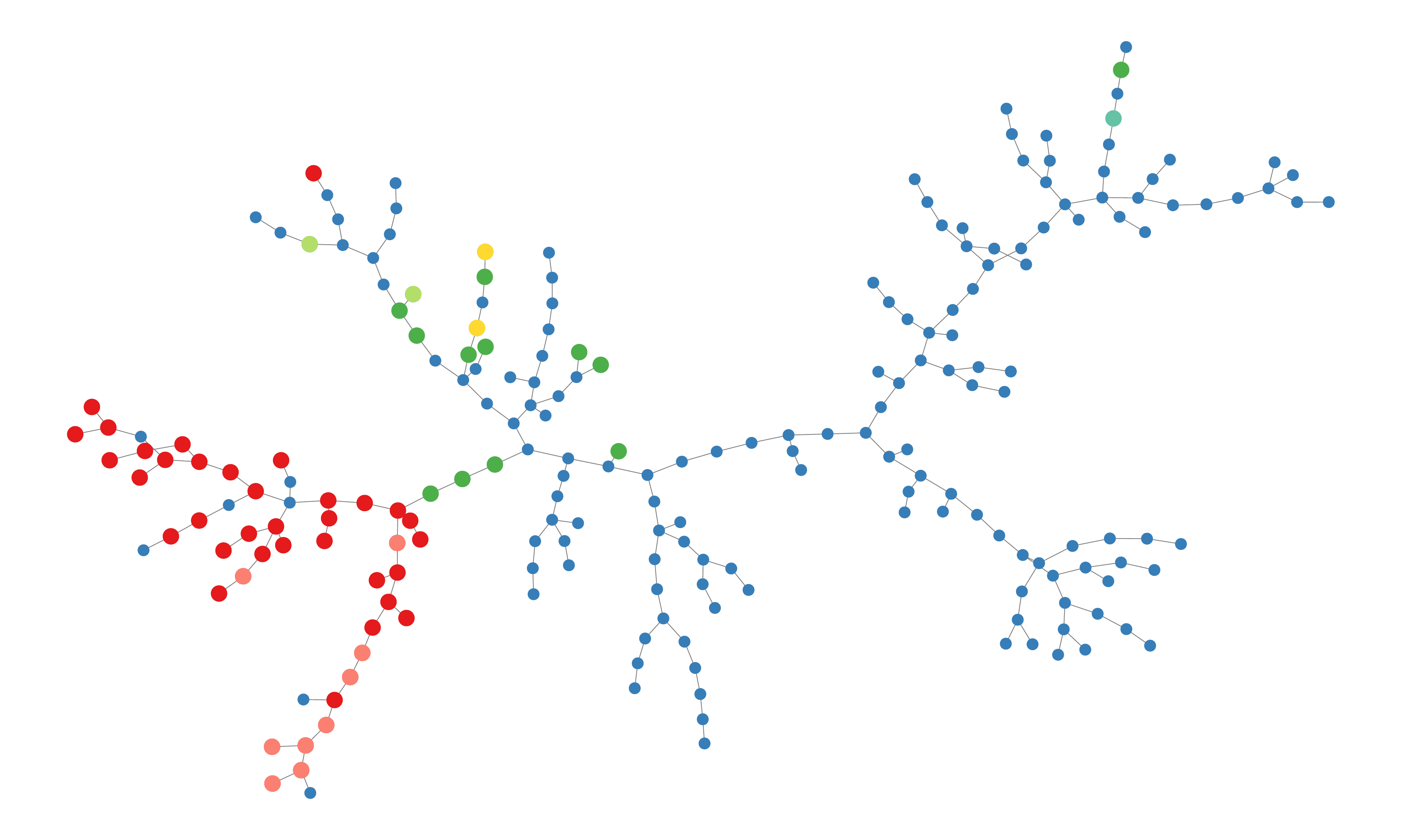}
\caption[The MST of the MSPs population.]{The MST of the binary pulsar population defined as $T(218, 217)$ based on the complete, undirected, and weighted graph $G(218, 23653)$ computed from the Euclidean distance among 10 scaled variables (or the equivalent 4 PCs that described their whole variance). 
Each node in the MST represents a pulsar. 
The MST notes separately confirmed BWs (red), RBs (green), and tMSPs J1023+0038 and J1227-
4853 (yellow), respectively. 
Also, the BWs and RBs in globular clusters (light red and light green, respectively) are highlighted.
The RBs J1622-0315 (green) and the RB candidate J1302-3258 (light teal) are also noted in the rightmost branch of the MST.
The unclassified ones appear in blue.
See Table~\ref{chapter_6_tab: BW_RW_tMSP} for more details.}
       \label{chapter_6_figures: MST_MSP}
\end{figure}

In Chapter~\ref{chapter2}, all the necessary concepts to calculate an MST are explained.
We define a Euclidean distance (see Equation~(\ref{chapter_2_eq: d_eucl})) over the variables or the PCs described in Section~\ref{chapter6: variables_PCA}. Note, as explained in Section~\ref{chapter4: variance}, that using the first four PCs ($\sim$ 100\% of the explained variance) produces the same MST as using all variables (and thus the results from the analysis that follows from the graph are the same) but is less demanding given the reduced dimensionality of the problem. With this Euclidean distance, we first compute a complete, undirected, and weighted graph $G(V, E)=G(218, 23653)$, with $|V|$=218 nodes and $|E|=23653$ edges, with a specific weight value defining each edge.
From that, we obtain the MST of this sample, $T(V, E')=(218, 217)$, with $|V|=218$ nodes and $|E'|=217$ edges, shown in Figure~\ref{chapter_6_figures: MST_MSP}, where each node represents a pulsar. 

\begin{figure}
\includegraphics[width=1\textwidth]{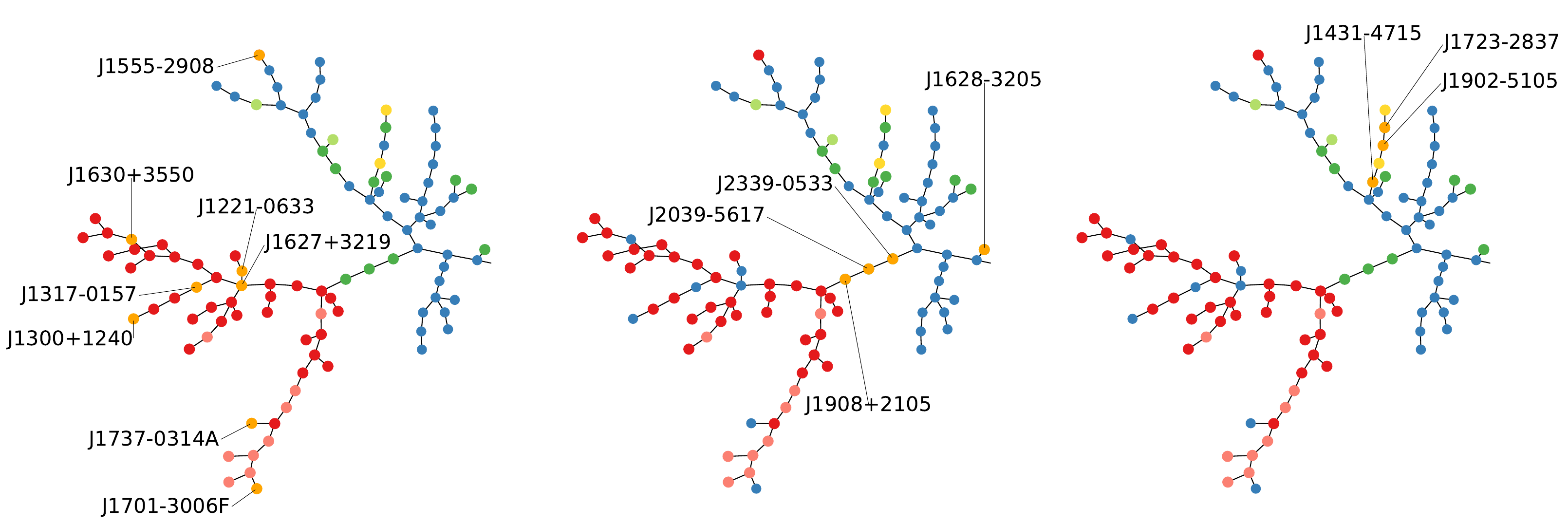}
  \centering
  \caption[Zoom of the leftmost part of the MST of the MSPs population.]{
Zoom of the leftmost part of the MST shown in Figure~\ref{chapter_6_figures: MST_MSP}, to observe more clearly the location of the pulsars (shown in orange) discussed in Section~\ref{chapter6: BWs_MST} (left panel), in Section~\ref{chapter6: RBs_MST} (central panel), and Section~\ref{chapter6: tMSPs_MST} (right panel).
The same color code of Figure~\ref{chapter_6_figures: MST_MSP} is shown, where the confirmed BWs appear in red, RBs in green, tMSPs in yellow, and the BWs and RBs in globular clusters in light red and light green, keeping the unclassified pulsars in blue.
 }
  \label{chapter_6_figures: MST_classes}
\end{figure}

The MST can separate BWs from RBs, with tMSPs appearing close in the same structure, as shown in Figure~\ref{chapter_6_figures: MST_MSP}. These tree´s regions are zoomed in Figure~\ref{chapter_6_figures: MST_classes}, where several pulsars that will be discussed next are highlighted.

\subsubsection{Black widows pulsars in the graph}
\label{chapter6: BWs_MST}

We will now focus on some specific pulsars following Figure~\ref{chapter_6_figures: MST_classes}.
Two of these are J1300+1240 and J1630+3550, which have not been explicitly classified as BWs yet but have been discussed by \citet{Yan2013, Sobey_2022}.
A possible formation scenario for J1300+1240, suggested by \citet{Yan2013}, is in low-mass, narrow-orbit bound binaries. Here, the initial point of departure is a low-mass binary pulsar with a very narrow orbit, such as the known BW J1959+2048. Due to their comparable velocity and rotation period to those of J1959+2048, J1300+1240 may also evolve in this scenario.
J1630+3550 is an MSP in a binary system of 7.6h orbiting a companion with a minimum mass of 0.0098 M$_{\odot}$. 
These parameters can indicate a BW as is stated in \citet{Lewis_2023}.
From the MST graph, we can see that they are located in the BW region. 
We can further consolidate their potential categorization as BWs by observing the ranking of the nearest pulsars based on the Euclidean distance derived from the defined variables, which can be obtained using the Pulsar Tree web\footnote{\url{https://www.pulsartree.ice.csic.es/}}.
Note that the ranking provides a distance-ordered list of one pulsar relative to all others. In contrast, the MST provides a global manner of optimizing the total pathway, considering all distances, to connect all pulsars.
Only the nearest neighbor to any given pulsar will have an immediately recognized position within the MST; as such, the closest neighbor will always be one of the pulsars connected to the one of interest in the MST.
For a deeper discussion and examples regarding the differences between the Ranking and the MST, see Appendix \ref{Appendix_4: nearness-comment} for a more detailed explanation.
J1300+1240 has nine pulsars among its 10 nearest neighbors, all of which have already been classified as BWs. The remaining one is J1630+3550, whose 10 nearest neighbors are also BWs.

Among the interesting sources highlighted in Figure~\ref{chapter_6_figures: MST_classes} are J1317-0157 and J1221-0633, classified as BW systems by \citet{Swiggum_2023}, having tight orbits, low-mass companions, and exhibiting eclipses.
In both cases, their nearest neighbor ranking contains nine confirmed BWs except for J1627+3219.
J1627+3219's position in the MST, underscored by its ranking, where eight of its ten nearest neighbors are confirmed BW pulsars, supports its categorization in this group. 
This assertion is further bolstered by \citet{Braglia2020}, who identified a candidate optical periodicity, necessitating verification through subsequent observations. 
Recent research by \citet{Corcoran2023} also underscores the potential BW nature of this pulsar. Still, the lack of detected radio pulsations highlights the need for further studies to fully understand its nature.

We also pinpoint J1737-0314A, which is likely a BW system according to the timing solution provided by \citet{Pan_2021b}.
Looking at its ranking, its categorization as a BW is favored since it contains nine confirmed BWs among its 10 nearest neighbors, the remaining being J1701-3006F.
In addition, the latter, J1701-3006F, considering its mass limit and orbital features, is also favored as a BW, as outlined in \citet{Freire_2005}, although eclipses have not been observed. 
However, this absence could be due to a low orbital inclination \citep{Vleeschower_2024}.
On the other hand, its nearest neighbor ranking shows six BWs, with the closest being J17001-3006F; see also \citet{King2005}.
Both J17001-3006E and J1701-3006F appear connected in the MST in Figure~\ref{chapter_6_figures: MST_classes}, filling the end of a branch that contains many of these cases.

Notably, J1555-2908 is the only confirmed BW away from the rest in the MST.
In the work by \citet{Kennedy_2022}, extensive analysis was performed using optical spectroscopy and photometry.
This study, combined with gamma-ray pulsation timing information alongside a companion mass of 0.06 M$_{\odot}$, concluded that J1555-2908 lies at the observed upper boundary of what is typically classified as a BW system.
Note that its nearest neighbor, J1835-3259B, is not a BW.

Of interest to all sources discussed in this section, we remark that no other non-BW pulsar, except J1908+2105 discussed in the next section, has more than five confirmed BWs in its ranking. 

\subsubsection{Redbacks pulsars in the graph}
\label{chapter6: RBs_MST}

Figure~\ref{chapter_6_figures: MST_classes} shows that RB pulsars do not appear clustered like BWs but are mainly close to each other, with a small fraction near the border of BWs.
The caveat here is the smaller number of RBs known so far. 

The pulsar J1908+2105 is an interesting case within the RB class.
This one is right on the frontier with the part of the MST mostly populated by BWs and could be classified as such, mainly due to its minimum companion mass of 0.06 M$_{\odot}$ and its short $P_B$ of 0.14 days \citep{strader_redbacks}.
However, the pronounced radio eclipses of this system align it more closely with RBs (also see, e.g., \citealt{Linares2023, Deneva_2021}).
Note that ranking the 10 nearest neighbor pulsars to J1908+2105 shows that 8 out of 10 are BWs. 
The fifth and ninth pulsars in the ranking, J2039-5617 and J0337+1715, respectively, are RB and uncategorized. 
This situation, where the ranking of a given pulsar thought to pertain to one class (RB) is dominated by pulsars belonging to the other, only arises in the case mentioned above.
Thus, the MST location and the ranking favor, or at the very least do not rule out, a BW classification for J1908+2105.

On the other hand, the gamma-ray source 3FGLJ2039-5617 (PSR J2039-5617) is almost certainly associated with an optical binary listed as an RB candidate by \citet{Strader_2019}. It validates its predicted nature of RB \citet{Corongiu_2021}, where they found clear evidence of eclipses of the radio signal for about half of the orbit, which they associate with the presence of intra-binary gas. It appears as a confirmed RB in \citet{Linares2023}.
Furthermore, J2039-5617 shares similarities with the confirmed RB J2339-0533, exhibiting a peak in gamma-ray emission close to a minimum in its optical emission, which contrasts with the expected phase of the intrabinary shock (IBS) as seen in \citet{An_2020}. 
The similarity between J2039-5618 and J2339-0533, as well as their position in the MST, highlights intriguing aspects of their behavior, prompting further analysis. 

The distinct positions of J1622-0315, J1302-3258, and J1628-3205, seen in Figure~\ref{chapter_6_figures: MST_MSP}, located far from the rest of the RBs in the MST, also capture attention. 
J1622-0315 is one of the lightest known RB systems, a companion mass of 0.15 M$_{\odot}$, with a relatively hot companion \citep{yap2023light, Sen_2024}. 
This system is notable for its $P_{B}$ of 0.16 days, marginally smaller than other RBs that typically have one of 0.2 days or more. 

On the other hand, J1302–3258 is classified as an RB candidate and reported to have a minimum companion mass of 0.15 M$_{\odot}$ \citep{Strader_2019}.
However, it is distinguished by the absence of an identified optical companion and the lack of published evidence of extensive radio eclipses. 
\citet{Linares2023} note that, unlike most known RBs and candidates, J1302-3258 does not have a \textit{Gaia} counterpart, a trend more typical of BWs, which generally possess more fabulous companion stars with fainter optical magnitudes. 
J1628-3205 is a RB in a 0.21~day-long orbit around and has a 0.17--0.24 M$_{\odot}$ companion star \citep{Ray2012, Li2014, strader_redbacks}. Its optical counterpart shows two peaks per orbital cycle. Such a shape is reminiscent of the ellipsoidal deformation of a star that nearly fills its Roche lobe in a high-inclination binary. It suggests that heating by the pulsar wind is not a major effect. In this context, J1628-3205 is an intermediate case among RBs, falling between systems where the companion star's emission is dominated by pulsar irradiation (e.g., J1023+0038) and those that are not (e.g., J1723-2837, J1622-0315, J1431-4715)(see \citealt{Li2014,yap2023light,deMartino2024} and references therein).

\subsubsection{Transitional pulsars in the graph}
\label{chapter6: tMSPs_MST}

The pulsars J1723-2837, J1902-5105, and J1431-4715 appear in the same part of the MST in Figure~\ref{chapter_6_figures: MST_classes} as the known tMSPs J1227-4853 and J1023+0038. 
In addition, these three are the only ones in the sample containing at least one known tMSP in their top three neighbors; see Table~\ref{chapter_6_tab: tMSP_ranking}, which also lists the closest neighbors of the two confirmed tMSPs.

\begin{table}
\caption[Ranking for the nodes in the region of the tMSPs.]{Ranking based on the Euclidean distances calculated over the 10 scaled variables considered for the nodes in the region of the tMSPs.}
\scriptsize
\centering
\begin{tabular}{ccccc}
\hline
            \textbf{J1023+0038} & \textbf{J1723-2837} & \textbf{J1902-5105} & \textbf{J1227-4853} & \textbf{J1431-4715} \\ 
\hline
                   J1723-2837 & J1902-5105 & J1723-2837 & J1902-5105 & J2205+6012  \\
                   J1227-4853 & J1023+0038 & J1227-4853 & J1431-4715 & J1227-4853  \\
                   J1902-5105 & J1227-4853 & J1431-4715 & J1723-2837 & J1342+2822B \\
                   
\hline
\end{tabular}
\label{chapter_6_tab: tMSP_ranking}
\end{table}

J1723-2837 is a nearby (d$\sim$1 kpc based on \textit{Gaia} parallax measure) 1.86~ms RB in a 15~h-period binary system \citep{Crawford2013}. With an X-ray luminosity of $10^{32}$erg s$^{-1}$, it ranks among the brightest RBs in that band \citep[see, e.g.][]{Lee2018}. Based on the similarity with the X-ray output of tMSPs in the rotation-powered pulsar state, \citet{Linares2014} suggested that it is one of the most promising candidates to observe a transition to an accretion state. The observed emission, in both soft \citep{Bogdanov2014, Hui2014} and hard X-rays \citep{Kong2017}, is modulated at the $P_{B}$, indicative of an origin in the IBS between the pulsar wind and mass outflow from the companion star. On the other hand, its optical counterpart shows no sign of significant irradiation with two peaks per orbital cycle \citep{Li2014}.

J1902–5105, a 1.74~ms radio MSP in a 2-day-period binary system, was discovered within the Parkes telescope surveys targeting unidentified \textit{Fermi-LAT} sources \citep{Kerr_2012ApJ}. It is a relatively bright MSP located at a distance of $\sim$1.2 kpc \citep{Camilo_2015ApJ}. It was suggested that the companion is a $0.2-0.3 \, \mathrm{M_{\odot}}$ white dwarf with a helium core \citep{Camilo_2015ApJ}. 

Finally, J1431-4715, discovered in the High Time Resolution Universe (HTRU) survey with the Parkes radio telescope, is an RB MSP with a $P$ of 2.01 ms in a 10.8 hr orbit with a companion mass of 0.20 M$_{\odot}$ \citep{bates2015mnras, MiravalZanon_2018J, deMartino2024}.

Results reported in \citet{MST-MSPs} indicate a lack of moding\footnote{Although J1723-2837, J1902-5105, and J1431-4715 are currently detected as radio MSP and are therefore not expected to exhibit the typical bimodality in the X-ray light curve, we analyzed \textit{XMM-Newton} archival observations of these sources (ObsID 0653830101 for J1723-2837, ObsID 0841920101 for J1902-5105, and ObsID 0860430101 for J1431-4715) to check for possible mode switching. This analysis is shown in Appendix \ref{Appendix_6: moding}.}, suggesting a lack of conclusive evidence regarding the transitional nature of J1723-2837, J1902-5105, and J1431-4715. 
Still, their position in the MST and neighbor ranking confirms them as subjects of interest as potential transitional systems, though they are currently in the radio pulsar state. 
The relative proximity of J1723-2837, particularly J1902-5105, renders them optimal candidates for potentially unveiling a future transition from a rotation-powered state to an accretion-powered one.

\section{MST clustering }
\label{chapter6: significance_branches}

\begin{figure}
\centering
\includegraphics[width=\textwidth]{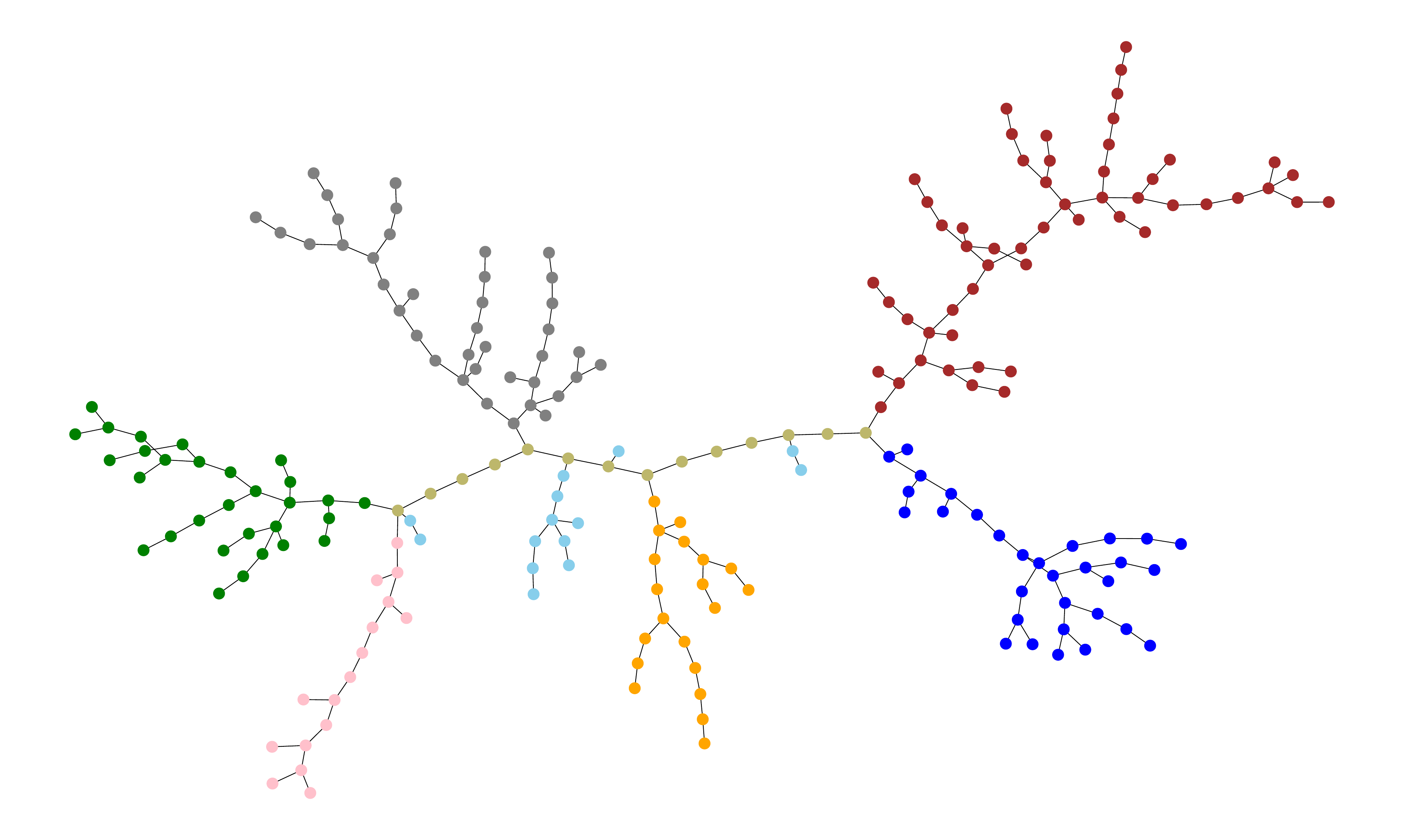}
\caption[The MST of the MSPs population separated into significant branches.]{The MST defined as $T(218, 217)$ of the MSPs population is separated into significant branches according to the algorithm described in the text (see Section~\ref{chapter6: significance_branches}), is shown.
The branches group a comparable number of pulsars: gray (39), orange (20), dark blue (31), brown (54), dark green (30), and pink (16).
The trunk (in dark khaki) has only 14 nodes, with 14 others (in sky blue) not considered to pertain to any significant branch or trunk, being the noise of the former structures.
}
       \label{chapter_6_figures: MST_siginificance_branches}
\end{figure}

\begin{figure}
\centering
\includegraphics[width=1\textwidth]{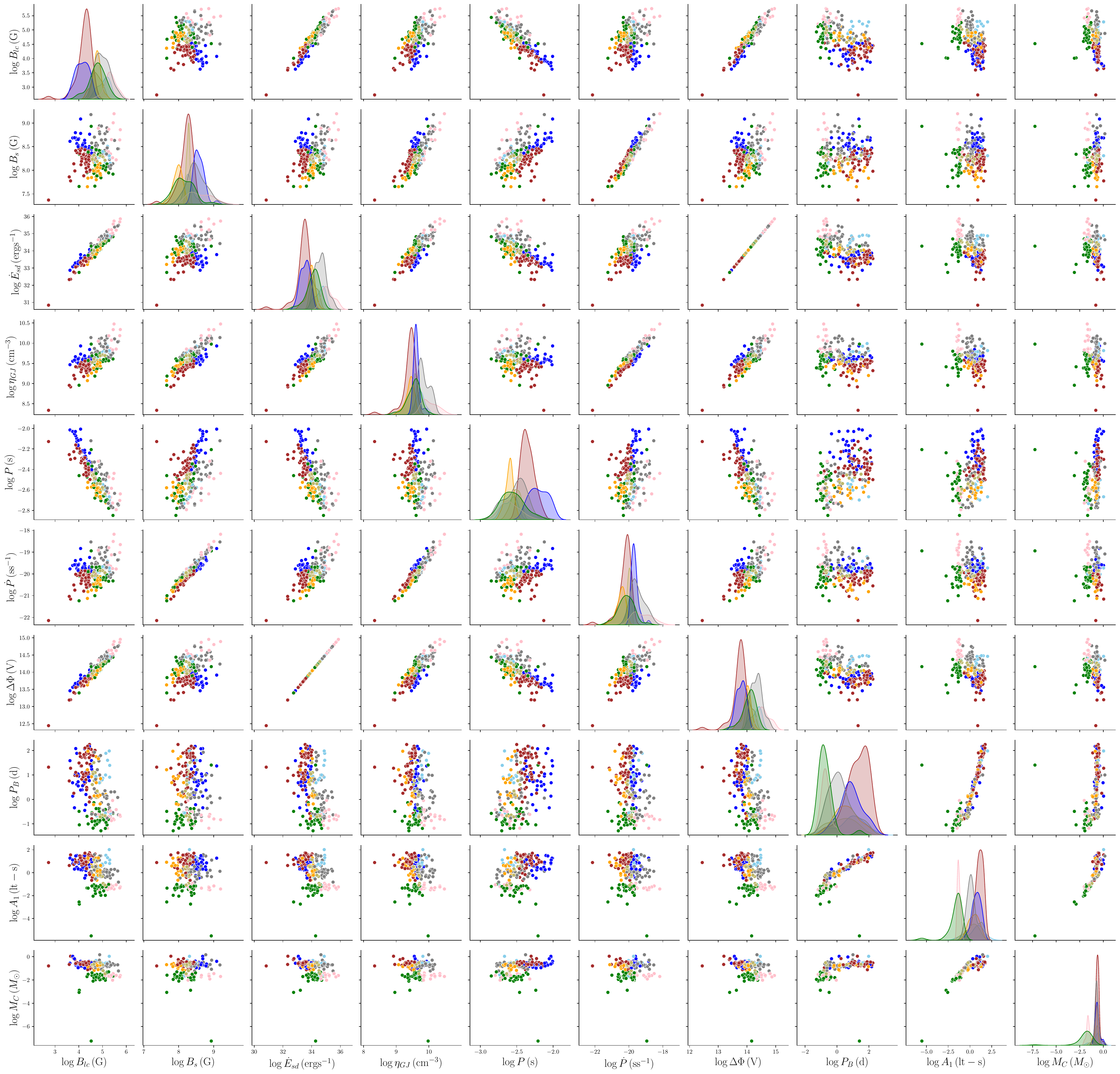}
\caption[Cross-correlation of the properties according to the significant branches.]{Cross-dependence and distributions of the 10 variables considered separated according to the branches in Figure~\ref{chapter_6_figures: MST_siginificance_branches}. 
}
       \label{chapter_6_figures: siginificance_branches_Vars}
\end{figure}

In this Section, we apply the clustering algorithm introduced in Chapter~\ref{chapter5} to separate an MST into distinct parts according to a quantitative prescription.
The methodology and specific details of this case are presented in the Appendix \ref{Appendix_6: significant_branches}.
Figure~\ref{chapter_6_figures: MST_siginificance_branches} shows separated branches, and Figure~\ref{chapter_6_figures: siginificance_branches_Vars} shows the distributions for each magnitude considering them. 
The MST visually represents the binary millisecond pulsar population in this way without making any prior assumptions about the nature of the nodes.

Two of the branches indicated by the method in Figure~\ref{chapter_6_figures: MST_siginificance_branches} (depicted in dark green and pink) contain all BWs except J1555-2908, which was already deemed as lying at the edge of the BWs' population as discussed in Section~\ref{chapter6: BWs_MST}.
In addition, one of these branches (pink) contains the most candidate or confirmed BWs seen in globular clusters. 
The BWs in globular clusters, together with the rest of the pulsars seen in the pink branch, show a $\dot{P}$ and $\dot{E}$ in general higher than the others; see Figure~\ref{chapter_6_figures: siginificance_branches_Vars}.
Note that these values cannot be reliably measured in a globular cluster due to the acceleration in the cluster's gravity well.
Similarly, tMSPs and most RBs fall in another branch (depicted in gray according to Figure~\ref{chapter_6_figures: MST_siginificance_branches}).
As we see in Figure~\ref{chapter_6_figures: siginificance_branches_Vars}, pulsars in the orange branch exhibit a $P_{B}$, $M_{C}$, and $A_{1}$ close to those in the gray branch.
However, in contrast, they show a lower $\dot{P}$ and thus a smaller $B_{s}$.
The rest of the population, particularly those pulsars in the dark blue and brown branches, which are significantly less different to each other, are markedly different from the other groups, as shown in Table~\ref{chapter_6_tab: significance_branches} (see Appendix \ref{Appendix_6: significant_branches} for details on the similarity criteria).
Figure~\ref{chapter_6_figures: siginificance_branches_Vars} shows that although these nodes are less obviously separated in the usual representations, such as the typical $P,\dot P$ diagram, they show larger $P_{B}$, $M_{C}$, $P$, and smaller $\dot E$ than the BW pulsars.

We can discuss differences among branches based on the distributions of the variables seen in the main diagonal of Figure~\ref{chapter_6_figures: siginificance_branches_Vars}.
We observe that $M_{C}$ exhibits sharp distributions around similar values, except for the (pink) branch, which primarily contains BWs in globular clusters, for which $M_{C}$ shows smaller values.  
The dark green branch, plagued by BWs, displays a broader distribution, always within small values.
This is reflected in $A_1$ and $P_B$, where the behavior is similar even with somewhat broader distributions, mainly for the branches that do not contain BWs. 
The widest distributions are observed in $P$ for all branches except the one depicted in orange, which has a high density of cases around low values. The dark blue branch contains the pulsars with more extended periods.
The $\dot{P}$ shows different trends for some branches. In that order, the orange, brown, and dark green branches exhibit sharp behaviors that enable us to distinguish the pulsars they contain. Most of them are not classified as BWs or RBs.
This is reflected in $\dot E$, $\eta_{GJ}$, $B_{lc}$, and $B_S$;  branches containing BWs and RBs show a wider distribution, shifted towards larger values for the dark green, gray, and pink branches, considering that the latter two contain different types of spider pulsars.
In addition, Figure~\ref{chapter_6_figures: siginificance_branches_Vars} can be compared with Figure~\ref{chapter_6_figures: pair_plotclasses11v}. The former is more informative, as the clustering technique separates the nodes in groups that are significantly different from one another (see Table~\ref{chapter_6_tab: significance_branches} in Appendix \ref{Appendix_6: significant_branches} for details on the similarity criteria) beyond the known RBs and RBs.

\subsection{Tracking the variables along the MST}
\label{chapter6: tracking_variables}
\begin{figure}
  \includegraphics[width=1\textwidth]{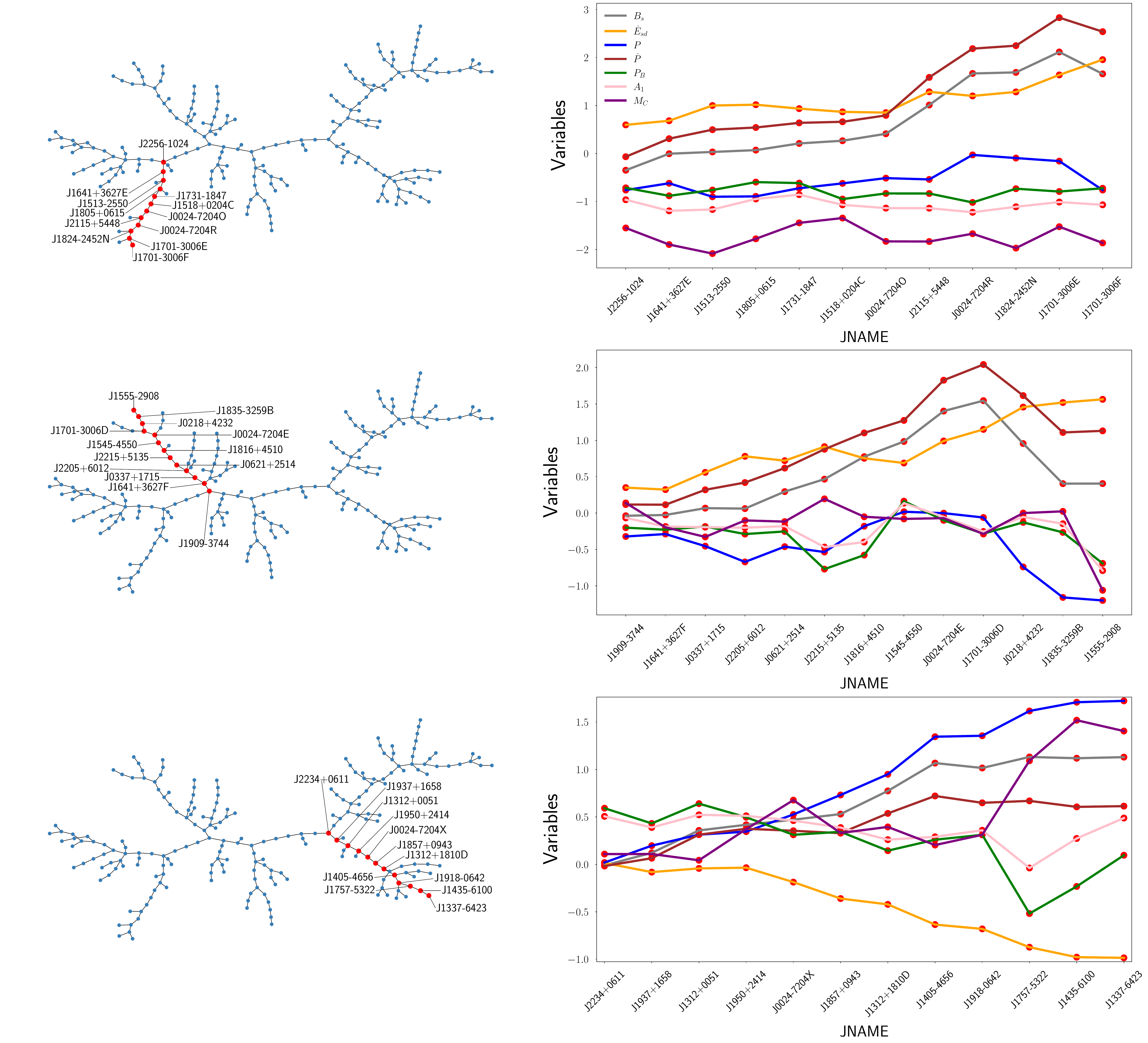}
  \centering
  \caption[Tracking the properties along a path in the MST.]{
    Examples of the values of $P$, $\dot P$, $B_{s}$, $\dot E$, $M_{C}$, $A_1$, and $P_{B}$ (the legend shown in the first row applies to all rows) along a path in the MST.
    Left column: Selected paths and nodes in the MST are highlighted in red.
    Right column: The y-axis shows the scaled variables. 
    The x-axis displays the pulsars along the selected path (nodes in red) by their identifier, according to the ATNF v2.1.1. The order shown responds to the path reading in the MST, which always goes from the central part (trunk) to the ends of the MST.
    } 
  \label{track_variables}
\end{figure}

In closing this Section, we note that the MST technique orders the pulsars according to one or several of the variables considered (as discussed in Section~\ref{chapter4: branching}).
Figure~\ref{track_variables} below serves as an example (data to construct similar figures for other paths are provided in the Pulsar Tree web that accompanies this work) of how the variables are ordered along a given path in the graph theory context.
Figure~\ref{track_variables} shows the values of $P$, $\dot P$, $B_{s}$, $\dot E$, $M_{C}$, $A_1$, and $P_{B}$ along a path in the MST.

\section{Localizing a new node}
\label{chapter6: predict_link}

We introduce an incremental algorithm for updating the MST when new nodes are considered. 
This can infer where a new node with given properties will fall in the MST.
When we add a new node and compute its distance from all other nodes in the graph, we preserve the known distances among the previously existing nodes.
In this process, we keep the scaling of the distance used in calculating the original MST, denoted by $T$, to which the new node will be added. 
This involves identifying potential cycles that would be formed with the existing $T$, adding new edges, and discarding them so that the resulting graph remains an MST.
While avoiding a complete recalculation of the MST, this approach provides the new node's location relative to the former ones, resulting in an updated $T$, which we shall refer to as $T'$.
This process follows the principles outlined in traditional graph theory, e.g., the correctness of Prim's algorithm as per the MST theorem and the MST lemma as the cycle property (see Chapter~\ref{chapter2} for technical details).

\subsection{IGR J18245-2452—PSR J1824-2452I}
\label{chapter6: IGRJ1824}

Since tMSP IGR J18245-2452 (or PSR J1824-2452I, see, e.g., \citealt{Papitto2013}) lacks a measured value of the $\dot{P}$, it is initially excluded from the sample used in this work (see Section~\ref{chapter6: sample}).
Considering its measured values for $P$, $P_{B}$, $M_{C}$, and $A_{1}$, we shall assume a value for the $\dot P$ and add this pulsar to $T$, in Figure~\ref{chapter_6_figures: MST_MSP}.
Based on the estimated $B_{s}\sim (0.7-35) \times 10^{8}$ G by \citet{Ferrigno2014}, and its known $P$, we can compute\footnote{For consistency, we assume $B_{s} = 3.2 \times 10^{19} \sqrt{P \, \dot{P}}$ G} a range of $\dot{P}\sim (1.2\times 10^{-21} - 3\times 10^{-18}) \mathrm{ss^{-1}}$.
We take 10 equispaced values in this range (and their concurrent values of all physical variables derived taking $\dot P$ into account, and we construct for each one a $T'$.

\begin{figure}
\centering
\includegraphics[width=\textwidth]{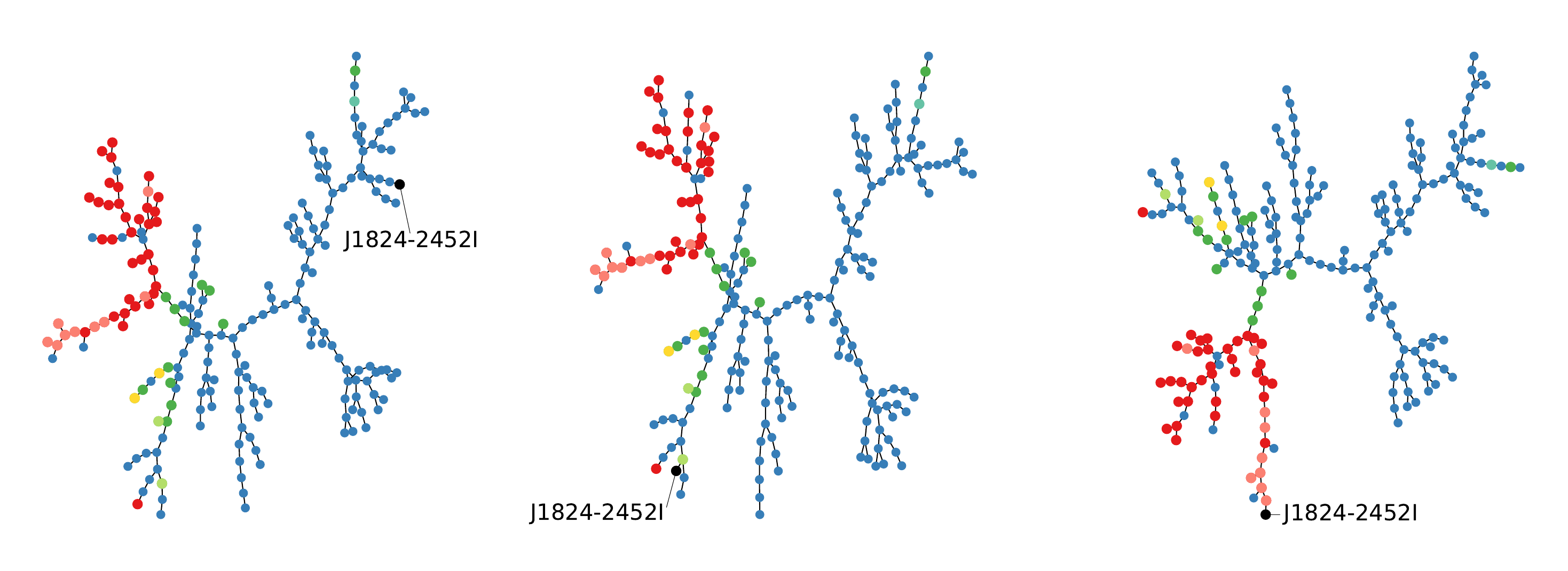}
\caption[The MSTs containing different realizations for $\dot P$.]{The MSTs containing different realizations for $\dot P$, 
$\dot{P}\sim 1.22\times10^{-20}\mathrm{ss^{-1}}$ (left panel);
$\dot{P}\sim 3.34\times10^{-19}\mathrm{ss^{-1}}$ (central panel);
$\dot{P}\sim 3\times10^{-18}\mathrm{ss^{-1}}$ (right panel),
of the tMSP J1824-2452I (in black). Each MST is defined as a $T'(219,218)$ and shown with the spider type of pulsars as seen in Figure~\ref{chapter_6_figures: MST_MSP}, where the confirmed BWs appear in red, RBs in green, tMSPs in yellow, the BWs and RBs in globular clusters in light red and light green, the RB candidate in light teal, and keeping the unclassified ones in blue.
}
\label{chapter_6_figures: MST_IGRJ181245-2452}
\end{figure}

When we assume the lowest values of the given range, $\dot{P}\lesssim 10^{-20}\mathrm{ss^{-1}}$, we find that the added pulsar is located along one of the rightmost branches (brown branch, see Figure~\ref{chapter_6_figures: MST_siginificance_branches}) of the MST, see the left panel in Figure~\ref{chapter_6_figures: MST_IGRJ181245-2452}.
As $\dot{P}$ increases to the range ($10^{-20}, 3\times10^{-19}) \, \mathrm{ss^{-1}}$, its MST location would climb until it reaches the end of the branch where the rest of the RBs, and therefore the known tMSPs, appear. 
It will remain at that end until it $\dot{P}\lesssim 6\times 10^{-19}\mathrm{ss^{-1}}$ is considered, see the central panel in Figure~\ref{chapter_6_figures: MST_IGRJ181245-2452}.
When this value of $\dot P$ is exceeded and up to the largest ones explored $\dot{P}\sim 3\times 10^{-18}\mathrm{ss^{-1}}$, the pulsar would be located in the BW branch, as we can see in the right panel of Figure~\ref{chapter_6_figures: MST_IGRJ181245-2452}.
Thus, as IGR J18245-2452 is an RB transitional pulsar, it would be reasonable to expect that it falls near the majority of the nodes of its class, implying we can limit searches for its $\dot P$ from $\dot{P}\sim (1.2\times 10^{-21} - 3\times 10^{-18})\, \mathrm{ss^{-1}}$ to just around $\dot{P}\sim (1\times 10^{-20}, 6\times 10^{-19}) \, \mathrm{ss^{-1}}$ or $B_{s}\sim (2\times 10^{8}, 1.55\times 10^{9})$G.

\subsection{tMSP candidate: 3FGL J1544.6-1125}
\label{chapter6: 3FGL}

3FGL J1544.6-1125 has shown variability with high and low modes exclusively seen in tMSPs, see \citet{Bogdanov_2015}. 
In \citet{Britt_2017}, the $P_{B}$ of this pulsar is estimated to be $P_{B} = 0.2415361(36) \, \mathrm{days} = 20868.72(31) \, \mathrm{s}$. The semi-amplitude of the radial velocity of the companion star $K_2 = 39.3 \pm 1.5 \, \mathrm{km \, s^{-1}}$ yields a companion mass $M_{2}\lesssim 0.7 \, M_\odot$.
To estimate the range of $A_{1}$, we start defining the center of mass location, relying on the fundamental relationship $ M_1 a_1 = M_2 a_2$  (see, e.g., \citealt{FundamentalAstronomy}), where $a_1$ and  $a_2$ represent the distances of the objects from the center of mass, and $M_1$ and $M_2$ are the masses of the neutron star and the companion star, respectively. 
Additionally, the radial velocity of the companion is defined as (see, e.g., \citealt{tauris2003formation}) $K_2 = \omega_{B} \cdot a_2 \cdot \sin(i)$, where $\omega_{B} = {2 \pi}/{P_{B}}$.
By using the above relationships, we derive \(A_{1} = a_1 \sin(i) = \frac{K_2 \cdot P_{B}}{2 \pi} \cdot \frac{M_2}{M_1} \).
We make several assumptions, where 
for $M_1$ we have
$ M_{1min}\sim 1.4 M_{\odot}$
and 
$ M_{1max}\sim 2 M_{\odot} $, 
and for $M_2$ we have
$M_{2min}\sim 0.05 M_{\odot}$
and 
$M_{2max}\sim 0.7 M_{\odot}$. 
Consequently, we calculate the range $A_1\sim(0.0088, 0.259)$ lt-s.
We consider values from $M_{2min}$ up to $M_{2max}$ for the range of $M_{C}$.
Also, as the pulsar has an unknown value of $P$, we impose no constraint on it and shall range it within $\sim (1,10)\times10^{-3}$ s.
Similarly, we shall also inspect the range between the minimum and maximum measurement of $\dot{P}$ seen in the sample according to $(7 \times  10^{-23}, 7 \times 10^{-19})$ ss$^{-1}$. 
We consider 10 values spanning each range for ($P$, $\dot{P}$, $A_1$, $M_{C}$), resulting in $10^4$ distinct combinations for which we apply the incremental algorithm. 

\begin{figure}
\centering
\includegraphics[width=\textwidth]{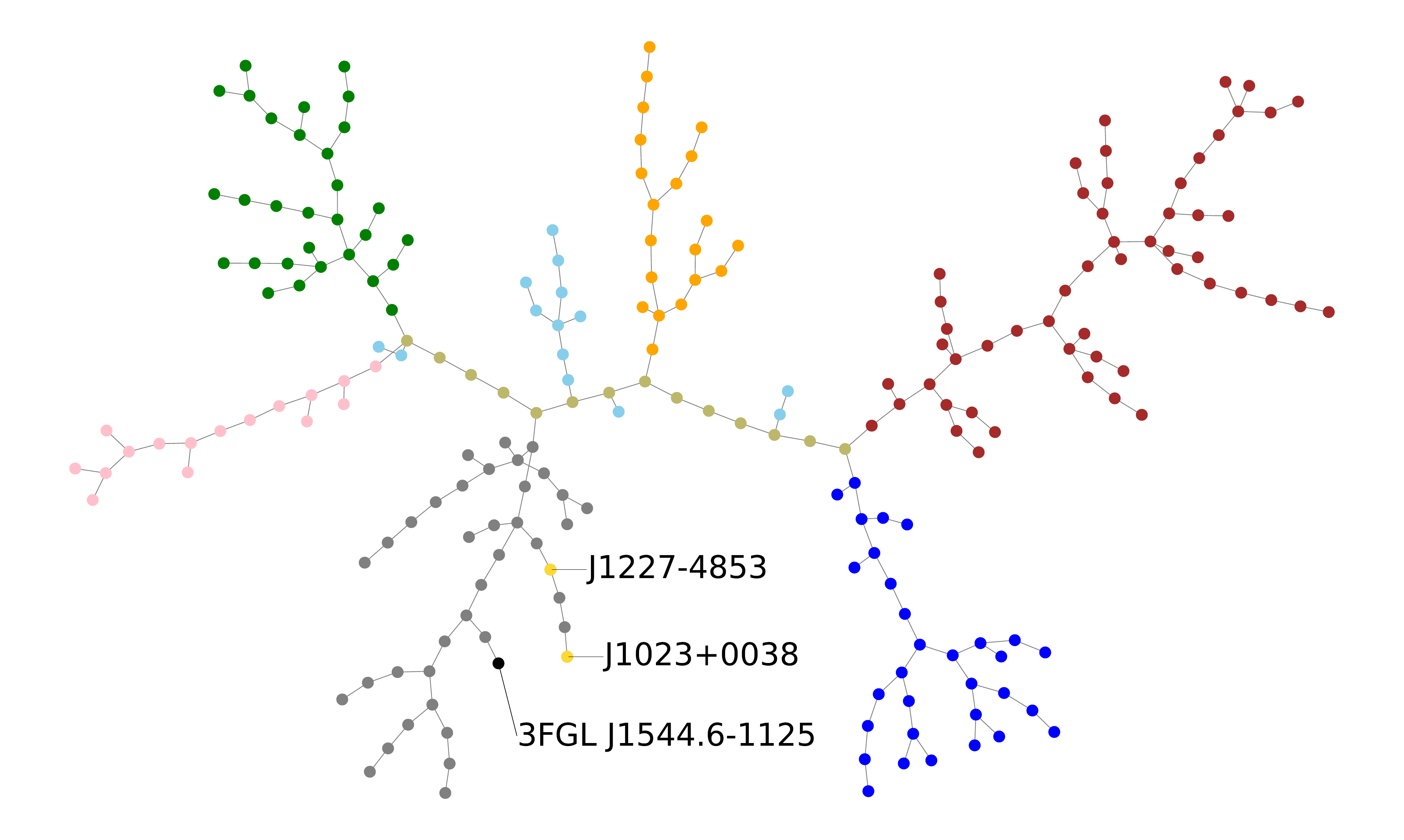}
\caption[The MST containing the 3FGL J1544.6-1125.]{The MST defined as $T'(219,218)$ containing the 3FGL J1544.6-1125 (in black) is shown with the known tMSPs (in yellow) labeled J1023+0038 and J1227-4853, as seen in Figure~\ref{chapter_6_figures: MST_MSP} and the color code of the significant branches, trunk, and the remaining pulsars as seen in Figure~\ref{chapter_6_figures: MST_siginificance_branches}.

}
       \label{chapter_6_figures: MST_3FGL1544}
\end{figure}

We observe that the structural stability\footnote{
The percentage of edges preserved between the original and updated MST after adding a new node. Although minor changes (even a single edge) can impact global properties, a high level of edge retention, such as the 99\%, serves as a useful indicator of consistency.} of $T'$ is extremely high compared to $T$, since in more than 99\% of the cases, they only differ on one edge. 
As seen in the previous analysis, this allows us to position 3FGL J1544.6-1125, considering known parts of $T$ as the regions with the marked classes of pulsars as in Figure~\ref{chapter_6_figures: MST_MSP} or the branches seen in Figure~\ref{chapter_6_figures: MST_siginificance_branches}.
In 67\% of the cases, the position of 3FGL J1544.6-1125 falls in a branch (gray), where the known tMSPs are located, as we show in Figure~\ref{chapter_6_figures: MST_3FGL1544}.
Despite the uncertainties of the variables, the MST promotes the tMSP classification of 3FGL J1544.6-1125. 

\subsection{SAX J1808.4-3658}
\label{chapter6: SAX}

\begin{figure}
\centering
\includegraphics[width=\textwidth]{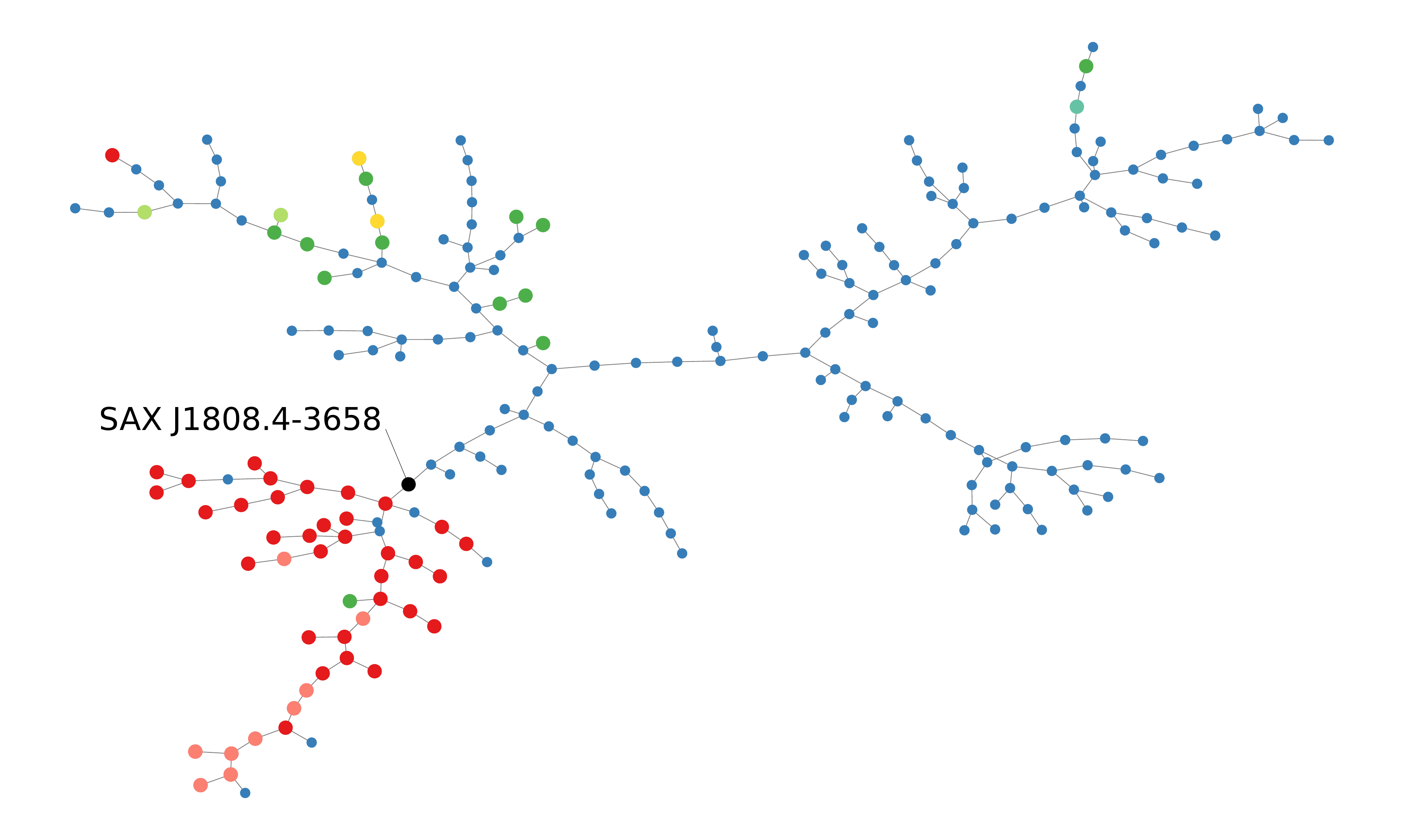}
\caption[The MST containing the AMXP SAX J1808.4-3658.]{The MST defined as $T'(219,218)$, containing the AMXP SAX J1808.4-3658 (in black) is shown. The confirmed BWs appear in red, RBs in green, tMSPs in yellow, the BWs and RBs in globular clusters in light red and light green, the RB candidate in light teal, and keeping the unclassified ones in blue, as seen in Figure~\ref{chapter_6_figures: MST_MSP}.

}
       \label{chapter_6_figures: MST_SAX J1808.4-3658}
\end{figure}

The location of SAX J1808.4-3658, an accreting millisecond X-ray pulsar (AMXP, see, e.g., \citealt{Patruno_Watts}), within the MST is also of considerable interest. 
It has exhibited approximately ten month-long outbursts, with a recurrence of about 2–3 years, making it the AMXP with the most numerous outbursts, which is suitable for in-depth investigation of its long-term timing properties, as discussed by \citet{Illiano2023}.
SAX J1808.4-3658 is likely a gamma-ray source \citep{deona2016}, and there is an ambiguity about its activity status as a rotation-powered millisecond pulsar during quiescent periods despite the lack of detected radio pulsations. 
Focusing on the long-term first derivative of the spin frequency, as detailed in Figure~2 and Section~3.2 of \citet{Illiano2023}, reveals a $\dot{\nu} = -1.152(56) \times 10^{-15} \, \mathrm{Hz \, s^{-1}}$,  in alignment with findings from previous works \citep{Patruno_2012, Sanna_2017, Bult_2020}. 
The variables taken are $P = 0.00249391976403(31)$ s, $A_1 = 0.0628033(57)$ lt-s, and the $P_B = 0.083902314(96)$ d (as listed in Table~1 of \citealt{Illiano2023}).
Consequently, we estimate $\dot{P} = - \dot{\nu}/\nu^2 \,  = - P^2 \, \dot{\nu} \sim 7.2 \times 10^{-21} \mathrm{s \, s^{-1}}$.
Past timing analysis indicates that the neutron star orbits a semi-degenerate companion of $M_{C}\sim0.05$ M$_\odot$ (see \citealt{Bildsten_2001}).
Considering the values computed for the other variables, given the known spin period and spin period derivative, we apply the incremental algorithm to obtain $T'$.
Using the BWs and RBs seen in Figure~\ref{chapter_6_figures: MST_MSP} as a reference, we show $T'$ in Figure~\ref{chapter_6_figures: MST_SAX J1808.4-3658} together with the AMXP SAX J1808.4-3658, which falls at the gate of the high-density zone of BWs.

\section{Concluding remarks}
\label{chapter6: conclusions}

Graph theory provides an elegant way to inspect a population.
A combination of the MST and the distance ranking underlying helps distinguish components and identify candidates for membership in a particular group.
Additionally, it is worth noting that no methodology has been established to identify candidates among radio-emitting pulsars that are not in a sub-luminous disk state, and the MST can contribute to this purpose.
In this work, we have focused on the MSP population and observed that its MST categorizes BWs and RBs into distinct categories. This separation is evident when plotting the $M_C$ for each system.
When the values of $M_C$ are not clear-cut between these populations, it is increasingly complex to make a safe classification without further studies.
Using the location in the MST, the individual distance ranking for a given pulsar, and the fact that in only very few cases, the ranking of a given pulsar thought to pertain to one class is dominated by pulsars belonging to another, suggests that 
\begin{itemize}

\item We promote the BW classification of J1300+1240, J1630+3550, J1317-0157, J1221-0633, J1627+3219, J1701-3006F, and J1737-0314A.

\item The MST location and the ranking favor, or at the very least do not rule out, a BW classification for J1908+2105.

\item The pulsars J1723-2837, J1902-5105, and J1431-4715 are the only ones in the sample containing at least one known
tMSP is in their top 3 neighbors. We promote further studies of these pulsars to observe them transitioning into an accretion-powered state, analogous to the known tMSPs.

\item The MST location of IGR J18245-2452/PSR J1824-2452I as an RB transitional pulsar implies that the yet unknown $\dot P$ is in the range $\dot{P}\sim (1\times 10^{-20}, 6\times 10^{-19}) \, \mathrm{ss^{-1}}$. Searches limited to this range could empower the techniques and lead to detection.

\item Spanning over the uncertainty in the variables of the putative transitional pulsar behind 3FGL J1544.6-1125, we see that it is located near the others in most cases. 

\item The AMXP SAX J1808.4-3658 is close to the group BWs and can be classified as such from its location in the MST. Indeed, \citet{diSalvo2008} proposed that the system is a hidden BW and that the source is ejecting matter in X-ray quiescence. Note that this fact alone would not necessarily imply its possible BW nature. 
    
\item If we use graph theory to cluster the population of MSPs via distinguishing branches (unsupervised methodology), we recover a clear distinction between BWs, RBs, and a significant number of other MSPs showing different features, larger $P_{B}$, $M_{C}$, $P$, and smaller $\dot E$ releases. 
    
\item This process also separates BWs in the field and these in globular clusters, placing them in different branches within the BW region.
    
\end{itemize}

\begin{center}
\scriptsize
\setlength\LTleft{-50pt}
\setlength\LTright{0pt}
\begin{longtable}{lrrrrrrrrrrr}
\caption[Properties of the MSPs population according to the ATNF v2.1.1.]{The truncated values according to the ATNF v2.1.1 (visit \url{https://www.atnf.csiro.au/research/pulsar/psrcat/} for updated versions) of the 10 variables considered for the confirmed BWs, RBs, and tMSPs (candidates for BWs or RBs are placed below the corresponding line).
The tMSPs are included in the RBs section and are denoted by an asterisk.
Pulsars belonging to globular clusters are marked with a star.
Note that the pulsar J1300+1240 (also known as PSR B1257+12, see \citealt{Wolszczan_1992, Wolszczan_1994, Wolszczan_2000}) is a planetary system, and its orbital parameters can be considered as lower limits depending on the mass of its planets (see, e.g., \citealt{Konacki_2002}).
} \label{chapter_6_tab: BW_RW_tMSP} \\
\toprule
JNAME & $P_{B}(\mathrm{d})$ &  $\dot{E}_{sd}(\mathrm{erg \, s^{-1}})$ &  $P(\mathrm{s})$ &  $B_{s}(\mathrm{G})$ &  $\dot{P}(\mathrm{s \,s^{-1}})$ & $M_{C}(\mathrm{M_{\odot}})$   &  $A_{1}(\mathrm{lt-s})$ &  $\Delta\Phi(\mathrm{V})$ &  $\eta_{GJ}(\mathrm{cm^{-3}})$ &  $B_{lc}(\mathrm{G})$ \\
\midrule
\endfirsthead

\multicolumn{11}{l}{\textbf{Table \thetable{} (continued):} Properties of the MSPs population.} \\
\toprule
JNAME & $P_{B}(\mathrm{d})$ &  $\dot{E}_{sd}(\mathrm{erg \, s^{-1}})$ &  $P(\mathrm{s})$ &  $B_{s}(\mathrm{G})$ &  $\dot{P}(\mathrm{s \,s^{-1}})$ & $M_{C}(\mathrm{M_{\odot}})$   &  $A_{1}(\mathrm{lt-s})$ &  $\Delta\Phi(\mathrm{V})$ &  $\eta_{GJ}(\mathrm{cm^{-3}})$ &  $B_{lc}(\mathrm{G})$ \\
\midrule
\endhead

\bottomrule
\endfoot

\bottomrule
\endlastfoot

\multicolumn{11}{c}{Black widows} \\ 
\hline
                 
 J0023+0923 & 0.13 & 1.58$\times 10^{34}$ & 0.0030 & 1.88$\times 10^{8}$ & 1.14$\times 10^{-20}$ & 0.018 & 0.034 &1.33$\times 10^{14}$& 4.28$\times 10^{9}$&  62401.91 \\
 J0024-7204O$^{\star}$ & 0.13 & 6.48$\times 10^{34}$ & 0.0026 & 2.86$\times 10^{8}$ & 3.03$\times 10^{-20}$ & 0.024 & 0.045 &2.69$\times 10^{14}$& 7.50$\times 10^{9}$& 145482.85 \\
 J0024-7204P$^{\star}$ & 0.14 & 5.41$\times 10^{35}$ & 0.0036 & 1.57$\times 10^{9}$ & 6.63$\times 10^{-19}$ & 0.019 & 0.038 &7.77$\times 10^{14}$& 2.98$\times 10^{10}$& 305080.04 \\
 J0024-7204R$^{\star}$ & 0.06 & 1.38$\times 10^{35}$ & 0.0034 & 7.27$\times 10^{8}$ & 1.48$\times 10^{-19}$ & 0.029 & 0.033 &3.93$\times 10^{14}$& 1.44$\times 10^{10}$& 161687.07 \\
 J0251+2606 & 0.20 & 1.82$\times 10^{34}$ & 0.0025 & 1.40$\times 10^{8}$ & 7.57$\times 10^{-21}$ & 0.027 & 0.065 &1.42$\times 10^{14}$& 3.82$\times 10^{9}$&  80162.82 \\
 J0312-0921 & 0.09 & 1.53$\times 10^{34}$ & 0.0037 & 2.73$\times 10^{8}$ & 1.97$\times 10^{-20}$ & 0.010 & 0.015 &1.30$\times 10^{14}$& 5.10$\times 10^{9}$&  50446.32 \\
 J0610-2100 & 0.28 & 8.45$\times 10^{33}$ & 0.0038 & 2.20$\times 10^{8}$ & 1.23$\times 10^{-20}$ & 0.024 & 0.073 &9.71$\times 10^{13}$& 3.95$\times 10^{9}$&  35950.28 \\
 J0636+5128 & 0.06 & 5.76$\times 10^{33}$ & 0.0028 & 1.00$\times 10^{8}$ & 3.44$\times 10^{-21}$ & 0.007 & 0.008 &8.02$\times 10^{13}$& 2.42$\times 10^{9}$&  39963.19 \\
 J0952-0607 & 0.26 & 6.66$\times 10^{34}$ & 0.0014 & 8.31$\times 10^{7}$ & 4.77$\times 10^{-21}$ & 0.022 & 0.062 &2.72$\times 10^{14}$& 4.06$\times 10^{9}$& 275776.46 \\
 J1124-3653 & 0.22 & 1.69$\times 10^{34}$ & 0.0024 & 1.21$\times 10^{8}$ & 6.01$\times 10^{-21}$ & 0.031 & 0.079 &1.37$\times 10^{14}$& 3.49$\times 10^{9}$&  81614.11 \\
 J1301+0833 & 0.27 & 6.64$\times 10^{34}$ & 0.0018 & 1.41$\times 10^{8}$ & 1.05$\times 10^{-20}$ & 0.027 & 0.078 &2.72$\times 10^{14}$& 5.29$\times 10^{9}$& 211158.03 \\
 J1311-3430 & 0.06 & 4.93$\times 10^{34}$ & 0.0025 & 2.34$\times 10^{8}$ & 2.09$\times 10^{-20}$ & 0.009 & 0.010 &2.34$\times 10^{14}$& 6.33$\times 10^{9}$& 130950.18 \\
 J1446-4701 & 0.27 & 3.66$\times 10^{34}$ & 0.0021 & 1.48$\times 10^{8}$ & 9.80$\times 10^{-21}$ & 0.021 & 0.064 &2.02$\times 10^{14}$& 4.67$\times 10^{9}$& 131662.94 \\
 J1513-2550 & 0.17 & 8.96$\times 10^{34}$ & 0.0021 & 2.16$\times 10^{8}$ & 2.16$\times 10^{-20}$ & 0.018 & 0.040 &3.16$\times 10^{14}$& 7.06$\times 10^{9}$& 213347.04 \\
J1518+0204C$^{\star}$ & 0.08 & 6.71$\times 10^{34}$ & 0.0024 & 2.57$\times 10^{8}$ & 2.60$\times 10^{-20}$ & 0.043 & 0.057 &2.73$\times 10^{14}$& 7.17$\times 10^{9}$& 157520.28 \\
 J1544+4937 & 0.12 & 1.09$\times 10^{34}$ & 0.0021 & 7.86$\times 10^{7}$ & 2.79$\times 10^{-21}$ & 0.019 & 0.032 &1.10$\times 10^{14}$& 2.51$\times 10^{9}$&  73224.85 \\
 J1555-2908 & 0.23 & 3.07$\times 10^{35}$ & 0.0017 & 2.85$\times 10^{8}$ & 4.45$\times 10^{-20}$ & 0.059 & 0.151 &5.86$\times 10^{14}$& 1.10$\times 10^{10}$& 468636.58 \\
J1641+3627E$^{\star}$ & 0.11 & 4.47$\times 10^{34}$ & 0.0024 & 2.10$\times 10^{8}$ & 1.74$\times 10^{-20}$ & 0.023 & 0.037 &2.23$\times 10^{14}$& 5.86$\times 10^{9}$& 128450.77 \\
 J1641+8049 & 0.09 & 4.27$\times 10^{34}$ & 0.0020 & 1.36$\times 10^{8}$ & 8.94$\times 10^{-21}$ & 0.046 & 0.064 &2.18$\times 10^{14}$& 4.65$\times 10^{9}$& 154502.68 \\
 J1653-0158 & 0.05 & 1.24$\times 10^{34}$ & 0.0019 & 6.95$\times 10^{7}$ & 2.40$\times 10^{-21}$ & 0.011 & 0.010 &1.17$\times 10^{14}$& 2.44$\times 10^{9}$&  85609.20 \\
J1701-3006E$^{\star}$ & 0.15 & 3.62$\times 10^{35}$ & 0.0032 & 1.01$\times 10^{9}$ & 3.10$\times 10^{-19}$ & 0.035 & 0.070 &6.36$\times 10^{14}$& 2.16$\times 10^{10}$& 281028.45 \\
 J1719-1438 & 0.09 & 1.63$\times 10^{33}$ & 0.0057 & 2.18$\times 10^{8}$ & 8.04$\times 10^{-21}$ & 0.001 & 0.001 &4.27$\times 10^{13}$& 2.60$\times 10^{9}$&  10546.33 \\
 J1731-1847 & 0.31 & 7.78$\times 10^{34}$ & 0.0023 & 2.46$\times 10^{8}$ & 2.54$\times 10^{-20}$ & 0.038 & 0.120 &2.94$\times 10^{14}$& 7.28$\times 10^{9}$& 179657.07 \\
 J1745+1017 & 0.73 & 5.77$\times 10^{33}$ & 0.0026 & 8.60$\times 10^{7}$ & 2.72$\times 10^{-21}$ & 0.015 & 0.088 &8.03$\times 10^{13}$& 2.24$\times 10^{9}$&  43265.82 \\
 J1805+0615 & 0.33 & 9.31$\times 10^{34}$ & 0.0021 & 2.22$\times 10^{8}$ & 2.27$\times 10^{-20}$ & 0.026 & 0.087 &3.22$\times 10^{14}$& 7.23$\times 10^{9}$& 216420.69 \\
 J1810+1744 & 0.15 & 3.97$\times 10^{34}$ & 0.0016 & 8.84$\times 10^{7}$ & 4.60$\times 10^{-21}$ & 0.049 & 0.095 &2.10$\times 10^{14}$& 3.68$\times 10^{9}$& 181229.94 \\
J1824-2452M$^{\star}$ & 0.24 & 4.42$\times 10^{34}$ & 0.0047 & 7.75$\times 10^{8}$ & 1.22$\times 10^{-19}$ & 0.012 & 0.032 &2.22$\times 10^{14}$& 1.12$\times 10^{10}$&  66406.37 \\
J1824-2452N$^{\star}$ & 0.19 & 1.66$\times 10^{35}$ & 0.0033 & 7.39$\times 10^{8}$ & 1.59$\times 10^{-19}$ & 0.021 & 0.049 &4.31$\times 10^{14}$& 1.52$\times 10^{10}$& 183891.70 \\
 J1833-3840 & 0.90 & 1.07$\times 10^{35}$ & 0.0018 & 1.84$\times 10^{8}$ & 1.77$\times 10^{-20}$ & 0.009 & 0.061 &3.46$\times 10^{14}$& 6.82$\times 10^{9}$& 265593.33 \\
J1836-2354A$^{\star}$ & 0.20 & 2.42$\times 10^{33}$ & 0.0033 & 8.92$\times 10^{7}$ & 2.31$\times 10^{-21}$ & 0.019 & 0.046 &5.20$\times 10^{13}$& 1.84$\times 10^{9}$&  22164.68 \\
 J1928+1245 & 0.13 & 2.40$\times 10^{34}$ & 0.0030 & 2.27$\times 10^{8}$ & 1.67$\times 10^{-20}$ & 0.010 & 0.018 &1.63$\times 10^{14}$& 5.21$\times 10^{9}$&  77477.09 \\
 J1959+2048 & 0.38 & 1.60$\times 10^{35}$ & 0.0016 & 1.66$\times 10^{8}$ & 1.68$\times 10^{-20}$ & 0.024 & 0.089 &4.23$\times 10^{14}$& 7.16$\times 10^{9}$& 375949.48 \\
 J2017-1614 & 0.09 & 7.80$\times 10^{33}$ & 0.0023 & 7.61$\times 10^{7}$ & 2.45$\times 10^{-21}$ & 0.030 & 0.043 &9.33$\times 10^{13}$& 2.27$\times 10^{9}$&  57631.66 \\
 J2047+1053 & 0.12 & 1.04$\times 10^{34}$ & 0.0042 & 3.02$\times 10^{8}$ & 2.08$\times 10^{-20}$ & 0.040 & 0.069 &1.07$\times 10^{14}$& 4.87$\times 10^{9}$&  35979.39 \\
 J2051-0827 & 0.09 & 5.48$\times 10^{33}$ & 0.0045 & 2.42$\times 10^{8}$ & 1.27$\times 10^{-20}$ & 0.030 & 0.045 &7.82$\times 10^{13}$& 3.72$\times 10^{9}$&  24801.53 \\
 J2052+1219 & 0.11 & 3.38$\times 10^{34}$ & 0.0019 & 1.16$\times 10^{8}$ & 6.70$\times 10^{-21}$ & 0.038 & 0.061 &1.94$\times 10^{14}$& 4.06$\times 10^{9}$& 139873.60 \\
 J2055+3829 & 0.12 & 4.32$\times 10^{33}$ & 0.0020 & 4.62$\times 10^{7}$ & 9.99$\times 10^{-22}$ & 0.025 & 0.045 &6.95$\times 10^{13}$& 1.53$\times 10^{9}$&  47537.98 \\
 J2115+5448 & 0.13 & 1.67$\times 10^{35}$ & 0.0026 & 4.46$\times 10^{8}$ & 7.49$\times 10^{-20}$ & 0.024 & 0.044 &4.32$\times 10^{14}$& 1.18$\times 10^{10}$& 237535.29 \\
 J2214+3000 & 0.41 & 1.91$\times 10^{34}$ & 0.0031 & 2.16$\times 10^{8}$ & 1.47$\times 10^{-20}$ & 0.015 & 0.059 &1.46$\times 10^{14}$& 4.81$\times 10^{9}$&  67001.37 \\
 J2234+0944 & 0.41 & 1.66$\times 10^{34}$ & 0.0036 & 2.73$\times 10^{8}$ & 2.01$\times 10^{-20}$ & 0.017 & 0.068 &1.36$\times 10^{14}$& 5.21$\times 10^{9}$&  53684.04 \\
 J2241-5236 & 0.14 & 2.60$\times 10^{34}$ & 0.0021 & 1.24$\times 10^{8}$ & 6.89$\times 10^{-21}$ & 0.013 & 0.025 &1.70$\times 10^{14}$& 3.93$\times 10^{9}$& 111420.40 \\
 J2256-1024 & 0.21 & 3.71$\times 10^{34}$ & 0.0022 & 1.63$\times 10^{8}$ & 1.13$\times 10^{-20}$ & 0.034 & 0.082 &2.03$\times 10^{14}$& 4.92$\times 10^{9}$& 126750.83 \\
 J2322-2650 & 0.32 & 5.54$\times 10^{32}$ & 0.0034 & 4.54$\times 10^{7}$ & 5.83$\times 10^{-22}$ & 0.0008 & 0.002 &2.48$\times 10^{13}$& 9.08$\times 10^{8}$&  10266.96 \\
 \hline
 J1221-0633	&0.38	& 2.87$\times 10^{34}$&	0.0019&	1.02$\times 10^{8}$&	5.26$\times 10^{-21}$& 0.015&    0.05&	1.79$\times 10^{14}$&3.65$\times 10^{9}$&	132280.06\\
 J1300+1240$^\dag$   &25.2   & 1.87$\times 10^{34}$& 0.0062&	8.53$\times 10^{8}$&	1.14$\times 10^{-19}$& 5.6$\times 10^{-8}$&  3$\times 10^{-6}$&1.44$\times 10^{14}$&9.49$\times 10^{9}$&	 33265.14\\
 J1317-0157	&0.08	& 8.82$\times 10^{33}$&	0.0029&	1.27$\times 10^{8}$&	5.49$\times 10^{-21}$& 0.020&    0.02&	9.92$\times 10^{13}$&3.04$\times 10^{9}$&	 48767.61\\
 J1627+3219	&0.16	& 2.07$\times 10^{34}$&	0.0021&	1.10$\times 10^{8}$&	5.47$\times 10^{-21}$& 0.025&    0.05&	1.52$\times 10^{14}$&3.50$\times 10^{9}$&	 99738.68\\
 J1630+3550	&0.31	& 2.44$\times 10^{34}$&	0.0032&	2.62$\times 10^{8}$&	2.08$\times 10^{-20}$& 0.011&    0.03&	1.65$\times 10^{14}$&5.62$\times 10^{9}$&	 73150.90\\
 J1701-3006F$^{\star}$ &0.20	& 7.25$\times 10^{35}$&	0.0022&	7.22$\times 10^{8}$&	2.22$\times 10^{-19}$& 0.024&    0.05&	9.00$\times 10^{14}$&2.17$\times 10^{10}$&	560490.08\\
 J1737-0314A$^{\star}$ &0.22	& 4.86$\times 10^{35}$&	0.0019&	4.40$\times 10^{8}$&	9.55$\times 10^{-20}$& 0.018&    0.04&	7.36$\times 10^{14}$&1.53$\times 10^{10}$&	531633.70\\
\hline
\multicolumn{11}{c}{Redbacks and transitional millisecond pulsars} \\
\hline
 J1023+0038$^{\ast}$& 0.19 & 5.68$\times 10^{34}$ & 0.0016 & 1.09 $\times 10^{8}$ & 6.92 $\times 10^{-21}$ & 0.15 & 0.34 &2.52$\times 10^{14}$&4.48$\times 10^{9}$& 213328.66 \\
 J1048+2339 & 0.25 & 1.16$\times 10^{34}$ & 0.0046 & 3.79 $\times 10^{8}$ & 3.00 $\times 10^{-20}$ & 0.35 & 0.83 &1.14$\times 10^{14}$&5.62$\times 10^{9}$&  35000.56 \\
 J1227-4853$^{\ast}$& 0.28 & 9.12$\times 10^{34}$ & 0.0016 & 1.38 $\times 10^{8}$ & 1.10 $\times 10^{-20}$ & 0.24 & 0.66 &3.19$\times 10^{14}$&5.67$\times 10^{9}$& 270534.19 \\
 J1431-4715 & 0.44 & 6.83$\times 10^{34}$ & 0.0020 & 1.70 $\times 10^{8}$ & 1.41 $\times 10^{-20}$ & 0.14 & 0.55 &2.76$\times 10^{14}$&5.86$\times 10^{9}$& 196263.68 \\
 J1622-0315 & 0.16 & 7.93$\times 10^{33}$ & 0.0038 & 2.12 $\times 10^{8}$ & 1.14 $\times 10^{-20}$ & 0.11 & 0.21 &9.41$\times 10^{13}$&3.81$\times 10^{9}$&  34977.90 \\
 J1628-3205 & 0.21 & 1.42$\times 10^{34}$ & 0.0032 & 1.98 $\times 10^{8}$ & 1.19 $\times 10^{-20}$ & 0.18 & 0.41 &1.26$\times 10^{14}$&4.27$\times 10^{9}$&  56116.06 \\
J1717+4308A$^{\star}$ & 0.20 & 7.65$\times 10^{34}$ & 0.0031 & 4.44 $\times 10^{8}$ & 6.11 $\times 10^{-20}$ & 0.18 & 0.39 &2.92$\times 10^{14}$&9.73$\times 10^{9}$& 132169.23 \\
 J1723-2837 & 0.61 & 4.65$\times 10^{34}$ & 0.0018 & 1.19 $\times 10^{8}$ & 7.54 $\times 10^{-21}$ & 0.27 & 1.22 &2.28$\times 10^{14}$&4.46$\times 10^{9}$& 175618.45 \\
J1740-5340A$^{\star}$ & 1.35 & 1.36$\times 10^{35}$ & 0.0036 & 7.92 $\times 10^{8}$ & 1.68 $\times 10^{-19}$ & 0.21 & 1.65 &3.90$\times 10^{14}$&1.50$\times 10^{10}$& 152737.57 \\
 J1803-6707 & 0.38 & 7.49$\times 10^{34}$ & 0.0021 & 2.00 $\times 10^{8}$ & 1.84 $\times 10^{-20}$ & 0.33 & 1.06 &2.89$\times 10^{14}$&6.51$\times 10^{9}$& 193679.03 \\
 J1816+4510 & 0.36 & 5.22$\times 10^{34}$ & 0.0031 & 3.75 $\times 10^{8}$ & 4.31 $\times 10^{-20}$ & 0.18 & 0.59 &2.41$\times 10^{14}$&8.13$\times 10^{9}$& 108100.42 \\
 J1908+2105 & 0.14 & 3.23$\times 10^{34}$ & 0.0025 & 1.90 $\times 10^{8}$ & 1.38 $\times 10^{-20}$ & 0.06 & 0.11 &1.90$\times 10^{14}$&5.14$\times 10^{9}$& 105961.15 \\
 J1957+2516 & 0.23 & 1.74$\times 10^{34}$ & 0.0039 & 3.33 $\times 10^{8}$ & 2.74 $\times 10^{-20}$ & 0.11 & 0.28 &1.39$\times 10^{14}$&5.82$\times 10^{9}$&  50306.20 \\
 J2039-5617 & 0.22 & 3.00$\times 10^{34}$ & 0.0026 & 1.96 $\times 10^{8}$ & 1.41 $\times 10^{-20}$ & 0.19 & 0.47 &1.83$\times 10^{14}$&5.11$\times 10^{9}$&  98678.12 \\
 J2215+5135 & 0.17 & 7.41$\times 10^{34}$ & 0.0026 & 2.98 $\times 10^{8}$ & 3.33 $\times 10^{-20}$ & 0.24 & 0.46 &2.87$\times 10^{14}$&7.91$\times 10^{9}$& 157526.43 \\
 J2339-0533 & 0.19 & 2.32$\times 10^{34}$ & 0.0028 & 2.04 $\times 10^{8}$ & 1.41 $\times 10^{-20}$ & 0.30 & 0.61 &1.60$\times 10^{14}$&4.89$\times 10^{9}$&  79742.09 \\
  \hline
 J1902-5105*& 2.01 & 6.86$\times 10^{34}$ & 0.0017 & 1.28$\times 10^{8}$ & 9.19$\times 10^{-21}$& 0.18 & 1.90 & 2.76$\times 10^{14}$ & 5.08$\times 10^{9}$ & 227053.37 \\
 J1302-3258 & 0.78 & 4.81$\times 10^{33}$ & 0.0037 & 1.58$\times 10^{8}$ & 6.54$\times 10^{-21}$& 0.17 & 0.92 & 7.33$\times 10^{13}$ & 2.91$\times 10^{9}$ &  27786.94 \\
 \hline
\hline
\end{longtable}
\end{center}

\chapter{Separating repeating fast radio bursts using the minimum spanning tree as an unsupervised methodology}
\label{chapter7}

\section{Introduction} 
\label{chapter7: intro}

Fast Radio Bursts (FRBs) are transient radio pulses, typically lasting on the order of milliseconds and originating from cosmological distances (see Section~\ref{chapter1: FRBs}).
They have attracted considerable interest and numerous physical interpretations; see, e.g., \citet{Katz2018, Popov2018, Cordes2019, Petroff2019, Platts2019, Zhang2020, Xiao2021, Petroff2022, Xiao2022, Zhang2023}. 
Their observed properties, such as their dispersion (DM) and rotation (RM) measures, provide clues into the physical characteristics of the intervening medium, including density and magnetic field strength. 
Their energy or brightness temperature also offers valuable hints regarding the underlying mechanisms driving their emission. 
After the discovery of the first repeater in 2016, FRB 20121102A \citep{Spitler2016, Scholz2016}, FRBs are broadly categorized into two groups based on their repetition properties: repeaters and (apparently) non-repeaters. 
However, synthesizing the physical causes behind these differences and escaping from often-not-fully-understood observational biases remains a significant challenge. See Chapter~\ref{chapter1} for a detailed discussion of FRBs.

In this chapter based on the work \citep{MST-FRBs}, we aim to explore a novel classification tool for FRBs, separating them into repeaters and non-repeaters (for other approaches, see, e.g., \cite{Palaniswamy2018, Caleb2019, Ai2021, ML-FRBI, ML-FRBII, FRBs-Sun25}). 
We utilize graph theory (see Chapter~\ref{chapter2}) as a framework for this methodology, and PCA (see Chapter~\ref{chapter3}) as an alternative to describe the properties of FRBs. 
In particular, we introduce the MST as an unsupervised learning approach with a supervised evaluation.
The MST, where each node will represent an FRB, provides a manageable and compact structure calculated in an $N$-dimensional space.
Aided by the MST's fundamental properties, we will explore its capacity to establish the variables with the best separation power and then indicate a group of likely repeating FRBs that have yet to be classified.
Using an unsupervised learning approach is particularly appealing when one does not control which properties within a data set drive a classification, as is the case here. 
In addition, even the labels themselves are subject to possible change in time, as non-repeaters could become repeaters. 
Also, such methods do not necessarily require sample splitting for proper application.
This is valuable since the sample is insignificant and contains at most 750 FRBs, so we are not dealing with big data. Randomly cutting the sample in a training set (say, 80\%/20\% in size) would imply that 150 FRBs are sufficient to test the pattern behind a repeater/non-repeater classification, which we consider a risk.
Finally, the MST, as an unsupervised method, provides transparency in identifying the variables that best distinguish repeaters from non-repeaters. This becomes less clear in supervised methods when techniques such as feature importance (see, e.g., \cite{Feature_importance_1, Feature_importance_2, Feature_importance_3, Feature_importance_4}) or Shapley Additive exPlanations (SHAP, see \cite{SHAP_values1, SHAP_values2} for more details) are applied for that purpose.

\section{Sample, variables, PCA, and MST}
\label{chapter7: sample_variables_PCA_MST}

We use the sample from the Canadian Hydrogen Intensity Mapping Experiment Fast Radio Burst (CHIME) catalog \citep{CHIME_2021, CHIME_2023}.
We consider their 750 FRBs, comprising 265 repeaters and 485 non-repeaters. 
The repeaters group corresponds to events from 70 localizations. Each FRB is considered individually as a distinct event, labeled as a repeater (as in CHIME). 
Following the CHIME catalog, sub-bursts can be attributed to repeaters and non-repeaters. They are also considered individual bursts, each with the corresponding label.
We exclude the six FRBs for which neither flux nor fluence has been measured.

To characterize an FRB, we shall consider the logarithm of the following variables\footnote{This is done just for convenience, as their values may differ by several orders of magnitude for different FRBs} \citep{CHIME_2021, CHIME_2023}:
\begin{itemize}
\item peak frequency $\nu_{c}$ (MHz), 
\item flux $S_{\nu}$ (Jy), 
\item fluence $F_{\nu}$ (Jy ms), 
\item boxcar width $\Delta t_{BC}$ (ms), 
\end{itemize}
and also derived parameters based on the DM-$z$ relation (as presented in \citet{ML-FRBI, ML-FRBII} and references therein), such as:
\begin{itemize}
\item redshift $z$, 
\item rest-frame frequency width $\Delta \nu$ (MHz), 
\item rest-frame width $\Delta t_{r}$ (ms), 
\item burst energy $E$ (erg), 
\item luminosity $L$ (erg/s), and 
\item brightness temperature $T_B$ (K).
\end{itemize}

As their distribution is not normal (nor is it that of the original variables without logarithm), we use the robust scaler to scale them (see Equation~(\ref{chapter_4_eq: robust_scaler})).

\begin{figure}
  \includegraphics[width=1\textwidth]{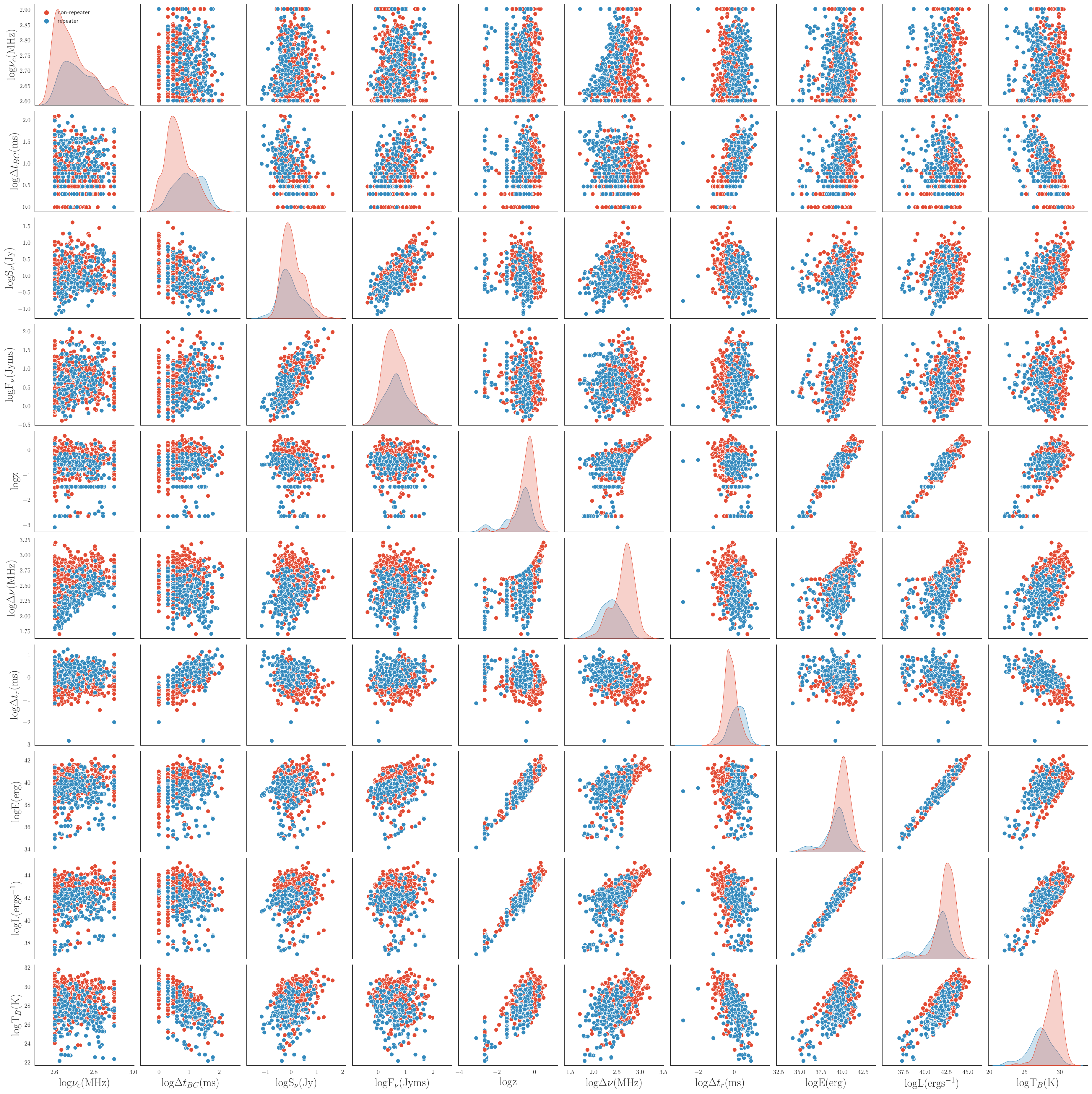}
  \centering
\caption[Cross-correlation of the magnitudes considered.]{
Cross-correlation of the logarithm of the 10 magnitudes considered.
Repeaters (in blue) and non-repeaters (in red) are separately noted.
The main diagonal shows the distribution for each variable.
}
  \label{chapter_7_figures: pairplot_classes_originalVariables}
\end{figure}

As we can observe from the cross-correlations shown in Figure~\ref{chapter_7_figures: pairplot_classes_originalVariables}, the physical properties exhibit different classification powers.
This can be corroborated by applying a simple separation method: for each of the magnitudes considered, we note the median of each distribution and count the number of repeaters positioned on each side of this central location.
The rest-frame frequency width provides the best separation, yielding 217 (82\%) grouped FRB repeaters (all with a width lower than 397.24 MHz).
This is far off from the random expectation: After producing 10$^5$ simulations in which repeater labels are randomly assigned, the average separation achieved is 52\%, that is, a separation consistent with an entirely random process in which labels are separated in equal numbers on both sides of the median.
This high separation power of the rest-frame frequency width will serve as a threshold to validate the performance of other algorithms or classifiers.
We shall look below for classifiers that will perform better than 82\% in separating repeaters from non-repeaters.

\begin{figure}
  \centering
  \includegraphics[width=1\textwidth]{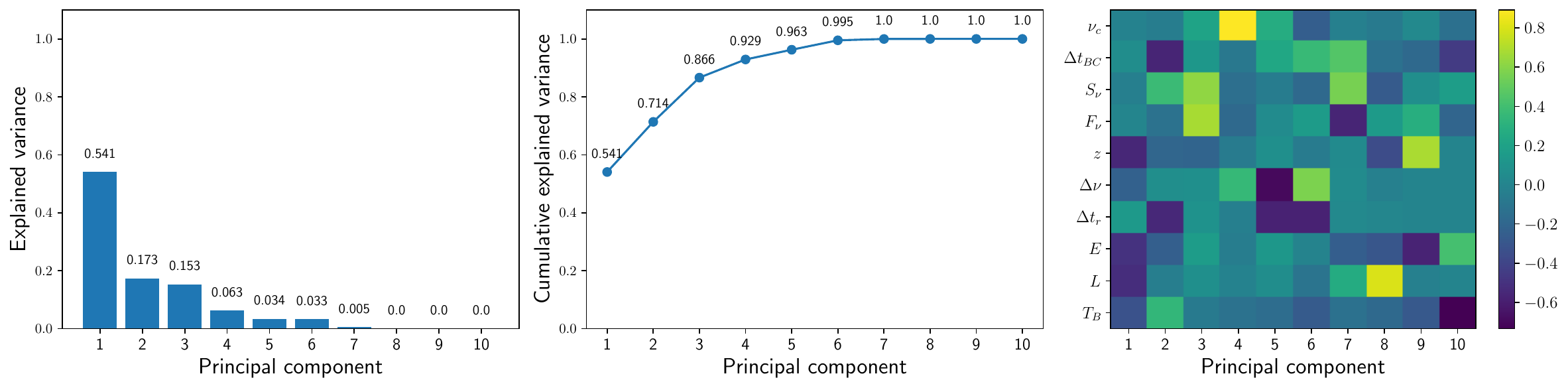}
  \caption[PCA results.]{PCA results for the logarithm of the set of variables defined for the population of 750 FRBs, see Section~\ref{chapter3: visual_tools} for further explanations about PCA definitions.

}
  
  \label{chapter_7_figures: PCA}
\end{figure}

Figure~\ref{chapter_7_figures: PCA} presents the outcome of conducting PCA over the logarithm of all the variables defined in this section, see Chapter~\ref{chapter3} for more details.
While the first six PCs contain the majority of variance, retaining over 99\% of its informational content, the first three PCs contain 87\% of it.
Full variance coverage needs nine PCs, so the dimensionality reduction is not extreme if it is to be entirely retained. 
Thus, the number of PCs needed to cover the total variance, combined with the relatively flat distribution of the loadings of the first PCs, renders its use a priori unappealing.
In other words, if almost all PCs are needed (9 out of 10), dimensionality is not significantly reduced. If the loadings are flat, no strong directional variance is identified, thus no dominant underlying structure appears. 
This justifies saying that PCA does not extract meaningful simplification or patterns.

Considering the MST, we shall introduce the Euclidean distance (see Equation~(\ref{chapter_2_eq: d_eucl})), using all or part of the defined physical variables mentioned above, or the PCs (employing the PCs that account for 100\% of the explained variance yields an MST identical to the one derived using all magnitudes). We initially generate a complete, undirected, weighted graph $G(V, E)$, characterized by a set $V$ of nodes (each of them is an FRB) and a set $E$ of edges to define the relationships between the FRBs. The latter are assigned to a specific weight $w$ equal to the Euclidean distance between the FRBs, using all or part of the variables or PCs, as described.
We will start from the graph $G(750,280875)$ for each combination of variables and derive an MST, $T(750,749)$. 

\section{Algorithm and results}
\label{chapter7: results}

The total number of combinations (without repetition) from a set of 10 variables is 
$\sum_1^{10} \binom
{10}{k} = 2^{10} - 1 = 1023$,
reflecting the number of MSTs we compute.
For each of these MSTs, we identify the node with the highest centrality of the graph through the betweenness centrality estimator (see Section~\ref{chapter2: bet_cen}).
This estimator determines the frequency at which a particular node appears on the shortest path between any two other nodes, thereby measuring the node's importance in terms of connectivity within the MST. 
Eliminating this node would separate $T$ into several connected parts (or branches).
The branch with the highest number of repeaters will be named the repeater branch.
All other branches will be considered as non-repeater branches.
A high density of repeaters in the repeater branch would qualify the variables under which the underlying Euclidean distance is built as a well-separated filter via the MST algorithm.
To judge the performance of each MST seen as a classifier, we use the metrics precision, recall, $F_{1}$, and $F_{2}$ score, plus the area under the curve (AUC or ROC-AUC), see Appendix \ref{Appendix_7} for a more detailed explanation.
Precision measures the ratio of correctly identified repeaters regarding the repeater branch's total FRBs (both repeaters and non-repeaters). At the same time, recall assesses the proportion of actual repeaters that are correctly identified. The $F_{1}$ score balances precision and recall equally, whereas the $F_{2}$ score places more on recall. AUC evaluates the classifier's overall discrimination ability.
In identifying high-density locations of repeaters, metrics such as recall and precision focus on correctly identifying actual repeaters and minimizing non-repeaters that are misclassified as repeaters. 
A high $F_{1}$ score indicates a good balance between recall and precision. 
On the other hand, the $F_{2}$ score, as seen, places more emphasis on recall than precision, effectively tolerating a certain level of misclassification of non-repeaters as repeaters.
Lastly, a high AUC suggests that MST predictions rank actual repeater instances higher than non-repeaters, which is crucial for identifying many repeaters among the whole sample.
To determine the best combination of variables, we look for a combination that offers a balanced trade-off between precision and recall (as reflected in the $F_{1}$ Score), also considering the $F_{2}$ Score (which weighs recall higher than precision) and the ROC-AUC value for overall performance.
This is, in practice, implemented by an overall rank (see, e.g., \cite{OverallRank2}), which averages all ranks across these evaluating metrics.

Armed with the MST/betweenness centrality methodology and the rankings, we find 25 combinations of variables and three combinations of PCs for which we obtain a separating, i.e., recall better than 82\%. 
Results are shown in Table~\ref{chapter_7_table: final_results}.
Recall values are notably high across the board, which is excellent if the primary concern is to minimize missing repeaters. 
In some combinations, precision is lower than recall, which is expected when recall is prioritized.
The $F_{1}$ scores are still reasonably good, indicating a balanced performance between precision and recall. 
The $F_{2}$ scores are also high, reinforcing the model's orientation towards recall, which is desirable in contexts where the consequences of missing repeaters outweigh those of non-repeaters seen as repeaters.
The ROC-AUC values are also far from what could be considered a random guess, showing a reasonable degree of separability.
Based on the overall rank, a combination of peak frequency, rest-frame frequency width, and brightness temperature (\#1) achieves the best balance.
Although this is not the best combination for recall, for which it shows 0.8528, it is not far from the best 0.8604 (a difference of just two repeaters) obtained with the combination (\#7).
Additionally, it achieves a precision of 0.5825, the second highest, which, combined with the recall above, yields the combination with the best $F_{1}$ score, $F_{2}$ score, and ROC value. 

Table~\ref{chapter_7_table: final_results} shows the results obtained using the PCs. 
Recalling that we need 9 PCs to define the entire variance, we consider $\sum_1^{9} \binom{9}{k} =2^{9} - 1 = 511$.
Only three combinations of PCs present results exceeding the recall threshold, although they are not particularly informative or ranked better than those obtained directly using physical variables.

We consistently observe high recall rates, exceeding 0.82 across all combinations, which highlights their effectiveness in accurately identifying repeaters.
The precision values, ranging from 0.51 to 0.55, and corresponding $F_{1}$ scores between 0.63 and 0.67, suggest a reasonable balance between the accuracy of robust predictions and the method's ability to identify actual repeaters.
Furthermore, the $F_{2}$ scores, centered around 0.75, emphasize the evaluation's focus on recall over precision, aligning with our classification goal to reduce misclassified repeaters.
The approximately 0.65 to 0.68 ROC-AUC values demonstrate the MST based on PCs' robust capability to distinguish between repeaters and non-repeaters, indicating significant discriminative power. 
Considering these points through the overall rank, the combination ($PC_1$, $PC_2$, $PC_3$) emerges as the most effective. 
This represents 87\% of the total variance and has the best separation power of all those using principal components. However, describing more variance with more PCs does not yield better separation power, implying that something other than the variables used to describe the sample here likely plays a role in the separation. 

\subsection{Significance testing via Monte Carlo label permutations}
\label{chapter7: MC_permutations}

To evaluate whether the observed separation power could arise by chance, we test the randomness of the process in two ways, employing a Monte Carlo permutation test considering the labels of the FRBs. 
First, we question whether, if there is no link between repeating/non-repeating labels and the variables, we could still find a group of 25 combinations out of 1023 possible ones that could lead to a separating power of over 82\%. 
We do so by randomly relabelling the FRBs in repeaters and non-repeaters, maintaining the sample proportions, and building 10$^{5}$ fake samples for each MST using the same variables as the original.

In any of the simulations performed, no combination of variables out of the 1023 possible exceeded the threshold of 82\% separation.
Second, for each combination selected in Table~\ref{chapter_7_table: final_results}, we compare the actual result (i.e., its real separation power) with the average recall of the fake samples using the Z-score. This statistical measure quantifies how many standard deviations a data point (in this case, the recall) is from the mean of the distribution (see the Appendix \ref{Appendix_7} for more details).
Table~\ref{chapter_7_table: final_results} shows that Z-scores obtained are at 9$\sigma$ or higher for recall values, along all combinations, compared to distributions from fake sets.
Additionally, considering the PCs, the high Z-scores of 10 and 12 across the combinations underscore the statistical significance of these results, affirming that the observed recall rates are not due to chance.
These randomness tests are a strong indicator of the effectiveness and robustness of this classification approach. 
In particular, the randomness of the best-ranked combination of variables (\#1), viewed in the Z-score of the first row in Table~\ref{chapter_7_table: final_results}, is one of the lowest, which proves a robust classification. 

\begin{table}
\scriptsize
\centering
\caption[Summary of the evaluating metrics.]{Summary of performance metrics of the combination of variables selected as those exceeding the threshold of 0.82 recall is seen in the rest-frame frequency width, sorted by overall rank.
The combinations seen in the 'variables' column are named as follows:
(1): $\mathrm{log}\nu_{c}$, 
(2): $\mathrm{log}\Delta t_{BC}$, 
(3): $\mathrm{log}S_{\nu}$, 
(4): $\mathrm{log}F_{\nu}$, 
(5): $\mathrm{log}z$, 
(6): $\mathrm{log}\Delta \nu$, 
(7): $\mathrm{log}\Delta t_{r}$, 
(8): $\mathrm{log}E$, 
(9): $\mathrm{log}L$, and 
(10):$\mathrm{log}T_B$.
}

\begin{tabular}{c|ccccccccccc}
\hline
\# & Variables & Precision & Recall & $F_{1}$ Score & $F_{2}$ Score & ROC-AUC & Z Score \\
\hline
\toprule
1 & [1, 6, 10] & 0.5825 & 0.8528 & 0.6922 & 0.7804 & 0.7773 & 14 \\
2 & [1, 5, 6, 7] & 0.5622 & 0.8528 & 0.6777 & 0.7729 & 0.7033 & 13 \\
3 & [1, 2, 5, 6, 10] & 0.5589 & 0.8415 & 0.6717 & 0.7642 & 0.7285 & 13 \\
4 & [1, 3, 4, 6, 7, 8, 9, 10] & 0.5784 & 0.8491 & 0.6881 & 0.7764 & 0.6834 & 13 \\
5 & [1, 5, 6, 8, 10] & 0.5878 & 0.8340 & 0.6895 & 0.7695 & 0.6961 & 14 \\
6 & [6] & 0.5722 & 0.8226 & 0.6749 & 0.7564 & 0.7522 & 19 \\
7 & [1, 2, 3, 6, 7, 8] & 0.5352 & 0.8604 & 0.6599 & 0.7672 & 0.6941 & 12 \\
8 & [1, 3, 5, 6, 9] & 0.5510 & 0.8566 & 0.6706 & 0.7711 & 0.6577 & 13 \\
9 & [1, 2, 3, 4, 6, 7, 9, 10] & 0.5457 & 0.8340 & 0.6597 & 0.7543 & 0.7007 & 12 \\
10 & [2, 5, 10] & 0.5598 & 0.8302 & 0.6687 & 0.7571 & 0.6999 & 13 \\
11 & [1, 3, 5, 6] & 0.5509 & 0.8377 & 0.6647 & 0.7587 & 0.6645 & 12 \\
12 & [1, 4, 5, 6, 9] & 0.5688 & 0.8264 & 0.6738 & 0.7578 & 0.6583 & 13 \\
13 & [5, 6, 9, 10] & 0.5198 & 0.8415 & 0.6427 & 0.7488 & 0.6883 & 11 \\
14 & [8, 10] & 0.5298 & 0.8377 & 0.6491 & 0.7505 & 0.6859 & 11 \\
15 & [2, 6, 8] & 0.5366 & 0.8302 & 0.6519 & 0.7483 & 0.7006 & 12 \\
16 & [1, 2, 4, 6, 7, 9, 10] & 0.5239 & 0.8264 & 0.6413 & 0.7409 & 0.7230 & 11 \\
17 & [1, 2, 4, 5, 6, 10] & 0.5201 & 0.8302 & 0.6395 & 0.7417 & 0.7118 & 11 \\
18 & [1, 4, 6, 9, 10] & 0.5116 & 0.8302 & 0.6331 & 0.7383 & 0.7013 & 10 \\
19 & [1, 2, 3, 6, 9, 10] & 0.5117 & 0.8226 & 0.6310 & 0.7335 & 0.7180 & 10 \\ 
20 & [1, 2, 4, 5, 6, 9, 10] & 0.5034 & 0.8302 & 0.6268 & 0.7348 & 0.7002 & 10 \\
21 & [1, 2, 3, 4, 6, 7, 8, 9] & 0.4966 & 0.8377 & 0.6236 & 0.7366 & 0.6590 & 10 \\
22 & [1, 2, 4, 5, 6] & 0.4944 & 0.8340 & 0.6208 & 0.7332 & 0.6866 & 10 \\
23 & [1, 6, 8, 9, 10] & 0.4846 & 0.8302 & 0.6120 & 0.7266 & 0.7026 & 9 \\
24 & [1, 5, 6, 9, 10] & 0.4944 & 0.8302 & 0.6197 & 0.7309 & 0.6987 & 10 \\
25 & [1, 4, 5, 6, 9, 10] & 0.4943 & 0.8226 & 0.6176 & 0.7262 & 0.6851 & 10 \\
\hline
1 & [$PC_1$,$PC_2$,$PC_3$] & 0.5504 & 0.8453 & 0.6667 & 0.7635 & 0.6858 & 12    \\ 
2 & [$PC_1$,$PC_7$] & 0.5178 & 0.8226 & 0.6356 & 0.7360 & 0.6716 & 11 \\ 
3 & [$PC_1$,$PC_3$,$PC_6$,$PC_7$,$PC_8$] & 0.5092 & 0.8377 & 0.6334 & 0.7420 & 0.6523 & 11  \\ 
\hline
\end{tabular}
\label{chapter_7_table: final_results}
\end{table}

\begin{figure}
  \centering
  \includegraphics[width=1\textwidth]{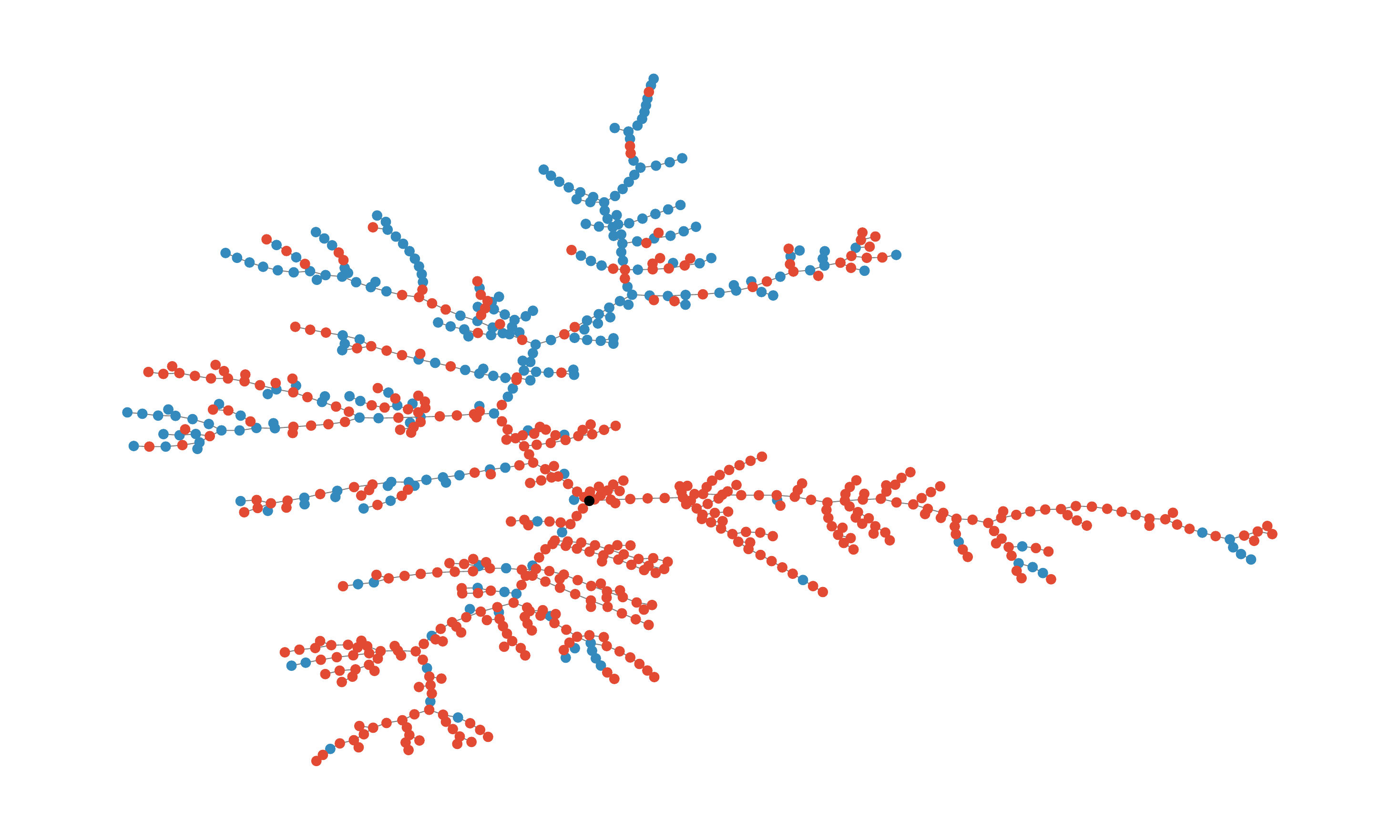}
  \caption[$T(750,749)$ based on the combination of $\mathrm{log}\nu_{c}$, $\mathrm{log}\Delta \nu$, and $\mathrm{log}T_B$.]{$T(750,749)$ computed from the Euclidean distance based on the combination of peak frequency, rest-frame frequency width, and brightness temperature. Repeaters are shown in blue, and non-repeaters in red. The most central node in terms of betweenness centrality appears in black.}
  \label{chapter_7_figures: MST_winner}
\end{figure}

Combination \#1 successfully identified a significant number of repeaters (226/265), ensuring that relatively few repeaters are missed (39/265). 
Some non-repeaters are located in the repeater branch of the MST (162/485).
If there is any physical link between the variables and the repeating property, some of them should be candidates to appear as repeaters in the next catalogs, depending on duty cycles. We will return to this below.
Figure~\ref{chapter_7_figures: MST_winner} shows the MST with the arrangement of repeaters and non-repeaters of combination \#1; it separates the two types of FRBs. 

\begin{sidewaystable}
\scriptsize
\centering
\caption[Properties of the FRBs candidates.]{Properties of the currently labeled non-repeater FRBs seen as repeaters for all combinations shown in Table~\ref{chapter_7_table: final_results}.
The "Sub" column shows the sub-burst number to which each FRB corresponds.
Candidate FRBs reported in \citet{ML-FRBI, ML-FRBII} are highlighted in black. 
Repeaters not classified as such in any of the selected combinations (outliers) are shown below the line.
}
\begin{tabular}{llcccccccccccc}
\hline
Name & Sub &
RA  & Dec   & $\nu_{c}$  & $\Delta t_{BC}$  & $S_{\nu}$  & $F_{\nu}$  & $z$ & $\Delta \nu$  & $\Delta t_{r}$  & $E$ & $L$  & $T_B$  \\
    &       & $(^\circ)$ &   $(^\circ)$ &  (MHz) &  (ms) & (Jy) &  (Jy ms) & &  (MHz) &  (ms) &  (erg) & (erg/s) &  (K) \\
\hline
\toprule
 FRB20180907E   &  0 & 167.88 & 47.09 & 400.20 & 11.80 & 0.73 & 6.90 & 0.3118 & 178.54 & 3.16 & $7.11        \times 10^{39}$ & $9.87 \times 10^{41}$ & $7.74 \times 10^{27}$ \\
 FRB20180920B   &  0 & 191.09 & 63.52 & 421.10 & 10.81 & 0.35 & 1.70 & 0.4007 & 116.53 & 1.66 & $3.10        \times 10^{39}$ & $8.93 \times 10^{41}$ & $6.71 \times 10^{27}$ \\
 FRB20180928A   &  0 & 312.95 & 30.85 & 400.20 & 2.95 & 1.34 & 2.50 & 0.0022 & 92.11 & 0.27 & $1.20          \times 10^{35}$ & $6.44 \times 10^{37}$ & $1.06 \times 10^{25}$ \\
\textbf{FRB20181017B}   &  0 & 237.76 & 78.50 & 593.20 & 12.78 & 1.06 & 6.50 & 0.2067 & 247.97 & 1.91 & $4.24\times 10^{39}$ & $8.35 \times 10^{41}$ & $1.86 \times 10^{27}$ 
 \\
 FRB20181022E  & 0 & 221.18  & 27.13 & 443.70 & 2.95 & 0.69 & 2.08 & 0.2073 & 193.28 & 0.33 & $1.02          \times 10^{39}$ & $4.09 \times 10^{41}$ & $4.09 \times 10^{28}$ \\
 FRB20181125A &  0 & 147.94 & 33.93 & 434.50 & 14.75 & 0.39 & 3.20 & 0.1710 & 156.33 & 1.09 & $1.04          \times 10^{39}$ & $1.48 \times 10^{41}$ & $6.49 \times 10^{26}$ \\
 FRB20181125A &  1 & 147.94 & 33.93 & 436.60 & 14.75 & 0.39 & 3.20 & 0.1710 & 177.76 & 1.23 & $1.04          \times 10^{39}$ & $1.49 \times 10^{41}$ & $6.43 \times 10^{26}$ \\
 FRB20181125A &  2 & 147.94 & 33.93 & 426.50 & 14.75 & 0.39 & 3.20 & 0.1710 & 141.34 & 1.35 & $1.02          \times 10^{39}$ & $1.45 \times 10^{41}$ & $6.73 \times 10^{26}$ \\
 FRB20181214A & 0 & 70.00   & 43.07 & 435.00 & 2.95 & 0.156 & 0.41 & 0.2308 & 116.19 & 0.43 & $2.47          \times 10^{38}$ & $1.15 \times 10^{41}$ & $1.20 \times 10^{28}$ \\
 FRB20181220A &  0 & 346.11 & 48.43 & 400.20 & 2.95 & 1.33 & 3.00 & 0.0022 & 196.64 & 0.43 & $1.44           \times 10^{35}$ & $6.39 \times 10^{37}$ & $1.05 \times 10^{25}$ \\
 FRB20181226E & 0 & 303.56  & 73.64 & 400.20 & 2.95 & 0.48 & 1.35 & 0.1779 & 186.23 & 0.99 & $4.37           \times 10^{38}$ & $1.83 \times 10^{41}$ & $2.55 \times 10^{28}$
 \\
 \textbf{FRB20181229B} &  0 & 238.37 & 19.78 & 445.50 & 20.64 & 0.42 & 4.90 & 0.3197 & 154.80 & 2.55 & $5.92 \times 10^{39}$ & $6.70 \times 10^{41}$ & $1.24 \times 10^{27}$ \\
 \textbf{FRB20190112A} & 0 & 257.98  & 61.20 & 697.70 & 9.83 & 1.40 & 16.20 & 0.3476 & 317.48 & 1.22 & $3.64 \times 10^{40}$ & $4.24 \times 10^{42}$ & $8.80 \times 10^{27}$ \\
 FRB20190128C &  0 &  69.80 & 78.94 & 491.60 & 15.73 & 0.71 & 5.90 & 0.1772 & 238.50 & 5.23 & $2.32          \times 10^{39}$ & $3.29 \times 10^{41}$ & $8.73 \times 10^{26}$ \\
 FRB20190206B & 0 & 49.76   & 79.50 & 506.40 & 19.66 & 0.95 & 9.60 & 0.2190 & 350.34 & 5.82 & $6.03          \times 10^{39}$ & $7.27 \times 10^{41}$ & $1.09 \times 10^{27}$ \\
 \textbf{FRB20190206A} &  0 & 244.85 &  9.36 & 534.50 & 5.90 & 1.40 & 9.10 & 0.0618 & 213.84 & 0.76 & $4.53  \times 10^{38}$ & $7.40 \times 10^{40}$ & $1.20 \times 10^{27}$ \\
 \textbf{FRB20190218B} &  0 & 268.70 & 17.93 & 588.00 & 17.69 & 0.57 & 5.90 & 0.4416 & 334.17 & 1.42 & $1.84 \times 10^{40}$ & $2.56 \times 10^{42}$ & $2.56 \times 10^{27}$ \\
 FRB20190223A & 0 & 64.72   & 87.65 & 444.80 & 3.93 & 0.47 & 1.58 & 0.2865 & 149.61 & 0.59 & $1.52           \times 10^{39}$ & $5.81 \times 10^{41}$ & $3.05 \times 10^{28}$ \\
 FRB20190308C &  0 & 188.36 & 44.39 & 453.40 & 21.63 & 0.47 & 4.80 & 0.4542 & 218.85 & 0.28 & $1.22          \times 10^{40}$ & $1.74 \times 10^{42}$ & $2.51 \times 10^{27}$ \\
 FRB20190308C &  1 & 188.36 & 44.39 & 449.00 & 21.63 & 0.47 & 4.80 & 0.4542 & 211.58 & 0.38 & $1.21          \times 10^{40}$ & $1.72 \times 10^{42}$ & $2.56 \times 10^{27}$ \\
 FRB20190323D &  0 &  56.88 & 46.93 & 400.20 & 12.78 & 0.37 & 2.49 & 0.5930 & 204.86 & 3.41 & $9.66          \times 10^{39}$ & $2.29 \times 10^{42}$ & $1.26 \times 10^{28}$ \\
 \textbf{FRB20190329A} &  0 &  65.54 & 73.63 & 432.30 & 11.80 & 0.52 & 2.24 & 0.0022 & 73.87 & 1.04 & $1.16  \times 10^{35}$ & $2.70 \times 10^{37}$ & $2.20 \times 10^{23}$ \\
 \textbf{FRB20190410A} &  0 & 263.47 & -2.37 & 515.70 & 6.88 & 1.59 & 5.80 & 0.0728 & 182.92 & 0.94 & $3.89  \times 10^{38}$ & $1.14 \times 10^{41}$ & $1.51 \times 10^{27}$ \\
 \textbf{FRB20190412B} &  0 & 285.65 & 19.25 & 400.20 & 42.27 & 0.68 & 12.80 & 0.0146 & 228.59 & 6.70 & $2.60\times 10^{37}$ & $1.40 \times 10^{39}$ & $1.11 \times 10^{24}$ \\
 \textbf{FRB20190423B} &  0 & 298.58 & 26.19 & 537.60 & 9.83 & 0.87 & 7.00 & 0.0031 & 159.79 & 2.48 & $8.54  \times 10^{35}$ & $1.06 \times 10^{38}$ & $6.48 \times 10^{23}$ \\
 \textbf{FRB20190423B} &  1 & 298.58 & 26.19 & 524.60 & 9.83 & 0.87 & 7.00 & 0.0031 & 148.96 & 8.47 & $8.33  \times 10^{35}$ & $1.04 \times 10^{38}$ & $6.81 \times 10^{23}$ \\
 \textbf{FRB20190429B} &  0 & 329.93 &  3.96 & 422.40 & 16.71 & 0.74 & 5.00 & 0.1944 & 50.64 & 5.34 & $2.05  \times 10^{39}$ & $3.62 \times 10^{41}$ & $1.32 \times 10^{27}$ \\
 FRB20190430A &  0 &  77.70 & 87.01 & 433.80 & 19.66 & 0.75 & 7.70 & 0.2278 & 214.13 & 2.75 & $4.50          \times 10^{38}$ & $5.38 \times 10^{41}$ & $1.27 \times 10^{27}$ \\
 \textbf{FRB20190527A} &  0 &  12.45 &  7.99 & 484.70 & 57.02 & 0.47 & 10.10 & 0.5367 & 205.46 & 1.74 & $3.87\times 10^{40}$ & $2.77 \times 10^{42}$ & $4.46 \times 10^{26}$ \\
 FRB20190527A &  1 &  12.45 &  7.99 & 449.10 & 57.02 & 0.47 & 10.10 & 0.5367 & 172.11 & 1.61 & $3.59         \times 10^{40}$ & $2.56 \times 10^{42}$ & $5.20 \times 10^{26}$ \\
 FRB20190601C &  0 &  88.52 & 28.47 & 517.00 & 5.90 & 1.32 & 5.80 & 0.1753 & 223.54 & 0.58 & $2.35           \times 10^{39}$ & $6.28 \times 10^{41}$ & $1.02 \times 10^{28}$ \\
 FRB20190601C &  1 &  88.52 & 28.47 & 502.20 & 5.90 & 1.32 & 5.80 & 0.1753 & 201.91 & 0.43 & $2.28           \times 10^{39}$ & $6.10 \times 10^{41}$ & $1.08 \times 10^{28}$ \\
 FRB20190617B &  0 &  56.43 &  1.16 & 459.30 & 13.76 & 0.99 & 9.20 & 0.1655 & 217.37 & 6.50 & $2.94          \times 10^{39}$ & $3.69 \times 10^{41}$ & $1.58 \times 10^{27}$ \\
 \hline
FRB20180910A &  0 & 354.83 & 89.01 & 417.60 & 0.98 & 6.50 & 5.60 & 0.6230 & 649.22 & 0.126 & $2.51 \times 10^{40}$ & $4.73 \times 10^{43}$ & $3.82 \times 10^{31}$ \\
FRB20190210C &  0 & 313.90 & 89.19 & 448.50 & 1.97 & 2.37 & 3.60 & 0.5798 & 631.93 & 0.181 & $1.50 \times 10^{40}$ & $1.56 \times 10^{43}$ & $2.58 \times 10^{30}$ \\
FRB20200726D &  0 & 294.75 & 59.40 & 684.00 & 4.92 & 0.76 & 3.48 & 1.3812 & 541.72 & 0.407 & $1.23 \times 10^{41}$ & $6.37 \times 10^{43}$ & $3.17 \times 10^{29}$ \\

\hline
\end{tabular}
\label{chapter_7_table: candidates}
\end{sidewaystable}

\subsection{Results considering selection effects}
\label{chapter7: results_effects}

As explained in \citet{CHIME_2021}, assessing selection effects is a challenge to be addressed in the study of FRBs. 
We have approached this question under the considerations outlined in Section~3.3 of \citet{CHIME_2023}. The following cuts have been imposed in the sample:

\begin{enumerate}
    \item Events measured by the \textit{bonsai}\footnote{FRB search code implemented to address computational challenges and generates SNR estimations, see \citet{Bonsai} and reference therein for more details.} real-time detection pipeline S/N $<$ 12 are excluded due to being more likely to be misclassified as noise.
    \item Events with DM $<$ 1.5 max(DM$_{\mathrm{NE2001}}$, DM$_{\mathrm{YMW16}}$) are excluded to minimize the possibility of having a wrong identification of rotating radio transients or radio pulsars as FRBs considering that a $\sim 50\%$ errors may be common in models used to estimate the Galactic DM.
    \item Events detected in the telescope's sidelobes (see \cite{Lin_sideLobes1, Lin_sideLobes2}) are excluded since understanding the primary beam shape at large zenith angles is limited. This makes it challenging to characterize events. 
\end{enumerate}

After considering such cuts, we are left with 459 FRBs. This is a significant reduction from the earlier sample (by about 40\%). In this sample, we find 135 FRBs classified as repeaters. 
Again, for each variable considered, we note the median of each distribution and count the number of repeaters positioned on each side of this central location. The rest-frame frequency width, which can separate 109 of the 135 repeaters (80\%), provides the best classification for this reduced sample.
Thus, we test whether our methodology provides a separating power beyond this threshold.
After applying it similarly to what has been done for the entire sample, we find 15 different combinations of variables 
for which the separating performance exceeds that provided by the rest-frame frequency width.

We note two aspects of interest: selection effects may force us to work with reduced samples whose properties are likely not well understood yet, due to low sample sizes and changes in total variance. As a result, the PCA analysis and the combination of variables with high separating power differ from those corresponding to the entire sample. 
However, it can be recognized that peak frequency, rest-frame frequency width, and brightness temperature (variables in \#1, see Table~\ref{chapter_7_table: final_results}) are among the variables that appear the most in all combinations with ample separating power, whereas, on the other hand, the burst energy is the least implied, offering little classification power. 

\section{Conclusions}
\label{chapter7: conclusions}

To our knowledge, this is the first application of the graph theory MST technique to separate the repetition properties of FRBs.
To some extent, this work is related to that of \citet{Bhatporia2023}, which utilizes topological data analysis to cluster the population of FRBs; however, performance metrics are not provided in their case.
The performance metrics seen in Table~\ref{chapter_7_table: final_results} indicate that the MST classifiers do a good job of separating repeaters among the FRBs, excelling especially in recall and thus minimizing false negatives, i.e., repeaters seen as non-repeaters. 
Furthermore, the clarity in its construction can contribute to the scientific understanding of FRBs by highlighting the importance of certain variables over others in determining what best describes a source's repetitive character.
Due to these motivations, this approach demonstrates the potential for practical application. 
Methodologically, we do not define clusters over the MST; instead, we compute the betweenness centrality for each node and identify the one with the highest value (the black node in Figure~\ref{chapter_7_figures: MST_winner}). 
This node acts as a bridge within the MST; we partition the graph into separate branches by removing it. We then use the labels to count the number of repeaters on each branch, which allows us to study the distribution of repeaters and non-repeaters. 
Since we do not cluster, we do not encounter issues such as chaining or unbalanced clustering, as seen, for example, in single linkage algorithms (see \cite{Everit_Landau_clusters} for a discussion). 
Importantly, we do not use the labels during the construction of the MST or the identification of the most central node; our method is unsupervised and requires no training. 
As a result, the MST works with the entire sample at once, without dividing it into smaller pieces, thereby minimizing the risk of losing relevant information.

Table~\ref{chapter_7_table: final_results} shows that while rest-frame frequency width and brightness temperature demonstrated strong discriminatory power per se via median separation of the sample, the frequent occurrence of peak frequency in the selected combinations highlights its potential contribution to the classification process. 
This information can be valuable for optimizing the current method and informing feature selection strategies in other classification methods, such as machine learning. 

The performance of this method is on par with or exceeds that of others, for instance, in \citet{Chen_2021, ML-FRBI, ML-FRBII}, where they utilize samples from \citet{CHIME_2021} under a machine learning approach.
In more detail, our methodology is applied under the same conditions as in \citet{ML-FRBI, ML-FRBII}, where the sample comprises 594 FRBs, consisting of 500 non-repeaters and 94 repeaters.  
Under these assumptions, the results are a 97\% recall and an $F_{2}$ score of 77\%, comparable to that obtained in the quoted papers.
The recall is higher than that seen in \citet{Chen_2021}, where the sample is the same, but they identify one repeater less (501 non-repeaters and 93 repeaters).
In this quoted work, the variable set adds more properties to deal with a total of 13 (10 observational and three model-dependent parameters, as they define), several of which are set.

Regarding the forecasting power of this methodology, in Table~\ref{chapter_7_table: candidates}, we show 33 currently labeled as non-repeaters FRBs that appear in the repeater branch, identified using the 25 selected combinations of variables that show a separating power exceeding a recall threshold of 82\%.
In the context of the MST strategy, these are considered strong candidates for repeating bursts.
Moreover, a few have been identified as well using machine learning methodologies in \citet{ML-FRBI, ML-FRBII}.
Also, Table~\ref{chapter_7_table: candidates} shows the FRBs denoted as outliers, known repeaters, but are consistently classified in the non-repeater branch of any of the 25 MSTs considered. 
Both candidates and outliers provide a handle on the technique's power and caveats, which should be monitored with future samples of FRBs.

Lastly, reducing false positives can be essential to gain robustness, and attempts could be made to improve precision without significantly compromising recall. 
To refine the classification, a methodology that involves dividing the graph into significant branches (as introduced in Chapter~\ref{chapter5}) could be helpful. 
If given the opportunity, we would like to consider the repetition rate of each of the repeating FRBs, which would be a noticeable enhancement.
In particular, we would like to see whether the MST grouping can distinguish a soft transition across the sample from the most commonly repeating FRBs to the non-repeating ones. 
However, this exercise is hindered by the current availability of data, as the rate should consider the optimal integration time for a particular field of view, which is not provided in the catalog information \citet{CHIME_2021, CHIME_2023}. 
We encourage providing such data in future catalog versions, as it would benefit this and similar analyses, comparing it with any modeling of the repeating mechanism.

\chapter{Dynamic Time Warping}
\label{chapter8}

\section{Introduction}
\label{chapter8: introduction}

The development of robust methods for time series alignment has been a long-standing area of study. 
The goal has been to develop techniques that are sufficiently stable to be adapted effectively.
Early efforts focused on the application of Euclidean distance \citet{Faloutsos1994, Korn1997, Das1998, Debregeas1998, Keogh_et_al2000, Yi2000, Keogh_et_al2001, Chan2003, Agrawal2005}. 
However, works such as \citet{Berndt_1994, Kadous_1999, Keogh_Pazzani_2000, Aach_2001, Bar-Joseph2002, Chu_2002, Vlachos_et_al_2002} have observed crucial weaknesses, including its rigidity in the alignment.

\begin{figure} 
\centering
  \includegraphics[width=1\textwidth]{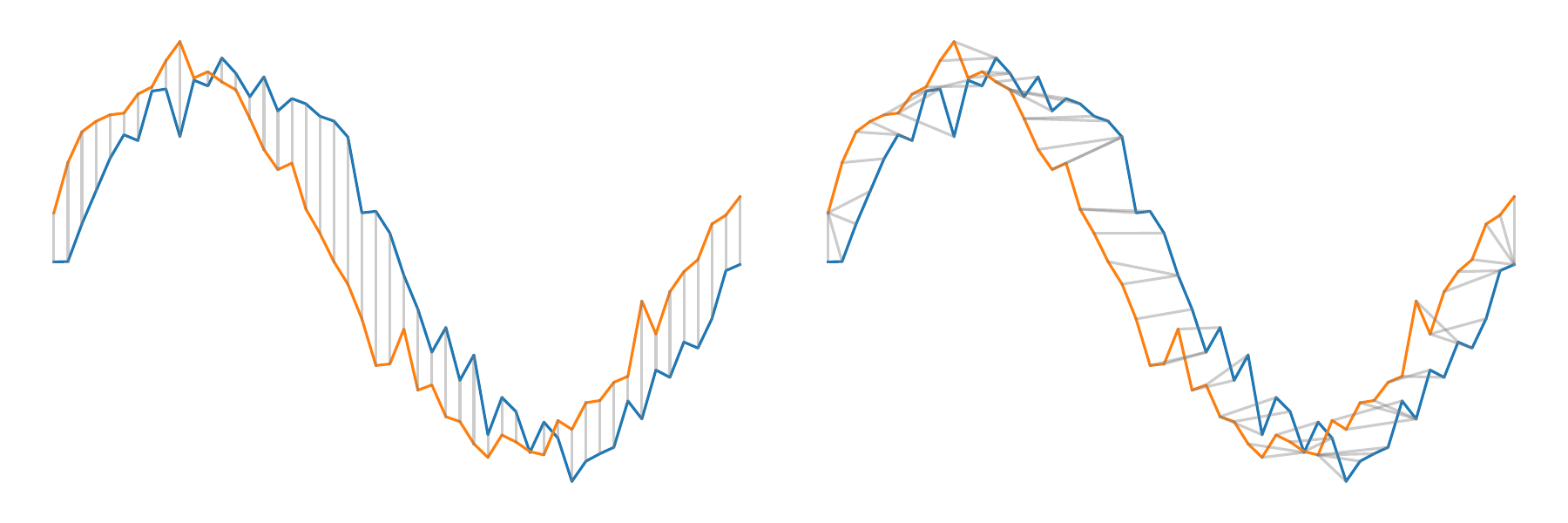}
  \caption[Alignments: Euclidean versus DTW.]{
  Left panel: A Euclidean alignment is shown where the elements of one time series (blue) are aligned (gray lines) point by point, vertically, with the elements of the other time series (orange).
  Right panel: shows DTW alignment on the same time series as the left panel, where points are not aligned vertically, allowing elements of one series to be aligned with different elements of the other series more than once.
  }
  \label{chapter_8_fig: dtw_vs_euclidean_raw}
\end{figure}

These caveats led to the search for methods that would allow flexibility when producing alignments, which would solve, for example, the issue of similar time series that were out of phase or shortened relative to each other.
Thus, based on the concept of dynamic programming, as seen in \citet{Bellman_1959, Bellman_1961} and widely used in speech recognition \citep{Vintsyuk_1968, Itakura_75, Sakoe1978, Myers_1980, Myers_1981, Godin_1989, Rabiner_1993}, \citet{Berndt_1994} introduced the idea of Dynamic Time Warping (DTW), see Figure~\ref{chapter_8_fig: dtw_vs_euclidean_raw}.

The goal of DTW is to account for differences in phase or time by finding a potentially nonlinear alignment between the elements of the two time series. Intuitively, this can be achieved by either skipping certain elements of a time series or using some of them multiple times. 
Methodologically, DTW identifies the optimal alignment between the two time series under specific constraints. 
Based on this alignment, the time series can be warped in a nonlinear manner to match each other (see \cite{Keogh_2005, Muller15} for more details).

The DTW approach is broadly utilized across various fields: it is used for human activity classification \citep{Holt_2007, Sempena_2011}, music and motion analysis \citep{Müller_2007}, signature or fingerprint recognition \citep{Munich_1999, Kovacs-Vajna_2000}, shape matching \citep{Bartolini_2005}, bioinformatics \citep{Aach_2001}, chemical engineering \citep{Dai_2011}, or medicine \citep{Kiani_2017, Gogolou_2019}. 
Beyond these applications, it has also been commonly used to analyze economic trends in stock markets and financial time series, particularly for predicting economic behavior \citep{Tasneem_2017, Franses_2020, Arya_2021}.
DTW has also recently started to be used in astrophysics:
Examples include the similarity between light curves of Gamma-ray bursts, which offers insight into their central engine activities \citep{Zhang_2016}, the reconstruction of gravitational wave signals from core-collapse supernovae \citep{Suvorova2019}, and the assessment of solar wind time series \citep{Samara2022}. Recently, \citet{Vohl2024} has employed an unsupervised method involving DTW to categorize radio pulsars by profile shape similarity using graph theory.

In this chapter, we introduce the fundamental aspects of DTW, as well as the methodological steps required for its application.
Additionally, we will examine the advantages of its implementation as a similarity measure compared to more conventional methods, such as Euclidean distance.

\section{Computing Dynamic Time Warping}
\label{chapter8: part_methods}

Consider two time series $X$ and $Y$ of length $n$ and $m$, 
\begin{equation}  
    \begin{aligned}
        X &= [x_1, x_2, \dots, x_i, \dots, x_n] \\
        Y &= [y_1, y_2, \dots, y_j, \dots, y_m]~.
    \end{aligned}
    \label{chapter_8_eq: time_series}
\end{equation}

The indices $(i,j)$ run along the elements of each time series, respectively ($i=1, \dots, n$, $j=1, \dots, m$).
To be considered a warping path, i.e., a set of connected pairs of points $(x_i, y_j)$ covering all the elements of both series in Equation~(\ref{chapter_8_eq: time_series}), we shall require three conditions that ensure efficient, meaningful, and temporally consistent alignments (see, e.g., \cite{Berndt_1994}):

\begin{enumerate} 
    \item Boundary condition. First, each of the time series's initial and final elements faces each other so that any warping path has constrained endpoints, that is, $i_{1}=1$, $j_{1}=1$, and $i_{k}=n$, $j_{k}=m$.
    \item Monotonicity condition. We shall also require that no elements of the series that break the temporal order be connected, i.e., no backward steps. Said mathematically, $i_{k-1} \leq i_{k}$ and $j_{k-1} \leq j_{k}$.
    \item Continuity condition. Finally, we shall also request that the elements of the series being connected do so without gaps or simply that all elements in the series are involved in the path, $i_{k}-i_{k-1} \leq 1$ and $j_{k}-j_{k-1} \leq 1$.
    
\end{enumerate}

Taking these conditions into account, the DTW is computed as (see, e.g., \cite{Hailin_2014}):

\begin{eqnarray}
\text{DTW}(X, Y) = 
\min_{W(X,Y)\in \mathcal{A}}
\Big(\sum_{(i,j)\in W}d(x_{i},y_{j})^{2}\Big)^{\frac{1}{2}}
,\label{chapter_8_eq: DTW_optimized}
\end{eqnarray}

where $W$ represents an explored warping path that pertains to the set $\mathcal{A}$ of all paths fulfilling the above-referred conditions (hereafter warping path conditions). The DTW value from Equation~(\ref{chapter_8_eq: DTW_optimized}) is the cost of the optimal warping path ($W_{\mathrm{o}}$), that is, the minimum value obtained after adding the distance $d$ between all pairs $(i,j)$ of $W_{\mathrm{o}}$.

\subsection{Dynamic programming approach}
\label{chapter8: dynamic_programming}

From the perspective of DTW calculation, dynamic programming plays a crucial role in the efficient computation of DTW. 
Following \citet{Kruskal_Liberman1983, Rabiner_1993, Keogh_2005, Muller15} as a reference, we compute the DTW on the time series seen in Equation~(\ref{chapter_8_eq: time_series}).
We construct a cost matrix ($E$) of size $n\times m$ where each cell ($i^{th}, j^{th}$) represents an alignment, and contains the distance between $x_i$ and $y_j$ seen as $d(x_i, y_j)=(x_i-y_j)^2$.
In this case, $W$ can be viewed as a series of $E$ cells that determine the alignment between $X$ and $Y$.
In such a way, we can define:

\begin{eqnarray}
W = w_{1}, w_{2},\dots,w_{k},\dots,w_{\mathrm{K}}
~,
\label{chapter_8_eq: warping_path_in_E}
\end{eqnarray}

where $w_{k}=(i,j)_{k}$ represents the $k^{th}$ of $W$ and contains the values seen in ($i^{th}, j^{th}$) of $E$, i.e., $d(x_i, y_j)$ defined above.
Equation~(\ref{chapter_8_eq: warping_path_in_E}) satisfies $\max(m,n) \leq \mathrm{K} < m+n-1$.
As discussed, there exists an $\mathcal{A}$ composed of an exponential number of $W$ satisfying the warping path conditions, so analogously to Equation~(\ref{chapter_8_eq: DTW_optimized}): 

\begin{eqnarray}
\text{DTW}(X, Y) = 
\min_{W\in \mathcal{A}}
\Bigg(\sqrt{\sum_{{w\in W}}w}\Bigg) = \Bigg(\sqrt{\sum_{k=1}^{K}w_{k}}\Bigg)_{W_{\mathrm{o}}}~.
\label{chapter_8_eq: DTW_optimized_E}
\end{eqnarray}

In Equation~(\ref{chapter_8_eq: DTW_optimized_E}), the value of DTW is given by $W_{\mathrm{o}}$ whom is defined by Equation~(\ref{chapter_8_eq: warping_path_in_E}).
However, the application of Equation~(\ref{chapter_8_eq: DTW_optimized_E}) on every $W$ is not feasible due to the enormous size of $\mathcal{A}$.
For this purpose, a recurrence relation is implemented, which is at the core of dynamic programming, to find $W_{\mathrm{o}}$.
The idea behind this process is to solve the problem by combining the solutions from the simplified subproblems, which were derived from the original problem.
In other words, an expression that defines a sequence based on conditions that gives a term as a function of the previous terms.
From this, we work out what is known as the accumulative cost matrix ($D$) computed by:

\begin{eqnarray}
D(i,j)=E(i,j)+ 
\min\big(D(i-1,j-1), D(i-1,j),D(i,j-1)\big)
~.\label{chapter_8_eq: D_cumulative}
\end{eqnarray}

As derives from Equation~(\ref{chapter_8_eq: D_cumulative}), $D$ is a matrix $(n \times m)$ where $D(1,1)=E(1,1)$ by definition.
Therefore, $W$, according to the warping path conditions, is constrained in $D$ as follows:

\begin{enumerate}
    \item The $W$ has to start and end on opposite diagonals of $D$: $w_1=(1,1)$ and $w_{k}=(m,n)$.
    \item The $W$ has to traverse $D$ evenly spaced in terms of time with no backward jump: let $w_k=(i,j)$, then $w_{k-1}=(i',j')$ with $i-i' \geq 0$ and $j-j' \geq 0$. 
    \item The $W$ has restricted advancement in $D$ to adjacent cells: let $w_k=(i,j)$, then $w_{k-1}=(i',j')$ with $i-i' \leq 1$ and $j-j' \leq 1$.
\end{enumerate}

\begin{figure} 
\centering
  \includegraphics[width=1\textwidth]{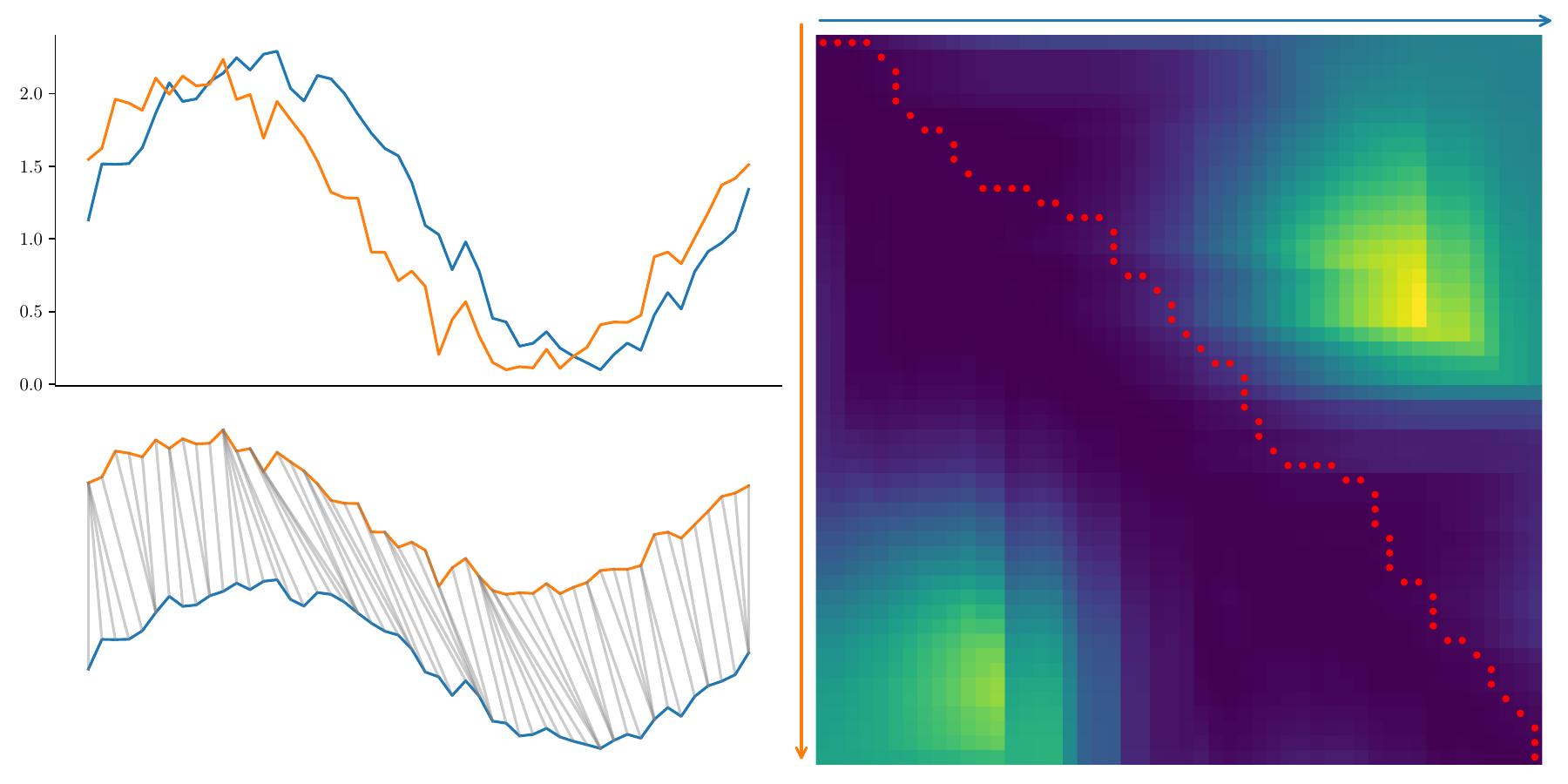}
  \caption[Optimal warping path.]{
  First column: two time series, $X$ in orange and $Y$ in blue, are shown unaligned (top panel), and aligned from $W_{\mathrm{o}}$ by DTW, where each gray line is a $w_{k}\in W_{\mathrm{o}}$ (bottom panel).
  Second column: shown is $D$ calculated over the time series in the first column, with orange and blue arrows indicating the temporal order, respectively. The path of $W_{\mathrm{o}}$ is shown in red. The values tending towards yellow represent the cells containing higher cumulative costs, while the darker cells represent those with lower costs.
  }
  \label{chapter_8_fig: dtw_cost_matrix}
\end{figure}

Thus, the value of DTW seen in Equation~(\ref{chapter_8_eq: DTW_optimized_E}) is given by the $W_{\mathrm{o}}$ found in $D$ whose cost is determined by $E$ (see Equation(\ref{chapter_8_eq: warping_path_in_E})), see Figure~\ref{chapter_8_fig: dtw_cost_matrix} to get an idea of this computation.

An example of the DTW application, along with its computation step by step, can be seen in Appendix \ref{Appendix_8: example}.

\subsection{Euclidean distance in the Dynamic Timer Warping context}
\label{chapter8: DTW_ED}

\begin{figure} 
\centering
  \includegraphics[width=1\textwidth]{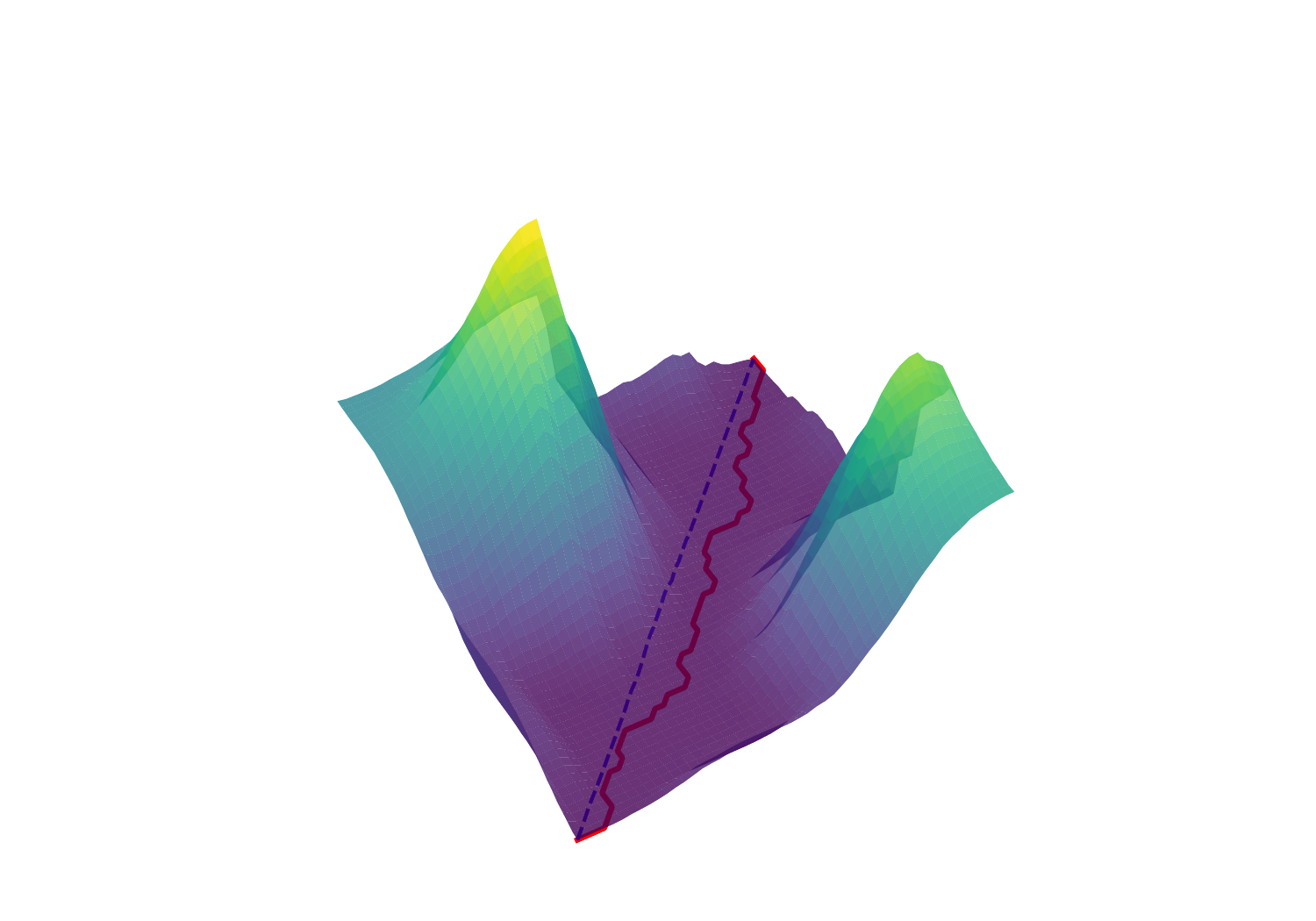}
  \caption[Euclidean and optimal warping path.]{
  The $D$ depicted in Figure~\ref{chapter_8_fig: dtw_cost_matrix} is represented as a three-dimensional surface, where the vertical z-axis indicates the cumulative cost of the alignment between the elements of $X$ and $Y$.
  The $W_{\mathrm{o}}$ is shown in red and the $W_{\mathrm{E}}$ in blue.
  
  }
  \label{chapter_8_fig: dtw_cost_matrix_realWorld_ED_DTW}
\end{figure}

The Euclidean distance (ED) can be viewed as a restricted case of DTW.
In such a way, let $W$ according to Equation~(\ref{chapter_8_eq: warping_path_in_E}), $w_{k}$ takes $k=i=j=$, defining what we denote as Euclidean path ($W_{E}$).
Therefore, $W_{\mathrm{E}}$ represents the main diagonal of $D$, so it can only be computed when $n=m$ in Equation~(\ref{chapter_8_eq: time_series}).
Furthermore, $W_{E}$, since it satisfies the warping path conditions, is always explored in Equation~(\ref{chapter_8_eq: DTW_optimized_E}). 
This leads to the conclusion that for any $X$ and $Y$ with $n=m$, we have $\text{DTW}(X, Y)\leq \text{ED}(X, Y)$.
Figure~\ref{chapter_8_fig: dtw_cost_matrix_realWorld_ED_DTW} highlights how $W{\mathrm{o}}$ deviates from $W_{\mathrm{E}}$, allowing flexibility in aligning time series looking for the minimum cost.

An example of $W_{\mathrm{E}}$ calculated step by step in comparison with $W_{\mathrm{o}}$ can be seen in Appendix \ref{Appendix_8: euclideanPath}.

\subsection{Global constraints}
\label{chapter8: global_constraints}

\begin{figure} 
\centering
  \includegraphics[width=1\textwidth]{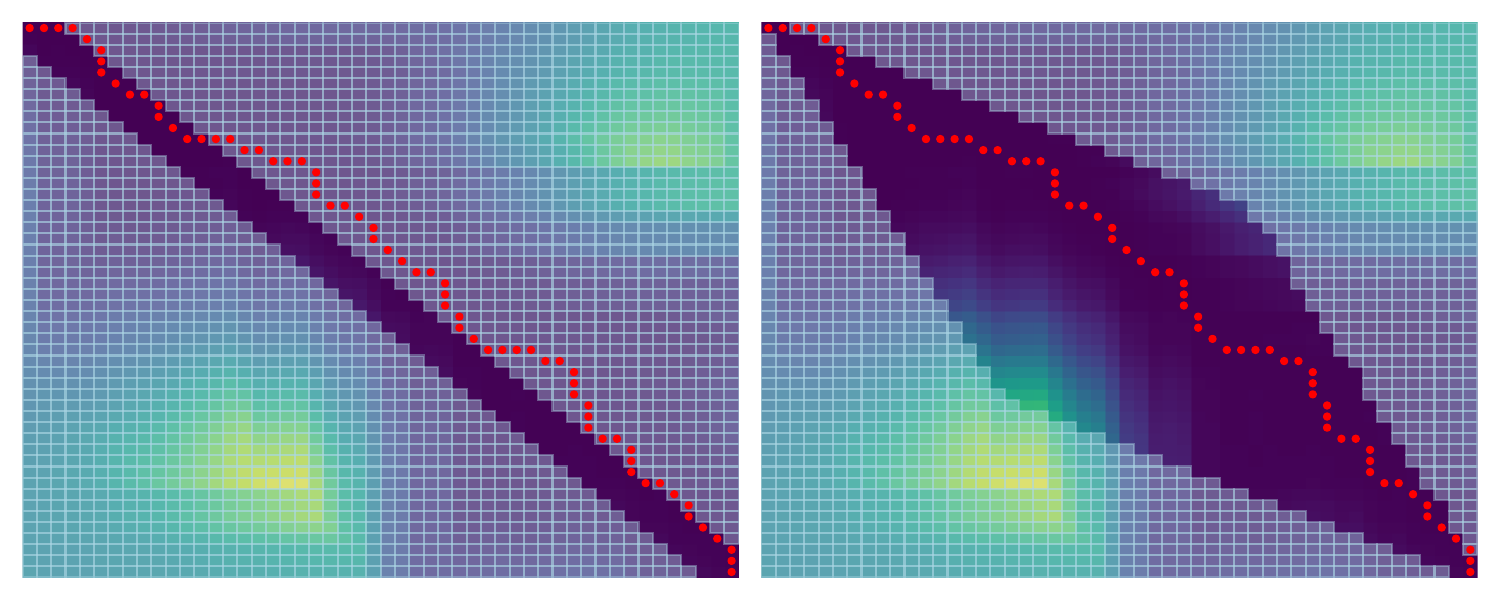}
  \caption[Global constraints in the warping path.]{
   The Sakue-Chiba band and the Itakura parallelogram constraints on the $D$ are shown in the right and left panels, respectively, highlighted as the only possible region for computing DTW. The $W_{\mathrm{o}}$ is marked in red when no constraint is applied, as seen in Figure~\ref{chapter_8_fig: dtw_cost_matrix} where the $D$ shown was computed.
  }
  \label{chapter_8_fig: global_constraints}
\end{figure}

In addition to the warping path conditions, other global constraints limit the number of $W$ seen in $A$.
This not only helps to speed up the DTW calculation but also provides more global control over the path of $W$ by avoiding what is known as "pathological" warpings, where few elements of $X$ are aligned with many elements of $Y$.
These types of constraints determine how far $W$ can be moved away from the diagonal of $D$ by determining what is known as a warping window or global constraint region. 
The two best-known global constraints are the Sakoe-Chiba band (\cite{Sakoe1978}) and the Itakura parallelogram (\cite{Itakura_75}).
Figure~\ref{chapter_8_fig: global_constraints} shows how the scope of $W$ is restricted to the highlighted area. 
While the Sakoe-Chiba band traverses the diagonal of $D$ with a width fixed by the global constraint area, the Itakura parallelogram restricts the slope of $W$ (see \cite{SakoeVsItakura} for a discussion about a comparison of their performance).

\section{Properties of DTW}
\label{chapter8: DTW_properties}

DTW can be defined as a similarity distance; however, it is not strictly a metric (\cite{VidalRuiz1985} for a discussion). Let $X$ e $Y$ as defined in Equation~(\ref{chapter_8_eq: time_series}), we can state the following properties:

\begin{itemize}
    \item Self-similarity: $\text{DTW}(X,X)=0$.
    \item Symmetry: $\text{DTW}(X, Y)=\text{DTW}(Y,X)$.
    \item Positivity: If $X \neq Y$, then $\text{DTW}(X, Y)\geq 0$.
\end{itemize}

However, to be considered a metric, DTW should also meet the triangular inequality: $\text{DTW}(X, Y)\leq \text{DTW}(X, Z)+\text{DTW}(Y, Z)$ with $Z$ as a time series with non restriction on length. 
This property is not always satisfied by DTW and has served as a starting point, among others, for the search of new and improved forms of its implementation; see, e.g., \citet{Keogh_2005, Athitsos_2008, Ding2008, Keogh_accelerating}.
Furthermore, it must be noted that due to the dynamic alignment performed by DTW, the following property may not hold: if $\text{DTW}(X, Y)=0$, then $X=Y$.  
However, considering this situation, we enter the realm of the concept of similarity since $X$ and $Y$ may differ due to $n\neq m$. 
For instance, a $\text{DTW}\sim 0$ would allow us to establish a strong similarity relation, which is highly useful in fields where time series represent physical phenomena observed with temporal differences but that somehow respond to similar phenomenological processes.

\section{Conclusions}
\label{chapter8: appl_astro}

The advent of DTW as a distance measure between time series stems from the weaknesses imposed by the usual distance measures, such as ED.
Methods such as ED limit the cases where they can be applied, as they require fundamental conditions, such as $n=m$ in Equation~(\ref{chapter_8_eq: time_series}). 
Moreover, they show less efficiency when, even under the same length condition, the aim is to find the difference between morphologies due to their static alignments, point by point.
This causes time series that are similar in terms of morphological characteristics, such as the number of peaks, the value of higher intensity, or the width of these, but out of phase, to show high values. 
However, a dynamic alignment such as the one carried out by DTW allows us to solve these problems. 

In Chapter~\ref{chapter9}, we use these premises to conduct a morphological comparison of the gamma-ray light curves of a sample of approximately 300 pulsars collected by the {\it Fermi}-LAT Third Pulsar Catalog (3PC, \cite{Fermi3PC}).
The high variability of these detections, as well as the different observation times that produce light curves of varying sizes, makes DTW an optimal approach for establishing a similarity distance between them. 
In this way, we have been able to establish a ranking of similarity between pulsars through their light curves, leading to a clustering of them and to suggestions about their emission processes and the factors influencing them.

\chapter{Quantitative exploration of the similarity of gamma-ray pulsar light curves}
\label{chapter9}

\section{Introduction} 
\label{chapter9: intro}

Comparing light curves of two or more gamma-ray pulsars has been chiefly an artisanal business. 
One would usually compare how similar two light curves are to the eye of the beholder, mentally disregarding the different flux levels.
However, this is not always possible, as the light curve count rates can vary from a few to more than a thousand for the same phase bin. 
To improve this, consider global light curve indicators like the number of peaks, their separation, width, and height. 
This is not as obvious as it seems for several reasons.
First, a solid definition of a light curve peak is lacking. 
Any characterization of light curves would inherently depend on this prior definition.
Secondly, light curves with two peaks of different heights can have equal relative heights, or light curves with a distinct peak separation can be identical in other respects.
It is clear, then, that the former indicators do not exhaust the need for a similarity assessment and do not provide a quantitative measurement for it.

Recent theoretical studies of light curve morphology, as seen in \citet{Iniguez2024, Cerutti2024, Iñiguez2025}, emphasize the importance of such similarity measurements.
Results from theoretical models suggest that light curve properties are primarily determined by geometry and do not correlate with other physical parameters of the pulsars or their magnetospheres.
\citet{Iniguez2024} has, in particular, demonstrated that pulsars with different timing and spectral parameters have similar radiation skymaps. 
This implies a likely similarity among their magnetospheres, with the geometry widely dominating over the spectra in shaping the light curves. 
If so, considering a sample of light curves generated for the same pulsar but for different geometries and observers should relate to exploring the population of all observed pulsars.
Thus, it is essential to advance our understanding of pulsar magnetospheres to know to what extent pulsars with different properties have similar light curves. 

This point is further supported by \citet{Iñiguez2025}, who, considering the same population and the same synchro-curvature model above,  presented a simultaneous fit of spectra and light curves.
The results obtained show that light curve morphology is largely insensitive to timing and spectral parameters, reinforcing the dominant role of geometry in shaping pulsar emission.

Section~\ref{chapter1: emission_models} briefly reviewed the development of gamma-ray emission models to give some theoretical context to the physical origin of the light curves analyzed in this chapter.

Dynamic time warping (DTW) is a well-established method for comparing time series and identifying similar behaviors, even when they exhibit variations in speed or other temporal differences. The fundamental concepts and the DTW implementation can be found in Chapter~\ref{chapter8}.

In this chapter, we present the DTW methodology proposed in \citet{DTW_I} for the first time to quantitatively assess the degree of similarity between two gamma-ray light curves and apply it to the whole set of light curves in the {\it Fermi}-LAT Third Pulsar Catalog (3PC, \cite{Fermi3PC}).

\section{{\it Fermi}-LAT 3PC Light curves }
\label{chapter9: part_set_data}

The set of time series considered for this work comprises 294 light curves associated with the pulsars reported in the 3PC.
In this sample, there are 143 millisecond pulsars (MSPs) with periods shorter than 10 ms ($P<10$ ms).

\begin{figure}
  \includegraphics[width=1\textwidth]{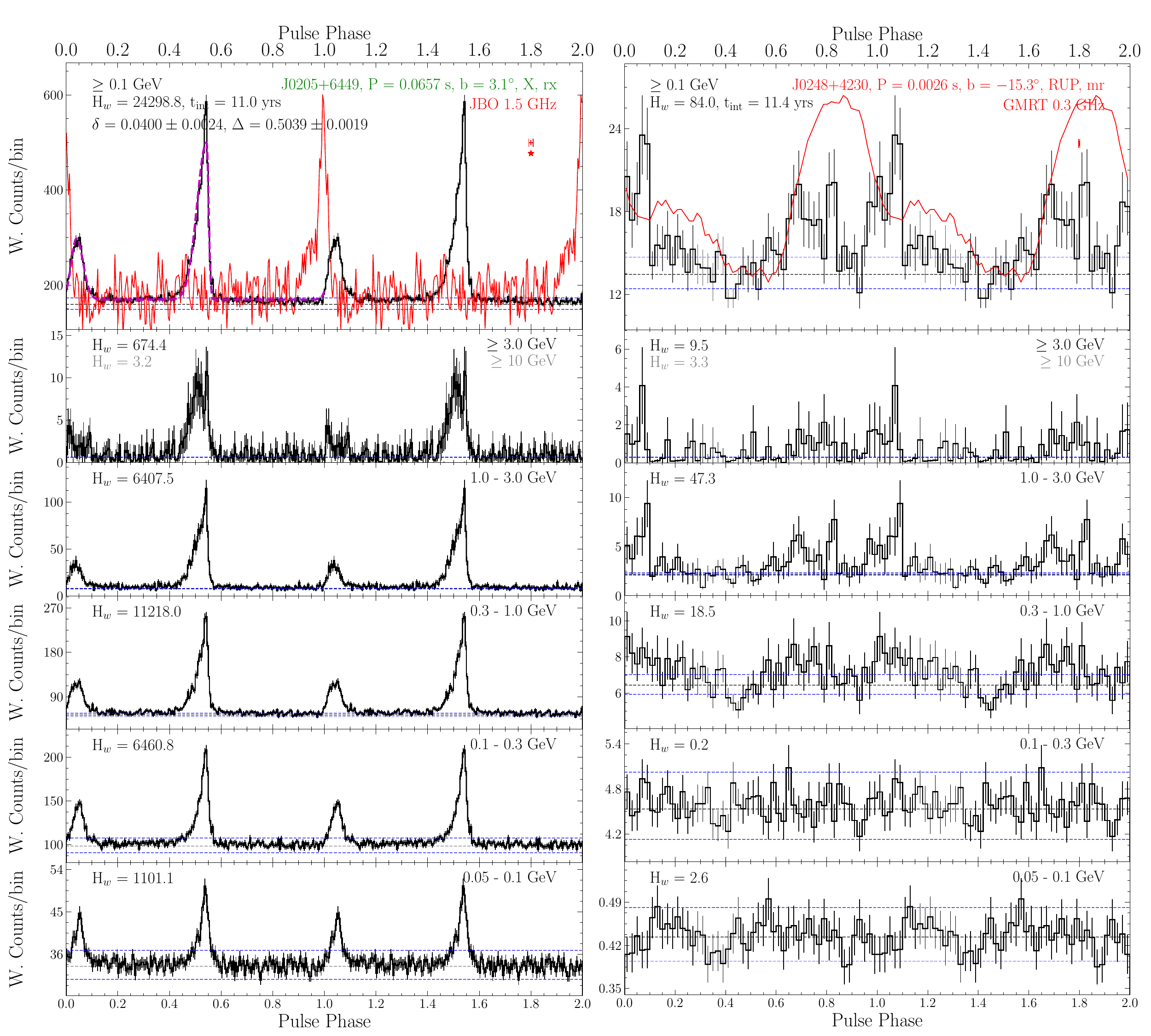}
  \centering
\caption[Example of light curves seen in the 3PC.]{Example of light curves seen in the 3PC (adopted from \url{https://fermi.gsfc.nasa.gov/ssc/data/access/lat/3rd_PSR_catalog/}) of J0205+6449 (first column) and J0248+4230 (second column) at different energies, as indicated in the different panels, are shown. See Figure~9 in \citet{Fermi3PC} for more information.
}

       \label{chapter_9_fig: 3PC_LightCurves}
\end{figure}

We consider the pulsars' weighted counts above $E \geq 100$ MeV (see, e.g., top panel in Figure~\ref{chapter_9_fig: 3PC_LightCurves}), subtracting the background level for each pulsar as reported in the 3PC (see, e.g., black dotted line in the top panel in Figure~\ref{chapter_9_fig: 3PC_LightCurves}).
When the resulting value is lower than zero, we set it to zero (no pulsar contribution above the background). 
As shown in Figure~\ref{chapter_9_fig: 3PC_LightCurves}, light curves can exhibit significantly different flux levels.
As we are interested in comparing the morphology of the light curves, they should have the same scale; thus, we consider the \textit{Min-Max scaler} after background subtraction, i.e., all light curves are transformed to lie between 0 and 1 via:

\begin{eqnarray}
X^{\dag}=\frac{x_i-x_{\text{min}}}{x_{\text{max}}-x_{\text{min}}}.
\label{chapter_9_eq: min-max_scaler}
\end{eqnarray}

Gamma-ray light curves have different sizes, i.e., are discretized in different bins ($N$).
The 3PC contains 48 light curves described with 25, 116 with 50, 98 with 100, 27 with 200, 4 with 400, and 1 with 800 bins.
These bins correspond to the rotational phase of a pulsar ($\phi \in (0,1)$).

\begin{figure}
  \includegraphics[width=1\textwidth]{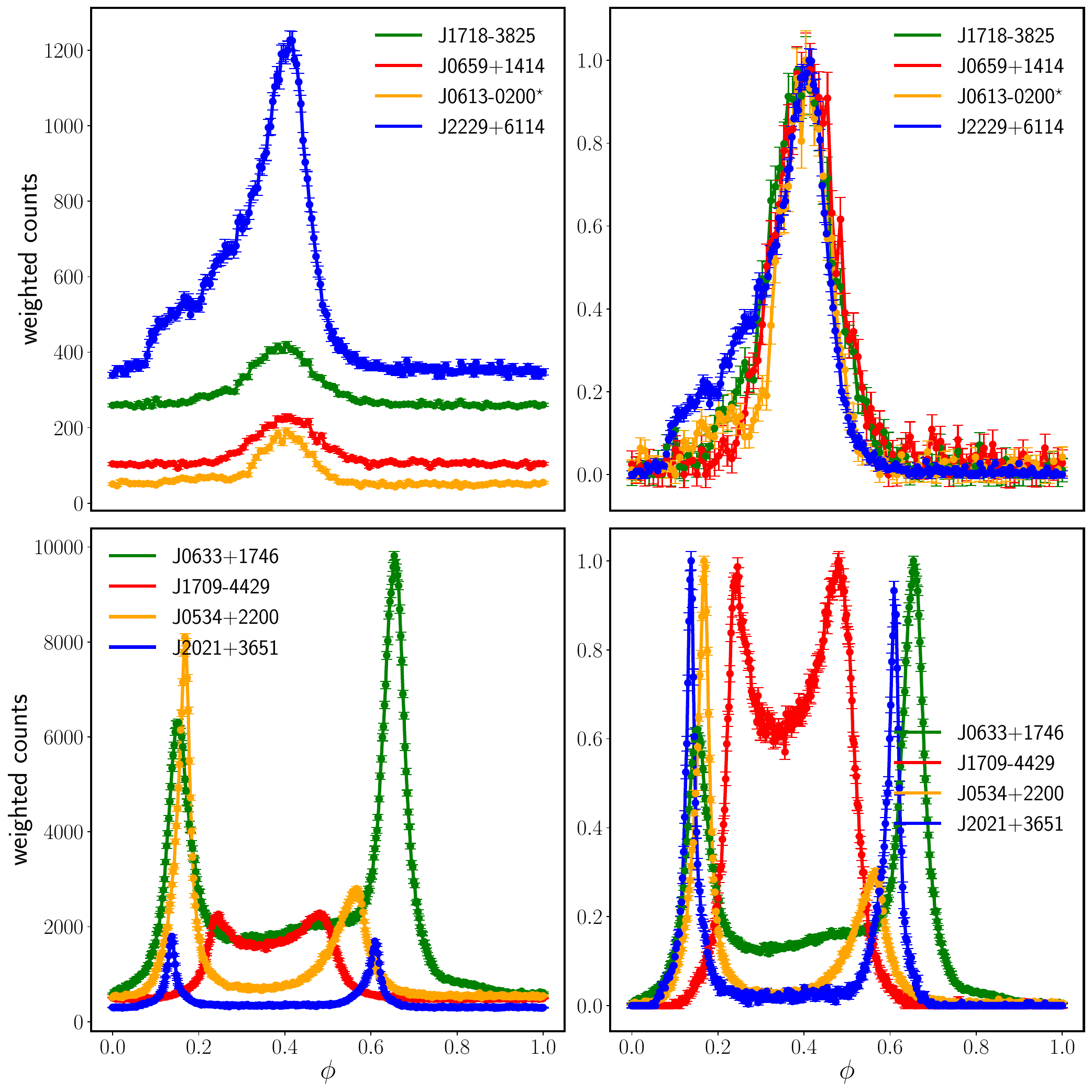}
  \centering
\caption[Light curves of the 3PC after being transformed.]{Light curves of the 3PC after being transformed by subtracting the background, phase rotating to achieve 
the best alignment between them, and applying Equation~(\ref{chapter_9_eq: min-max_scaler}). First column: light curves taken from the 3PC. Second column: resulting light curves after transformation. 

}

       \label{chapter_9_fig: LCs_transformed}
\end{figure}

One example of the transformation performed by subtracting the background, rotating the light curves to achieve the best alignment between them (see Section~\ref{chapter9: results} for further explanation), and applying Equation~(\ref{chapter_9_eq: min-max_scaler}) is shown in Figure~\ref{chapter_9_fig: LCs_transformed}. We note in the first row that morphological similarities are involved after this process; however, this is not always strictly the case, as observed in the second row in Figure~\ref{chapter_9_fig: LCs_transformed}.

\section{Results} 
\label{chapter9: results}

Considering the definitions in Section~\ref{chapter8: part_methods}, we will apply DTW to the 3PC light curves.
The arrangement of its weighted counts in representation of its flux levels can be seen according to Equation~(\ref{chapter_8_eq: time_series}).
Then, Equation~(\ref{chapter_9_eq: min-max_scaler}) is applied after subtracting the corresponding background. 
The result of this light curve processing enables us to effectively apply Equation~(\ref{chapter_8_eq: DTW_optimized}) and obtain a DTW value that shows the similarity between the given light curves. 
However, as $\phi$=0 is arbitrary, we shall consider all possible phase shifts, getting a DTW for each using Equation~(\ref{chapter_8_eq: DTW_optimized}).
The minimum value of the DTW measures how similar two light curves are, given the most favorable phase shift between them.
Finally, in addition to comparing light curves using DTW, we shall also directly compute the Euclidean distance (ED) when the lengths in the two-time series are the same or have been correspondingly rebinned to the same. 
The Appendix \ref{Appendix_8: example} provides a worked-out example that will help generate intuition regarding these applications.

\subsection{Comparing light curves with the same number of bins via ED}
\label{chapter9: ed_part_dtw}

\begin{figure}
  \includegraphics[width=1\textwidth]{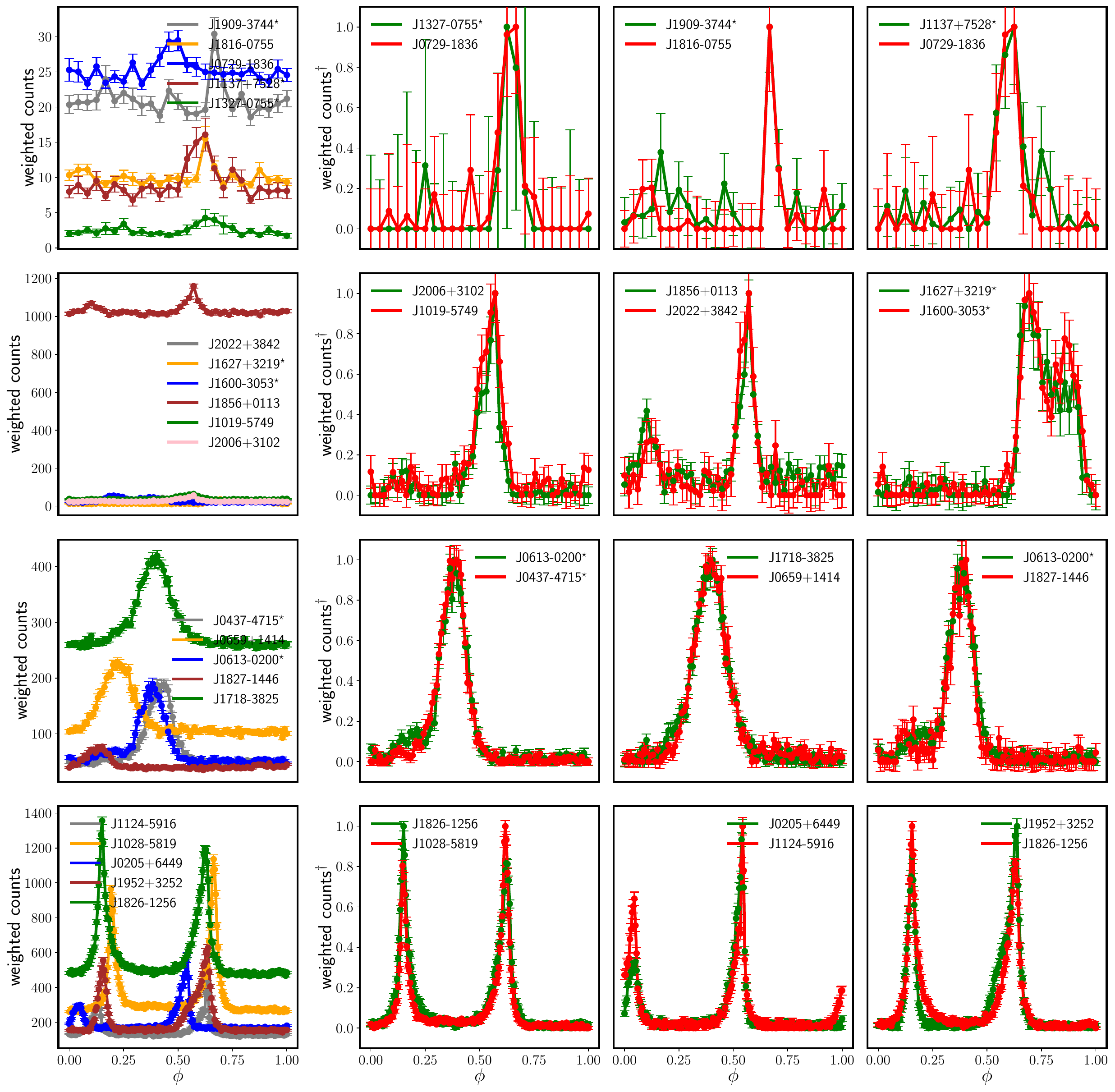}
  \centering
\caption[Ranking light curves according to $\mathrm{ED}$.]{Each row shows the three most similar pairs of light curves according to $\mathrm{ED}$ for light curves having 25, 50, 100, and 200 bins.
The plot in the first column of each row shows the 3PC data directly before processing.
MSPs are denoted with a star on the pulsar name. }

       \label{chapter_9_fig: ranking_ED}
\end{figure}

Figure~\ref{chapter_9_fig: ranking_ED} shows the three most similar pairs of light curves quantified by ED. 
Figure~\ref{chapter_9_fig: ranking_ED} also shows how these light curves look as read from the 3PC before processing, i.e., before background subtraction, before the Equation~(\ref{chapter_9_eq: min-max_scaler}) is used to compare the shape disregarding different flux levels, and before rotations are performed to produce the best alignment leading to the minimum value of ED. 
The morphological similarity of these pairs of light curves is striking. 
In particular, MSPs and young pulsars share detailed light curve morphology.
The pulsar spin period ($P$) and the spin period derivative ($\dot{P}$), and the quantities derived from them using the rotating dipole model (see Section~\ref{chapter1: Spin_periods}), such as the magnetic field at the surface ($B_{s}$), the magnetic field at the light cylinder ($B_{lc}$), and the spin-down energy loss rate ($\dot{E}_{sd}$), can be very different from one pulsar to another, even when their light curves are very similar. 
Differences up to a factor of 70 in $\dot{E}_{sd}$, of 4 in $B_{s}$, and of 30 in $B_{lc}$ can be found even between young pulsar pairs sharing light curve morphology. 
The differences in parameters, when the pair mixes MSPs and young pulsars, can be much larger, e.g., reaching up to a factor of more than $3\times 10^4$ in
$B_{s}$.
The gamma-ray emission properties, as reported in the 3PC, e.g., phase-averaged integral energy flux in the 0.1 to 300 GeV energy band, $G_{100}$ ($F_{E>100\mathrm{MeV}}$), and the energy of the maximum peak of the SED ($E_p$) also span an extensive range, representing very different gamma-ray spectra.
Light curve similarity is also immune to such differences.

\subsection{Comparing light curves of the same size via DTW}
\label{chapter9: ed_dtw}

\begin{figure}
  \includegraphics[width=1\textwidth]{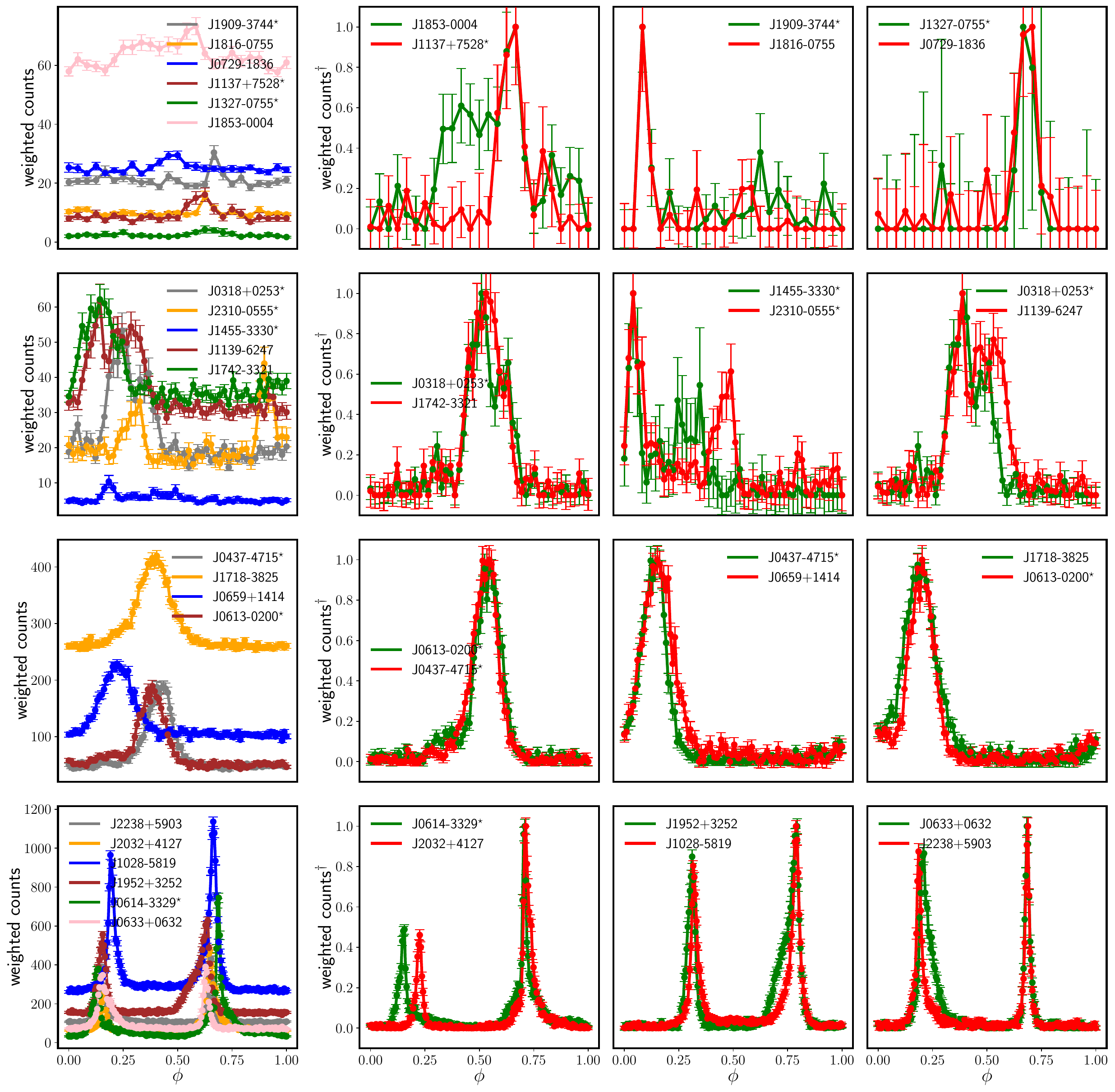}
  \centering
\caption[Ranking light curves according to DTW limited by size.]{The three most similar pairs of light curves according to DTW for light curves having 25, 50, 100, and 200 bins. The plot in the first column of each row shows the 3PC data directly before processing.
MSPs are denoted with a star on the pulsar name. }
       \label{chapter_9_fig: ranking_dtwbyed}
\end{figure}

We also employ the DTW methodology to determine whether it offers any advantages over ED.
Results for similar, equal-size light curves under DTW are shown in Figure~\ref{chapter_9_fig: ranking_dtwbyed}.
We recall that the cumulative cost matrix in DTW ($D$, see details in Equation~(\ref{chapter_8_eq: D_cumulative})) includes the ED comparison.
Thus, all paths considered via ED are also considered when using DTW. 
In some cases, the DTW ranking of the three most similar pairs for each subset of equal-$N$ light curves is the same as those in ED. This is not the case; see Section~\ref{chapter8: DTW_ED} for more details.

The similarity of the pulsar light curves showcased by DTW goes beyond simple visual inspection (which is what ED matches more closely).
It does not focus on, e.g., having similar peaks separated by the same phase interval, which our eyes would do directly if we could rotate all light curves to consider all possible shifts. Instead, it focuses on the global morphology. 
Consider, for example, the case of the group of $N=200$ bins. 
The DTW ranking selects the most similar pair, J0614-3329 and J2032+4127, which is not even among the first three pairs ranked by ED. 
The pulsar pairs selected by ED are visually appealing to the eye because the $\phi$ separation is the same, despite the height of each peak being different. Under DTW, however, these do not rank as highly, as this method prioritizes overall morphological similarity rather than strict phase alignment.
One example is the top pair ranked by DTW, which exhibits a slightly different phase separation between peaks (which, for DTW, is not a problem but would significantly increase ED) and an almost exact match of peak heights. When comparing these light curves, DTW emphasizes peak values more, whereas ED focuses more on phasing.
As before, they involve pulsars of all kinds with significantly different physical and spectral parameters.

\subsection{Caveats of ED and rebinning the whole sample}
\label{chapter9: caveats_ED}

ED has a restricted application to light curves of equal size. 
This restriction would exclude the best-described light curves from most similarity comparisons unless they were rebinned to a lower resolution, resulting in a loss of details. 
Rebinning, thus, (unless rebinning to the worst-resolved light curve, with $N=25$) will produce the loss of significant portions of the sample.
DTW remains stable despite rebinning.
For instance, over the sample of light curves rebinned to 100 bins (formed by 130 light curves, as those with $N<100$ bins are ignored), DTW rankings of the five most similar light curves from any given one retain three pulsars ranked also when no rebinning is applied in 85\% of the cases.
If we look at the two best matches, at least one is reproduced by both situations (with and without rebinning) in 88\% of the cases.
In addition to reducing the sample, we have demonstrated that ED is capable of producing large measures (signaling dissimilarity) even in cases where the overall structure remains the same.
This will occur mainly when two or more peaks are not precisely aligned, there is a slight difference in phase separation, or the peak widths differ slightly.
When rebinning, as the resolution at which the comparison will be made will, at most, be that of the lesser-sampled light curve, these issues will appear more often.
DTW solves both problems simultaneously.

\subsection{Using DTW to compare all light curves}
\label{chapter9: part_dtw}

Since the DTW methodology does not require reapplication over time series of equal length, we now consider all light curves without rebinning. 
We obtain 294 ordered rankings, one for each pulsar; we have (294 $\times$ 293)/2 = 43071 DTW values. 

\begin{figure}
  \includegraphics[width=1\textwidth]{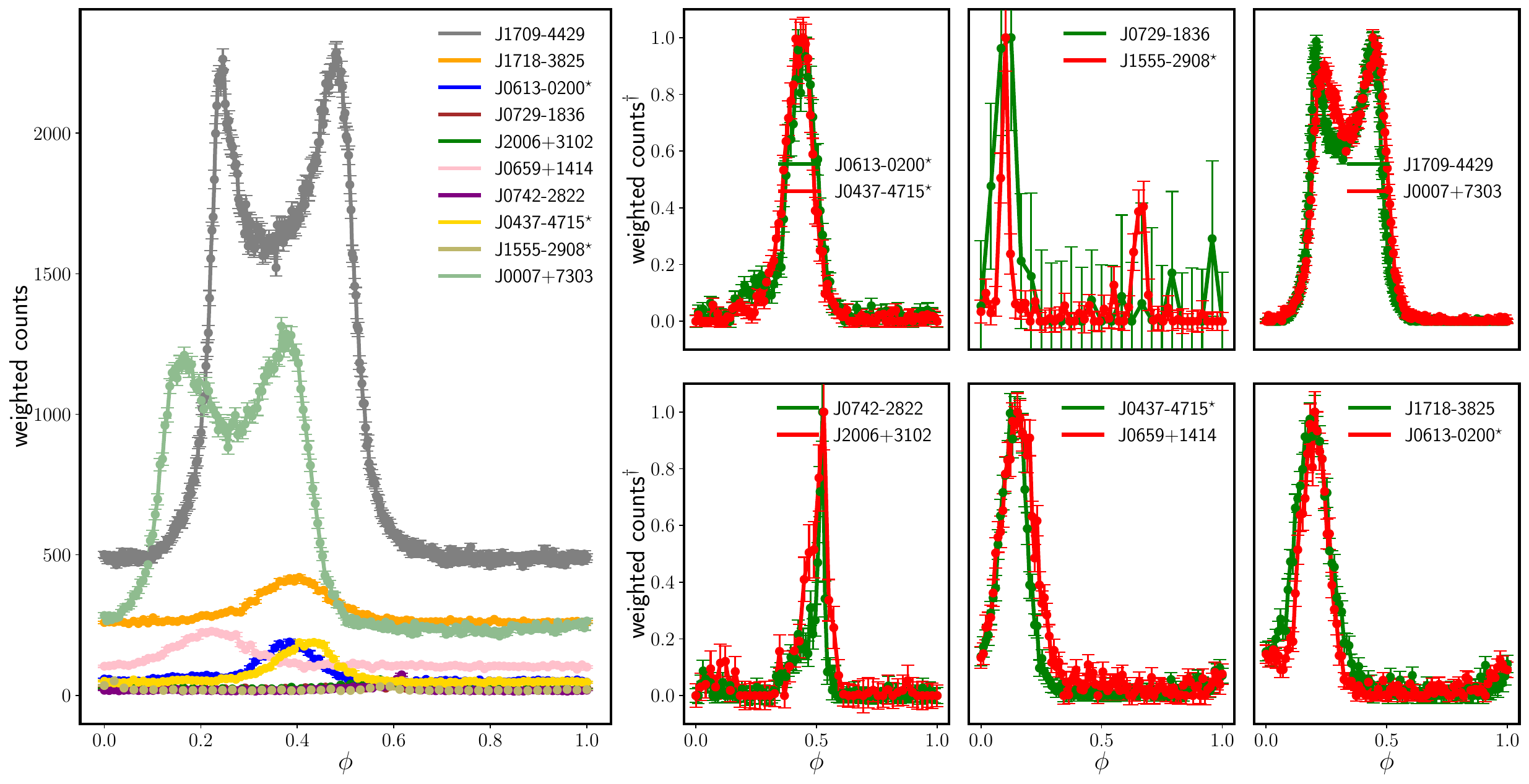}
  \centering
\caption[Ranking light curves according to DTW.]{The six most similar pairs of light curves according to DTW are displayed from left to right, from top to bottom. The plot in the first column shows the 3PC data directly before processing.
MSPs are denoted with a star on the pulsar name. 
}
       \label{chapter_9_fig: ranking_DTW}
\end{figure}

Therefore, we derive a total ranking by comparing all the light curves using DTW. In Figure~\ref{chapter_9_fig: ranking_DTW}, we display the six most similar light curves identified from this DTW ranking. 

In addition to associating morphological structures, even if they do not occur in the same phase, as we saw in Section~\ref{chapter9: ed_dtw}, DTW can be applied to light curves with any number of bins without rebinning (see Section \ref{chapter9: caveats_ED}). Figure~\ref{chapter_9_fig: ranking_DTW} shows that DTW identifies within its first six closest light curve pairs J0729-1836 and J1555 with 25 and 50 bins, respectively; J1709-4429 and J0007+7303 with 400 and 200 bins; and J0742-2822 and J2006+3102 with 100 and 50 bins.

\begin{figure}
  \includegraphics[width=0.4\textwidth]{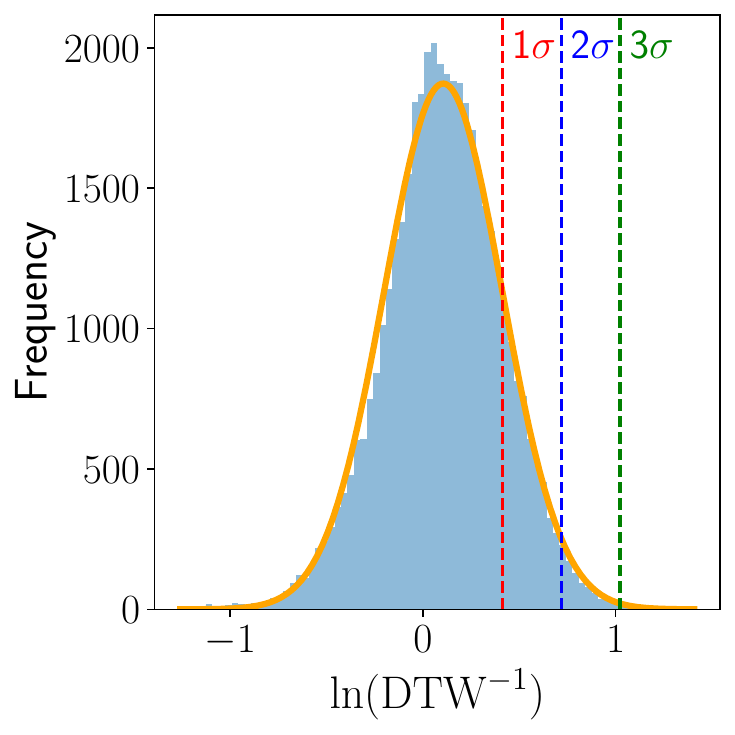}
  \centering
    \caption[Distribution of the natural logarithm of $\mathrm{DTW}^{-1}$ values.]{
    Distribution of the natural logarithm of $\mathrm{DTW}^{-1}$ values, $\ln (\mathrm{DTW}^{-1})$. 
    The orange line shows the normal distribution identified as the best fit for the data. The dashed vertical lines denoted with $1\sigma$ (red), $2\sigma$ (blue), and $3\sigma$ (green) represent the intervals of the distribution according to the empirical rule.
    }
       \label{chapter_9_fig: distribution_dtw}
\end{figure}

The log-normal distribution is identified as the best fit for both $\mathrm{DTW}$ and $\mathrm{DTW}^{-1}$ values (we take the inverse values to facilitate looking at the most similar light curve pairs in the far end tail of the distribution).
Thus, considering the distribution of $\ln (\mathrm{DTW}^{-1})$ values, as shown in Figure~\ref{chapter_9_fig: distribution_dtw}, allows us to define confidence intervals in the usual way, as a Gaussian is the best fit for this distribution. 
We use the mean $\mu$ and the standard deviation ($\sigma$) of the distribution of $\ln (\mathrm{DTW}^{-1})$, equal to 0.105 and 0.306, respectively, to obtain a value of $3\sigma$ deviation equal to 2.78. That is, a value of $\mathrm{DTW}^{-1}=2.78$ represents two pairs of light curves whose similarity measured by DTW is 3$\sigma$ beyond the mean.
Beyond this value, 70 pairs of light curve comparisons are found, involving 63 pulsars (13 MSPs).

\begin{figure}
  \includegraphics[width=0.9\textwidth]{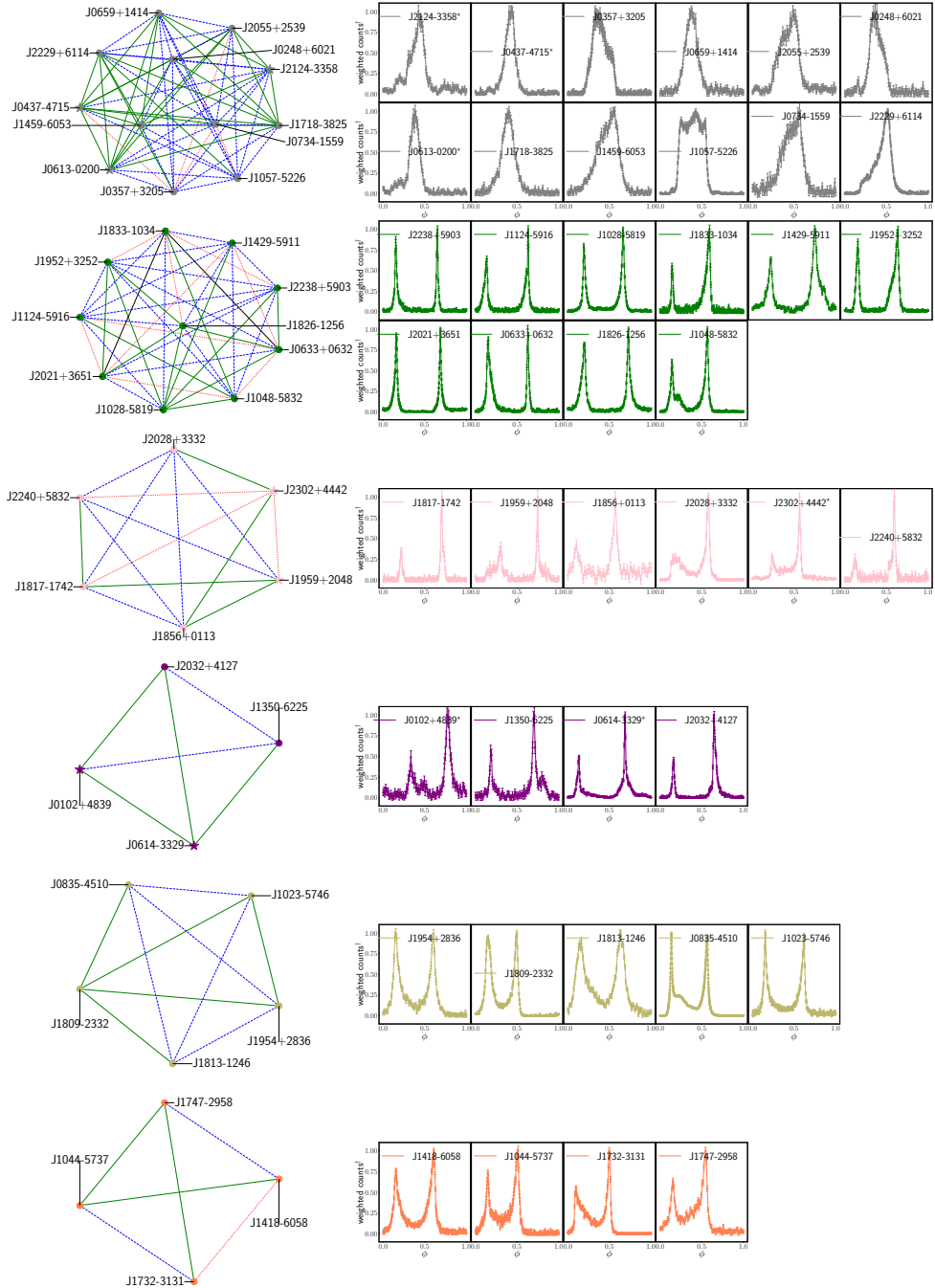}
  \centering
\caption[Clusters at a 3$\sigma$ significance level.]{Clusters with four or more pulsars and all members related to another by at least one similarity at a 3$\sigma$ significance level. Each node represents a pulsar, and each colored edge connecting the nodes represents the similarity significance of the corresponding light curves (black for $<1\sigma$, red for cases between 1 and $2\sigma$, blue for cases between 2 and $3\sigma$, and green for $>3\sigma$).
The light curves associated with each cluster are shown next.
MSPs are denoted with a star-shaped node. }
       \label{chapter_9_fig: clusters_3sigma}
\end{figure}

We can use the DTW distribution to partition the light curve sample into clusters. 
If we request that clusters contain at least four pulsars, we find six clusters, and members are related by at least one similarity at the 3$\sigma$ level. 
We can depict these connections in a graph.
Figure~\ref{chapter_9_fig: clusters_3sigma} shows these six clusters, where 41 pulsars are displayed as the graph nodes. 
The properties of these pulsars are shown in Table~\ref{chapter_9_tab: pulsar_parameters}.
The different similarity levels according to the distribution of $\mathrm{DTW}$ values are shown by the color-coded edges in the corresponding complete cluster graphs (where all nodes are connected). 
Figure~\ref{chapter_9_fig: clusters_3sigma} also shows the light curves of the members of each cluster.
The complete graphs show that the vast majority of the light curves within a given cluster also belong in the tail of the distribution, with similarities beyond $>1$ and $>2 \sigma$.

\begin{table}
\centering
\scriptsize
\caption[Physical and gamma-ray properties of the clusters seen in DTW.]{The columns show the pulsar timing, derived properties using the rotating dipole model, and gamma-ray properties of the clusters, including color coding, shown in Figure~\ref{chapter_9_fig: clusters_3sigma}. MSPs are denoted with a star on the pulsar name, and their corresponding $\dot{P}$ has an additional factor $10^{-6}$ multiplying the column heading. J1856$+$0113 has a Test Statistic (TS) below 25, so no spectral fit results are provided. See the 3PC for more details. 
}
\begin{tabular}{l | ccccc | cc }
\hline
\textbf{ PSRJ}  & $P$ & $\dot P$  & $\dot E_{sd}$  & $B_s$ & $B_{lc}$  & $F_{E>100 {\rm MeV}}$  & $E_p$ \\
                &  [s] &  [s/s]   &  [erg/s]       &  [G]  & [G]       &  [erg/cm$^2$/s]        & [GeV] \\
                &      &   [$\times 10^{-14}$]       & [$\times 10^{35}]$& [$\times 10^{10}$] & [$\times 10^{4}$] & [$\times 10^{-11}$]\\

\hline

\textcolor{Gray}{J0248$+$6021} & 0.217 & 5.52 &   2.13 & 350.35 &  0.32 &  2.94 &           0.774 \\
\textcolor{Gray}{J0357$+$3205} & 0.444 & 1.31 &   0.06 & 244.02 &  0.03 &  6.01 &           0.869 \\
\textcolor{Gray}{J0437$-$4715$^{\star}$} & 0.005 & 5.72 &   0.12 &   0.06 &  2.80 &  1.76 & 0.700 \\
\textcolor{Gray}{J0613$-$0200$^{\star}$} & 0.003 & 0.96 &   0.13 &   0.02 &  5.57 &  3.82 & 1.156 \\
\textcolor{Gray}{J0659$+$1414} & 0.384 & 5.50 &   0.38 & 465.38 &  0.08 &  2.65 &           0.156 \\
\textcolor{Gray}{J0734$-$1559} & 0.155 & 1.25 &   1.32 & 140.95 &  0.35 &  4.57 &           0.623 \\
\textcolor{Gray}{J1057$-$5226} & 0.197 & 0.58 &   0.30 & 108.54 &  0.13 & 29.59 &           1.213 \\
\textcolor{Gray}{J1459$-$6053} & 0.103 & 2.53 &   9.08 & 163.32 &  1.37 & 12.27 &           0.372 \\
\textcolor{Gray}{J1718$-$3825} & 0.074 & 1.32 &  12.49 & 100.35 &  2.22 & 10.37 &           0.564 \\
\textcolor{Gray}{J2055$+$2539} & 0.319 & 0.41 &   0.05 & 115.85 &  0.03 &  5.32 &           1.215 \\
\textcolor{Gray}{J2124$-$3358$^{\star}$} & 0.004 & 2.06 &   0.07 &   0.03 &  2.48 &  3.88 & 1.949 \\
\textcolor{Gray}{J2229$+$6114} & 0.051 & 7.53 & 215.58 & 199.49 & 13.33 & 24.01 &           0.740 \\

\hline

\textcolor{green}{J0633$+$0632} & 0.297 &   7.96 &   1.19 &  492.24 &  0.17 &  9.56 &1.427 \\
\textcolor{green}{J1028$-$5819} & 0.091 &   1.42 &   7.34 &  115.28 &  1.39 & 24.51 &0.888 \\
\textcolor{green}{J1048$-$5832} & 0.124 &   9.55 &  19.92 &  347.86 &  1.69 & 18.52 &0.968 \\
\textcolor{green}{J1124$-$5916} & 0.136 &  75.15 & 119.15 & 1021.18 &  3.78 &  6.11 &0.362 \\
\textcolor{green}{J1429$-$5911} & 0.116 &   2.39 &   6.06 &  168.30 &  1.00 & 11.28 &0.540 \\
\textcolor{green}{J1826$-$1256} & 0.110 &  12.12 &  35.72 &  369.86 &  2.54 & 41.35 & 0.870 \\
\textcolor{green}{J1833$-$1034} & 0.062 &  20.20 & 336.24 &  357.81 & 13.89 &  8.91 &0.305 \\
\textcolor{green}{J1952$+$3252} & 0.040 &   0.58 &  37.23 &   48.56 &  7.24 & 14.77 &0.993 \\
\textcolor{green}{J2021$+$3651} & 0.104 &   9.48 &  33.53 &  317.40 &  2.62 & 49.21 &1.064 \\
\textcolor{green}{J2238$+$5903} & 0.163 &   9.69 &   8.87 &  401.71 &  0.86 &  6.64 &0.673 \\

\hline

\textcolor{Lavender}{J1817$-$1742} & 0.150 &  2.06 & 2.42 & 177.62  &  0.49 & 2.85            &  1.123 \\
\textcolor{Lavender}{J1856$+$0113} & 0.267 & 20.59 & 4.25 & 750.91  &  0.36 &  -              &     - \\
\textcolor{Lavender}{J1959$+$2048$^{\star}$} & 0.002 &  1.68 & 1.60 &   0.02  & 36.73 & 1.57  & 1.038 \\
\textcolor{Lavender}{J2028$+$3332} & 0.177 &  0.49 & 0.35 &  93.74  &  0.16 & 5.71            & 1.511 \\
\textcolor{Lavender}{J2240$+$5832} & 0.140 &  1.53 & 2.20 & 147.82  &  0.50 & 0.98            & 1.138 \\
\textcolor{Lavender}{J2302$+$4442$^{\star}$} & 0.005 &  1.33 & 0.04 &   0.03  &  1.75 & 3.90  & 2.355 \\
\hline

\textcolor{Plum}{J0102$+$4839$^{\star}$} & 0.003 & 1.17 & 0.17 &   0.02 & 6.68 &  1.50  & 1.700 \\
\textcolor{Plum}{J0614$-$3329$^{\star}$} & 0.003 & 1.78 & 0.22 &   0.02 & 7.06 & 11.46  & 2.835 \\
\textcolor{Plum}{J1350$-$6225} & 0.138 & 0.89 & 1.33 & 112.10 & 0.39 &  3.60            & 1.951 \\
\textcolor{Plum}{J2032$+$4127} & 0.143 & 1.16 & 1.56 & 130.56 & 0.41 & 14.24            & 2.606   \\

\hline

\textcolor{olive}{J0835$-$4510} & 0.089 & 12.23 &  67.64 &  334.51 & 4.31 & 929.32 &  1.267 \\
\textcolor{olive}{J1023$-$5746} & 0.111 & 37.99 & 108.21 &  658.53 & 4.37 &  14.60 &  0.621 \\
\textcolor{olive}{J1809$-$2332} & 0.147 &  3.44 &   4.29 &  227.33 & 0.66 &  42.48 &  1.285 \\
\textcolor{olive}{J1813$-$1246} & 0.048 &  1.76 &  62.39 &   92.96 & 7.70 &  24.78 &  0.483 \\
\textcolor{olive}{J1954$+$2836} & 0.093 &  2.12 &  10.48 &  141.71 & 1.64 &  10.70 &  1.172 \\
\hline

\textcolor{orange}{J1044$-$5737} & 0.139 &  5.46 &  8.02 & 278.71  & 0.95 & 11.40 & 0.650 \\
\textcolor{orange}{J1418$-$6058} & 0.111 & 17.10 & 49.92 & 440.00  & 3.00 & 29.70 & 1.001 \\
\textcolor{orange}{J1732$-$3131} & 0.197 &  2.80 &  1.46 & 237.56  & 0.29 & 17.93 & 2.089 \\
\textcolor{orange}{J1747$-$2958} & 0.099 &  6.11 & 24.98 & 248.61  & 2.37 & 15.91 & 0.526 \\

\hline
\end{tabular}
\label{chapter_9_tab: pulsar_parameters}
\end{table}

\section{Conclusions}
\label{chapter9: conclu}

This chapter provides a quantitative assessment of gamma-ray light curve similarity that extends beyond estimating global properties, such as the number of peaks, peak width, or peak separation. 
Studying via DTW the similarity of all light curves (more than 43000 pairs, each with their corresponding rotations arising by considering that phase zero is arbitrary), a well-behaved distribution is found, from which similarity can be assessed.
Referring to that distribution of DTW values, we could define confidence intervals and, for the first time, determine whether a pair of pulsars has light curve similarity above or below a given threshold. 
This methodology lacks the caveats affecting the point-by-point comparison provided by the Euclidean distance between two time series.
The latter can only be applied to light curves described with equal resolution. Rebinning not only loses details of the morphology in cases that are better sampled but also requires cutting the sample unless light curves with the worst resolution drive the morphological comparison.
Even if we rebin and cut the sample accordingly, light curves that, for instance, have two very similar peaks with similar heights and widths but with slightly different peak separations would never be considered identical by the Euclidean distance estimator.
Instead, DTW would mark those as such, since it examines the global light curves.
Using this methodology, we could cluster the light curve sample quantitatively.

The variety of physical and spectral properties in light curves that are morphologically similar is large.
Having the whole distribution of the DTW, we have explored whether the similarity of any two light curves, as measured with the DTW technique, is correlated with nearness in any other quantity of interest. 
In particular, we tested for correlations between the DTW values for any two pulsars of the sample and the Euclidean distance under each one of the pulsar properties considered above.
For instance, we considered the $(\mathrm{DTW})^{-1}$ measure between two given pulsars versus the correctly normalized difference of their periods or their magnetic field at the light cylinder. 
No correlation arises for any of the physical or spectral parameters.
Similar light curves do not imply that pulsars are also identical in terms of their timing properties and can mix MSPs with young ones. 
Even for pulsars of the same kind, similar morphologies do not necessarily imply identical period derivatives, magnetic fields, or spin-down energies; they can all vary significantly.
This suggests that the exact mechanism in all pulsars radiates gamma-ray emission and that geometry dominates the morphology of the light curve.

\chapter{General Conclusions and Future Work}
\label{chapter10}

In this chapter, I summarize the main points of the thesis and discuss some avenues for future exploration.

\section{General Conclusions}
\label{chapter10: Conclusions}

This thesis has concentrated on advanced mathematical techniques, with a special focus on graph theory, PCA, and DTW, to analyze pulsar and FRB populations.

The population of pulsars has been extensively described in Chapter~\ref{chapter1}, from their origin to the physical properties that best characterize their evolution and emission, highlighting the caveats they present. In this chapter, we have also contextualized FRBs, as well as their possible origin and their more than likely connection to pulsars, specifically magnetars.

In Chapters~\ref{chapter2} and \ref{chapter3}, we have presented the mathematical concepts that comprise the frameworks of this thesis, along with Chapter~\ref{chapter8}, where we do the same for DTW.

The innovative classification, grouping, and prediction methodologies developed are supported in these more conceptual chapters.

The MST was brought in Chapter~\ref{chapter4} as a tool to visualize the pulsar population collected in the ATNF, demonstrating that the traditional $P\dot P$ diagram does not fully capture its variance. Using PCA, we showed that physical properties derived must be considered to define as much information as possible. 
We have demonstrated that the pulsar MST (or simply, the Pulsar Tree) encodes physical meaning within its structure, extending beyond being a mathematical approach.
This work led to the development of the Pulsar Tree web, an interactive platform for exploring pulsar populations from a graph theory perspective.

Building on this approach, Chapter~\ref{chapter5} shows the application of the MST to automatically identify pulsar subpopulations, utilizing betweenness centrality estimators and non-parametric statistical tests, as the KS test, to define significant branches. This clustering method provided information about the historical evolution of pulsars. 
The latter is quantified through the implementation of \textit{growth} and \textit{togetherness}, two measures designed to capture the stability of significant branches over time. 
This algorithm also represents a new member of the family of clustering methods, which has the advantage of relying on the underlying topology of the population, making it a rigorous approach.

Furthering the MST application, Chapter~\ref{chapter6} focused on MSPs in binaries seen in the ATNF, incorporating binary parameters alongside physical properties. Based on their MST proximity to confirmed systems, we identified candidates for BWs, RBs, and tMSPs. This approach allowed us to estimate missing parameters and the possible nature of specific pulsars. 
Using the clustering method introduced in Chapter~\ref{chapter5}, we found that the clusters organize according to notably distinct physical and binary properties. Some align with known classes, such as spider pulsars, while others suggest new avenues for exploration.
These advancements were integrated into the Pulsar Tree web, expanding its capabilities for MSPs.

In a separate application, Chapter~\ref{chapter7} employed MST as an unsupervised classification method to distinguish between repeating and non-repeating FRBs observed by CHIME. We developed a graph-based methodology that outperforms traditional machine learning classifiers by computing multiple MSTs based on observational and derived properties. The process identified new repeater candidates and highlighted the most discriminative features of FRB separation.
This suggests that the MST could serve as a valuable preprocessing tool for complex, computationally intensive machine learning methods that rely on robust, well-validated inputs.

Shifting our focus to time-series analysis, we introduced the DTW method for gamma-ray pulsar light curves from the \textit{Fermi}-LAT 3PC in Chapter~\ref{chapter9}. 
The mathematical context and advantages of this approach are discussed in Chapter~\ref{chapter8}.
This work represents the first quantitative clustering of gamma-ray pulsars based on light curve morphology. It defines similarity thresholds and demonstrates that pulsars with nearly identical light curves can exhibit distinct timing properties. These findings suggest that geometry dominates gamma-ray emission rather than intrinsic properties.

The results presented throughout this thesis reveal physical insights that support the establishment of these techniques as valid approaches for studying high-energy astrophysics.

\section{Future Work}
\label{chapter10: future_work}

The reach of the methodologies developed in this thesis for graph theory and DTW applications in high-energy astrophysics is far from over.
While the MST has provided insights into pulsar populations and FRBs, further analysis could reveal evolutionary trends, transitional sources, new classes, and hidden connections. 
One example is how specific parameters evolve along branch paths in the MST, which could provide insight into pulsar evolution and classification. 
Another example is an ongoing project we are currently pursuing, which utilizes DTW to quantify the similarity between confirmed and candidate tMSPs through their light curves. 
This approach could be extended to different time series, such as Gamma Ray Bursts (GRBs) or magnetar flares, which exhibit various light curve structures, often with complex variability. 
There are also opportunities to combine graph theory and DTW. 
The MST could be used to cluster pulsars based on the DTW distances of their gamma-ray light curves and explore the underlying behaviour of the clusters. 
A natural extension of this thesis is the application of Graph Neural Networks (GNNs), which enable the direct encoding and learning of relational information between astrophysical sources. 
Having developed algorithms based on graph structures, the context is well-prepared to extend these concepts into dynamic learning frameworks.


\appendix 


\chapter{Graph centrality}  
\label{Appendix_2}

\section{Betweenness centrality estimator: Practical example}
\label{Appendix_2: practical_example}

Here, we provide a simple example of the betweenness centrality estimator. To calculate the coefficient $C_B$ by applying Equation~(\ref{chapter_2_eq: betweenness_centrality}) on a graph, we recall to take into account the following considerations:

 \begin{itemize}
    \item Select all pairs of nodes $(v_s,v_t)$. One can exclude the adjacent ones, as there cannot be another node between them. These define all possible paths that may include $v_j$ as an intermediate node.
    \item Select all shortest paths within the graph joining $(v_s,v_t)$, eliminating duplicities (i.e., $\sigma_{s,t} = \sigma_{t,s}$). 
    If the graph is unweighted, shortest paths are defined by the minimum number of edges that must be traversed. If weighted, shortest paths minimize the total edge weight.
    If the shortest path can be reached by different paths due to cycles, $\sigma_{s,t}$ takes a value equal to the number of such paths; otherwise, it equals 1.
    \item Each time, $v_j$, appearing on these paths will be assigned the numerical value 1, i.e., $\sigma_{s,t}(v_j)$=1.
    \item Make the total sum of the values according to the previous steps. We will normalize by multiplying the result obtained by $2/(N-1)(N-2)$.
 \end{itemize}

\begin{figure}
  \includegraphics[width=1\textwidth]{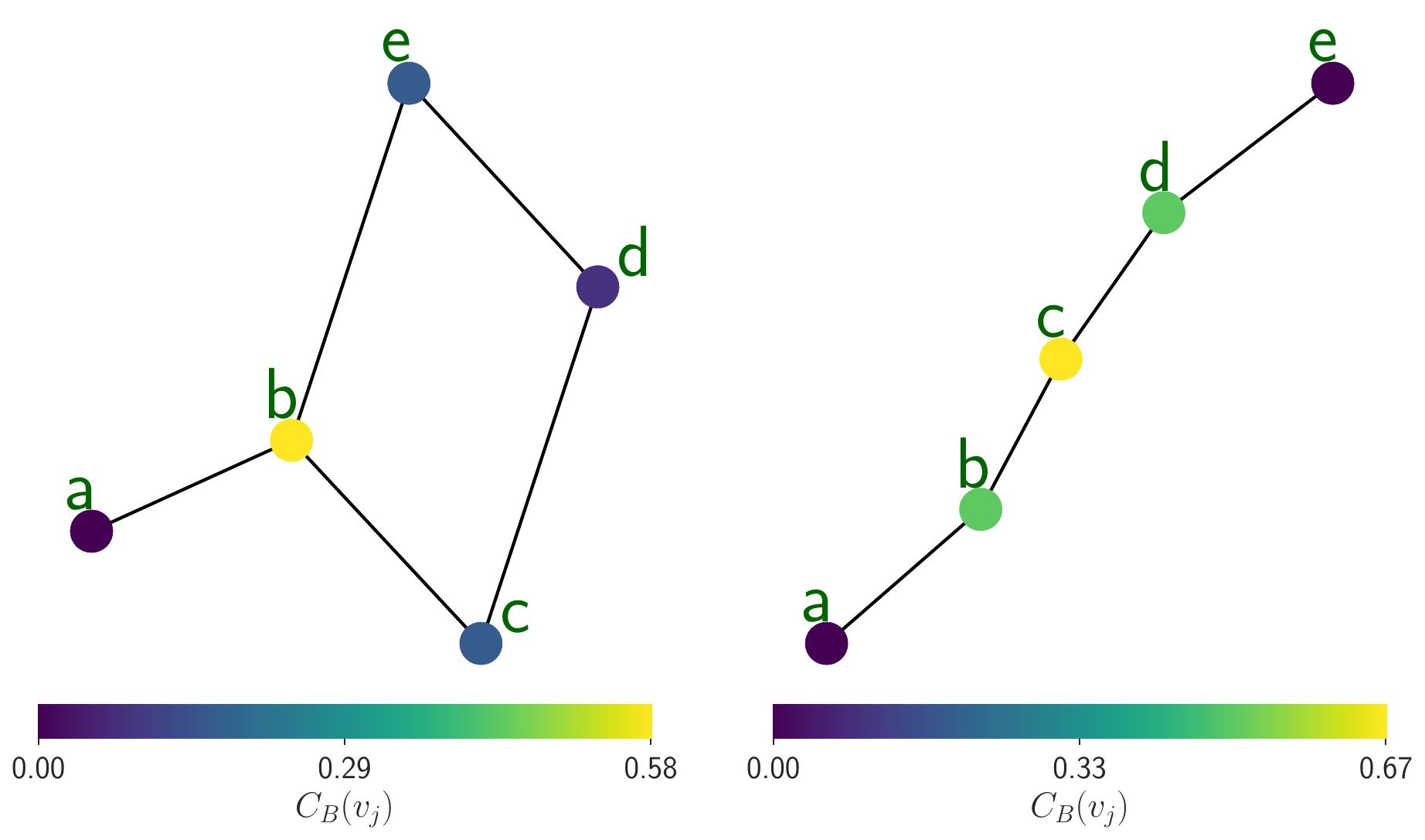}
  \centering
  \caption[Example of Betweenness Centrality estimator.]{
  Left: A graph $G(5,5)$ with a cycle composed of nodes $(b, c, d, e)$. The node with the largest $C_B$ is the node $b$, whose value would be 0.58, while the node $a$ would take the value of 0. 
  Right: A graph $T(5,4)$ is the MST from $G$. The node with the largest $C_B$ is the node $c$, whose value is 0.67, while the nodes $a$ and $e$ would take the value of 0. Similar to Figure~\ref{chapter5_fig: BetCen_ModelVersion}, the color map represents the highest $C_B$ values in yellow colors with increasing darkness for values toward zero.
}
  \label{Appendix2: CB_appendix}
\end{figure}

We apply these steps to two different examples, as shown in Figure~\ref{Appendix2: CB_appendix}, to illustrate how the arrangement of nodes in the graph affects $C_B$. Note that for simplicity, both graphs are considered unweighted when calculating $C_B$. The left panel shows a graph $G(5,5)$, which is not complete but connected, whose node $b$ has the highest $C_B$. To calculate this value, we consider that $b$ appears in the shortest paths $\sigma_{a,c}(b)$, $\sigma_{a,d}(b)$, $\sigma_{a,e}(b)$ and $\sigma_{c,e}(b)$, and we assign the value 1. Of those pairs, on the other hand, $\sigma_{a,d}$ and $\sigma_{c,e}$ have a shortest path reached by two different paths; therefore, a value of 2 will be assigned to each of them. Concurrently, $\sigma_{a,c}$ and $\sigma_{a,e}$ will equal one because there is only one shortest path to link the corresponding pairs. 
Note that $\sigma_{a,d}(b)$ passes through $b$ in both its shortest paths, so its final contribution is 2. So the sum total for node $b$, following the paths $(a, c), (a, d), (a, e), (c, e)$, is: $\frac{1}{1}+\frac{2}{2}+\frac{1}{1}+\frac{1}{2}$=3.5 and if we normalize ($N=5$) the final value is $C_B(b)$=0.58. The second example (right panel of Figure~\ref{Appendix2: CB_appendix}) represents an MST, $T(5,4)$, of the graph $G$. An MST does not contain loops, so the shortest distance between a pair of nodes can only be reached through a unique path. This path will, therefore, always be the shortest one so that $\sigma_{s,t}$ in Equation~(\ref{chapter_2_eq: betweenness_centrality}) will always be 1. Then, $\sigma_{s,t}(v_j)$ will, at most, be equal to 1 for each node $v_j$. Focusing again on node $b$, it appears in the following paths $\sigma_{a,c}(b)$, $\sigma_{a,d}(b)$ and $\sigma_{a,e}(b)$, which will all contribute 1 to the sum. The final sum, then 3, normalized as defined above, is 0.5.

\section{Closeness centrality estimator: Practical example}
\label{Appendix_2: practical_example2}

Here, we provide a simple example of the closeness centrality estimator. To calculate the coefficient $C_C$ by applying Equation~(\ref{chapter_2_eq: closeness_centrality}) on a graph, we recall to take into account the following considerations:

 \begin{itemize}

    \item Select the node $v_j$ for which you want to compute the closeness centrality.
    \item Compute the shortest path length $d_G(v_j, v_t)$ from node $v_j$ to every other node $v_t$ in the graph, with $v_j \ne v_t$. 
    If the graph is unweighted, shortest paths are defined by the minimum number of edges that must be traversed. If weighted, shortest paths minimize the total edge weight. If the shortest path can be reached by different paths due to cycles, $d_G(s,t)$ takes the value of the minimum number of edges (or the minimum total edge weight) that defines the shortest path.
    \item Sum all shortest path distances from $v_j$ to the other nodes: $\sum_{j\ne t}d_G(v_j, v_t)$.
    \item Take the reciprocal of the step above. We will normalize by multiplying the result obtained by $N-1$.
 \end{itemize}

\begin{figure}
  \includegraphics[width=1\textwidth]{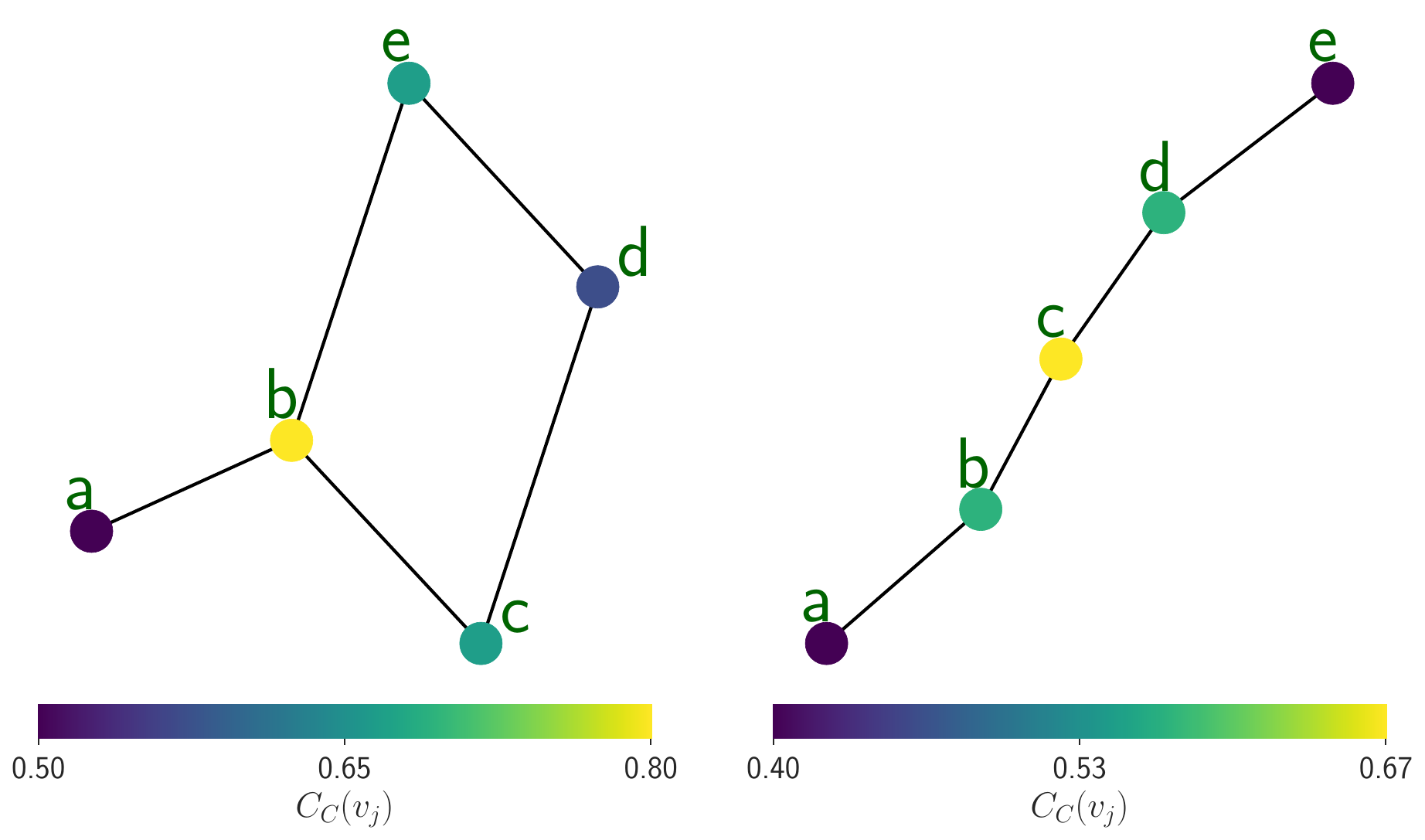}
  \centering
  \caption[Example of Closeness Centrality estimator.]{
  Left: A graph $G(5,5)$ with a cycle composed of nodes $(b, c, d, e)$. The node with the largest $C_C$ is the node $b$, whose value would be 0.8, while the node $a$ would take the value of 0.5. 
  Right: A graph $T(5,4)$ is the MST from $G$. The node with the largest $C_C$ is the node $c$, whose value is 0.67, while the nodes $a$ and $e$ would take the value of 0.4. 
  Similar to Figure~\ref{Appendix2: CB_appendix}, the color map represents the $C_C$ values.
}
  \label{Appendix2: CC_appendix}
\end{figure}

Following Appendix~\ref{Appendix_2: practical_example}, we apply these steps to two different examples, as shown in Figure~\ref{Appendix2: CC_appendix}, to illustrate how the arrangement of nodes in the graph affects $C_C$.
Note that for simplicity, both graphs are considered unweighted when calculating $C_C$.
The left panel shows a graph $G(5,5)$, which is not complete but connected, whose node $b$ has the highest $C_C$.
The distances from node $b$ to the rest are considered the number of edges in the shortest path:
$
d_G(b, a) = 1\text{, } 
d_G(b, c) = 1\text{, }
d_G(b, d) = 2, \text{ and }
d_G(b, e) = 1~.
$

The total distance is: $
\sum_{b \ne t} d_G(b,t) = 1 + 1 + 2 + 1 = 5~.
$
The $C_C$ value is then given by Equation~(\ref{chapter_2_eq: closeness_centrality}): $
C_C(b) = \frac{4}{5} =0.8~.
$
The second example (right panel of Figure~\ref{Appendix2: CC_appendix}) represents an MST, $T(5,4)$, of the graph $G$. An MST does not contain loops, so the shortest distance between a pair of nodes can only be reached through a unique path. Focusing again on node $b$: $
d_T(b, a) = 1\text{, } 
d_T(b, c) = 1\text{, }
d_T(b, d) = 2, \text{ and }
d_T(b, e) = 3~.
$
The total distance is: $
\sum_{b \ne t} d_T(b,t) = 1 + 1 + 2 + 3 = 7~.
$
The $C_C$ value is then given by Equation~(\ref{chapter_2_eq: closeness_centrality}): $
C_C(b) = \frac{4}{7} =0.57~.
$

\chapter{PCA using EVD and SVD}
\label{Appendix_3}

We illustrate PCA step by step using a small dataset with $n = 4$ data points and $N = 2$ variables. Let us consider the dataset:

\begin{equation}
X = \begin{bmatrix}
3 & 2 \\
4 & 2 \\
1 & 1 \\
0 & 4 \\
\end{bmatrix}~,
\label{Appendix_3: data_matrix}
\end{equation}

where each row corresponds to a data point with two features.

\section{Computing PCA using EVD}

\subsubsection{Step 1: Center the data}

PCA requires the data to be mean-centered. We compute the mean of each column:

\begin{equation}
\mu = \frac{1}{4} \sum_{i=1}^4 X_i =
\begin{bmatrix}
2 \\
2.25
\end{bmatrix}~.
\end{equation}

The centered data matrix $\tilde{X}$ is obtained by subtracting $\mu$ from Equation~(\ref{Appendix_3: data_matrix}):

\begin{equation}
\tilde{X} = X - \mu^T =
\begin{bmatrix}
1  & -0.25\\
2  & -0.25\\
-1 & -1.25\\
-2 &  1.75
\end{bmatrix} ~.
\label{Appendix_3: centered_data_matrix}
\end{equation}

\subsubsection{Step 2: Compute the Covariance matrix}

The covariance matrix $C$ is computed as described in Equation~(\ref{chapter_3_eq: cov_matrix_EVD}) using Equation~(\ref{Appendix_3: centered_data_matrix}):

\begin{equation}
\tilde{X}^T \tilde{X} =
\begin{bmatrix}
10 & -3 \\
-3 & 4.75
\end{bmatrix}~.
\label{Appendix_3: tilde_XTX}
\end{equation}

Then: 
\begin{equation}
C = \frac{1}{3}
\begin{bmatrix}
10 & -3  \\
-3 & 4.75
\end{bmatrix}
=
\begin{bmatrix}
3.3 & -1  \\
-1 & 1.58
\end{bmatrix}~.
\label{Appendix_3: covariance_Matrix}
\end{equation}

\subsubsection{Step 3: Apply Eigenvalue Decomposition}

We compute the eigenvalues $\lambda$ and eigenvectors $\vec{v}$ of $C$, see Equation~(\ref{chapter_3_eq: EVD_derived}), through the characteristic polynomial:

\begin{equation}
    \det(C - \lambda I) = 0~.
    \label{Appendix_3: solving_EVD}
\end{equation}

Substituting Equation~(\ref{Appendix_3: covariance_Matrix}) in Equation~(\ref{Appendix_3: solving_EVD}), and the identity matrix $I$, we obtain:

\begin{equation}
\det\left(
\begin{bmatrix}
3.33 - \lambda & -1 \\
-1 & 1.58 - \lambda
\end{bmatrix}
\right)
=
(3.33 - \lambda)(1.58 - \lambda) - 1 = 0
~.
\label{Appendix_3: solving_EVD_expanded}
\end{equation}

Expanding Equation~(\ref{Appendix_3: solving_EVD_expanded}):

\begin{equation}
\lambda^2 - (3.33 + 1.58)\lambda + (3.33)(1.58) - 1 = 0 \Rightarrow \lambda^2 - 4.91\lambda + 4.26 = 0~.
\label{Appendix_3: solving_EVD_expanded_quadratic}
\end{equation}

Solving Equation~(\ref{Appendix_3: solving_EVD_expanded_quadratic}), we obtain:

\begin{equation}
\lambda_1 \approx 3.78, \quad \lambda_2 \approx 1.12~,
\label{Appendix_3: eigenvalues}
\end{equation}

where $\lambda_1$ and $\lambda_2$ are the eigenvalues. To find the corresponding $\vec{v}$, we have to solve: 

\begin{equation}
(C - \lambda_i I) \vec{v}_i = 0~.
\label{Appendix_3: eigenvalues_corresponding_eigenvectors}
\end{equation}

Considering Equation~(\ref{Appendix_3: eigenvalues_corresponding_eigenvectors}), for each $\lambda$ seen in Equation~(\ref{Appendix_3: eigenvalues}), we obtain:

\begin{equation}
\vec{v}_1 \approx
\begin{bmatrix}
- 0.91 \\
  0.41
\end{bmatrix}~, \quad
\vec{v}_2 \approx
\begin{bmatrix}
-0.41 \\
-0.91
\end{bmatrix}~.
\label{Appendix_3: eigenvectors}
\end{equation}

Equation~(\ref{Appendix_3: eigenvectors}) shows the final results after being normalized by $||\vec{v}_i||$.

\subsubsection{Step 4: Project the data}

To transform the original data into the new basis defined by the PCs, we project $\tilde{X}$ onto $\vec{v}$:

\begin{equation}
Z = \tilde{X} \cdot V
\label{Appendix_3: projectionZ}
\end{equation}

where $V$ is the matrix of $\vec{v}$ seen as $V=[\vec{v}_1 \quad \vec{v}_2]$. Applying Equation~(\ref{Appendix_3: projectionZ}) and considering Equation~(\ref{Appendix_3: eigenvectors}), we obtain:

\begin{equation}
Z =
\begin{bmatrix}
 -1.01 & -0.19\\
 -1.92 & -0.60\\
  0.39 &  1.55\\
  2.54 & -0.77
\end{bmatrix}~.
\label{Appendix_3: projectionZ_values}
\end{equation}

where each row in Equation~(\ref{Appendix_3: projectionZ_values}) gives the new coordinates, PC scores, of $X$ seen in Equation~(\ref{Appendix_3: data_matrix}). 
In other words, each column in Equation~(\ref{Appendix_3: projectionZ_values}) is a PC, which can be calculated as we saw in Equation~(\ref{chapter_3_eq: PC_space}):
\begin{equation} 
    \begin{aligned}
        PC_1 &=  -0.91 \cdot \tilde{X}_{:,1} + 0.41 \cdot \tilde{X}_{:,2}~\\
        PC_2 &= -0.41 \cdot \tilde{X}_{:,1} - 0.91 \cdot \tilde{X}_{:,2}~.
    \end{aligned}
    \label{Appendix_3: PC1_PC2_EVD}
\end{equation}

In Equation~(\ref{Appendix_3: PC1_PC2_EVD}) we consider Equation~(\ref{Appendix_3: eigenvectors}) and Equation~(\ref{Appendix_3: data_matrix}), where $\tilde{X}_{:,1}$ and $\tilde{X}_{:,2}$ represent each column of $\tilde{X}$, respectively.

\section{Computing PCA using SVD}

Given the $X$ seen in Equation~(\ref{Appendix_3: data_matrix}), the SVD factorizes it as according to Equation~(\ref{chapter_3_eq: SVD_methodology}), where:
\begin{itemize}
    \item $ U \in \mathbb{R}^{4 \times 2} $ contains the left singular vectors,
    \item $ \Sigma \in \mathbb{R}^{2 \times 2} $ is a diagonal matrix of singular values $ \sigma_i \geq 0 $,
    \item $ V \in \mathbb{R}^{2 \times 2} $ contains the right singular vectors.
\end{itemize}

To calculate $U$, $\Sigma$, and $V$, according Equation~(\ref{chapter_3_eq: cov_SVD_methodology}), we have to use Equation~(\ref{Appendix_3: tilde_XTX}).

\subsubsection{Step 1: Apply Eigenvalue Decomposition}

Let's solve the characteristic polynomial:

\begin{equation}
    \det(\tilde{X}^{T}\tilde{X} - \lambda I) = 0~.
    \label{Appendix_3: solving_SVD}
\end{equation}

From Equation~(\ref{Appendix_3: solving_SVD}):

\begin{equation}
\det\left(
\begin{bmatrix}
10 - \lambda & -3 \\
-3 & 4.75 - \lambda
\end{bmatrix}
\right)
=
(10 - \lambda)(4.75 - \lambda) - 9 = 0
~.
\label{Appendix_3: solving_SVD_expanded}
\end{equation}

Expanding Equation~(\ref{Appendix_3: solving_SVD_expanded}):

\begin{equation}
\lambda^2 - (10 + 4.75)\lambda + (10)(4.75) - 9 = 0 \Rightarrow \lambda^2 - 14.75\lambda + 38.5 = 0~.
\label{Appendix_3: solving_SVD_expanded_quadratic}
\end{equation}

Solving Equation~(\ref{Appendix_3: solving_SVD_expanded_quadratic}), we obtain:

\begin{equation}
\lambda_1 \approx 11.36, \quad \lambda_2 \approx 3.39~,
\label{Appendix_3: eigenvalues_SVD}
\end{equation}

where $\lambda_1$ and $\lambda_2$ are the eigenvalues of $\tilde{X}^{T}\tilde{X}$. 

To find the corresponding $\vec{v}$, we have to solve: 

\begin{equation}
(\tilde{X}^{T}\tilde{X} - \lambda_i I) \vec{v}_i = 0~.
\label{Appendix_3: eigenvalues_corresponding_eigenvectors_SVD}
\end{equation}

Considering Equation~(\ref{Appendix_3: eigenvalues_corresponding_eigenvectors_SVD}), for each $\lambda$ seen in Equation~(\ref{Appendix_3: eigenvalues_SVD}), we obtain:

\begin{equation}
\vec{v}_1 \approx
\begin{bmatrix}
- 0.91 \\
  0.41
\end{bmatrix}~, \quad
\vec{v}_2 \approx
\begin{bmatrix}
-0.41 \\
-0.91
\end{bmatrix}~.
\label{Appendix_3: eigenvectors_SVD}
\end{equation}

Equation~(\ref{Appendix_3: eigenvectors_SVD}) shows the final results after being normalized by $||\vec{v}_i||$.

\subsubsection{Step 2: Singular values ($\Sigma$)}

We compute singular values considering Equation~(\ref{Appendix_3: eigenvalues_SVD}):

\begin{equation}
    \sigma_1 = \sqrt{11.36} \approx 3.37, \quad \sigma_2 = \sqrt{3.39} \approx 1.84~.
    \label{Appendix_3: eigenvalues_singular_values}
\end{equation}

Then, considering Equation~(\ref{Appendix_3: eigenvalues_singular_values}):

\begin{equation}
\Sigma = \begin{bmatrix}
3.37 & 0 \\
0 & 1.84 \\
\end{bmatrix}~.
\label{Appendix_3: SIGMA_SVD}
\end{equation}

Note applying Equation~(\ref{chapter_3_eq: lambdas_SVD_methodology}), we recover Equation~(\ref{Appendix_3: eigenvalues}) based on $C$.

\subsubsection{Step 3: Right singular vectors ($V$)}

Using Equation~(\ref{Appendix_3: eigenvectors_SVD}), we get:

\begin{equation}
    V = 
\begin{bmatrix}
-0.91 & -0.41 \\
0.41 &  -0.91 \\
\end{bmatrix}~.
\label{Appendix_3: V_SVD}
\end{equation}

\subsubsection{Step 4: Left singular vectors ($U$)}

Using Equation~(\ref{chapter_3_eq: PCs_SVD_methodology}), we have:

\begin{equation}
   U = \tilde{X} V \Sigma^{-1}~.
   \label{Appendix_3: U_SVD}
\end{equation}

Then, computing $\Sigma^{-1}$ considering Equation~(\ref{Appendix_3: SIGMA_SVD}):

\begin{equation}
\Sigma^{-1} \approx \begin{bmatrix}
0.30 & 0 \\
0 & 0.54 \\
\end{bmatrix}~,
\label{Appendix_3: SIGMA_SVD_inverse}
\end{equation}

and considering Equations~(\ref{Appendix_3: centered_data_matrix}) and (\ref{Appendix_3: V_SVD}):

\begin{equation}
    \tilde{X} V \approx
\begin{bmatrix}
 -1.01 & -0.19 \\
 -1.92 & -0.60 \\
  0.39 &  1.55\\
  2.54 & -0.77 \\
\end{bmatrix}~,
\label{Appendix_3: tildeXV_SVD_PCs}
\end{equation}

following Equation~(\ref{Appendix_3: U_SVD}), we obtain:

\begin{equation}
U = \tilde{X} V \Sigma^{-1} \approx
\begin{bmatrix}
-0.30& -0.10\\
-0.58& -0.32\\
 0.12 & 0.84 \\
 0.76 &-0.41 \\
\end{bmatrix}~.
\label{Appendix_3: U_SVD_values}
\end{equation}

\subsubsection{Step 5: Project the data}

The PC scores can be obtained via Equation~(\ref{Appendix_3: tildeXV_SVD_PCs}) or equivalently, using Equation~(\ref{Appendix_3: U_SVD_values}) and Equation~(\ref{Appendix_3: SIGMA_SVD}):

\begin{equation}
Z = \tilde{X} V = U \Sigma \approx
\begin{bmatrix}
 -1.01 & -0.19 \\
 -1.92 & -0.60 \\
  0.39 &  1.55 \\
  2.54 & -0.77 \\
\end{bmatrix}~.
\label{Appendix_3: projectionZ_values_SVD}
\end{equation}

Equation~(\ref{Appendix_3: projectionZ_values_SVD}) and Equation~(\ref{Appendix_3: projectionZ_values}) give the same results, and are equivalent ways of calculating PCs.
\chapter{Details on algebra with principal components, distance, and rendering of the MST}  
\label{Appendix_4}

\section{Algebra with $PC_1$ and $PC_2$ }
\label{Appendix_4: algebra_PC1_PC2}

The variables used depend on powers of $P$ and $\dot{P}$, and therefore, once the logarithm is taken, the expressions defining each variable are linear. For example, $\log B_{s} = K + \ logP + \log \dot{P}$, where $K$ is a constant, geometrically speaking, the Equation of a plane, $Ax+By+Cz+D=0$. 
The eigenvalues associated with the PCs beyond $PC_1$ and $PC_2$ will be strictly null, and therefore, all the explained variance is accumulated in the first two PCs. 
Equation~(\ref{chapter_4_eq: PC1_PC2_8V}) is expressed as a function of the normalization of the logarithm of the variables, $(V_{l})^{\dag}$, where the sub-index $l$ indicates the logarithm of the generic variable $V$. Here, given that any of the variables we are considering can be written as $V(P,\dot {P}) = KP^{a} \dot P^{b}$, $V_{l} = \log V = K_{l} + aP_{l} + b\dot{P}_{l}$, the normalization via Equation~(\ref{chapter_4_eq: robust_scaler}) is:

\begin{eqnarray}
        (V_{l})^{\dag} = \frac{V_{l} - Q_{2,V_{l}}}{IQR_{V_{l}}}=
        \frac{K_{l} + aP_{l} + b\dot{P}_{l} - Q_{2,V_{l}}}{IQR_{V_{l}}}~.
    \label{Appendix_4_eq: norm_log_var}
\end{eqnarray} 

Once Equation~(\ref{Appendix_4_eq: norm_log_var}) is applied to all the variables, a summation will be obtained, which leads to Equation~(\ref{chapter_4_eq: PCA_2V}). 
In Figure~\ref{chapter_4_figure: figure3}, we show how a circle in the $P\dot{P}$ diagram would appear in the plane defined by the PCs. 
The simulated points located along a circle in the $P\dot P$ diagram, defined by its logarithmic coordinates ($P_{l}, \dot P_{l}$), can be described by a fixed value $r^{2} = \Delta P_{l}^{2} + \Delta \dot{P}_{l}^{2}$ where $\Delta $ is used to represent that the circle can be displaced from the origin. 
Considering Equation~(\ref{chapter_4_eq: PCA_2V}), we get:

\begin{eqnarray}
r^{2} = 0.702\Delta PC_{1}^{2} + 0.148\Delta PC_{2}^{2} - 0.363 \Delta PC_{1}\Delta PC_{2}
~.\label{Appendix_4_eq: ellipse}
\end{eqnarray}
Equation~(\ref{Appendix_4_eq: ellipse}) describes an ellipse (i.e., $Ax^{2}+Bxy+Cy^{2}+Dx+Ey+F=0$) with an angle of rotation $\cot(2\theta) = (A-C)/{B}$. Starting counter-clockwise from the positive axis of $PC_{1}$ this leads to $\theta = 73.38^{\circ}$.

\section{Nearness in the distance context and the MST}
\label{Appendix_4: nearness-comment}

\begin{figure}
\centering
  \includegraphics[width=0.6\textwidth]{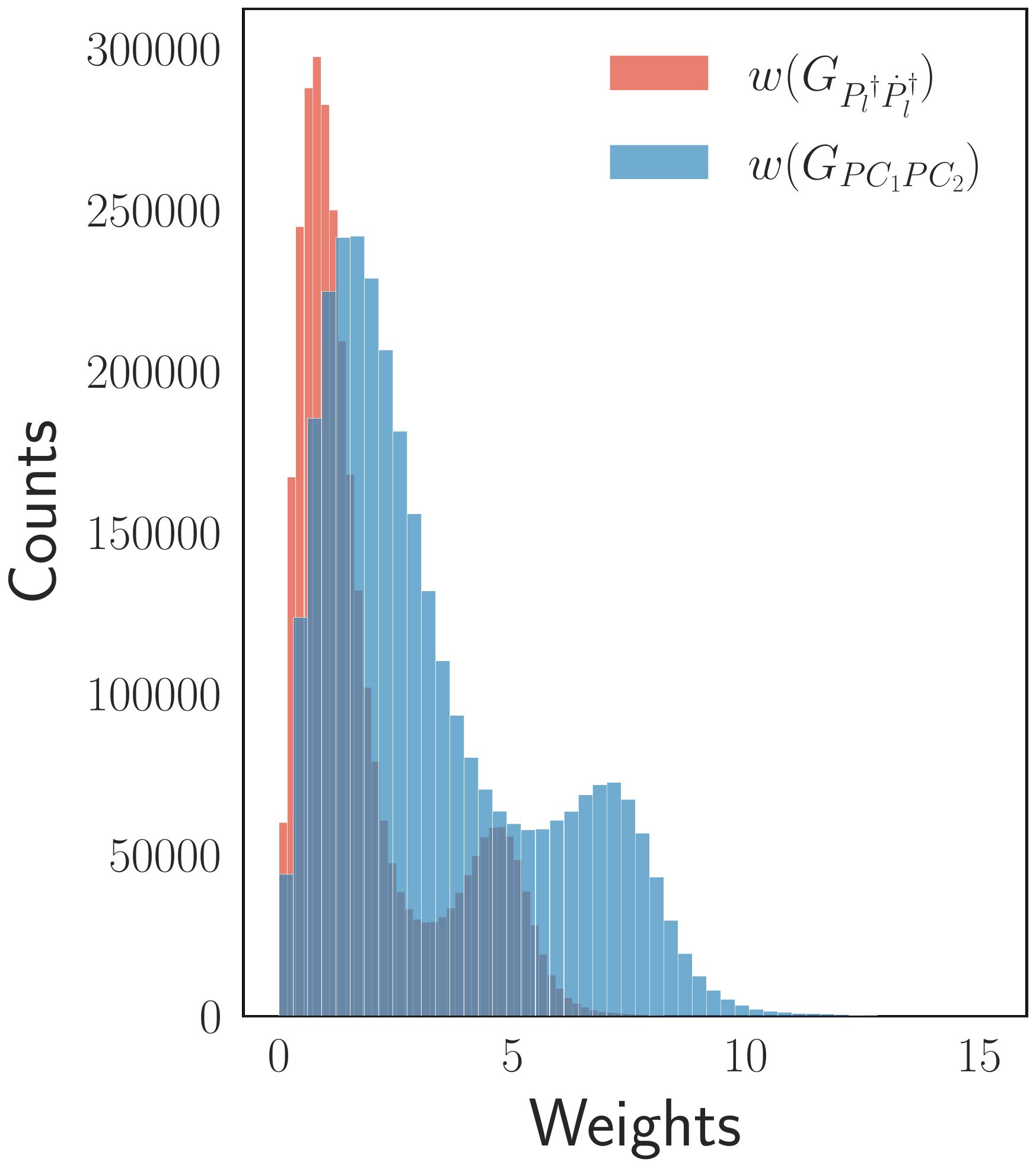}
  \caption[Distributions of the weights associated with a complete graph.]{Distributions of the weights associated with the complete, undirected, and weighted graph for two different distance definitions. We show in red the distribution of distances based on the logarithms of the variables $P$ and $\dot{P}$, once they have been normalized, and in blue, the distribution of distances based on $PC_1$ and $PC_2$. 
}
  \label{Appendix_4_figure: figure12}
\end{figure}

The nearness between two pulsars depends on the definition of distance and primarily on the variables considered to compute it; thus, the more comprehensive this distance is, the more accurate it becomes. One way to see this is by comparing the distributions of the weights associated with a complete, undirected, and weighted graph, considering only $P,\dot P$ with those obtained from a similar graph but considering the set of 8 variables described in Section~\ref{chapter4: variables} (equivalently, the first two PCs). 
This is illustrated in Figure~\ref{Appendix_4_figure: figure12}, where the distributions differ, which relates to the discussion derived from the bottom panels of Figure~\ref{chapter_4_figure: figure3}. In addition, from the set of weights associated with the graph, it is possible to know the nearest neighbors of any randomly chosen pulsar. 
Considering the ranking, a distance-ordered list of one pulsar relative to all others, it can be seen that neighboring pulsars differ according to each distance definition. For instance, for every pulsar, we identify the nearest three neighbors using a distance based on $P$ and $\dot{P}$, and compare these with the ones obtained using the complete set of variables of interest. We find that 45\% of the population incorporates a new pulsar.
Even when the same pulsars are concerned, the ranking positions may change in many cases: 40\% of the population has at most one pulsar in the same position.

Due to MST properties, we know that the first neighbor in the distance ranking from a given source is indeed one of the nodes attached to the MST. However, the second neighbor and beyond in the distance ranking do not necessarily reflect vis-à-vis the visual appearance of the MST. Thus, visual nearness from a given source in the MST is not a one-to-one overall translation of the distance ranking from that source. The MST minimizes the distance to a given node and the total distance required to visit the entire population. This introduces a global perspective in selecting how particular sources are connected to the rest, ultimately serving as a basis for clustering analysis techniques. 

\section{Representing the MST: how rendering is done}
\label{Appendix_4: representing_MST}

Given that an MST lacks a fixed set of axes, i.e., it lives in a non-Euclidean space, its rendering (angles among branches, orientation, branch direction) does not hold any particular meaning; only the connections among nodes do. A different representation could be chosen, while conserving the fundamental characteristic —i.e., preserving the exact distance definition —which will keep each node linked to the same neighbors and in the same order.
Despite the appearance difference, the nearness and connections described in one MST would be the same as in the other. 
An obvious example is to take one of the branches that go rightwards in the Pulsar Tree (see Figure~\ref{chapter_4_figure: figure4}) and force it to go leftwards. This would produce the same MST (the same adjacency matrix, in technical terms, see Chapter~\ref{chapter2: basics_graph}), and everything we have said looking at the MST with the same branch pointing rightwards would also apply to the new rendering. It is not the appearance of the overall graph that is important, but rather the properties of the graph. 
For the MST representations in this work, we utilize the {\sc neato} program, which employs the Graphviz Python library \citep{graphviz} to minimize a pseudo-energy and select the orientation of the different components. 

\chapter{Details on significant branches and the
MST for the MSP population}  
\label{Appendix_6: significant_branches}

Here, we present a concise and focused explanation of how the clustering algorithm introduced in Chapter~\ref{chapter4} is applied to the population of MSPs. We apply the betweenness centrality estimator (see Section \ref{chapter2: bet_cen}) to identify the most central nodes of the graph $T$. Each value identified as an outlier through the betweenness centrality will be referred to as a potential trunk node (pTN).
The trunk is defined as a path, i.e., a sequence of non-repeated nodes. 
At the end of the trunk, the pTNs that delimit are called contour nodes (CNs).
These CNs must be nodes of degree greater than 2, acting as articulation points for the graph. Removing these nodes partitions the graph into connected components.
The connected components originating in the CNs are called contour branches (CBs).
This leaves a clustered structure with a minimum of two groupings at each extreme of the trunk. 
To avoid limiting the trunk in nodes of degree 2 or larger but from which the departing structures are of non-representative size (i.e., we aim to prevent noise near the trunk), we define a significant threshold ($\alpha_{s}$) so that the number of nodes in the CBs is requested to exceed this lower limit. 
We set the threshold at 5\%, implying that branches will contain at least 11 pulsars of the total population.
Once all the possible CNs are identified, we calculate all the possible trunks as the paths that pair these CNs.
The number of trunks obtained:
$\# Trunks = \mathrm{CNs} \times (\mathrm{CNs}-1)/2
~.\label{eq: number_trunks}
$
For each possible trunk, we have a set of branches constituted by the connected components starting from every node of a degree larger than 2 (including the CBs) and having $\alpha_{s}$ as a lower limit to the number of nodes each one must contain.
This results in six sets of branches and trunks.
We choose the set that provides more excellent uniformity in the size of the resulting groups, which contributes to the robustness
of a comparison analysis, such as a non-parametric test, such as the Kolmogorov-Smirnov (KS) statistics \citep[see, e.g.,][]{KS-test3, KS-test4, KS-test5}. 
In this case, the null hypothesis ($H_0$) is that the distribution of the variables of the nodes of two given branches is consistent with them coming from the same parent distribution. 
If we reject this $H_0$, say at the 95\% confidence level (CL) or better, we may conclude that two branches may be formed by different pulsars or at different evolutionary stages.
Table~\ref{chapter_6_tab: significance_branches} shows the KS test results for these branches, where the $H_0$ column indicates which branches differ the most under the above assumptions.
It is observed that most branches are distinguished by most individual variables.

\begin{table}
\scriptsize
\centering
\caption[Comparison of significant branches.]{
Comparison of the branches in Figure~\ref{chapter_6_figures: MST_siginificance_branches}.
We denote with a number 1 when two branches, according to the KS-test for a 95\% CL, reject $H_0$ for the analyzed variable.
Instead, we denote the opposite case with a number of 0, where we can not reject $H_0$.
The last column shows the accumulated sum for each confrontation between branches, counting for which variables these branches reject $H_0$.
For visualization reasons, we denote with "*" those colors in which dark is suppressed, alluding to the dark green and dark blue branches.
}
\setlength{\tabcolsep}{5pt}
\begin{tabular}{l rrrrrrrrrr | r}
\hline
Branches    &  $B_{lc}$ &  $B_{s}$ &  $\dot{E}$ &  $\eta_{GJ}$ &  $P$ &  $\dot{P}$ &  $\Delta\Phi$ &  $P_{B}$ &   $A_{1}$ & $M_C$ & $H_{0}$ \\
\hline
Green* - Gray & 0 & 1 & 1 & 1 & 0 & 1 & 1 & 1 & 1 & 1 & 8 \\
Green* - Pink & 1 & 1 & 1 & 1 & 0 & 1 & 1 & 0 & 0 & 0 & 6 \\
Green* - Blue* & 1 & 1 & 1 & 0 & 1 & 1 & 1 & 1 & 1 & 1 & 9 \\
Green* - Orange & 0 & 1 & 1 & 1 & 1 & 1 & 1 & 1 & 1 & 1 & 9 \\
Green* - Brown & 1 & 1 & 1 & 1 & 1 & 0 & 1 & 1 & 1 & 1 & 9 \\
Gray - Pink & 1 & 0 & 1 & 1 & 0 & 0 & 1 & 1 & 1 & 1 & 7 \\
Gray - Blue* & 1 & 1 & 1 & 1 & 1 & 1 & 1 & 1 & 1 & 0 & 9 \\
Gray - Orange & 1 & 1 & 1 & 1 & 1 & 1 & 1 & 0 & 0 & 0 & 7 \\
Gray - Brown & 1 & 1 & 1 & 1 & 1 & 1 & 1 & 1 & 1 & 1 & 10 \\
Pink - Blue* & 1 & 0 & 1 & 1 & 1 & 1 & 1 & 1 & 1 & 1 & 9 \\
Pink - Orange & 1 & 1 & 1 & 1 & 0 & 1 & 1 & 1 & 1 & 1 & 9 \\
Pink - Brown & 1 & 1 & 1 & 1 & 1 & 1 & 1 & 1 & 1 & 1 & 10 \\
Blue* - Orange & 1 & 1 & 1 & 1 & 1 & 1 & 1 & 0 & 0 & 0 & 7 \\
Blue* - Brown & 0 & 1 & 0 & 1 & 1 & 1 & 0 & 1 & 0 & 0 & 5 \\
Orange - Brown & 1 & 1 & 1 & 0 & 1 & 1 & 1 & 1 & 1 & 1 & 9 \\
\hline
\end{tabular}
\label{chapter_6_tab: significance_branches}
\end{table}

\chapter{No moding in identified tMSP candidates in the radio state}  
\label{Appendix_6: moding}
\begin{figure}
\centering
\includegraphics[width=0.99\textwidth]{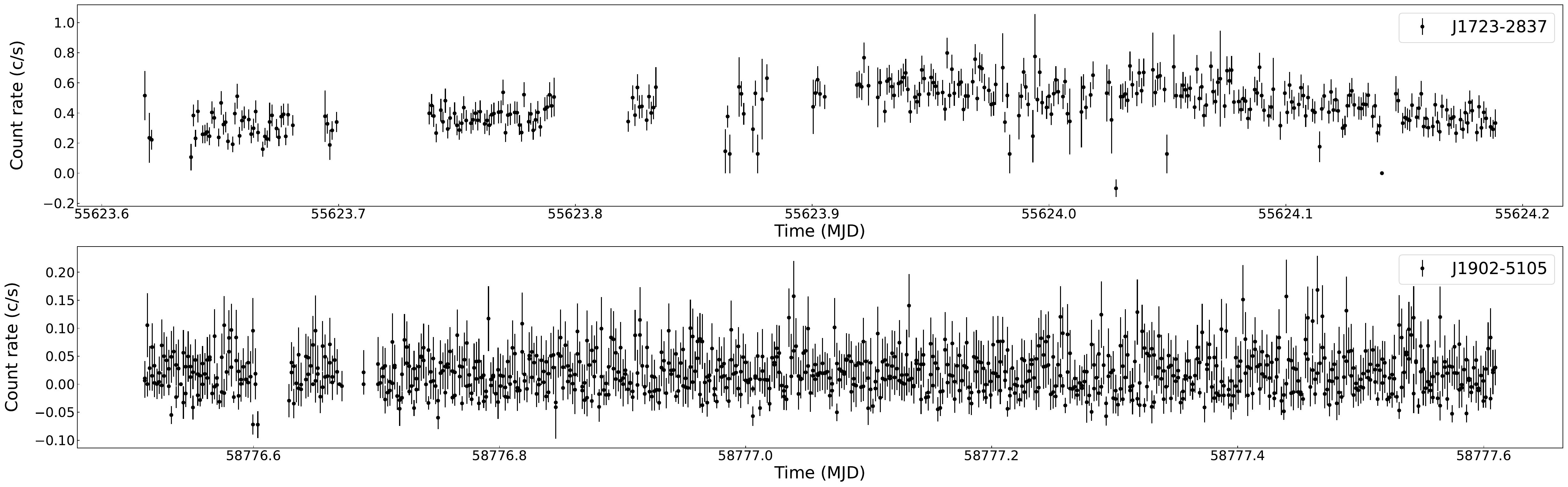}
\caption[\textit{XMM-Newton} 0.3-10 keV background-subtracted light curves.]{\textit{XMM-Newton} 0.3-10 keV background-subtracted light curves over the time interval in which the three EPIC instruments collected data simultaneously for J1723-2837 (top panel) and J1902-5105 (bottom panel). The light curves are binned with a time resolution of 100 s. For a detailed analysis of J1431-4715 see \cite{deMartino2024, Linares2023}.} 
\label{Appendix_6_fig: XMM_light_curves}
\end{figure}

The EPIC-pn \citep{Struder_2001A&A} operated with a time resolution of 73.4 $\mathrm{ms}$ (full frame mode) for 1723-2837 and J1902-5105, and 47.4 $\mathrm{ms}$ (large window mode) for J1431-4715. The two EPIC-MOS \citep{Turner_2001A&A} were employed with a time resolution of 2.6 s (full frame mode) in the first two cases and 0.9 s (large window mode) for J1431-4715. We processed and analyzed the data using the Science Analysis Software (SAS; v.21.0.0). All three observations exhibited high background activity in the 10–12 keV light curve, so we excluded the contaminated data intervals.
For the EPIC-pn and each MOS, we extracted source photons within a circular region centered on the source position with a 40$^{\circ}$ radius, and background photons from an 80$\arcsec$-wide, source-free circular region.
Using the \texttt{epiclccorr} task, 0.3-10 keV background-subtracted light curves were extracted from the three EPIC instruments over the simultaneous coverage time interval, binned with a time resolution of 100 s. 
As previously noted, we recall that the X-ray flux of J1723-2837 (top panel of Figure~\ref{Appendix_6_fig: XMM_light_curves}) is variable with the orbital phase as discussed in \cite{Bogdanov2014}.

We searched the \textit{XMM-Newton} light curves for the typical bimodality between "high" and "low" intensity modes (see Figure~\ref{Appendix_6_fig: XMM_light_curves}; see \cite{deMartino2024}, see also \citep{Linares2023} for a detailed analysis of J1431-4715). We experimented with different bin sizes but did not observe the expected bimodal feature. Further analysis of the count rate distributions confirmed the absence of this bimodality. A deep search for pulsation has identified PSR J1431-4715 as a gamma-ray pulsar \cite{Fermi3PC}; its gamma-ray emission shows variable signature as those found for prototypical tMSPs \cite{Torres2017}.
Note that the prototype for tMSPs, J1023+0038, is located at $\sim$1.37 kpc \citep{Deller_2012ApJ}, while the estimated distances for J1723-2837 and J1902–5105 are $\sim$0.75 kpc \citep{Bogdanov2014} and $\sim$1.2 kpc \citep{Camilo_2015ApJ}, respectively. Assuming comparable luminosities, if a similar bimodal behavior were present, we would expect to observe it in these closer sources.
Moreover, distinct high and low modes are evident in the faint candidate tMSP 3FGL J1544.6-1125 \citep{Bogdanov_2015}. We have re-analyzed the \textit{XMM-Newton} archival observation ObsID 072080101 as a check for the latter source. The bimodal pattern was detected in the EPIC/MOS1 0.3-10 keV background-subtracted light curve, binned with a time resolution of 100 s and exhibiting an average count rate comparable to that of J1723-2837. Thus, should it be present in J1723-2837, it should have become similarly apparent.
\chapter{Evaluating metrics}  
\label{Appendix_7}
We summarize the key statistical metrics used to evaluate the separation between repeater and non-repeater FRBs seen in Chapter~\ref{chapter7}. The evaluating metrics provided are precision, recall, F-scores, the Receiver Operating Characteristic (ROC) analysis, and the Z-score used for assessing statistical significance relative to randomized baselines (see, e.g., \cite{han2011data, Bishop2007, Geron2019}). These tools enable an interpretable assessment of how well the selected properties, obtained through the MST approach, distinguish between the determined FRB categories in a manner consistent with machine learning methods.

\textit{Precision} quantifies the ratio of repeaters that were correctly identified compared to the total in the branch, while \textit{recall} (also known as \textit{sensitivity}) measures the proportion of actual repeaters that were correctly classified.
The \textit{F measure}, defined as
\begin{equation}
F_{\beta} = (1 + \beta^2)  \frac{Precision \,\, Recall}{(\beta^2 \,\, Precision) + Recall}
\end{equation} 
evaluates the predictive performance by combining precision and recall through a weighted harmonic mean. 
The importance of precision ($\beta < 1$) versus recall ($\beta > 1$) is modulated by the $\beta$ parameter.
When $\ beta=1$, precision and recall are equally weighted. 
The $F_{1}$ score is obtained when $\beta = 1$, resulting in equal emphasis on precision and recall.
The $F_{2}$ score, alternatively, is derived when $\beta = 2$, giving more importance to recall over precision. 
The \textit{Receiver Operating Characteristic (ROC)} curve is a graphical representation of a binary classification model's performance across different threshold settings. 
It plots the true positive rate (sensitivity) against the false positive rate (1 - specificity) at various threshold values. 
Here, Specificity, also known as the true negative rate, measures the proportion of actual negative instances (non-repeater FRBs) correctly identified as negative.
The ROC curve helps assess the trade-off between sensitivity and specificity, providing insights into the classifier's ability to discriminate between positive and negative instances.
The Area Under the Curve (AUC) value quantifies the overall performance of a binary classification model represented by the ROC curve. 
It measures the area under the ROC curve, with values ranging from 0 to 1. 
AUC values closer to 1 indicate the classifier's better discrimination ability, while values closer to 0.5 suggest poor discrimination (similar to random guessing).
    
The labels assigned to repeaters and non-repeaters according to the repeater branch are as follows:
\begin{itemize}
  \item \textbf{True Positive (TP)}: Repeaters correctly classified as repeaters 
  \item \textbf{False Positive (FP)}: Non-repeaters incorrectly classified as repeaters.
  \item \textbf{True Negative (TN)}: Non-repeaters correctly classified as non-repeaters.
  \item \textbf{False Negative (FN)}: Repeaters incorrectly classified as non-repeaters.
\end{itemize}

The scores are defined as
  \begin{eqnarray}
    Precision &=& \frac{TP}{TP + FP}\\
    Recall &=& \frac{TP}{TP + FN} \\
    Specificity &=& \frac{TN}{TN + FP}\\
    F_{1}\; Score &=& \frac{2 \,\, Precision \,\, Recall}{Precision + 
    Recall}\\
    F_{2}\; Score &=& \frac{5 \,\, Precision \,\, Recall}{4 \,\, Precision + 
    Recall}
  \end{eqnarray}
The \textit{Z-score}, also known as the standard score or z-value, is a statistical measure that quantifies how many standard deviations a data point is from the mean of a dataset. It is defined as 
\begin{eqnarray}
    Z = \frac{x - \mu}{\sigma}.
\end{eqnarray}

Here, $x$ is the value of the data point (the classifier's recall), $\mu$ and $\sigma$ are the mean and the standard deviation of the dataset (distribution from the fake cases), respectively, see Section \ref{chapter7: MC_permutations} for a proper application. 
A Z-score of 0 indicates that the data point is strictly at the mean of the dataset, while positive and negative Z-scores suggest that the data point is above and below the mean, respectively.
A Z-score higher than 2$\sigma$ or 3$\sigma$ in absolute value is significant at the 5\% or even 1\% level.
\chapter{Working with DTW}  
\label{Appendix_8}

\section{Optimal warping path computation}
\label{Appendix_8: example}

\begin{figure}
\centering
  \includegraphics[width=0.45\textwidth]{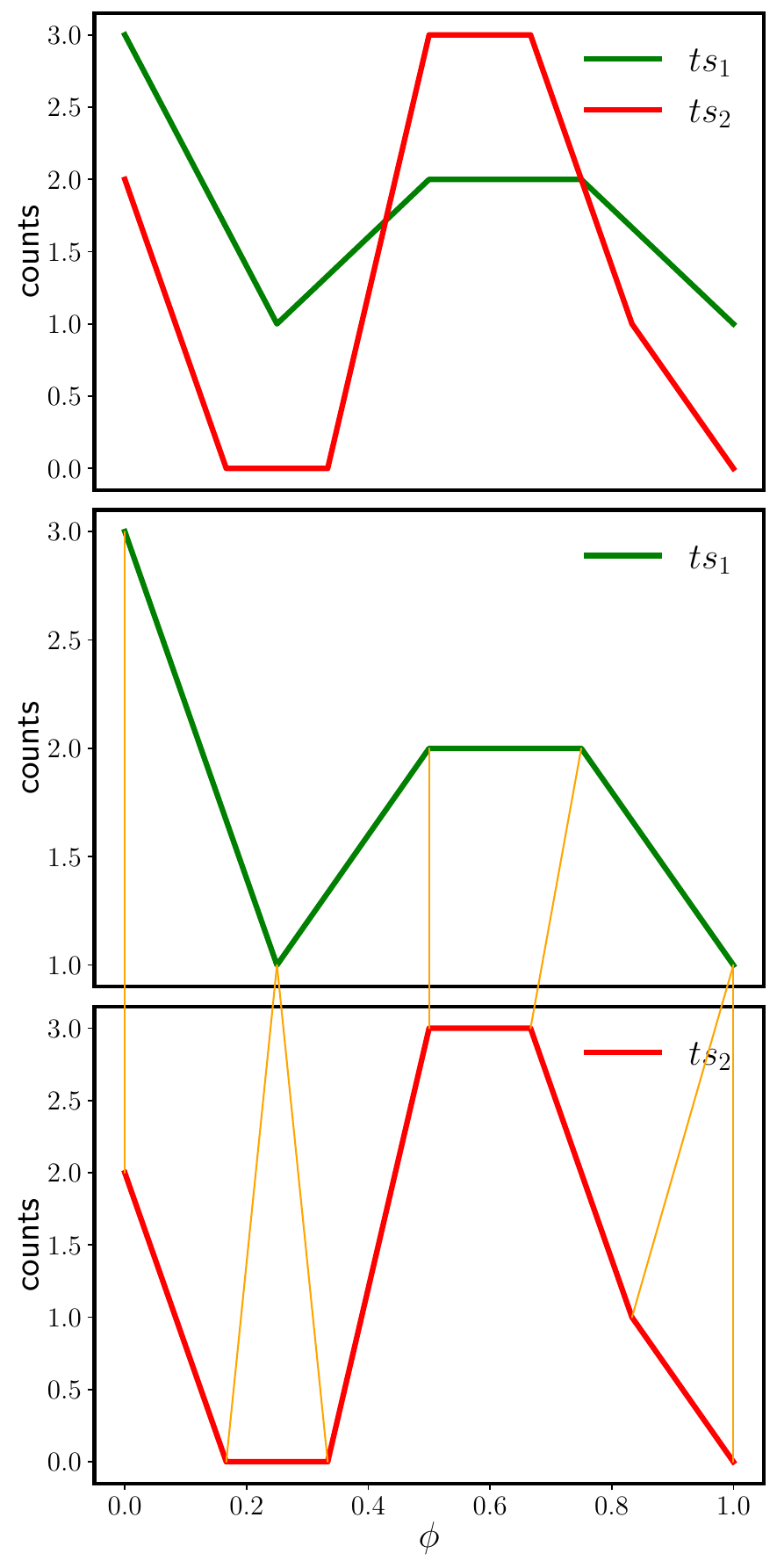}
  \caption[Example of DTW application.]{Top panel: 
Time series $ts_1$ (green) and $ts_2$ (yellow) used in the DTW example.
According to Equation~(\ref{Appendix_8_eq: toy_example}), the y-axis represents an arbitrary number of counts.
The x-axis represents the phases for each value, from 0 to 1, defined by the number of values in each time series.
Bottom panel:
Application of DTW. 
The orange lines represent $W_{\mathrm{o}}$, the connection between the pairs of the time series (i.e., the chosen ($i, j$) pairs from Equation~(\ref{Appendix_8_eq: MatrixED})).
}
  \label{Appendix_8_fig: appendix1_tseries_DTW}
\end{figure}

Consider, as an example, the following two-time series, see Figure~\ref{Appendix_8_fig: appendix1_tseries_DTW}, one with five and another with seven instances
\begin{equation}
  \begin{aligned}
     ts_1&=& 3, 1, 2, 2, 1\\
     ts_2&=& 2, 0, 0, 3, 3, 1, 0.
    \label{Appendix_8_eq: toy_example}
\end{aligned}  
\end{equation}

We calculate the Euclidean distance matrix ($E$, see, eg., \cite{EuclideanDistance_matrix}), as
\begin{equation}
E=E(d(i,j)^{2})=\begin{bmatrix}
        1 & 9 & 9 & 0 & 0 & 4 & 9 \\
        1 & 1 & 1 & 4 & 4 & 0 & 1 \\
        0 & 4 & 4 & 1 & 1 & 1 & 4 \\
        0 & 4 & 4 & 1 & 1 & 1 & 4 \\
        1 & 1 & 1 & 4 & 4 & 0 & 1 \\
\end{bmatrix}
~.\label{Appendix_8_eq: MatrixED}
\end{equation}
Here, each value is the square of the difference between the elements of the series located in ($i, j$), that is, $d = (ts_{1}(i)-ts_{2}(j))^{2}$, where $i$ is an index running on $ts_1$, from 1 to 5, and $j$ runs on $ts_2$, from 1 to 7.
In the next step, we compute the cumulative distance matrix ($D$) through the following recurrence relation seen in the application of 
Equation~(\ref{chapter_8_eq: D_cumulative}):
\begin{equation}
D= D(i,j) = \begin{bmatrix}
    \textcolor{orange}{1} & 10 & 19 & 19 & 19 & 23 & 32 \\
    2 & \textcolor{orange}{2}  &  \textcolor{orange}{3} &  7 & 11 & 11 & 12 \\
    2 &  6 &  6 &  \textcolor{orange}{4} &  5 &  6 & 10 \\
    2 &  6 & 10 &  5 &  \textcolor{orange}{5} &  6 & 10 \\
    3 &  3 &  4 &  8 &  9 &  \textcolor{orange}{5} &  \textcolor{orange}{6} 
    \label{Appendix_8_eq: MatrixDcum}
\end{bmatrix}
\end{equation}
For instance, the value seen in row 2 and column 2 in Equation~(\ref{Appendix_8_eq: MatrixDcum}) is obtained as $D(2,2)=E(2,2)+ \min\big(D(1,2), D(2,1), D(1,1)\big)=1+\min(2,10,1)=2$.
The aim will be to find the optimal path (the one with the minimal cost), referred to as $W_{\mathrm{o}}$, that relates the two series.
This can be found by tracing backward (backtracking) in $D$, taking the previous cells' lowest cumulative values from the initial cell.
The optimization will be subject to the following conditions.
First, it needs to connect the beginning and the end of each time series, i.e., it will be a path, $W$, that starts at cell (5, 7) and ends at cell (1, 1). 
Additionally, we must ensure that the path $W$ respects the sequential order of the time series, so it should move through the cells either to the left or upward, but never to the right or downward. For instance, from cell (5, 7), the next lowest value is 5, which occupies the cell (5, 6).
Finally, to avoid skipping any time series element, each step in $W$ is constrained to reach only neighboring cells. 
For instance, the next cell to get from (5, 6) can only be the cells (4, 5), (4, 6), or (5, 5), whose values, respectively, are 5, 6, and 9, thus $W$ goes through (4, 5) which is the cell with the minimum cost.
When more than one possible cell with the same value exists, the diagonal cell is prioritized as the most natural one-to-one alignment.
In this case, the colored values seen in Equation~(\ref{Appendix_8_eq: MatrixDcum}) represent $W_{\mathrm{o}}$, which connects the elements (1, 1), (2, 2), (2, 3), (3, 4), (4, 5), (5, 6) and, (5, 7).
Its final cost, i.e., the DTW value, is the square root of the sum of each of the above-selected elements, as per Equation~(\ref{Appendix_8_eq: MatrixED}) and yields 2.45. 
Visually, $W_{\mathrm{o}}$ is shown in the bottom panel of Figure~\ref{Appendix_8_fig: appendix1_tseries_DTW}, where each orange line represents the connection of each of the elements extracted from Equation~(\ref{Appendix_8_eq: MatrixDcum}).

\section{The Euclidean path}
\label{Appendix_8: euclideanPath}

Consider the following time series, having the same length
\begin{eqnarray}
\begin{aligned}
     ts_3&=&3, 1, 2, 2, 1\\
     ts_4&=&2, 0, 0, 3, 3
    \label{Appendix_8_eq: toy_example2}
\end{aligned}
\end{eqnarray}

Calculating $E$ and $D$, as shown before we get
\begin{equation}
E= \begin{bmatrix}
    1 & 9 & 9 & 0 & 0 \\
    1 & 1 & 1 & 4 & 4 \\
    0 & 4 & 4 & 1 & 1 \\
    0 & 4 & 4 & 1 & 1 \\
    1 & 1 & 1 & 4 & 4 \\
\end{bmatrix}
~,\label{Appendix_8_eq: MatrixED2}
\end{equation}
and
\begin{equation}
D = \begin{bmatrix}
    \colorbox{blue}{\textcolor{orange}{1}} & 10 & 19 & 19 & 19 \\
    2 &  \colorbox{blue}{\textcolor{orange}{2}} &  \textcolor{orange}{3} &  7 & 11 \\
    2 &  6 &  \colorbox{blue}{6} &  \textcolor{orange}{4} &  5 \\
    2 &  6 & 10 &  \colorbox{blue}{\textcolor{orange}{5}} &  5 \\
    3 &  3 &  4 &  8 &  \colorbox{blue}{\textcolor{orange}{9}} 
\end{bmatrix}
~.\label{Appendix_8_eq: MatrixDcum2}
\end{equation}

\begin{figure} 
\centering
  \includegraphics[width=0.79\textwidth]{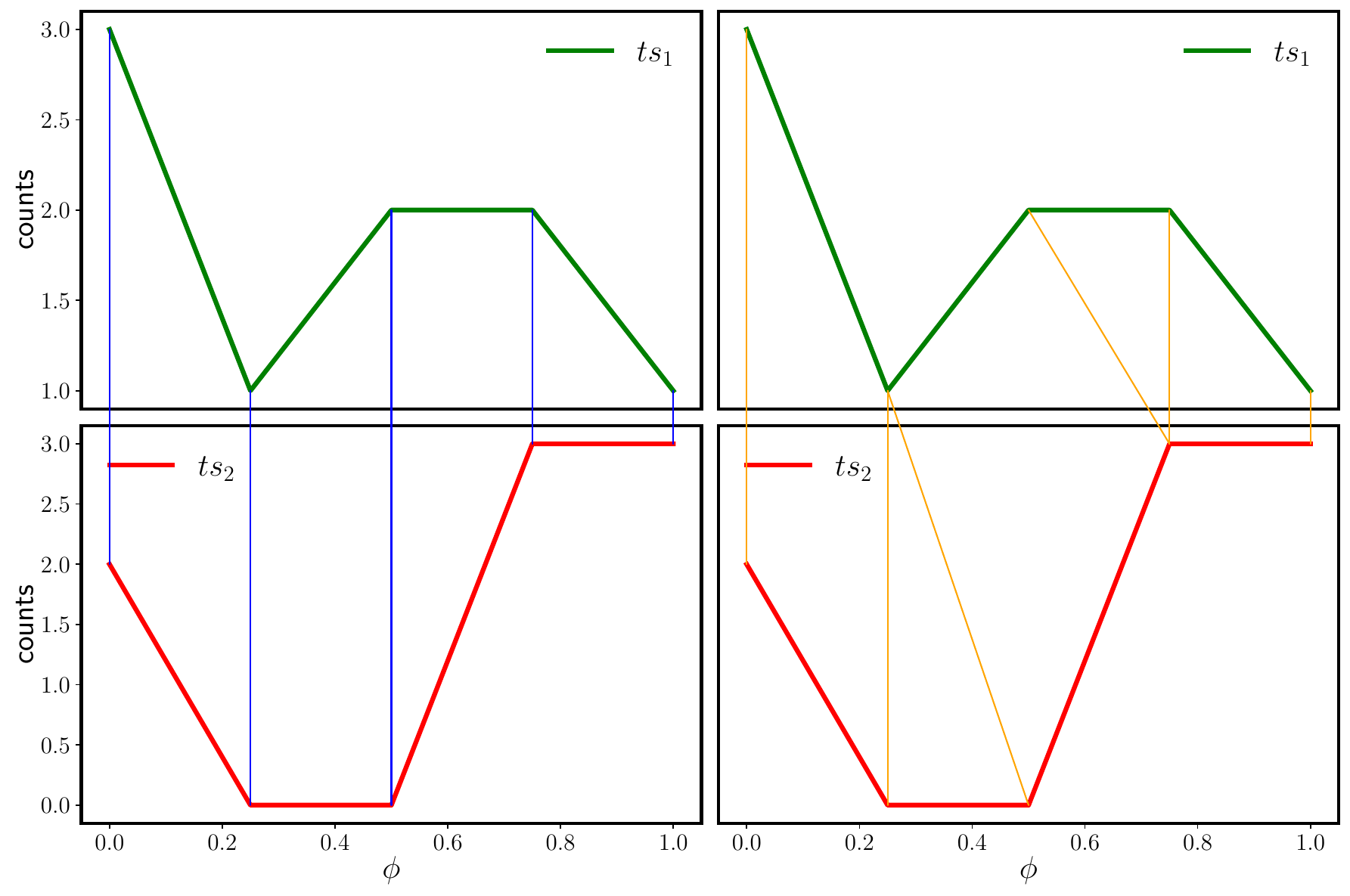}
  \caption[The Euclidean path in DTW.]{Time series $ts_3$ (green) and $ts_4$ (yellow) seen in Equation~(\ref{Appendix_8_eq: toy_example2}) connected via the Euclidean alignment (left) or DTW (right).
}
  \label{Appendix_8_fig: appendix2_DTW_ED}
\end{figure}

The main diagonal in Equation~(\ref{Appendix_8_eq: MatrixDcum2}) (noted in blue) represents a $W$ that has been explored in computing Equation~(\ref{chapter_8_eq: DTW_optimized}), arising by the connection of all $(i,j)$ with $i=j$; a point-by-point comparison of the time series, referred as $W_{\mathrm{E}}$
According to Equation~(\ref{Appendix_8_eq: MatrixED2}), its cost would be 3.31.
However, when applying Equation~(\ref{chapter_8_eq: DTW_optimized}), we see that another $W$, that would involve a different set of pairs $(i,j)$, and has a lower cost, being $W_{\mathrm{o}}$ in Equation~(\ref{Appendix_8_eq: MatrixDcum2}) (noted in orange).
According to Equation~(\ref{Appendix_8_eq: MatrixED2}), the cost of the orange path would be 3.00.
In Figure~\ref{Appendix_8_fig: appendix2_DTW_ED}, the alignment through the set ($i,j$) following $W_{\mathrm{E}}$ and $W_{\mathrm{o}}$, corresponding to both methods, is shown with the same color code used in Equation~(\ref{Appendix_8_eq: MatrixDcum2}).
Therefore, if the time series have the same length, the direct alignment between them (also referred to as the Euclidean alignment or just the Euclidean distance of the time series) will be computed and compared in the search for the minimal value.
This may or may not be the $W_{\mathrm{o}}$; in this example, it is not.

\chapter{Implementation and computational cost}  
\label{Appendix_10}

All the codes of this thesis are built on \textit{Python} v 3.11.2 \citep{python}, in which we implement the \textit{Psrqpy} package \citep{psrqpy} to deal with the population of pulsars from the ATNF catalogue. 
For the application of the PCA, we used the \textit{Scikit-learn} library \citep{scikit}. The previous libraries and packages contain as requirements other well-known libraries such as \textit{NumPy} \citep{numpy}, \textit{Pandas} \citep{pandas}, and \textit{Matplotlib} \citep{matplotlib}, through which, in addition, we have been able to develop those parts of the code that were necessary to obtain the results seen. 
The Pulsar Tree web is done using \textit{Bokeh} \citep{Bokeh}.
All computations, except those shown in Section~\ref{Appendix_10: compute_DTW}, were performed on a laptop equipped with a 2.0 GHz Quad-Core Intel Core i5 processor, 16 GB of 3733 MHz LPDDR4X memory, and Intel Iris Plus Graphics (1536 MB). 

\section{Computing MSTs}
\label{Appendix_10: compute_MSTs}

To create the graphs, we use the \textit{NetworkX} \citep{networkx} and the \textit{Graphviz} \citep{graphviz} libraries. 
On the other hand, the \textit{SciPy} library \citep{scipy} contains Kruskal's algorithm, which we have used to calculate the MST. 
The number of edges is a quickly growing function of the number of nodes, and so are the computational time requirements. 
For example, the computational time to calculate the MST with eight variables, and 2509 nodes from a complete, undirected, and weighted graph with 2509 nodes and 3146286 edges can be addressed from two approaches:
\begin{itemize}
    \item A brute-force method that builds the complete graph explicitly, computing all pairwise Euclidean distances and storing them as edge weights in a fully connected graph using \textit{NetworkX}. This graph is then processed using \textit{NetworkX}'s \texttt{minimum\_spanning\_tree} function to extract the MST. This method yields all the necessary edges: 3146286 edges for the given dataset of 2509 nodes. In practice, this implementation took $\sim$300 seconds per execution.
    \item This method computes the MST directly from the pairwise distance matrix using \texttt{scipy.sparse.csgraph.minimum\_spanning\_tree}. It avoids creating the full graph with all edges and instead uses a more efficient internal representation. The MST for the given dataset is computed in about 1.2 seconds.
\end{itemize}

\section{Computing Significant Branches}
\label{Appendix_10: compute_sig_branches}

The clustering algorithm designed to identify significant branches in a given MST begins by computing betweenness centrality using \textit{NetworkX}'s \texttt{betweenness\_centrality} function.
This allows us to detect the most central nodes, which are considered candidates for forming the main trunk of the tree. 
From these nodes, the graph is then partitioned around these paths to identify the branches. In the second stage, the algorithm evaluates the statistical significance of each branch using the two-sample KS test, \texttt{scipy.stats.ks\_2samp}, applied to physical properties across branch pairs. 
For the dataset of 2509 pulsars, computing the significant branches and storing the potential ones takes approximately 35 seconds. 

\section{Computing DTW}
\label{Appendix_10: compute_DTW}

The work in Chapter~\ref{chapter9} was conducted utilizing the \textit{dtaidistance} 2.3.10 package (\cite{DTW_package}) for the DTW application. 
The number of DTW values for each light curve comparison, DTW($lc_1$, $lc_2$), depends on the $N$ of each light curve.
Due to the applied rotation being $N_1$, and $N_2$, the size corresponding to $lc_1$ and $lc_2$, respectively, to this comparison is $N_{1}\times N_{2}$. 
Considering the 43071 possible pairs, the $N$ of each light curve, the number of computations is higher than $3\times 10^8$.
The implementation has been carried out on the Hidra High-Performance Computing (HPC) Cluster, which is hosted at the Institute of Space Sciences (ICE, CSIC).
For example, in a comparison involving light curves of 25 bins and 50 bins, the time usage required for a single (bin-to-bin) DTW computation is $\sim$0.0014~s, while the total time considering the shift is 1.407 s. In this case, the memory usage is negligible, 0.78 MB.
For comparison, taking light curves of 100 bins and 200 bins, the time usage for a single DTW is $\sim$0.0190 s, with a total of 371.186~s if looking at the complete rotation set. 
Let $lc1, lc2$=400 bins, the number of DTW computations performed for a complete rotation set is $\sim$160000.
The total time taken for all calculations, assuming the complete sample was no less than 72 hours, was mainly driven by the $lc$ of 400 bins.
The memory usage is around 3.75 MB. The computations were distributed across 16 cores using multiprocessing for a more efficient process.


\printbibliography[heading=bibintoc]


\end{document}